\documentclass[11pt]{article}
\usepackage{graphicx,amssymb}
\newcommand{\beq} {\begin{equation}}
\newcommand{\eeq} {\end{equation}}
\newcommand {\car}{$^{12}$C }
\newcommand {\oxy}{$^{16}$O }
\newcommand {\caI}{$^{40}$Ca }
\newcommand {\caII}{$^{48}$Ca }
\newcommand {\lead}{$^{208}$Pb }

\newcommand{\ton} {t_1}
\newcommand{\ttw} {t_2}
\newcommand{\tth} {t_3}
\newcommand{\balpha}{\mbox{\boldmath$\alpha$}}
\newcommand{\btau}{\mbox{\boldmath$\tau$}}
\newcommand{\bsigma}{\mbox{\boldmath $\sigma$}}

\newcommand{\bqu}{{\bf q}}
\newcommand{\bk}{{\bf k}}
\newcommand{\br}{{\bf r}}
\newcommand{\rot}{r_{12}}
\newcommand{\half}{\frac{1}{2}}
\begin{document}
%
%\begin{frontmatter}
%
\begin{titlepage}
\thispagestyle{empty}

%\vspace*{2cm}

\begin{center}
{\Large \bf 
Renormalized Fermi hypernetted chain 
approach in medium-heavy nuclei
}

\vspace{1.5cm}
{\large  F. Arias de Saavedra$ ^{\,1}$, 
C. Bisconti$ ^{\,2}$,
G. Co'$ ^{\,\,2}$,
and A. Fabrocini$^{\,3}$} \\ 
\vspace{1.cm}
{$^{1)}$ Departamento de F\'{\i}sica
At\'omica, Molecular y Nuclear, \\
Universidad de Granada, 
E-18071 Granada, Spain} \\
\vspace{.5cm}
{$^{2)}$ Dipartimento di Fisica,  Universit\`a del Salento  \\
and 
Istituto Nazionale di Fisica Nucleare  sez. di Lecce, 
\\ I-73100 Lecce, Italy} \\
\vspace{.5cm}
{$^{3)}$ Dipartimento di Fisica, Universit\`a di Pisa\\
and 
Istituto Nazionale di Fisica Nucleare  sez. di Pisa, \\
 I-73100 Pisa, Italy} \\
\end{center}
\vskip 1.5 cm 
\begin{abstract}
  The application of the Correlated basis function theory and of the
  Fermi hypernetted chain technique, to the description of the ground
  state of medium-heavy nuclei is reviewed.  We discuss how the
  formalism, originally developed for symmetric nuclear matter, should
  be changed in order to describe finite nuclear systems, with
  different number of protons and neutrons.  This approach allows us
  to describe doubly closed shell nuclei by using microscopic
  nucleon-nucleon interactions. We presents results of numerical
  calculations done with two-nucleon interactions of Argonne type,
  implemented with three-body forces of Urbana type. Our results
  regard ground-state energies, matter, charge and momentum
  distributions, natural orbits, occupation numbers, quasi-hole wave
  functions and spectroscopic factors of \car, \oxy, \caI, \caII and
  \lead nuclei.
\end{abstract} 
\vskip 1.cm
PACS: 21.60.-n; 21.10.Dr; 21.10.Ft; 21.10.Gv; 21.10.Jx; 21.90+f 
\end{titlepage}
\newpage
%\tableofcontents
%
\section{Introduction}
\label{sect:introduction}          
Aim of the many-body theories is the description of composite systems
in terms of their elementary components.  In our present understanding
of nature, only leptons, quarks, and gauge bosons are considered to
be elementary. In principle, any composite system should be described
in terms of these entities.  In reality, pragmatical difficulties
hinder the accomplishment of such an ambitious program. For example,
the difficulties in dealing with the non perturbative features of 
Quantum Chromodinamics (QCD) complicates the description even
of the individual hadrons. The description of even more complex
systems in terms of quarks and gluons is evidently not practicable, at
least at present.

A more pragmatical, and fruitful, approach for the study of composite
systems abandons quarks and leptons and uses effective degrees of
freedom and interactions. For nuclear systems, the most convenient
choice is to consider the nucleon as the basic degree of freedom. The
nucleon-nucleon (NN) interaction, even if modeled in various manners,
is always fixed to reproduce the NN elastic scattering data and the
deuteron properties \cite{mac87,mac89,sto93,sto94,wir95}.

After choosing the basic degrees of freedom and their interaction it
is necessary to define the theoretical framework to use. In the case
of nuclear systems, if one is interested in the ground state
properties and in low energy phenomenology (we mean here energies well
below a GeV) the role of antiparticles can be neglected. Therefore a
good description of these systems can be provided by the Schr\"odinger
equation.

In the last decade the Schr\"odinger equation has been solved without
approximations by using Green function Montecarlo (GFMC) techniques,
for several light nuclei up to A=12 \cite{pud95,pud97,pie01,pie05}.
In these calculations the two-body interaction has been implemented
with a three-body force chosen to reproduce the triton binding energy.
The great success in describing binding energies and low-lying
spectrum of these nuclei, is the pragmatical demonstration of the
validity of the non relativistic many-body approach.

Various techniques to solve the many-body Schr\"odinger equation
without approximations have been developed.  Some of them are tailored
to describe only few-body systems \cite{kam01}.  Others, formulated to
handle any number of nucleons, like the GFMC, are limited for
computational reasons to deal with light nuclei. Recently, a new
Montecarlo approach, called Auxiliary Field Diffusion Montecarlo
(AFDMC) \cite{gan06}, has been developed and it shows potentialities
to be applied to the description of medium-heavy nuclear systems.

The many-body theories try to overcome the difficulties in solving the
many-body Schr\"odinger equation by using approximations which
simplify the problem, but still provide a proper description of the
relevant physics effects.  Because of the strong repulsion in the
scalar channel of the NN interaction at short internucleonic distances
the use of the most traditional, and simplest, approximations, such as
the mean-field approximation, fails badly. For example, the use of
microscopic interactions in Hartee-Fock (HF) calculations leads to
unbound nuclei \cite{mut00b}. Dealing with the strong repulsion at
short distances is the major issue of the nuclear many-body theories.

Loosely speaking the various nuclear many-body theories can be
classified in two categories depending upon how they treat the
short-range repulsion problem.  There are theories acting on the
interaction, and others working on the trial many-body wave function.
The Brueckner theory, and all the theories constructing effective
interactions from microscopic ones, belong to the first category. One
of the theories belonging to the second category is the Correlated
Basis Function (CBF) theory whose most recent extensions and
applications to medium-heavy nuclei will be presented in this report.
Strictly speaking, our approach is the lowest order approximation of
the CBF theory \cite{kro02}.

The starting point of the CBF approach is the solution of the
many-body Schr\"odinger equation by means of the variational
principle:
\begin{equation}
\delta E[\Psi]=\delta\frac{<\Psi|H|\Psi>}{<\Psi|\Psi>}=0.
\label{eq:in-varp}
\end{equation}
The search for the minimum is done by using 
trial wave functions of the form:
\begin{equation}
\Psi(1,...,A)=F(1,...,A)\Phi(1,....,A) \,\,.
\label{eq:in-trial}
\end{equation}
In the above equation $\Phi$ describes the system as a set of $A$
particles moving independently from each other. We call Independent
Particles Model (IPM) this picture, which, in our calculations, is
modified by the correlation function $F$.  In its easiest form, we use
for $F$ the expression \cite{jas55}:
\begin{equation}
F(1,....,A)=\prod_{j>i=1}^{A}f(r_{ij}) \,\,,
\label{eq:in-jastrow}
\end{equation}
where $f(r_{ij})$ is a scalar function of the distance between two
particles of the system. 

The peculiarity of our approach consists in the technique used to
calculate the expectation values of Eq. (\ref{eq:in-varp}).  This
technique is inspired by the cluster expansion method used in
statistical mechanics to describe liquids \cite{may40}. The particles,
correlated by the function $f$, form clusters.  A topological study of
the various clusters shows that it is possible to construct a set of
integral equations which allows one to sum in a closed form the
contributions of all the clusters with some specific topological
properties.  This set of integral equations, called HyperNetted Chain
(HNC) equations, can be used to describe both classical and bosonic
systems \cite{lee59}.

In the mid 1970's, the cluster expansion techniques were extended to
include also the Pauli exclusion principle, and the Fermi HyperNetted
Chain (FHNC) equations were formulated \cite{gau71,fan74,fan75}. The
complexity of the NN interactions requires the use of correlation
functions that are more complex than those of Eq.
(\ref{eq:in-jastrow}).  These new correlations contain operator
dependent terms which commute neither with the nuclear hamiltonian,
nor among them.  Also for this reason it became necessary to extend
the FHNC equations to deal with this new type of correlations
\cite{fan78}.  The computational difficulties require the use of an
approximation called Single Operator Chain (SOC). The resulting set of
equations is called FHNC/SOC \cite{pan79}, and it has been
successfully applied to describe infinite systems
\cite{fan84,fan87,wir88,akm98}.

In this review, we are concerned about the application of the FHNC/SOC
computational scheme to medium-heavy nuclei.  The extension of the
FHNC theory to finite Fermi systems was introduced by Fantoni and
Rosati in the late 1970s \cite{fan79a}.  In their works they have
shown that a cluster expansion with an infinite numbers of terms can
be formulated even for finite systems. Consequently, the basic set of
FHNC equations, can be used also for finite systems.  However, one has
to consider that the loss of translational invariance in these systems
produces the so-called vertex corrections.  We shall refer to the new
set of equations as Renormalized Fermi HyperNetted Chain (RFHNC)
equations.

The results of the first numerical application of the RFHNC equations
to finite nuclear systems were presented in Ref. \cite{co92}.  In
that article, model nuclei were described. Protons and neutrons
wave functions were produced by a unique mean field potential, and in a
$ls$ coupling scheme.  The NN interactions considered had only central
terms, and the correlations were scalar functions.  This simplified
situation was used to test the theoretical, and numerical, feasibility
of the approach. Results for binding energies of \oxy and \caI model
nuclei were presented in \cite{co92} while the momentum distributions
where shown in a following article \cite{co94}.

A more realistic description of doubly closed shell nuclei was given
in \cite{ari96}, where proton and neutrons were separately treated,
and the single particle wave functions were expressed in a $jj$
coupling scheme. The RFHNC equations required a non trivial
reformulation.  Binding energies, matter densities and momentum
distributions, have been calculated for various doubly magic nuclei up
to \lead.  However, also in this case, simple central interactions and
scalar correlations were used.

In a following step, the RFHNC equations were extended to treat the
correlation terms commuting neither with the hamiltonian, nor among
themselves. This involved the extension of the SOC approximation.
Because of the technical difficulties the RFHNC/SOC equations have
been first formulated to deal with spin and isospin saturated nuclei,
and with single particle wave functions in a $ls$ coupling scheme.
Again only \oxy and \caI nuclei could be treated. The results of these
calculations have been presented in Refs. \cite{fab98,fab00,fab01}.

A formulation of the RFHNC/SOC equations general enough to handle
separately protons and neutrons in the more realistic $jj$ coupling
scheme was finally done. Binding energies and density distributions
have been shown in Ref. \cite{bis06} for the \car, \oxy, \caI, \caII
and also \lead nuclei. Here fully realistic microscopic interactions,
with tensor and spin-orbit terms were used. The hamiltonian
included also three-body interactions.

In the literature, there are various reviews regarding the FHNC/SOC
formalism applied to infinite nuclear systems
\cite{pan79,cla79,ros82,fab87,pol02}, but there is a void regarding
finite nuclei. For the sake of brevity in writing journal articles,
the formalism presented in the papers quoted above is incomplete.  The
aim of the present article is to provide a complete, coherent, and
self-contained presentation of the FHNC/SOC formalism for finite
nuclear systems, and to review the most recent results.

We recall in Sect. \ref{sec:infinite} the HNC, FHNC and FHNC/SOC
equations for infinite systems. They are important, not only because
we want to give a self-contained presentation, but especially because
the RFHNC/SOC formulation for the finite systems is constructed by
modifying that of infinite systems.  The RFHNC/SOC set of equations
will be presented in Sect. \ref{sec:finite}, and it will be applied in
Sect. \ref{sec:ene} to evaluate the energy of the system.  A selected,
but significant, set of recent numerical results will be presented,
and discussed, in Sect. \ref{sec:results}. In Sect. \ref{sec:per} we
provide a short overview of the possible extensions of the formalism.
Conclusions are presented in Sect. \ref{sec:conclusions}.

To improve the readability of the paper we present many technical
details of the derivation of the various expressions in the
Appendices. Furthermore, because of the large use of acronyms and
symbols, we list them in the Appendices \ref{sec:acronyms} and
\ref{sec:symbols}, respectively.

\section{Infinite systems}
\label{sec:infinite}
In this section we present the HNC and FHNC equations for infinite
systems. This presentation does not aim to substitute, or update, the
excellent review articles describing in detail the derivations of the
various expressions, see for example Refs.
\cite{pan79,cla79,ros82,fab87,pol02}. Our purpose is to recall the
main ideas and to emphasize those details which should be reconsidered
in the description of finite systems.

\subsection{Bosons}
\label{sec:inf-boson}
We start to present the CBF approach by describing a system composed
of $A$ bosons contained in a volume, $V$.  We are interested in getting
an infinite system by using the 
thermodynamic limit, i.e. $A$ and $V$ go to infinity keeping
the density, $\rho=A/V$, constant. We consider a homogeneous
and translationally invariant system, with a constant density, 
$\rho$.  The wave
function describing the system when the interaction between the
particles is switched off is:
\begin{equation}
\Phi(x_1,x_2,\ldots,x_A)={\cal S}
\Bigl( \phi_{1}(x_1) \cdots \phi_{A}(x_A) \Bigr) \,\,,
\label{eq:inf-phibose}
\end{equation} 
where we have indicated with ${\cal S}$ the symmetrization operator,
with $\phi_{i}(x_i)$ the single particle wave functions, and with
$x_i$ the generalized coordinate of the $i$-th particle.

In this IPM description of the ground
state of the system, all the bosons occupy the lowest single particles
state. We consider spin zero bosons, and 
because of the translational invariance of the system, 
the single particle wave functions are eigenfunctions of the
momentum $\bk$, and they can be expressed as:
\begin{equation}
\label{eq:inf-singlep1}
\phi_{j}(x_j)= \frac{1}{\sqrt{V}}\,e^{i \bk_j \cdot \br_j} \,\,.
\end{equation}
In this case, the generalized coordinate $x$ corresponds to $\br$. 

The density of the system can be obtained by using Eqs.
(\ref{eq:inf-phibose}) and (\ref{eq:inf-singlep1}),
\begin{equation}
\label{eq:inf-denszerob}
\rho_{0}(x)=A\phi^{*}(x)\phi(x)=\frac{A}{V}=\rho \,\,,
\end{equation}
which is constant, as expected.

As we have already discussed in the introduction, the idea is to solve
the Schr\"odinger equation by means of the variational principle by
using a  trial wave function of the form:
\beq
\Psi(x_1,...,x_A)=F(x_1,....,x_A) \Phi(x_1,...,x_A) \,\,,
\label{eq:inf-trial}
\eeq
where, in this case, the expression of $ \Phi$ is that of Eq.
(\ref{eq:inf-phibose}). 

For this specific bosonic case we describe the many-body correlation
function $F(x_1,...x_A)$ by using the so called  Jastrow ansatz 
\cite{din49,jas55}:
\beq
F(x_1,....,x_A)=\prod_{j>i=1}^{A}f(r_{ij}) \,\,,
\label{eq:inf-jastrow} 
\eeq
where the two-body correlation function (TBCF), $f(r_{ij})$, is a 
scalar function of the distance between the $i$-th and $j$-th particles.

In the calculation of the  energy functional
\beq
E[\Psi]= \frac{<\Psi|H|\Psi>}{<\Psi|\Psi>} \,\,,
\label{eq:inf-energy}
\eeq
it is very useful to employ 
the two-body distribution function (TBDF) defined as:
\begin{equation}
\label{eq:inf-tbdf}
g(x_1,x_2)=\frac{ A(A-1)
{\displaystyle 
\int dx_3\ldots dx_A
\Psi^{*}(x_1,\ldots,x_A)\Psi(x_1,\ldots,x_A)}}
{ {\displaystyle 
\rho^{2} \int dx_1dx_2\ldots dx_A
\Psi^{*}(x_1,\ldots,x_A)\Psi(x_1,\ldots,x_A)}} \,\,.
\end{equation}
The expectation value of any two-body operator, such as the two-body 
interaction, is obtained by integrating the TBDF on the two
coordinates $x_1$ and $x_2$:
\begin{equation}
\label{eq:inf-tbmv}
<O>=\frac 1 2 \rho^2 \int dx_1dx_2 \, g(x_1,x_2) \, O(x_1,x_2)
\,\,.
\end{equation}
The evaluation of the TBDF allows the calculation of the many-body
effects independently from the explicit expression of the operator.

By using Eqs. 
(\ref{eq:inf-phibose}), (\ref{eq:inf-singlep1}), 
(\ref{eq:inf-denszerob}) and the expressions (\ref{eq:inf-trial}) and 
(\ref{eq:inf-jastrow}) the numerator and the denominator of Eq. 
(\ref{eq:inf-tbdf}) can be written respectively as:
\begin{equation}
{\cal N}=(A-1)\frac{\rho^{A-2}}{A^{A-1}} \int dx_3dx_4...dx_A \, 
\, \prod_{i<j}f^{2}(r_{ij}) \,\,,
\label{eq:inf-nbos}
\end{equation}
and
\begin{equation}
{\cal D}= \frac{\rho^A}{A^A} \, \int dx_1dx_2...dx_A  
 \, \prod_{i<j}f^{2}(r_{ij}) \,\,.
\label{eq:inf-dbos}
\end{equation}

The cluster expansion is done by defining a new function $h(r_{ij})$
such as: 
\begin{equation}
f^{2}(r_{ij})=1+h(r_{ij})   \,\,.
\label{eq:inf-hdef}
\end{equation}

The product of $f^2$ factors can be rewritten by collecting all the
terms with the same number of  $h$-functions.  Let's first consider the
denominator ${\cal D}$, Eq. (\ref{eq:inf-dbos}), which can be written
as:
\begin{eqnarray}
\nonumber
{\cal D}=\frac{\rho^A}{A^A} \int
dx_1dx_2...dx_A &~&
\Big[ 1 
+\sum_{i<j}h(r_{ij})+3\sum_{i<j<k}h(r_{ik})h(r_{kj}) 
\\
&~& 
+\sum_{i<j<k<l}h(r_{ij})h(r_{kl})
+\ldots \Big]  \,\,.
\label{eq:inf-dbos1}
\end{eqnarray}

A convenient way of investigating the structure of the various terms
of Eq. (\ref{eq:inf-dbos1}) is to use the graphical representation
introduced by Yvon and Mayer \cite{may40}. In this formalism the
integrated points $x_i \equiv \br_i$, which are called {\em internal}
points, are represented by solid circles, and the $h$-functions by
dashed lines. The expression of Eq. (\ref{eq:inf-dbos1}) is obtained
by associating to each integrated point the contribution of the
density.  In the present case the density is constant, therefore its
contributions can be factorized out of the integral. This will not be
the case for finite systems.

%%%%%%%%%%%%%%%%%%%%%%%%%%%%%%
%
% figure bose denominator
%
%%%%%%%%%%%%%%%%%%%%%%%%%%%%%%%
\vskip 0.5 cm
\begin{figure}[htb]
\begin{center}
\includegraphics[scale=0.4]{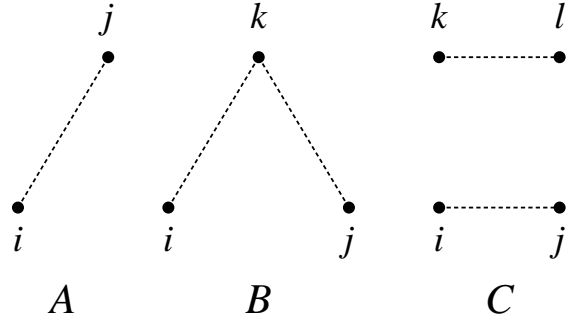}
\end{center}
\caption{
\small  Graphical representation of some terms contributing 
to Eq.(\ref{eq:inf-dbos1}). 
The dashed line represents the correlation function $h$. 
The black dots represents the integration points.
}
\label{fig:inf-bose1}
\end{figure}
\vskip 0.5 cm
%%%%%%%%%%%%%%%%%%%%%%%%%%%%%%%

The first sum of Eq. (\ref{eq:inf-dbos1}) is represented by the
diagram $A$ of Fig. \ref{fig:inf-bose1}. The second sum of 
Eq. (\ref{eq:inf-dbos1}) is represented by the diagram B. In this case, 
the point $k$ is in common with the two $h$-functions of the sum. 
The total contribution of this type of term is:
\beq
\frac{1}{2}\frac{(A-1)(A-2)}{A^2} \, \rho^3 \,
\int dx_idx_jdx_kh(r_{ik})h(r_{kj})  \,\,,
\label{eq:inf-diag2a}
\eeq
where the $(A-1)(A-2)$ factor is due to the fact that the sums on the
$i,j$ and $k$ indexes are limited to $i<j<k$.

In the third sum the two $h$-functions involves four different
points. Its contribution is represented by the diagram C of Fig.
\ref{fig:inf-bose1}, and it is given by:
\beq
\frac {A!}{4! (A-4)!}
\,\frac {\rho^4} {A^4}\,\int
dx_idx_jdx_kdx_l h(r_{ij})h(r_{kl})  \,\,,
\label{eq:inf-diag2b}
\eeq
%
%%%%%%%%%%%%%%%%%%%%%%%%%%%%%%
%
% figure bose numerator
%
%%%%%%%%%%%%%%%%%%%%%%%%%%%%%%%
\vskip 0.5 cm
\begin{figure}[htb]
\begin{center}
\includegraphics[scale=0.5]{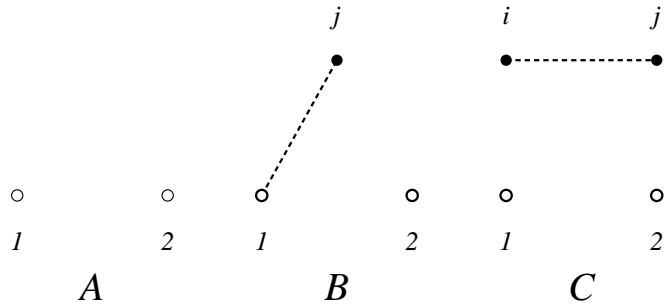}
\end{center}
\caption{
\small  Graphical representation of some terms contributing 
to Eq. (\protect\ref{eq:inf-nbos1}).
}
\label{fig:inf-bose2}
\end{figure}
\vskip 0.5 cm 
%%%%%%%%%%%%%%%%%%%%%%%%%%%%%%%

The use of Eq. (\ref{eq:inf-hdef}) in the numerator of
Eq. (\ref{eq:inf-tbdf}), allows us to obtain the expression:
\begin{eqnarray}
\nonumber
{\cal N}&=&f^{2}(r_{12})\frac{(A-1)}{A}\Biggl[ 1+
2\frac{\rho}{A} (A-2) \sum_{j>2}
\int dx_jh(r_{1j})
\\&~&
+ \frac{(A-2)(A-3)}{2}\frac{\rho^2}{A^2} \sum_{j>i>2}
\int dx_idx_jh(r_{ij})+\ldots \Biggr]  \,\,.
\label{eq:inf-nbos1}
\end{eqnarray}
A new symbol is required for the graphical representation of the 
numerator, since it is necessary to indicate the two coordinates which
are not integrated.  These coordinates are called {\em external}
points, and we have labeled them 1 and 2. The external points are
indicated by white circles as it is shown in Fig. \ref{fig:inf-bose2},
where we represent the lowest order terms of Eq. (\ref{eq:inf-nbos1}).
The uncorrelated term is represented by the A diagram. The B diagram
represents the terms of the first sum of Eq. (\ref{eq:inf-nbos1}),
where the $h$-function connects an external and an internal
point. Also the second sum of Eq. (\ref{eq:inf-nbos1}) contains only a
single $h$-function but it connects, in this case, only internal
points. The contribution of this sum is represented by the diagram C
of Fig. \ref{fig:inf-bose2}.

The numerator and the denominator of the TBDF (\ref{eq:inf-tbdf}) are
expressed by Eqs. (\ref{eq:inf-dbos1}) and (\ref{eq:inf-nbos1}) as
sums of terms characterized by the number of the $h$-functions, and by
that of the external, and internal points. Each term of these sums
forms a cluster of particles, and can be described by a diagram. We
proceed now by doing a topological classification of the various
diagrams.

The C diagrams of Fig. \ref{fig:inf-bose1} 
can be written by factorizing the non connected terms:
\begin{eqnarray*}
& \frac {A!}{4! (A-4)!} \,\frac {\rho^4} {A^4}\,\int
{\displaystyle \int
dr_idr_jdr_kdr_l h(r_{ij})h(r_{kl})}=&\\
&\frac{1}{4!}(1-\frac{6}{A}+\frac{11}{A^2}-\frac{6}{A^3})\rho^2
{\displaystyle \int
dr_idr_jh(r_{ij})\cdot\rho^2\int dr_kdr_l h(r_{kl})}&  \,\,.
\end{eqnarray*} 
Any diagram that can be factorized in two or more independent pieces
is called {\em unlinked}. Also the C diagram of Fig.
\ref{fig:inf-bose2} is unlinked. The diagrams that cannot be expressed
as a product of independent parts as the diagram B of
Fig. \ref{fig:inf-bose1}, are called {\em linked}.

%
%%%%%%%%%%%%%%%%%%%%%%%%%%%%%%
%
% figure reducible
%
%%%%%%%%%%%%%%%%%%%%%%%%%%%%%%%
\vskip 0.5 cm
\begin{figure}[htb]
\begin{center}
\includegraphics[scale=0.5]{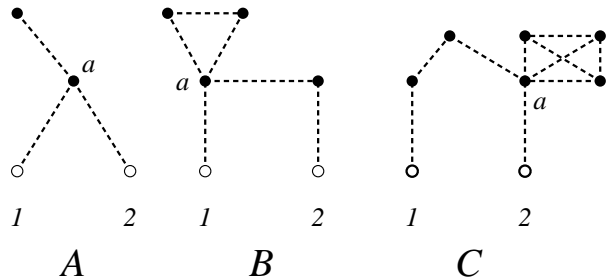}
\end{center}
\caption{
\small  Examples of reducible diagrams in the numerator of the TBDF, 
Eq. (\protect\ref{eq:inf-nbos1}).
}
\label{fig:inf-reducible}
\end{figure}
\vskip 0.5 cm 
%%%%%%%%%%%%%%%%%%%%%%%%%%%%%%%

The linked diagrams of Fig. \ref{fig:inf-reducible}, have the
properties of being {\sl reducible}. From the graphical point of view
the reducible diagrams are characterized by the presence of, at least,
one point linking a part of the diagram containing the external point,
and another part containing internal points only. Because of the
translational invariance of the system, the contributions of these two
parts can be factorized.  In general, every linked diagram whose
contribution to the TBDF can be expressed as a product of independent
integrals is called {\em reducible}.  In bosonic systems both
reducible and unlinked diagrams are factorizable. These factorizable 
diagrams of the numerator simplify, up to the $1/A$ order, all the
diagrams of the denominator.  The rigorous proof of this property is
given in Ref.  \cite{fan74}.

So, in the expression (\ref{eq:inf-tbdf}) of the TBDF, the denominator
diagrams compensate the contribution of the unlinked and of the reducible
diagrams of the numerator. Therefore, the TBDF can be expressed as the
sum of all the irreducible linked diagrams containing the two
external points 1 and 2:
\begin{equation}
\label{eq:inf-tbdfirr}
g(r_{12})=f^{2}(r_{12})\sum_{all\: orders}Y_{irr}(r_{12})
         =f^{2}(r_{12})\left(1+S(\rot)+C(\rot) \right)  \,\,.
\end{equation}
The translational invariance of the infinite system makes the TBDF
dependent only on the relative distance between the external points,
$r_{12}$.  A further topological classification of these irreducible
diagrams, divides them into {\em simple} and {\em composite}, and in
the above equation we have called $S(\rot)$ and $C(\rot)$ the
corresponding contributions to the TBDF (\ref{eq:inf-tbdfirr}).

%%%%%%%%%%%%%%%%%%%%%%%%%%%%%%%%
%
% figure simple plus composite 
%
%%%%%%%%%%%%%%%%%%%%%%%%%%%%%%%%
\vskip 0.5 cm 
\begin{figure}[htb]
\begin{center}
\includegraphics[scale=0.5]{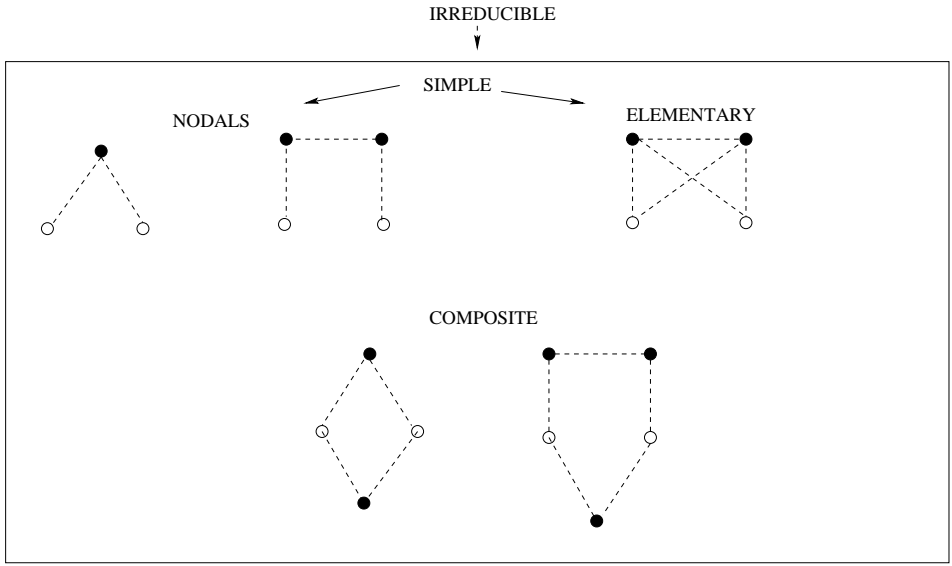}
\end{center}
\caption{
\small  Example of irreducible diagrams, classified as composite and
simple. This latter type of diagrams is sub-divided into nodal and
elementary ones. 
}
\label{fig:inf-irreducible}
\end{figure}
\vskip 0.5 cm 
%%%%%%%%%%%%%%%%%%%%%%%%%%%%%%%%

The composite diagrams are those composed by parts that are connected
only through the two external points 1 and 2, see Fig.
\ref{fig:inf-irreducible}. They can be expressed in terms of simple
diagrams.  Since there is no integration on the external points, the
contribution of a composite diagram is given by the product of the
simple diagrams connected to the external points. For example, the
contribution of all the composite diagrams which are formed by two
simple diagrams only, is $S^{2}(\rot)$.  Since the exchange of all the
particles of one subdiagrams with those of the other one, produces the
same composite diagram, we have to divide by $2$ to avoid double
counting. Repeating the same procedure we find that the contribution
of the composite diagrams formed by three simple diagrams is
$S^{3}(\rot) / 3!$, and so on.  The total sum of composite diagrams
can be written as:
\begin{equation}
C(\rot) =
\frac{S^{2}(\rot)}{2!}+\frac{S^{3}(\rot)}{3!}+
\frac{S^{4}(\rot)}{4!}+\ldots \,\,.
\label{eq:inf-composite}
\end{equation}
The TBDF, (\ref{eq:inf-tbdfirr}), can be rewritten as:
\begin{eqnarray}
\nonumber
g(r_{12}) &=& f^{2}(r_{12}) \left[
1+S(\rot)+\frac{S^{2}(\rot)}{2!}+\frac{S^{3}(\rot)}{3!}+
\ldots \right] \\
&=& f^{2}(r_{12})\,\exp[S(r_{12})] \,\,,
\label{eq:inf-bgr12a}
\end{eqnarray}
where the last equality appears because our system has an infinite
number of particles and is called hypernetted connection.

The above equation expresses the TBDF in terms of simple diagrams
only, which are further classified as {\em nodal} and {\em elementary}
ones. In a nodal diagram there is at least one point where all the
paths going from one external point to the other one have to pass.
This point is called a node. In the literature, the diagrams without
nodes are called {\em elementary} or {\em bridge} diagrams.  We shall
always use the adjective {\em elementary}.  Some examples of the type
of diagrams we have just defined can be found in Fig.
\ref{fig:inf-irreducible}.

If we call $N$ the contribution of all the nodal diagrams, and $E$
that of the elementary ones, we can write the TBDF as:
\begin{eqnarray}
\label{eq:inf-hnc1}
g(r_{12})&=& f^{2}(r_{12})\, \exp[N(r_{12})+E(r_{12})] \\
\nonumber 
&=&\left[ 1+h(\rot) \right]
\left[1 + N(\rot) + E(r_{12}) + \ldots \right] \\ 
&=& 1 + N(\rot) + X(\rot) \,\,.
\label{eq:inf-hnc2}
\end{eqnarray}
The above equation defines the diagrams contained in $X(\rot)$, which
are usually named {\sl non-nodal} diagrams since they have no nodes.  

%%%%%%%%%%%%%%%%%%%%%%%%%%%%%%%%%%%%
%
% figure nodal diagram
%
%%%%%%%%%%%%%%%%%%%%%%%%%%%%%%%%%%%%%5
\vskip 0.5 cm
\begin{figure}[htb]
\begin{center}
\includegraphics[scale=0.5]{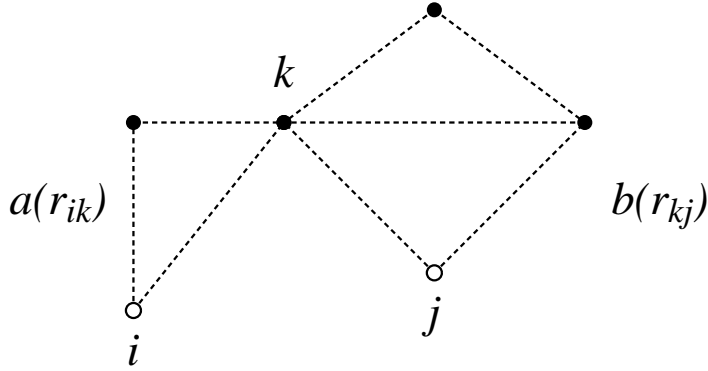}
\end{center}
\caption{
\small  Example of nodal diagram. We indicate with $a(r_{ik})$ the
contribution of the part of diagram to the left of the 
nodal point $k$, and with $b(r_{kj})$ the contribution 
of the right part.  
}
\label{fig:inf-nodal}
\end{figure}
\vskip 0.5 cm 
%%%%%%%%%%%%%%%%%%%%%%%%%%%%%%%%%%%%%5

A nodal diagram can be considered composed by parts which are linked at
the nodal point. Every nodal diagrams can be obtained by integrating
on the node the two functions representing the parts of the diagram.
Let us consider, for example, the nodal diagram of Fig.
\ref{fig:inf-nodal} having $i$ and $j$ as external points and $k$ as
node. If we call $a(r_{ik})$ and $b(r_{kj})$ the two functions
describing the two parts of the diagrams, the contribution of this
diagram to the TBDF (\ref{eq:inf-tbdfirr}) is:
\begin{equation}
\int d{\br}_{k}a(r_{ik})b(r_{kj})  \rho(r_k) 
= \rho \int d{\br}_{k}a(r_{ik})b(r_{kj})  
\equiv \Big(a(r_{ik})\Big|\rho(\br_k) b(r_{kj})\Big) \,\,,
\label{eq:inf-folprod}
\end{equation}
where a density function $\rho(\br_k)$ has been associated to the $\br_k$
integration point in order to recover the proper normalization,   
and since the density is constant in this case, it has been 
factorized out of the integral. The last term defines the symbol 
$\Big( \, \Big| \, \Big)$ we shall use henceforth to indicate  the folding
product or chain connection.  

By using the above considerations we can write a closed expression
which allows the evaluation of all the nodal diagrams.  The global
contribution $N(r_{ij})$ of all the nodal diagrams between the two
points $i$ and $j$ can be obtained as a folding product at the node
$\br_k$ of all the irreducible diagrams which can be constructed
between $i$ and $k$ and $k$ and $j$:
\begin{equation}
N(r_{ij})=\Big(X(r_{ik})\Big|\rho(\br_k) [N(r_{kj})+X(r_{kj})] \Big)
\,\,.
\label{eq:inf-hnc3}
\end{equation}
Every nodal diagram has at least one node and any path between its
external points $i$ and $j$ must pass through all the nodes. The above
equation tells us that the part of the diagram between $i$ and the
first node $k$, which is a non-nodal diagram, has to be folded to: i)
the non-nodal diagrams producing in this case nodal diagrams with only
one node and, ii) the nodal diagrams producing nodal diagrams with
more than one node.  The folding of two nodal diagrams at the $\br_k$
point is forbidden since it would produce many times the same diagram.

The set of Eqs.  (\ref{eq:inf-hnc1}), (\ref{eq:inf-hnc2}) and
(\ref{eq:inf-hnc3}) are known as HyperNetted Chain (HNC) equations.
Equation (\ref{eq:inf-hnc1}) allows one to express the TBDF in terms
of the simple diagrams after summing in a closed form the composite
diagrams and (\ref{eq:inf-hnc3}) allows the evaluation of the
contribution of all the nodal diagrams in a closed form.  However,
there is no closed expression to evaluate the contribution of the
elementary diagrams which must be calculated one by one. Calculations
of the TBDM without the contribution of the elementary diagrams are
labelled as HNC/0. When the contribution of the first elementary
diagram is included the calculation of the TBDF is called HNC/4, since
this diagram, shown in Fig. \ref{fig:inf-irreducible}, has four
particles.  These equations are usually solved with an iterative
procedure starting from the ansatz $N(r_{12})=E(r_{12})=0$, then
$X(r_{12})=f^{2}(r_{12})-1$ and we can get new nodals using
(\ref{eq:inf-hnc3}).

\subsection{Fermions}
\label{sec:inf-fermion}
In the description of a system of fermions we have to deal with the Pauli
exclusion principle. The IPM wave function $\Phi$ to be used in the
trial wave function (\ref{eq:in-trial}), is now a Slater determinant
of single particle wave functions $\phi$:
\begin{equation}
\Phi(x_1,....,x_A)=\frac{1}{\sqrt{A!}} \, \, 
\begin{tabular}{|cccc|}
$\phi_{1}(x_1)$&$\phi_{1}(x_2)$&\ldots&$\phi_{1}(x_A)$\\
$\phi_{2}(x_1)$&$\phi_{2}(x_2)$&\ldots&$\phi_{2}(x_A)$\\
\vdots & \vdots & $\ddots$ & \vdots \\
$\phi_{A}(x_1)$&$\phi_{A}(x_2)$&\ldots&$\phi_{A}(x_A)$
\label{eq:inf-slater}
\end{tabular}
\,\,,
\end{equation}
For an infinite system we can write the single particle wave functions
as:
\begin{equation}
\phi_{a}(x_j)=\frac{1}{\sqrt{V}}\,e^{i\bk_a\cdot\br_j}
\, \chi_{s_a}(j) \, \chi_{t_a}(j) \,\,.
\label{eq:inf-spwave}
\end{equation}
where we have indicated with $s$ and $t$ the projections on the $z$
axis of the spin and isospin and with $\chi_s$ and $\chi_t$ the
Pauli spinors.  In the fermions case, the generalized coordinate
$x$ indicates position $\br$, spin and isospin third components, in
addition to the total spin and isospin values.

Before attacking the problem of the calculation of the TBDF
(\ref{eq:inf-tbdf}) we discuss some property of
$|\Phi|^{2}$ which we write as:
\begin{equation}
|\Phi(1,2,\ldots,A)|^2=
\begin{tabular}{|cccc|}
$\rho_0(x_1,x_1)$&$\rho_0(x_1,x_2)$&\ldots&$\rho_0(x_1,x_A)$\\
$\rho_0(x_2,x_1)$&$\rho_0(x_2,x_2)$&\ldots&$\rho_0(x_2,x_A)$\\
\vdots & \vdots & $\ddots$ & \vdots \\
$\rho_0(x_A,x_1)$&$\rho_0(x_A,x_2)$&\ldots&$\rho_0(x_A,x_A)$
\end{tabular}
\,\,,
\label{eq:inf-phi2}
\end{equation}
where we have defined the various elements of the above determinant
as: 
\begin{equation}
\rho_0(x_i,x_j)=\sum_a\phi_a^{*}(x_i)\phi_a(x_j)
\,\,.
\label{eq:inf-uobdm}
\end{equation}
In the above expression the sum runs over all the occupied single
particle states of the system. We have defined in Eq.
(\ref{eq:inf-uobdm}) the uncorrelated One-Body Density Matrix (OBDM)
which is the basic ingredient of the calculation of the TBDF in the
fermion case. A fundamental property of the uncorrelated OBDM, due to
the orthonormality of the single particle wave functions, is:
\begin{equation}
\int d{x}_{j}\rho_0(x_i,x_j)\rho_0(x_j,x_k)=\rho_0(x_i,x_k)
\,\,,
\label{eq:inf-contrho}
\end{equation}
where in the above integral sign we include both the space
integration and the sum on the spin and isospin third components,
their {\em trace}.

We define the sub-determinant as:
\begin{equation}
\label{eq:inf-subdet}
\Delta_{p}(1,...,p)=
\begin{tabular}{|cccc|}
$\rho_0(x_1,x_1)$&$\rho_0(x_1,x_2)$&\ldots&$\rho_0(x_1,x_p)$\\
$\rho_0(x_2,x_1)$&$\rho_0(x_2,x_2)$&\ldots&$\rho_0(x_2,x_p)$\\
\vdots & \vdots & $\ddots$ & \vdots \\
$\rho_0(x_p,x_1)$&$\rho_0(x_p,x_2)$&\ldots&$\rho_0(x_p,x_p)$
\end{tabular}
\,\,,
\hspace{6mm} p\le A \,\,.
\end{equation}
Because of the property (\ref{eq:inf-contrho}) of the uncorrelated
OBDM the sub-determinants have the property:
\begin{equation}
\label{eq:inf-subdetp1}
\int d{x}_{p+1}\Delta_{p+1}(1,....,p+1)=(A-p)\Delta_{p}(1,...,p)
\,\,,
\end{equation}
and, by iterating it, we obtain:
\begin{equation}
\label{eq:inf-subdetp2}
\int d{x}_{p+1}...d{x}_{A}\Delta_{A}(1,....,A)=
(A-p)!\Delta_{p}(1,...,p)
\,\,.
\end{equation}
The above expression implies that:
\begin{equation}
\label{eq:inf-subdetp3}
\Delta_{p}=0 \,\,,\hspace{3cm} p>A \,\,.
\end{equation} 
The property (\ref{eq:inf-subdetp3}) will be extremely useful in the
application of the cluster expansion technique to both finite and 
infinite fermion systems.

The properties of the uncorrelated OBDM and of the sub-determinats we
have just presented, depend only on the orthonormality of the single
particle wave functions, and not on their explicit expressions.  For
this reason, they will remain valid also in the case of finite
fermions systems.  The expression (\ref{eq:inf-spwave}) of the single
particle wave functions has been chosen to describe a infinite and
homogeneous system. In this case, we obtain for the uncorrelated OBDM
the expression:
\begin{equation}
\rho_0(x_i,x_j)= \frac \rho \nu \ell (k_F r_{ij}) 
\sum_{s,t} \chi_{s}^+(i) \chi_{t}^+(i) \chi_{s}(j) \chi_{t}(j)
\,\,.
\label{eq:inf-defro}
\end{equation}
In the above equation we have indicated with $\nu$ the spin-isospin
degeneration of the system, $4$ in the nuclear matter case, and with
$k_F=(6\pi^2 \rho /\nu)^{1/3}$ the Fermi momentum.  In the literature
the function $\ell(x)$ is called Slater function \cite{fan79}, and has
the following explicit expression:
\begin{equation}
\ell (x)
=\frac 3 {x^3} (\sin{x}-x\cos{x}) \,\,.
\label{eq:inf-slatfun}
\end{equation}

In the description of fermion systems, it is necessary to include in
the Mayer diagrams a new graphical symbol identifying the presence,
and the role, of $\rho_0(x_i,x_j)$, which, in the calculation of the
TBDF, forms closed non overlapping loops.  This is an oriented line
connecting the two points $x_i$ and $x_j$. These lines are called {\em
  statistical} correlations to distiguish them from the {\em
  dynamical} correlations, $f(r_{ij})$. In the calculation of the TBDF
for the infinite system a term $-\ell(k_F r_{ij})/\nu$ should be
considered for each statistical line joining the $i$ and $j$ points,
and a factor $-\nu$ for every closed statistical loop which is related to 
the spin and isospin trace \cite{ros82}.

There is a basic difference between dynamical and statistical correlations.
While any number of dynamical lines may arrive at a given point only none
or two statistical lines may arrive at that point.

By using the trial wave function (\ref{eq:in-trial}) with the Jastrow
ansatz (\ref{eq:in-jastrow}) and the definition (\ref{eq:inf-hdef})
of the  $h$-function we write the TBDF (\ref{eq:inf-tbdf}) as:
\begin{eqnarray}
\nonumber
&~& g(x_1,x_2) \\
\nonumber
&=& \frac{ A(A-1) 
{\displaystyle \int d{x}_{3}....d{x}_{A}
(1+\sum_{i<j}h_{ij}+\sum_{i<j<k}h_{ij}h_{jk}+....)|\Phi(x_1,....,x_A)|^{2}}}
{{\displaystyle \rho^{2} \int d{x}_{1}....d{x}_{A}(1+\sum_{i<j}h_{ij}+
\sum_{i<j<k}h_{ij}h_{jk}+....)|\Phi(x_1,....,x_A)|^{2}}}
\,\,,
\label{eq:inf-tbdff1}
\end{eqnarray}
with $h_{ij} \equiv h(r_{ij})$.
By using the definition of sub-determinat (\ref{eq:inf-subdet}) the
numerator and the denominator of the above equation can be expressed
as sums of terms identified by the number of
$h$-functions:
\begin{eqnarray}
\nonumber
{\cal N} = \frac{A(A-1)}{\rho^2} f^2(r_{12}) &\int&
d{x}_{3}....d{x}_{A} \\
&~&
\left(1+\sum_{i<j}h_{ij}+\sum_{i<j<k}h_{ij}h_{jk}+....\right)\Delta_{A}
\,\,,
\label{eq:inf-nferm}
\end{eqnarray}
\begin{equation}
{\cal D}=\int d{x}_{1}....d{x}_{A}\left(1+\sum_{i<j}h_{ij}+
\sum_{i<j<k}h_{ij}h_{jk}+....\right)\Delta_{A}
\,\,.
\label{eq:inf-dferm}
\end{equation}
We rewrite the expressions of ${\cal N}$ and ${\cal D}$
by grouping the terms with the same number of points, $p$, and we
indicate them as $X^{(p)}(1,2,3,..,p)$. For example
\[
X^{(3)}(1,2;i) = h_{1i}+h_{2i}+h_{1i}h_{2i}
\,\,.
\]
The expression of the TBDF we obtain is:
\begin{eqnarray*}
g(x_1,x_2)&=&
\frac{A(A-1)}{\rho^2}f^{2}(r_{12})\int d{x}_{3}...d{x}_{A}
\Delta_{A}\Bigg[1+\\&&
\sum_{p=3}^A\frac{(A-2)!}{(p-2)!(A-p)!}X^{(p)}(1,2;\ldots,p)\Bigg]\\
&&\Bigg[\int d{x}_{1}...d{x}_{A}
\Delta_{A}\Bigg(1+\sum_{p=2}^A\frac{A!}{p!(A-p)!}X^{(p)}(1,\ldots,p)
\Bigg)\Bigg]^{-1}
\,\,.
\end{eqnarray*}
The factorials factors which multiply the $X^{(p)}$ functions, take
into account the fact that permutations of the $p$ internal points do
not change the value of the diagram.

By using the property (\ref{eq:inf-subdetp2}) of the sub-determinants,
we can integrate the above expression of the TBDM on all the
coordinates not involved by the correlations, i. e. not present in the
$X^{(p)}$ functions. So we obtain for the numerator and the
denominator of the TBDF, the expressions:
\begin{eqnarray}
{\cal N} &=& A!
\frac{f^2(r_{12})}{\rho^2}\sum_{p=2}^{A}
\frac{1}{(p-2)!}\int dx_3...dx_pX^{(p)}(1,2;...,p)\Delta_p(1,...,p)
\,,
\label{eq:inf-nferm1}
\\
{\cal D} &=& A!
\sum_{p=0}^{A}\frac{1}{p!}\int dx_1...dx_p X^{(p)}(1,...,p)
\Delta_p(1,...,p)
\,\,.
\label{eq:inf-dferm1}
\end{eqnarray}

We extend up to infinity the upper limits of all the sums of the above 
expression by using the property (\ref{eq:inf-subdetp3}) of the
sub-determinants. 
Each cluster term (diagram) can be divided in linked and unlinked
parts. Let us call ${\cal L}_n(1,2,i_3,...,i_n)$ the linked parts of
the various cluster terms containing the external points $1$ and $2$.
In these diagrams each internal point $i_3,..,i_n$ is connected to the
points $1$ and $2$ by at least one continuous path of dynamical and/or
statistical correlations. We call ${\cal U}_{p-n}(i_{n+1},...,i_p)$
the unlinked parts of the cluster terms. In this case none of the
$p-n$ points is connected to $1$ and $2$, or to another point of
${\cal L}_n$. The contribution of ${\cal L}_{n}$ does not change for a
permutation of some of its internal points. The same property holds for
${\cal U}_{p-n}$ and its internal points.  For this reason every
diagram of ${\cal N}$ separated in ${\cal L}_{n}$ and ${\cal U}_{p-n}$
parts, give $(p-2)!/ (n-2)!(p-n)!$ times the same contribution. We can
then express the numerator if we define $q=p-n$ as:
\begin{eqnarray}
\nonumber
{\cal N}&=& A!
\frac{f^2(r_{12})}{\rho^2}\sum_{n=2}^{\infty}
\frac{1}{(n-2)!}\int dx_3...dx_n\,{\cal L}_{n}(1,2;...,n)
\\
&&\Bigg[\sum_{q=0}^{\infty}\frac{1}{q!}\int dx_{1}...dx_{q} \,
{\cal U}_{q}(1,...,q)\Bigg]
\,\,.
\label{eq:inf-nferm2}
\end{eqnarray}
We extend the above considerations to the denominator
(\ref{eq:inf-dferm1}). Since in this case there are no external
points, the diagrams we have defined as linked ones, are
not present. Only the ${\cal U}_n$ diagrams contribute to the
denominator:
\beq
{\cal D} = A!
%f^2(r_{12})
\Bigg[\sum_{n=0}^{\infty}\frac{1}{n!}
\int dx_1...dx_n \, {\cal U}_{n}(1,...,n)\Bigg]
\,\,.
\label{eq:inf-dferm2}
\eeq
This expression is identical to that giving the contribution of the
unlinked terms of the numerator.  In the calculation of the TBDF the
denominator compensates all the unlinked diagrams of the numerator, and we
can write:
\begin{eqnarray}
\nonumber
&~&g(x_1,x_2) = g(r_{12}) \\
&~&=\frac{f^2(r_{12})}{\rho^2}\Bigg[\Delta_2(1,2)
+\sum_{p=3}^{\infty}\frac{1}{(p-2)!}\int dx_3...dx_p
\,{\cal L}_{p}(1,2;...,p)\Bigg]
\,\,.
\label{eq:inf-tbdfferm1}
\end{eqnarray}

%%%%%%%%%%%%%%%%%%%%%%%%%%%%%%%%%
% figure fermions reducible diagram
%%%%%%%%%%%%%%%%%%%%%%%%%%%%%%%%%
%
\begin{figure}[htb]
\begin{center}
\includegraphics[scale=0.5]{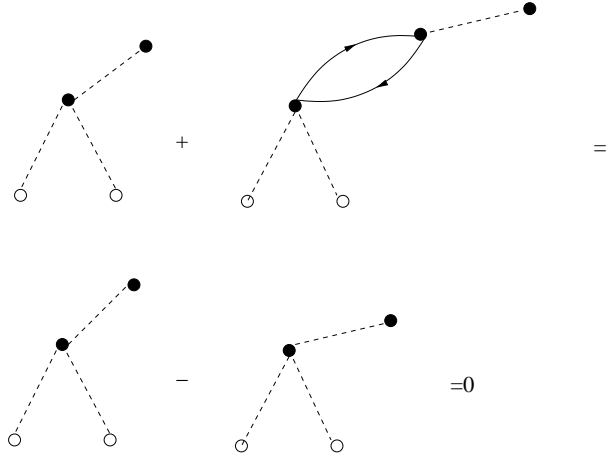}
\end{center}
\caption{
\small  Example of cancelation between two FHNC diagrams. The
statistical loop in the upper part produces a minus sign, and
therefore the total contribution is zero.
}
\label{fig:inf-canc}
\end{figure}
%%%%%%%%%%%%%%%%%%%%%%%%%%%%%%%%%

The above result shows that the TBDF can be obtained by calculating
linked cluster terms, only. As in the case of bosons, we define the
reducible diagrams as those linked diagrams containing a point, the
articulation point, which allows us to write the total contribution in
two or more separated contributions. An example of reducible diagrams
is given in Fig. \ref{fig:inf-canc}. Like in the bosonic case, the
factorization of the reducible diagrams in two or more subdiagrams is
due to the translational invariance of the system.  Also in the
fermionic case it is possible to show that the reducible diagrams do not
contribute to the calculation of the TBDF \cite{fan74}. However, in
the fermionic case, the mechanism which allows the elimination of the
contribution of the reducible diagrams, is very different from that of
the boson case. Furthermore, the cancelation of the reducible diagrams
is exact, not limited to $1/A$ power terms. The rigorous proof of this
cancellation is given in \cite{fan74}, and we present here only the
basic idea of how the cancellation mechanism works. This discussion
will become useful to present the vertex corrections
in the finite fermion systems case.
Let us consider, as example, the case of the diagrams shown in
the upper part of Fig. \ref{fig:inf-canc}.  These diagrams differ only
because the second diagram has an additional statistical loop.
Because the system is translationally invariant, and for the
properties of the Slater function (\ref{eq:inf-slatfun}), the
contribution of the the two diagrams is identical but with different
sign. Therefore, as is shown in the lower part of the figure, the
global contribution of the two diagrams is zero.  

%%%%%%%%%%%%%%%%%%%%%%%%%%%%
% fhnc diagram
%%%%%%%%%%%%%%%%%%%%%%%%%%%%
\begin{figure}[htb]
\begin{center}
\includegraphics[scale=0.65]{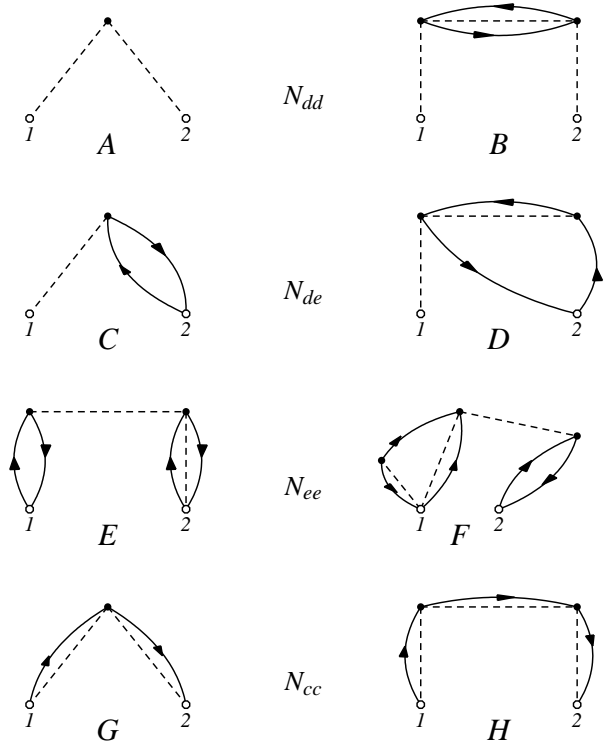}
\end{center}
\caption{
  \small The various types of nodal diagrams required by the FHNC
  equations.  The subindexes classify the diagrams with
  respect to the type of correlations reaching the external points $1$
  and $2$.
  }
\label{fig:inf-fhnc1}
\end{figure}
%%%%%%%%%%%%%%%%%%%%%%%%%%%%%%

The fermionic TBDF (\ref{eq:inf-tbdfferm1}) can be calculated by
considering the contribution of the irreducible diagrams only, in
analogy to Eq.  (\ref{eq:inf-tbdfirr}) for the bosons.  Again in
strict analogy with what has been done for the bosonic case, see
Eqs.(\ref{eq:inf-tbdfirr} - \ref{eq:inf-bgr12a}), it is possible to
show that the contribution of all the composite diagrams can be
obtained by considering simple diagrams only, which are classified in
nodal and elementary ones. The elementary, and nodal, diagrams in the
fermionic case are defined in analogy to those of the bosonic case,
but both statistical and dynamical correlations should be considered.
The presence of the statistical correlations hinders the possibility
of writing a single integral equation which allows the evaluation in
closed form of the contribution of all the nodal diagrams, as Eq.
(\ref{eq:inf-hnc3}) is doing. However, it is possible to find a set
of integral equations relating the contribution of the nodal diagrams
characterized by the type of correlations reaching the external points
$1$ and $2$ \cite{fan74,fan79}.

Graphical examples of the type of diagrams required to obtain the
various integral equations of interest are given in Fig.
\ref{fig:inf-fhnc1}. In the A and B diagrams only dynamical
correlations reach the external points. These diagrams are labeled
with the $dd$ (dynamical-dynamical) subscripts ($N_{dd}$). The C and D
diagrams have only dynamical correlations reaching the external point
$1$ and two statistical correlation lines reaching the external point
$2$. In this case, we label the nodal diagram with a $de$
(dynamical-exchange) subscript ($N_{de}$). The E and F diagrams are
labelled with a $ee$ (exchange-exchange) subscript since to both
external points arrive two statistical lines. Up to build those $ee$
diagrams with the external points in the same statistical loop, it is
convenient to define diagrams where a statistical correlation starts
from the external point $1$ and arrives to the external point $2$,
forming an open loop. We label these diagrams with the $cc$
(cyclic-cyclic) subscript and we remark that they do not contribute
directly to the TBDF.

As discussed in the bosonic case for Eq. (\ref{eq:inf-hnc3}), also in
this case the total contribution of the nodal diagrams can be obtained
by doing the folding product of various parts of the diagrams at the
nodal point. However, the Pauli exclusion principle, prohibits some of
the possible folding products. It is not possible to fold $cc$
diagrams with diagrams of different type and if one of the diagrams to
fold has the $e$ type at the nodal point the other one has to be $d$
type at this point.  This restriction is caused by the afore mentioned
fact that only two statistical lines may arrive at a point, in this
case the nodal point.

In analogy to the bosonic case, we call $N$ the sum of all the nodal
diagrams and $X$ the sum of all the irreducible non-nodal
diagrams. Of course now $N$ and $X$ are classified by the subindexes
$dd,de,ee$ and $cc$, and, for the nodal diagrams, we obtain the
following set of equations \cite{fan74,fan79}:
\begin{eqnarray}
\nonumber 
N_{dd}(r_{12})&=&\Big(X_{dd}(r_{13})+X_{de}(r_{13})\Big| \rho(\br_3) 
[ N_{dd}(r_{32})+X_{dd}(r_{32}) ] \Big)
\\
\nonumber
              &+ &\Big(X_{dd}(r_{13})\Big|  \rho(\br_3) 
[ N_{ed}(r_{32})+X_{ed}(r_{32}) ] \Big)
\,\,,
\\
\nonumber
N_{de}(r_{12})&=&\Big(X_{dd}(r_{13})+X_{de}(r_{13}) \Big| \rho(\br_3) 
 [ N_{de}(r_{32})+X_{de}(r_{32}) ] \Big)\\
\nonumber
              &+ &\Big(X_{dd}(r_{13}) \Big|  \rho(\br_3) 
[ N_{ee}(r_{32})+X_{ee}(r_{32}) ] \Big)
\,\,,
\\
\nonumber
N_{ee}(r_{12})&=&\Big(X_{ed}(r_{13})+X_{ee}(r_{13}) \Big|  \rho(\br_3) 
[ N_{de}(r_{32}) + X_{de}(r_{32}) ] \Big)\\ \nonumber
              &+ &\Big(X_{ed}(r_{13}) \Big|  \rho(\br_3) 
[ N_{ee}(r_{32})+X_{ee}(r_{32}) ] \Big)
\,\,,
\\
N_{cc}(r_{12})&=&\Big(X_{cc}(r_{13}) \Big|  \rho(\br_3)  
[ N_{cc}(r_{32})+X_{cc}(r_{32})
                 -\ell(k_{F}r_{32})/\nu ]\Big)
\,\,.
\label{eq:inf-fhncnod}
\end{eqnarray}
The equations for the non-nodal diagrams are:
\begin{eqnarray}
\nonumber
X_{dd}(r_{12})&=&g_{dd}(r_{12})-N_{dd}(r_{12})-1
\,\,,
\\
\nonumber
X_{de}(r_{12})&=&g_{dd}(r_{12})[N_{de}(r_{12})+E_{de}(r_{12})]-N_{de}(r_{12})
\,\,,
\\
\nonumber
X_{ee}(r_{12})&=&g_{dd}(r_{12})\{N_{ee}(r_{12})+E_{ee}(r_{12})+
[N_{de}(r_{12})+E_{de}(r_{12})]^{2}\\
\nonumber
&&-\nu[N_{cc}(r_{12})+E_{cc}(r_{12})-\frac{1}{\nu}\ell(k_{F}r_{12})]^{2}\}
-N_{ee}(r_{12})
\,\,,
\\ \nonumber
X_{cc}(r_{12})&=&g_{dd}(r_{12})[N_{cc}(r_{12})+E_{cc}(r_{12})-\frac{1}{\nu}
               \ell(k_{F}r_{12})] \\
&-& N_{cc}(r_{12})+\frac{1}{\nu}
               \ell(k_{F}r_{12})
\,\,.
\label{eq:inf-fhncnonod}
\end{eqnarray}
Finally, the partial TBDF are defined as: 
\begin{eqnarray}
\nonumber
g_{dd}(r_{12})& = & f^{2}(r_{12})\exp[N_{dd}(r_{12})+E_{dd}(r_{12})] 
\,\,,
\\
g_{de}(r_{12})&=& N_{de}(r_{12})+X_{de}(r_{12})\nonumber 
\,\,,
\\
g_{ed}(r_{12})&=&g_{de}(r_{12}) \nonumber 
\,\,,
\\
g_{ee}(r_{12})&=&N_{ee}(r_{12})+X_{ee}(r_{12}) \nonumber 
\,\,,
\\
g_{cc}(r_{12})&=&N_{cc}(r_{12})+X_{cc}(r_{12})-\frac{1}{\nu}\ell(k_Fr_{12})
\,\,. 
\label{eq:inf-fhnctbdf}
\end{eqnarray}

The total TBDF can be written in terms of the partial ones as:
\begin{equation}
g(r_{12})=g_{dd}(r_{12})+g_{ed}(r_{12})+g_{de}(r_{12})+g_{ee}(r_{12})
\label{eq:inf-fhnctot}
\end{equation}

%%%%%%%%%%%%%%%%%%%%%%%%%%%%%%%%%%%%
%
% Figure elementary fhnc diagram
%
%%%%%%%%%%%%%%%%%%%%%%%%%%%%%%%%%%%%
\vskip 0.5 cm
\begin{figure}[htb]
\begin{center}
\includegraphics[scale=0.5]{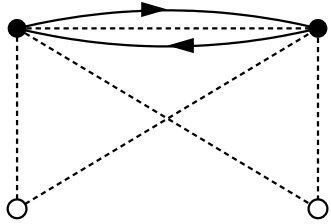}
\end{center}
\caption{\small  
Graphical representation of an elementary diagram.
}
\label{fig:inf-elementary}
\end{figure}
\vskip 0.5 cm
%%%%%%%%%%%%%%%%%%%%%%%%%%%%%%%%%%%%

The set of Eqs. (\ref{eq:inf-fhncnod}), (\ref{eq:inf-fhncnonod}),
(\ref{eq:inf-fhnctbdf}) and (\ref{eq:inf-fhnctot}) forms the Fermi
HyperNetted Chain (FHNC) equations. As we have already mentioned in
the case of bosons, also the FHNC equations allow the evaluation of
the contribution of all the composite and nodal diagrams in a closed
form. Again, the contributions of the elementary diagrams, such as
that shown in Fig. \ref{fig:inf-elementary}, should be included by
calculating them one by one. In analogy to the HNC case, it is common
practice to use the nomenclature FHNC/0, FHNC/4 etc. to indicate the
elementary diagrams included in the calculations.

\subsection{The operator dependent correlations}
\label{sec:inf-stdc}
The use of purely scalar correlations, as it is done in the Jastrow
ansatz (\ref{eq:in-jastrow}), is not adequate to deal with the
complicated structure of the nucleon-nucleon interaction.  For
example, the tensor terms of the interaction do not contribute to the
energy expectation value if only scalar correlations are used.  For
this reason the Jastrow ansatz has been extended by introducing
correlations which have the same operator structure of the NN
interaction, and are called in the literature \cite{pan79}
state-dependent correlations.  The general expression of these
type of correlations is:
\begin{equation}
{\cal F}(1,...,A)={\cal S}\Big( \prod_{j>i=1}^A F_{ij} \Big)={\cal
S}\Big(\prod_{j>i=1}^A\sum_{p=1}^{6}f_{p}(r_{ij})O^{p}_{ij} \Big)
\,\,.
\label{eq:inf-stcorr}
\end{equation} 
The operators $O^{p}_{ij}$ are defined as:
\beq
O^{p=1,6}_{ij} = 1,{\btau}_i\cdot{\btau}_j,
{\bsigma}_i\cdot{\bsigma}_j,
({\bsigma}_i\cdot{\bsigma}_j)({\btau}_i\cdot{\btau}_j),
S_{ij},S_{ij}({\btau}_i\cdot{\btau}_j)
\,\,,
\label{eq:inf-operators}
\eeq
where 
\beq
S_{ij} \equiv 
3(\bsigma_i\cdot\hat{\br}_{ij})(\bsigma_j\cdot\hat{\br}_{ij})
-\bsigma_i\cdot\bsigma_j
\,\,,
\label{eq:inf-tensor}
\eeq 
is the tensor operator. The symmetry operator ${\cal S}$ is required
to guarantee the antisymmetrization of the wave function
$\Psi(1,...,A)$ since, in general, the operators do not commute.  In
Eq. (\ref{eq:inf-operators}) we have indicated only the channels up to
$p=6$ since this is the correlation we have used in our numerical
calculations. State-dependent correlations constructed by considering
a larger number of channels have been used in nuclear matter
\cite{akm97} and in variational Monte Carlo calculations \cite{pie92}.

The order that we have introduced in the operators will be useful
for the finite system since we can write:
\begin{equation}
O_{ij}^{2k-1+l}=P_{ij}^k ({\btau}_i\cdot{\btau}_j)^l \,\,,
\label{eq:inf-separ}
\end{equation}
with $l=0,1$, $k=1,2,3$ and $P_{ij}^k=1,{\bsigma}_i\cdot{\bsigma}_j,S_{ij}$.
This allows us to separate clearly the spin and isospin parts of the
operators.  

The evaluation of the energy functional (\ref{eq:inf-energy}) requires
the calculation of the expectation value of two-body operators related
to the NN interaction which are written in terms of the operators
(\ref{eq:inf-operators}). In general, these operators can be expressed
as:
\begin{equation}
B(1,\ldots,A)= \sum_{j>i=1}^A \left( \sum_{p=1}^6 B ^p(r_{ij})
O^p_{ij} \right)
\,\,,
\end{equation}
and this suggest to define state-dependent TBDFs as:
\begin{equation}
g_p (\br_1,\br_2) =\frac{{\displaystyle A(A-1) \int
d{x}_3\ldots d{x}_A\Psi^{*}(1,\ldots,A) O^p_{12}
\Psi(1,\ldots,A)}}{
{\displaystyle \rho^2 \int
d{x}_1d{x}_2\ldots d{x}_A\Psi^{*}(1,\ldots,A) 
\Psi(1,\ldots,A)}}  
\,\,,
\label{eq:inf-sdtbdf}
\end{equation}
where we understand that all the spin and isospin traces are
done, including those of the external points 1 and 2. With the above
definition, the expectation value of $B$ can be calculated as:
\begin{equation}
<B> =\frac 1 2 \rho^2 \sum_{p=1}^6 \int
d{\br}_1d{\br}_2 B^p(r_{12}) g_p (\br_1,\br_2)
\,\,.
\label{eq:inf-expb}
\end{equation}

With the help of the sub-determinants (\ref{eq:inf-subdet}) we express
the state-dependent TBDF as:
\begin{eqnarray}
\nonumber
&~& g_p (\br_1,\br_2) = \\
&~&\frac{
{\displaystyle  A(A-1)  \int
d{x}_3\ldots d{x}_A {\cal S} \Big( \prod_{j>i=1}^A F_{ij} \Big) O^p_{12}
{\cal S}\Big( \prod_{j>i=1}^A F_{ij} \Big) 
\Delta_A (1,\ldots,A)}}{
{\displaystyle \rho^2 \int
d{x}_1d{x}_2\ldots d{x}_A{\cal S} \Big( \prod_{j>i=1}^A F_{ij} \Big) {\cal S}
\Big( \prod_{j>i=1}^A F_{ij} \Big) 
\Delta_A (1,\ldots,A)}}
\,\,.
%\label{eq:inf-sdtbdf}
\end{eqnarray}

In the calculation of the TBDF, we find it convenient to rewrite the
correlation function as:
\begin{equation}
\hspace*{-2mm}
F_{ij} =  \sum_{p=1}^{6}f_{p}(r_{ij})O^{p}_{ij}= f_1(r_{ij})
\left( 1+\sum_{p=2}^{6}\frac{f_{p}(r_{ij})}{f_{1}(r_{ij})}
O^{p}_{ij} \right)  =  f_1(r_{ij}) 
\left( 1 + H_{ij} \right)
\,.
\label{eq:inf-defhij}
\end{equation}

Because of the non commutativity of the operator dependent terms, in
the cluster expansion we have to consider also the ordering of the
various terms. Only the scalar term $p=1$, which commutes with all the
other ones, can be treated as we have indicated in the previous
section. By using the commutativity property of the scalar term we can
rewrite the correlation function as:
\begin{equation}
{\cal F}(1,...,A)={\cal S}\Big( \prod_{j>i=1}^A F_{ij} \Big)=
\left(\prod_{j>i=1}^A f_1(r_{ij}) \right)
{\cal S}\left[ \prod_{j>i=1}^A \big( 1 + H_{ij} \big) \right]
\,\,.
\label{eq:inf-correl}
\end{equation} 
This expression shows that each operator dependent term $H_{ij}$ can
be multiplied by any contribution from the central correlation
functions, $f_1$, without changing the operator structure of the
correlation.  In the many-body jargon when we incorporate into the
operator terms all the contributions from the central correlation, we
say that the Jastrow correlations dress the operator terms.

The general treatment of the state dependent correlations for nuclear
matter was first proposed in \cite{pan79}. This is the basic reference
for the interested reader. In the following we shall recall the basic
steps of the procedure, and we point out the features of interest for
the treatment of finite nuclear systems.

Although the notation in the demonstration would be  more involved 
than in the purely Jastrow case, it is still possible to show that 
the compensation between the unlinked diagrams of the numerator and 
all the diagrams of the denominator holds for the state-dependent 
correlations.

%%%%%%%%%%%%%%%%%%%%%%%%%%%%%%%%%%%%%%%%%%%%%%%%%%
% Figure of cancellation diagram with SOC
%%%%%%%%%%%%%%%%%%%%%%%%%%%%%%%%%%%%%%%%%%%%%%%%%%
\begin{figure}[htb]
\begin{center}
\includegraphics[scale=1.0]{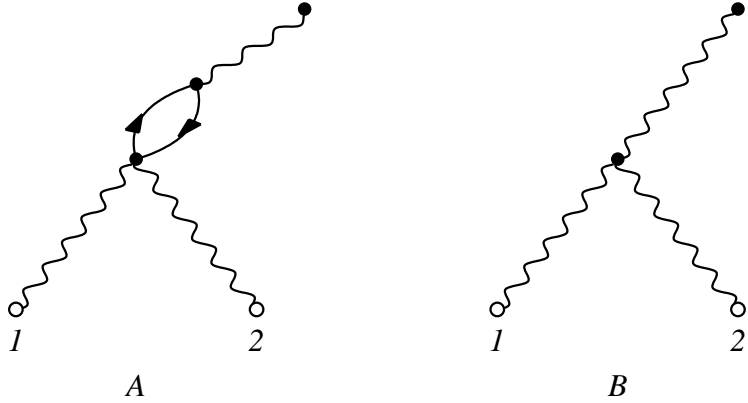}
\end{center}
\caption{
\small Diagrams analogous to those of Fig. \ref{fig:inf-canc} with
state dependent correlations.  
}
\label{fig:inf-canc2}
\end{figure}
%%%%%%%%%%%%%%%%%%%%%%%%%%%%%%%%%%%%%%%%%%%%%%%%%%

We have shown in the previous section that the second step in
obtaining the FHNC equations consisted in eliminating the contribution
of the reducible diagrams.  In the present case, this is no longer 
possible.  We explain the problem by using the example of Fig.
\ref{fig:inf-canc2} which is analogous to Fig. \ref{fig:inf-canc} but
with the scalar correlations substituted by state dependent
correlations, indicated by the wavy lines. We should remark that in the
graphical representation, the wavy lines indicate a generic operator
dependent term of the correlation. Diagrams with various wavy lines,
consider all the possible combinations and ordering of operators.  
Back to the case of Fig. \ref{fig:inf-canc2}, we should consider that 
all the statistical lines in a closed loop, but one, carry a spin-isospin 
exchange operator:
\beq
\Pi^{\sigma,\tau}(i,j) \equiv
\frac 1 4 
(1+\bsigma_i\cdot\bsigma_j)(1+\btau_i\cdot\btau_j)=
\frac 1 4 \sum_{p=1}^6 \Gamma^p O^p_{ij}
\,\,,
\label{eq:inf-siexch}
\eeq
on the corresponding pair of particles \cite{pan79} and with
$\Gamma^p$ given in Tab. \ref{tab:inf-delta} of 
Appendix \ref{sec:app-matrix}.
The spin and isospin dependent part of this operator is linear in
$\bsigma$ and/or $\btau$, therefore its trace is zero when only scalar
correlations are used, as in the case of Fig. \ref{fig:inf-canc}. In
that case, the contributions of the two diagrams were identical, with
a sign difference, therefore, the global result was zero.  In the
present case, the trace of the operator dependent part of the exchange
term is not always zero, but its value depends on the type of
operators linked to the points at the edges of the exchange loop. For
this reason, the traces of the A and B diagrams of Fig.
\ref{fig:inf-canc2} are in general different, therefore the global
result is, in general, different from zero.

This loss of irreducibility in the expansion of the TBDF joined to
the non commutativity among the operators makes it not
possible to calculate all the contributions of state-dependent 
correlations. This obliges us to use approximations. 
The difficulty in eliminating the reducible diagrams is overcome by
using an approximation, consisting in considering operator dependent
diagrams with specific topological properties. These diagrams are
classified as Single Operator Chain diagrams. Since the calculations of
these diagrams requires the evaluation of traces of non commuting
operators, we present first the technique used to calculate these
traces, and then the FHNC equations in the SOC approximation. 

\subsubsection{Traces}
\label{sec:inf-traces}
In the calculation of the TBDF, or of the energy functional, we have
to calculate expectation values of operators between the trial wave
functions (\ref{eq:in-trial}). These calculations require the
integration on the space coordinates and the sum on the spin and
isospin coordinates. We have called trace this last operation. There
are no specific strategies for evaluating the space integrals, which
 are done numerically. On the contrary, there are
strategies to obtain the spin and isospin traces. Here we shall
present them for the specific case of the infinite symmetric nuclear
matter, and, later, we shall generalize them for finite nuclei. We
have emphasized the characteristics of the system under discussion,
because we want to point out that it is a spin and isospin saturated
system, therefore the traces of terms linear in $\bsigma$ and/or 
$\btau$ operators are zero.

We start our discussion with the simplest possible case, the matrix
element between two points only, $1$ and $2$, having a single operator
$O_{12}^p$ acting between them. Because of the Pauli principle we have
direct and exchange terms. The direct term is:
\begin{eqnarray*}
&~&O_{12}^p \rho_0(1,1) \rho_0(2,2)  =  \\
&~&  \rho^2 \frac 1 {\nu^2}
\sum_{s_1,s_2,t_1,t_2} \chi_{s_1}^+(1)\chi_{s_2}^+(2)
\chi_{t_1}^+(1)\chi_{t_2}^+(2) O_{12}^p
\chi_{t_1}(1)\chi_{t_2}(2)\chi_{s_1}(1)\chi_{s_2}(2) \\
&~& =  \rho^2 C(O_{12}^p) 
\,\,,
\end{eqnarray*}
where we have used the expression (\ref{eq:inf-defro}) for $\rho_0$
and the fact that the limit of the Slater function (\ref{eq:inf-slatfun})  
when its argument goes to zero, is one. In the above expression we
have indicated with $C(O_{12}^p)$ the value of spin and isospin trace
relative to the operator $O_{12}^p$ divided by the number of states, 
$\nu^2$ in this case. We call it the $C$-trace. 
The exchange term for the case under study is:
\begin{eqnarray*}
&~&O_{12}^p \rho_0(1,2) \rho_0(2,1) = \rho^2 \frac {\ell^2
    (k_Fr_{12})}{\nu^2} \\
&~& 
\sum_{s_1,s_2,t_1,t_2} \chi_{s_2}^+(1)\chi_{s_1}^+(2)
\chi_{t_2}^+(1)\chi_{t_1}^+(2) O_{12}^p
\chi_{t_1}(1)\chi_{t_2}(2)\chi_{s_1}(1)\chi_{s_2}(2) \\
&~& = \rho^2\ell^2 (k_Fr_{12}) \, 
C\left(\frac 1 4 
(1+\bsigma_1\cdot\bsigma_2)(1+\btau_1\cdot\btau_2)
O_{12}^p \right) 
\,\,,
\end{eqnarray*}
where the term multiplying $O_{12}^p$ in the $C$-trace 
is the spin and isospin exchange operator which has to be
on the left of the rest of the operators.

The operators (\ref{eq:inf-operators}) are built to be scalar in the
Fock space formed by the product of configuration and spin, and
isospin, spaces, therefore they are constructed as scalar product of
spin, coordinates and isospin operators. For this reason, their
contributions can be evaluated by using the Pauli identity:
\begin{equation}
\label{eq:inf-pauliid}
({\balpha}_1\cdot{\bf A})({\balpha}_1\cdot{\bf B})=
{\bf A}\cdot{\bf B}+i{\balpha}_1\cdot({\bf A}\times{\bf B})
\,\,,
\end{equation}
where $\balpha=\bsigma, \ \btau$ and {\bf A} and {\bf B} are generic
vector operators. By using this identity, we can isolate the terms
linear in $\btau$ or $\bsigma$ which do not contribute in infinite
and symmetric nuclear matter, as we have already stated.

The two examples we have discussed are the easieast ones to calculate.
In evaluation of the TBDF, or of the energy functional, we have to
deal with more complicated situations. Following Ref. \cite{pan79} we
consider three type of situations.
\begin{enumerate}
\item[a)] Products of operators acting on the same pair, such as:
\[
O^{p>1}_{ij}O^{q>1}_{ij}\cdot\cdot\cdot O^{r>1}_{ij}
\,\,,
\]
\item[b)] 
 Products of operators acting on different connected points
 forming a ring, such as:
\[
O_{12}^{p>1}O_{23}^{q>1}.....O_{n-1n}^{r>1}O_{n1}^{s>1}
\,\,.
\]
We call this situation  Single-Operator Ring (SOR)
\item[c)] The situation when more than two operators act on 
an internal point. These are multipole operators terms. 
\end{enumerate}

{\em \ref{sec:inf-traces}.a Products of operators}\\
We analyze the trace algebra of the products of operators $O_{12}^{p}$ 
acting on the same pair of coordinates $1$ and $2$.

The $C$-trace of a single operator is:
\begin{equation}
C(O_{12}^p) = \delta_{p,1}
\label{eq:inf-ctrace1}
\end{equation}
since all operator linear in $\bsigma$ and/or $\btau$ have
zero $C$-trace in a spin and isospin saturated system, as we have
already mentioned. 

The $C$-traces of the product of two operators
$O^{p>1}_{12}O^{q>1}_{12}$ are obtained by using the 
relations 
\begin{eqnarray}
\nonumber
(\balpha_1\cdot\balpha_2)^2&=&
3-2\balpha_1\cdot\balpha_2  
\,\,,
\\
\nonumber
S_{12}\bsigma_1\cdot\bsigma_2&=&S_{12}
\,\,,
\\
S_{12}^2&=&6+2\bsigma_1\cdot\bsigma_2-2S_{12}
\,\,,
\label{eq:inf-spinisorel}
\end{eqnarray}
calculated by using the Pauli identity (\ref{eq:inf-pauliid})
and with $\balpha=\bsigma,\btau$. The values of the $C$-traces 
in this case can be summarized as:
\begin{equation}
C\Big(O^{p}_{12}O^{q}_{12}\Big)=B^{p}\delta_{pq} 
\,\,,
\label{eq:inf-ctrace2}
\end{equation}
where the values of $B^p$ are given in Tab. \ref{tab:inf-ap}  
of Appendix \ref{sec:app-matrix}. In \cite{pan79}, these are called
$A^p$ but we shall use this name for their spin parts.

The knowledge of the values of the $C$-traces of one operator and of
the product of two operators 
is enough to calculate the values of the 
$C$-traces for the product of any number of operator. The relations 
(\ref{eq:inf-spinisorel}) indicate that the product of two operators 
$O^{p>1}_{12} O^{q>1}_{12}$ can be written as sum of operators
$O^{r}_{12}$ multiplied by a coefficient. We can write:
\begin{equation}
\label{eq:inf-kmatrix}
O^{p}_{12}O^{q}_{12}=\sum_{r=1}^6 K^{pqr}O^{r}_{12}
\,\,,
\end{equation}
where the values of the matrix $K^{pqr}$ are given in Tab.
\ref{tab:inf-kpqr} of Appendix \ref{sec:app-matrix}. For example for
three operators we have: 
\begin{eqnarray*}
C\Big(O^{p}_{12}O^{q}_{12}O^{r}_{12}\Big)&=&\sum_{m=1}^6K^{pqm}C\Big(
O^m_{12}O^{r}_{12} \Big) = K^{pqr}B^r
\,\,.
\end{eqnarray*}

We would like to point out that since the operators $O^{p=1,6}$ acting
on the same points commute, their ordering does not matter in the
calculation of the $C$-trace. This means that:
\[ 
K^{pqr}B^{r}=K^{prq}B^{q}=K^{qrp}B^{p}....
\,\,.
\]

%%%%%%%%%%%%%%%%%%%%%%%%%%%%%%%%%%%%
% Figure SOR
%%%%%%%%%%%%%%%%%%%%%%%%%%%%%%%%%%%%
\vskip 0.5 cm 
\begin{figure}[htb]
\begin{center}
\includegraphics[scale=1.0]{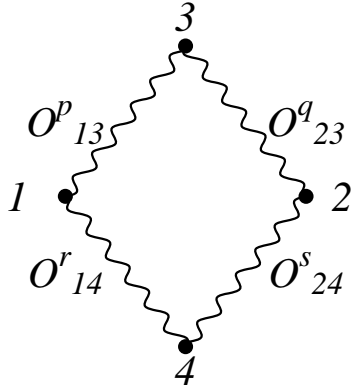}
\end{center}
\caption{
\small  Example of Single Operator Ring (SOR). 
}
\label{fig:inf-sor}
\end{figure}
\vskip 0.5 cm 
%%%%%%%%%%%%%%%%%%%%%%%%%%%%%%%%%%%%

{\em \ref{sec:inf-traces}.b Single operator rings} \\
In Fig. \ref{fig:inf-sor} we show an example of Single Operator
Ring (SOR). 
In the SOR diagram each point is reached by two operators
only. The ordering of these operators does not change the value of the
$C$-trace of the SOR. This is because, due to 
the Pauli identity
(\ref{eq:inf-pauliid}), the non commuting terms are linear in $\bsigma$
or $\btau$, therefore their trace is zero.

The basic step to evaluate of the $C$-trace of a SOR it is to
calculate the $C$-trace of two operators acting on a common point.
We call $O^{p}_{12}$ and $O^{q}_{23}$ the two operators, and $2$ is the
common point. All the variables relative to the common point $2$
should be summed or integrated. We sum on all the spin and isospin
third components and integrate on the azimuthal angle $\phi$:
\begin{equation}
\sum_{\sigma_2\tau_2}\int d\phi_2 \,O^{p}_{12} \, O^{q}_{23}=
\sum_{r=1}^6\int d\phi_2 \, \xi^{pqr}_{123} \, O^{r}_{13}
\,\,.
\label{eq:inf-xidef}
\end{equation}
The $\xi^{pqr}_{123}$ functions depend on the angles of the triangle
formed by the $1,2$ and $3$ points, and have the following properties:
\begin{eqnarray}
\nonumber
&~&
\xi^{2k_1-1+l_1 \,\, 2k_2-1+l_2 \,\, 2k_3-1+l_3}_{123} =
\zeta^{k_1 k_2 k_3}_{123} \delta_{l_1 l_2}\delta_{l_1 l_3}
\,\,,
\\
\nonumber
\zeta^{k_1 k_2 k_3}_{123} & = & \delta_{k_1 1}\delta_{k_2 1}\delta_{k_3 1}+
\delta_{k_1 2}\delta_{k_2 2}\delta_{k_3 2}
\\
\nonumber
&+&
P_2(\hat{r}_{13}\cdot\hat{r}_{23}) (\delta_{k_1 2}-\delta_{k_1 3}) 
\delta_{k_2 3}\delta_{k_3 3} 
\\
\nonumber
&+&
P_2(\hat{r}_{12}\cdot\hat{r}_{13}) \delta_{k_1 3}
(\delta_{k_2 2}-\delta_{k_2 3})\delta_{k_3 3}
\\ 
\nonumber
&+& 
P_2(\hat{r}_{12}\cdot\hat{r}_{23}) \delta_{k_1 3}
\delta_{k_2 3}(2\delta_{k_3 2}- \delta_{k_3 3}) 
\\
&-& 
\frac 1 2 \Big(9(\hat{r}_{13}\cdot\hat{r}_{23})
(\hat{r}_{12}\cdot\hat{r}_{13})(\hat{r}_{12}\cdot\hat{r}_{23})+ 1
\Big)\delta_{k_1 3}\delta_{k_2 3}\delta_{k_3 3} \,\,,
\label{eq:inf-xiprop}
\end{eqnarray}
with $P_2(x)=(3x^2-1)/2$ the Legendre polynomial of second degree and
we have used the separation in spin and isospin parts of the operators
presented in Eq. (\ref{eq:inf-separ}).
The global contribution of the SOR is calculated by using Eqs. 
(\ref{eq:inf-xidef}) and (\ref{eq:inf-xiprop}) for all the points of
the ring.

%%%%%%%%%%%%%%%%%%%%%%%%%%%%%
% Figure of Multiple operators 
%%%%%%%%%%%%%%%%%%%%%%%%%%%%%
\vskip 0.5 cm 
\begin{figure}[htb]
\begin{center}
\includegraphics[scale=1.0]{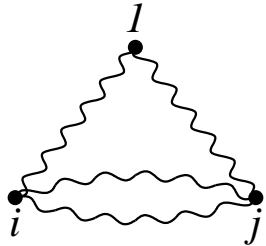}
\end{center}
\caption{
\small  Example of a multipole operator diagram 
where more than one operator acts on the same two points.
}
\label{fig:inf-mod}
\end{figure}
\vskip 0.5 cm 
%%%%%%%%%%%%%%%%%%%%%%%%%%%%%

{\em \ref{sec:inf-traces}.c Multiple-operators diagrams} \\
The last type of diagrams we discuss is represented by the diagram of
Fig. \ref{fig:inf-mod}. In this diagram two operators act on the same
points $i$ and $j$. The diagram  of the figure represents 
the product of the operators: $O^{p}_{ij}$, $O^{q}_{ij}$, 
$O^{r}_{1i}$ and $O^{s}_{1j}$.
In this case, the value of the $C$-trace depends on the ordering of 
the operators. It is possible to prove that this value is 
unchanged by a cyclic permutation of the operators of the same type
\cite{pan79}. That is:
\[
C \Big(O^{p}_{kl} (O^{q}_{ij} O^{r}_{ij} ...... ) \Big) =
C \Big( (O^{q}_{ij} O^{r}_{ij} ...... ) O^{p}_{kl} \Big) 
\,\,.
\]

As a consequence, for the evaluation of the $C$-trace 
we have to consider only two different orderings of
operators. A first one where $O^{p}_{ij}$ and
$O^{q}_{ij}$ are close to each other, and a second one, 
where these two operators are separated by another operators of the
type $O^{r}_{i1}$ or $O^{s}_{1j}$.
In the first case, by using 
Eqs. (\ref{eq:inf-kmatrix}) and (\ref{eq:inf-xiprop}) we obtain:
\beq
\int d\phi_1 C\Big(O^{p}_{ij}O^{q}_{ij}O^{r}_{1i}O^{s}_{1j}\Big)=
\sum_{t=1}^6K^{pqt}B^{t}\int d\phi_1 \xi^{rst}_{i1j}
\,\,.
\eeq
For the second case we have:
\begin{equation}
\int d\phi_1 C\Big(O^{p}_{ij}O^{r}_{1i}O^{q}_{ij}O^{s}_{1j}\Big)=
\sum_{t=1}^6 L^{pqt}\int d\phi_1 \xi^{rst}_{i1j}
\,\,,
\label{eq:inf-lpqrdef1}
\end{equation}
where we have defined:
\begin{equation}
L^{pqt}=\pm K^{pqt}B^{t}
\,\,.
\label{eq:inf-lpqrdef2}
\end{equation}
The $+$ sign is assigned if 
\[
C\Big(O^{p}_{ij}[O^{q}_{ij},O^{r}_{1i}]O^{s}_{1j}\Big)=0
\,\,,
\]
and the $-$ sign if 
\[
C\Big(O^{p}_{ij}\{O^{q}_{ij},O^{r}_{1i}\}O^{s}_{1j}\Big)=0
\,\,,
\]
where we have indicated with the symbols  
$[,]$ and $\{,\}$ the commutator and anticommutator respectively. 
The values of matrix $L^{pqr}$ are given in Tab. \ref{tab:inf-lpqr}
of Appendix \ref{sec:app-matrix}.  

Another important trace is that of
\[
{\cal C}\Big(O^{s}_{1i}O^{r}_{1i}O^{p}_{ij}O^{q}_{ij}\Big)
\,\,,
\]
which represents two SOR's linked at the point $i$. Also in this case
the result depends on the ordering of the operators and we
distiguish the case when 
$O^{p}_{ij}$ and $O^{q}_{ij}$ are close together or not. In the first 
case, we obtain:
\[
C\Big(O^{p}_{ij}O^{q}_{ij}O^{s}_{1i}O^{r}_{1i}\Big)=
B^p \delta_{p,q} B^{s} \delta_{s,r}
\,\,.
\]
To evaluate the $C$-trace of the second case we consider 
the fact that:
\begin{equation}
\label{eq:inf-dpqdef}
\sum_{\sigma_j\tau_j}O^{p}_{ij}O^{s}_{1i}O^{q}_{ij}=
\delta_{p,q}B^{p}(1+E_{ps})O^{s}_{1i}
\,\,,
\end{equation}
where, in the case of tensor operators, the above equation assumes an
integration over the angle between $\br_{ij}$ and $\br_{1i}$
\cite{pan79}. Then we obtain:
\begin{equation}
C\Big(O^{p}_{ij}O^{s}_{1i}O^{q}_{ij}O^{r}_{1i}\Big)=
B^p \delta_{p,q} (1+E_{ps}) B^{s} \delta_{s,r}
\,\,,
\end{equation}
with the values of  $E$, given in Tab. \ref{tab:inf-dpq} 
of Appendix \ref{sec:app-matrix}. In \cite{pan79} these
are called $D$ but we shall use this name for their spin parts.

\subsection{The Single Operator Chain (SOC) equations}
\label{sec:inf-soc}
\begin{figure}[htb]
\begin{center}
\includegraphics[scale=0.75]{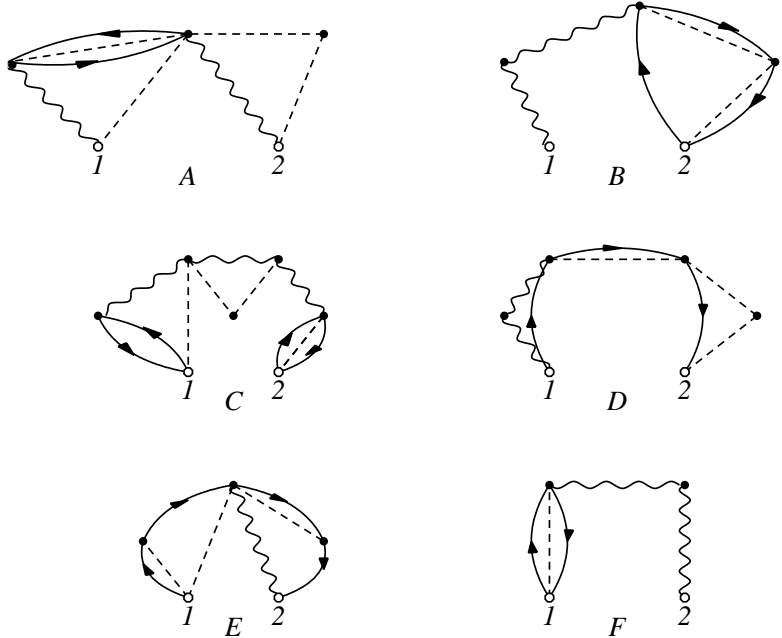}
\end{center}
\caption{
\small  Some nodal diagrams considered in the FHNC/SOC compuational
scheme. 
}
\label{fig:inf-soc}
\end{figure}

The strategy to attack the problems arising when state-dependent
correlations are used, consists in separating the purely scalar,
Jastrow, terms from those depending on the operators $O^p$ for $p>1$.
The Jastrow part is treated by using the set of FHNC equations
(\ref{eq:inf-fhncnod}) , (\ref{eq:inf-fhncnonod}) and
(\ref{eq:inf-fhnctbdf}).  With respect to the operator dependent part,
we have learnt that we need at least two operators arriving at a given
point to get a $C$-part different from zero. As the operators are
dynamical correlations, there is no limitation in the number of them
that can arrive at every point. An increasing number of operators
makes more complicated the evaluation of the traces so the Single
Operator Chain (SOC) approximation is adopted.  This supposes that
only a pair of operators arrive at every internal point of the
diagrams, this makes the operators form closed single chains and it
allows the formulation of closed expressions to calculate all the
nodal operators and those composite ones within the approximation.
The reliability of the SOC approximation is tested afterwards, by
controlling the validity of sum rules exhaustion. Examples of SOC
diagrams, of nodal type, are given in Fig.  \ref{fig:inf-soc}.

The SOC diagrams do not have limitations in the number of particles.
The discussion made in Sect. \ref{sec:inf-traces} has clarified that
the contribution of the SOC diagrams to the TBDF's and/or to the
energy, is independent of the ordering of the operators.  The single
operator between two points of the SOC diagram, can come from the
correlation, from the hamiltonian, or from an exchange line, whose
contribution is considered by inserting the spin-isospin exchange
operator (\ref{eq:inf-siexch}).

%%%%%%%%%%%%%%%%%%%%%%%%%%%%%%%%%%
% Figure SOC
%%%%%%%%%%%%%%%%%%%%%%%%%%%%%%%%%%
\vskip 0.5 cm 
\vskip 0.5 cm 
%%%%%%%%%%%%%%%%%%%%%%%%%%%%%%%%

The choice of considering SOC diagrams only, eliminates
the problem of the reducible diagrams, since all the SOC diagrams are
irreducible.  The equations we should consider contain only 
irreducible diagrams. Also in this case it is possible to show
\cite{pan79} that the composite diagrams can be obtained in terms of
simple diagrams classified again in nodal and elementary ones.

As in the FHNC case, the contribution of a nodal diagram
$N_{mn,p}(r_{ij})$, can be obtained by doing the folding product of
diagrams at the nodal point, as can be deduced by observing the
examples given in Fig.\ref{fig:inf-soc}.  Here we used $mn$ to
indicate the type of diagram $mn=dd,\ de, \ ed, \ ee, \ cc$.  As
already discussed, the folding product in the $\br_k$ point should be
done between the irreducible 
non nodal diagrams $X_{mn,p}(r_{ik})$ and
$X_{mn,q}(r_{kj}) + N_{mn,q}(r_{kj})$.  From the discussion made in the
previous section, it appears clear that not all the possible
combinations of operators acting on the integration point provide
contributions different from zero. The allowed combinations are ruled
by the values of $\xi^{pqr}$ given in Eqs.(\ref{eq:inf-xiprop}).

The sequence of Eqs.(\ref{eq:inf-fhncnod}) giving the nodal diagrams,
is generalized for the state-dependent terms as:
\begin{equation}
\hspace*{-5mm}
N_{mn,r}(r_{12})=\sum_{m',n'}\sum_{p,q=1}^6
\Big(X_{mm',p}(r_{13}) \, \xi^{pqr}_{132} \Big| \rho(\br_3)
 [X_{n'n,q}(r_{23})+N_{n'n,q}(r_{23})] \Big)
,
\label{eq:inf-nodalop}
\end{equation}
for $m,n=d,e$ and with $m'n'=dd,ed,de$. 
If we neglect the contribution of the elementary diagrams, i.e. in
approximation FHNC/0, we obtain the following expresssions for the 
TBDFs for $p> 1$:
\begin{eqnarray}
g_p(r_{12})&=&g_{dd,p}(r_{12})+2g_{de,p}(r_{12})+g_{ee,p}(r_{12}) 
\,\,,
\label{eq:inf-gp1} \\
g_{dd,p}(r_{12})
&=& h_{p}(r_{12})h_{c}(r_{12})=X_{dd,p}(r_{12})+N_{dd,p}(r_{12})
\,\,,
\label{eq:inf-gp2}\\
\nonumber
g_{de,p}(r_{12})&=&\Big(h_{p}(r_{12})N_{de}(r_{12})+f_{1}^2(r_{12})
N_{de,p}(r_{12})\Big)h_{c}(r_{12}) \\ 
&=&X_{de,p}(r_{12})+N_{de,p}(r_{12})
\,\,,
\label{eq:inf-gp3}\\ 
\nonumber
g_{ee,p}(r_{12})&=&\Big[h_{p}(r_{12})\Big(N_{de}(r_{12})N_{ed}(r_{12})+
N_{ee}(r_{12})\Big)
\\
\nonumber
&&+f_{1}^2(r_{12}) \Big(-\nu L^{2}(r_{12})\Gamma^{p}+N_{ee,p}(r_{12})
\,\,,
\\
\nonumber
&&+2N_{de,p}(r_{12})N_{de}(r_{12})\Big)\Big]h_{c}(r_{12}) \\
&=&X_{ee,p}(r_{12})+N_{ee,p}(r_{12})
\,\,,
\label{eq:inf-gp4}
\end{eqnarray}
where the $N$ and $X$ factor without index of the operator channel
indicate the scalar $p=1$ term, $\Gamma^p$ is given by Eq. 
(\ref{eq:inf-siexch}), and we have defined:
\begin{equation}
h_{p}(r_{12})=2f_{p}(r_{12})f_{1}(r_{12})+f_{1}^{2}(r_{12})N_{dd,p}(r_{12})
\,\,,
\label{eq:inf-hp}
\end{equation}
\begin{equation}
h_{c}(r_{12})=\exp[N_{dd}(r_{12})]
\,\,,
\label{eq:inf-hc}
\end{equation}
\begin{equation}
L(r_{12})=N_{cc}(r_{12})-\ell(k_Fr_{12})/\nu 
\,\,,
\label{eq:inf-bigl}
\end{equation}

For the calculation of the cyclic nodal diagrams within the SOC
approximation, it is convenient to distinguish two cases
\cite{pan79}. Since all statistical lines but one carry one
spin-isospin exchange operator, we have to add a dynamical operator to
complete the operator chain. This no exchange operator may be added to
the left or to right of the chain.  Example of these two cases is
given by the D and E diagrams of Fig. \ref{fig:inf-soc}.  In the D
diagram the point $1$ is connected with an operator dependent
correlation while the point $2$ is connected by the spin-isospin
exchange operator.  We label $L$ this case. The situation is reversed
in the case of the E diagram, and we call $R$ this case. By using this
classification we define for the cyclic case the quantities:
\beq
X^{L,R}_{c,p}(r_{12})
=\left[
h_{p}(r_{12})L(r_{12})+f_{1}^{2}(r_{12})N^{L,R}_{c,p}(r_{12})\right]
h_{c}(r_{12})
-N^{L,R}_{c,p}(r_{12})
\,\,,
\eeq
\beq
X_{cc}(r_{12}) =
\Big[f_{1}^{2}(r_{12})h_{c}(r_{12})-1\Big]L(r_{12})
\,\,,
\eeq
\beq
N_{cc,r}(r_{12}) =N^{L}_{c,r}(r_{12})+N^{R}_{c,r}(r_{12})
\,\,,
\eeq
\beq
N^{L}_{c,r}(r_{12})
=\sum_{p,q=1}^6\Big(X^L_{c,p}(r_{13})\xi^{pqr}_{132}
\Gamma^{q} \Big| \rho(\br_3) [X_{cc}(r_{23})+L(r_{23})] \Big)
\,\,,
\eeq
\beq
N^{R}_{c,r}(r_{12})
=\sum_{p,q=1}^6\Big( \Gamma^{p}X_{cc}(r_{13})
\xi^{pqr}_{132} \Big|  \rho(\br_3) 
[ X^R_{c,q}(r_{23})+N^R_{c,q}(r_{23}) ] \Big)
\,\,.
\eeq

The set of equations we have presented is called FHNC/SOC. In this
case the contribution of the elementary diagrams is neglected.  Some
example of diagrams considered by these equations is given in Fig.
\ref{fig:inf-soc}.  The diagram A is a $N_{dd,p}(r_{12})$ nodal
diagram; the diagrams B and F are examples of $N_{de,p}(r_{12})$
diagrams and the diagram C of the $N_{ee,p}(r_{12})$ diagram.
Finally, the diagrams D and E are of $N_{cc,p}(r_{12})$ type.

\section{Finite nuclear systems}
\label{sec:finite}

Two of the basic hypotheses done in the previous section, infinite
number of particles and translational invariance, are no longer valid
in the description of finite nuclei. In the literature the extension
of the FHNC theory to finite nuclear systems was first done by
considering systems with equal number of protons and neutrons with
single particle wave functions produced by a unique Mean-Field (MF)
potential, within the $ls$ coupling scheme \cite{co92,co94,fab98,fab00}.
This situation allowed a straightforward use of the spin and isospin
trace techniques developed to describe symmetric nuclear matter
\cite{pan79}.  The treatment of nuclei not saturated in isospin and
described in the $jj$ coupling scheme, was done in following works 
\cite{ari96,bis06}. Here we do not follow the historical development
of the theory but we present directly the formulation of the FHNC
theory for double closed shell nuclei not saturated in isospin and in
the $jj$ coupling scheme.

The changes of the OBDM expression, due to the loss of the
translational invariance, are presented in Sect. \ref{sec:fin-jj}.  We
should point out that we consider doubly magic nuclei only, which are
spherically symmetric.  In Sect. \ref{sec:fin-vertex} we present the
calculation of the TBDF and we introduce vertex corrections and the
Renormalized FHNC (RFHNC) equations, for purely scalar correlations.
The extension of the theory when state dependent correlations are
used, is presented in Sect.  \ref{sec:fin-stdc}.

\subsection{The single particle basis}
\label{sec:fin-jj}

The nuclear system under study has $Z$ protons, $N$ neutrons and,
therefore, $A=Z+N$ nucleons.  The set of single particles wave
functions used to describe this system is produced by solving the
one-body Schr\"{o}dinger equation:
\begin{equation}
h_{i}^t\phi_i^t(x_i)=\epsilon_i^t\phi_i^t(x_i)
\,\,,
\label{eq:fin-obschr}
\end{equation}
where the one-body hamiltonian is composed by the kinetic energy term
and a spherical mean-field potentials different for protons ($t=1/2$)
and neutrons ($t=-1/2$):
\begin{equation}
h_{i}^{t=\pm 1/2}=-\frac{\hbar^2}{2m_t}\nabla^{2}_{i}+
U^{t}(r_i)
\,\,.
\label{eq:fin-obham}
\end{equation}
where we have indicated with $m_t$ the nucleon mass. Our calculations
have been done with single particle wave functions generated by 
a Wood-Saxon potential of the form:
\begin{eqnarray}
\nonumber
U^t(r) &=&
\frac{V^t_0}
{1+ \exp { \left[ \Big( r-R^t_0 \Big) / a^t_0 \right] }} \\
&+&
\left[\frac{\hbar c}{m_{\pi}c^2}\right]^2 \,
V^t_{ls} \,
\frac{\exp{\left[ \Big( r-R^t_{ls} \Big) / a^t_{ls} \right] }}
{ \left\{ 
1+ \exp \left[ \Big( r-R^t_{ls} \Big) / a^t_{ls} 
      \right] \right\}^2 } 
\,\, {\bf l}\cdot \bsigma \, - V^t_{C}(r)
\,\,,
\label{eq:fin-wspot}
\end{eqnarray} 
where $m_{\pi}$ is the pion mass and the Coulomb term $V^t_C(r)$,
active only for protons, is that produced by a
homogeneous charge distribution. 
\begin{eqnarray}
V^{t=1/2}_{C}(r)&=& \left\{\begin{tabular}{cc}
$(Z-1)e^2/r$ & $r\geq R_C$\\
$\displaystyle \frac{(Z-1)e^2}{2R_C}
\left[3-\frac{r^2}{R_C^2}\right]$ & $r\leq R_C$ 
\label{eq:fin-coulpot}
\end{tabular}
\right.
\,\,.
\end{eqnarray}
The values of the parameters $V^t_{0}$, $V^t_{ls}$, $a^t_{0}$,
$a^t_{ls}$, $R^t_{0}$, $R^t_{ls}$ and $R_C$ are fixed by the
variational principle (\ref{eq:in-varp}). In an infinite system, the
variational parameter related to the single particle basis is the
density of the system.

The eigenfunctions of the hamiltonian (\ref{eq:fin-obham}) are also
eigenfunction of ${\bf j}^2$ and $j_z$ operators, where we have
indicated with ${\bf j}$ the total angular momentum of the single
nucleon.  The single particle wave functions are conveniently
expressed as:
\begin{eqnarray}
\nonumber 
\phi_{nljm}^t(x_i) & = & R_{nlj}^t(r_{i})
                \sum_{\mu,s}<l\mu \frac{1}{2}s|jm>
                Y_{l\mu}(\Omega_{i})\chi_{s}(i)\chi_t(i)
\\
& = & \phi_{nljm}^t(\br_i)\chi_t(i)= R_{nlj}^t(r_{i}) {\bf Y}^m_{lj}
(\Omega_{i})\chi_t(i)
\,\,,
\label{eq:fin-spwf}
\end{eqnarray}
where we have indicated with the symbol $<|>$ the Clebsh-Gordan
coefficients, and with $Y_{l\mu}(\Omega_{i})$ the spherical harmonics.
Here we used the symbol $\Omega_{i}$ to indicate both polar angles
$\theta_i$ and $\varphi_i$, characterizing the position of the nucleon
with respect to a fixed center of coordinates chosen to be the center
of the spherical nucleus. We have also defined the spin spherical
harmonics ${\bf Y}^m_{lj}(\Omega_{i})$ \cite{edm57}.

The uncorrelated OBDM (\ref{eq:inf-uobdm}) can be written as:
\begin{equation}
\rho_{0}(x_i,x_j)=\sum_{s,s',t}\rho_{0}^{ss't}({\bf r}_i,{\bf r}_j)
              \chi_{s}^+(i)\chi_{s'}(j)
              \chi_{t}^+(i)\chi_{t}(j)
\,\,,
\label{eq:fin-obdm}
\end{equation}
where the spatial part is defined as:
\begin{eqnarray}
\nonumber
\rho_0^{ss't}({\bf r}_i,{\bf r}_j) &=&\sum_{nlj}R_{nlj}^t(r_i)
                    R_{nlj}^t(r_j) \\
&~&
                    \sum_{\mu\mu' m}<l\mu \frac{1}{2}s|jm>
                    <l\mu ' \frac{1}{2}s'|jm>
                     Y_{l\mu}^{*}(\Omega_{i}) Y_{l\mu '}(\Omega_{j})
\,\,.
\label{eq:fin-obdmr}
\end{eqnarray}

We find it useful to consider separately the uncorrelated OBDMs of pairs
of particles with parallel or antiparallel third components of their
spins. For these OBDMs we obtain respectively the expressions:
\begin{eqnarray}
\nonumber
\rho_0^{\half \half \, t}({\bf r}_i,{\bf r}_j) & \equiv &
\rho_{0}^t({\bf r}_{i},{\bf r}_{j}) \\
&=&
               \frac{1}{8\pi}\sum_{nlj}(2j+1)R^t_{nlj}(r_{i})
                R^t_{nlj}(r_{j})P_{l}(\cos \theta_{ij})
\,\,,
\label{eq:fin-ropar}
\\
\nonumber
\rho_0^{\half \, -\half \, t}({\bf r}_i,{\bf r}_j) & \equiv &
\rho_{0j}^t({\bf r}_i,{\bf r}_j) 
\\
&=& \frac{1}{4\pi}
\sum_{nlj}(-1)^{j-l-1/2}R^t_{nlj}(r_i)R^t_{nlj}(r_j)
                \sin\theta_{ij}P'_{l}(\cos \theta_{ij})
\,\,,
\label{eq:fin-roantipar}
\end{eqnarray}
where we have called $\theta_{ij}$ the angle between $\br_i$ and
$\br_j$, and we have indicated with $P_{l}(x)$ the Legendre polynomial
of $l$th degree, and with $P'_{l}(x)$ its first derivative with
respect to $x$. Some useful symmetry properties of these OBDMs are:
\begin{eqnarray}
\rho_{0}^t({\bf r}_i,{\bf r}_j)&=&
        \rho_0^{\half \, \half \, t}({\bf r}_i,{\bf r}_j)=
       \rho_0^{-\half\, -\half \, t}({\bf r}_i,{\bf r}_j)
\,\,,
\label{eq:fin-rhoz}
\\
\rho_{0j}^t({\bf r}_i,{\bf r}_j)&=&
        \rho_0^{\frac{1}{2} \, -\frac{1}{2} \, t}({\bf r}_i,{\bf r}_j)=
       -\rho_0^{-\frac{1}{2}\, \frac{1}{2} \, t}({\bf r}_i,{\bf r}_j)
\,\,,
\label{eq:fin-rhozj}
\\
\rho_{0}^t({\bf r}_i,{\bf r}_j)&=&
       \rho_{0}^t({\bf r}_j,{\bf r}_i)
\,\,,
\label{eq:fin-rhoP}
\\
\rho_{0j}^t({\bf r}_i,{\bf r}_j)&=&
       -\rho_{0j}^t({\bf r}_j,{\bf r}_i)
\,\,. 
\label{eq:fin-rhoA}
\end{eqnarray}

The uncorrelated OBDM's describing finite nuclei do not depend only on
$r_{ij} = |\br_i-\br_j|$, as in the infinite systems case.  However,
the properties (\ref{eq:inf-contrho}), (\ref{eq:inf-subdetp1}) and
(\ref{eq:inf-subdetp2}), relevant for the construction of the FHNC
equations, remain valid.

The $ls$ coupling can be recovered by switching off the spin-orbit
term $V_{ls}^t=0$ in Eq. (\ref{eq:fin-wspot}). In this case, the
single particle energies, $\epsilon_i$, and the radial functions do
not depend on $j$, therefore the OBDM can be expressed as:
\begin{equation}
\rho_{0}(x_i,x_j)=\sum_{s,t}\rho_{0}^{t}({\bf r}_i,{\bf r}_j)
              \chi_{s}^+(i)\chi_{s}(j)
              \chi_{t}^+(i)\chi_{t}(j)
\,\,,
\end{equation}
with 
\begin{equation}
\rho_{0}^{t}({\bf r}_{1},{\bf r}_{2}) =
               \frac{1}{4\pi}\sum_{nl}(2l+1)R^t_{nl}(r_{1})
                R^t_{nl}(r_{2})P_{l}(\cos \theta_{12})
\,\,,
\end{equation}
showing that only the parallel spin OBDM survives.  The calculations
of Refs. \cite{co92,co94,fab98,fab00} have been done by using an $ls$
coupling scheme and by assuming equal number of protons and neutrons
moving in a unique MF potential. With these assumptions the expression
of the OBDM can be further simplified as:
\begin{equation}
\rho_{0}^\half({\bf r}_{1},{\bf r}_{2}) =
\rho_{0}^{-\half}({\bf r}_{1},{\bf r}_{2}) =
               \frac{1}{4\pi}\sum_{nl}(2l+1)R_{nl}(r_{1})
                R_{nl}(r_{2})P_{l}(\cos \theta_{12})
\,\,,
\end{equation}
i.e. the spatial part of the uncorrelated OBDM is also independent 
of the value of the isospin.

\subsection{The vertex corrections}
\label{sec:fin-vertex}

The construction of the FHNC equations for the finite systems follows
the steps used for the infinite systems. The minimization of the
energy functional, Eq. (\ref{eq:in-varp}), with the ansatz
(\ref{eq:in-trial}) on the wave function and (\ref{eq:in-jastrow}) on
the correlation, leads to the requirement of evaluating the TBDF
(\ref{eq:inf-tbdf}).  The loss of translational invariance obliges us
 to calculate also the One-Body Distribution Function (OBDF),
which, in the present case, depends on the isospin third component:
\begin{eqnarray}
\nonumber
\rho^{\ton}(\br_1) = \frac{{\cal N}_{\ton}}{<\Psi|\Psi>} 
&~& \int dx_2  \ldots dx_A \,\, 
\Psi^*(x_1,x_2,\ldots,x_A) \,P^{\ton}_1 \, \\
 &\times& 
\Psi(x_1,x_2,\ldots,x_A) \,\,\,,
\label{eq:fin-obd} 
\end{eqnarray}
where ${\cal N}_t$ indicates the number of protons ($t=1/2$) or of
neutrons ($t=-1/2$), and the projector operator $P^t$ selects the
particle with isospin third component $t$. By using the above
definitions, we can express the operator dependent TBDF as:
\begin{eqnarray}
\nonumber
&~&
\rho^2 g_q^{\ton \ttw}(\br_1,\br_2) \equiv 
\rho^{q,\ton \ttw}_{2}(\br_1,\br_2) 
= 
\frac {{\cal N}_{\ton} ({\cal N}_{\ttw}-\delta_{\ton \ttw})} 
{<\Psi|\Psi>} 
\\
&~&\times
\int dx_3...dx_A \, 
\Psi^*(x_1,x_2,\ldots,x_A) 
%\\
P^{\ton}_1 O^q_{12} P^{\ttw}_2
\Psi(x_1,x_2,\ldots,x_A) ,
\label{eq:fin-tbdm}
\end{eqnarray}
where the $O^q$ operators have been defined in Eq.
(\ref{eq:inf-operators}).  In the remaining part of this section we
shall be concerned only with the calculation of the scalar TBDF,
$q=1$. The evaluation of the other operator dependent TBDFs is treated
in Sect.  \ref{sec:fin-stdc}.

The first steps to be done to calculate the one- and two-body density
functions defined above, are analogous to those used in the infinite
systems case. We start by defining an $h$-function as in Eq.
(\ref{eq:inf-hdef}) and we use it to make the cluster expansion of the
numerator and the denominator of both OBDF and TBDF.  The various
terms of the cluster expansions can be analyzed by using Mayer
diagrams. The topological analysis of these diagrams is done in
analogy to what we have discussed in the case of infinite systems.

The arguments used in Sects. \ref{sec:inf-boson} and
\ref{sec:inf-fermion} to show that the contributions of the unlinked
diagrams of the numerator are simplified by the denominator, can be
repeated also in the finite systems case \cite{fan79a}.  The
demonstration is done by formally extending up to infinity all the
sums of the various cluster terms, since the property
(\ref{eq:inf-subdetp2}) of the sub-determinant $\Delta_p$ ensures that
diagrams containing a number of particles greater than the number of
particles forming the system, do not contribute.

In the infinite systems case, the next step was the elimination of the
reducible diagrams. We have already said that this elimination is only
approximated for boson systems, up the $1/A$ order, but it is exact
for infinite fermion systems.  The basic point of the demonstration
for this latter case, was the possibility to associate to each
reducible diagram, another diagram containing only one additional
exchange loop.  The contributions of these two diagrams to the TBDF
differ only by a sign, therefore they cancel each other.  This
cancellation mechanism is produced by two specific characteristics of
the infinite system. The fact that for a given reducible diagram it is
always possible to find another diagram having one additional
particle, and one additional exchange loop, is ensured by the presence
of an infinite number of particles. The translational invariance is
instead responsible for the fact that the additional exchange loop
contributes only an overall minus sign.  In the finite nuclei the
number of particles is limited, and the translational invariance is
lost, therefore there is no cancellation of the reducible diagrams.

\begin{figure}[]
\begin{center}
\includegraphics[scale=0.5]{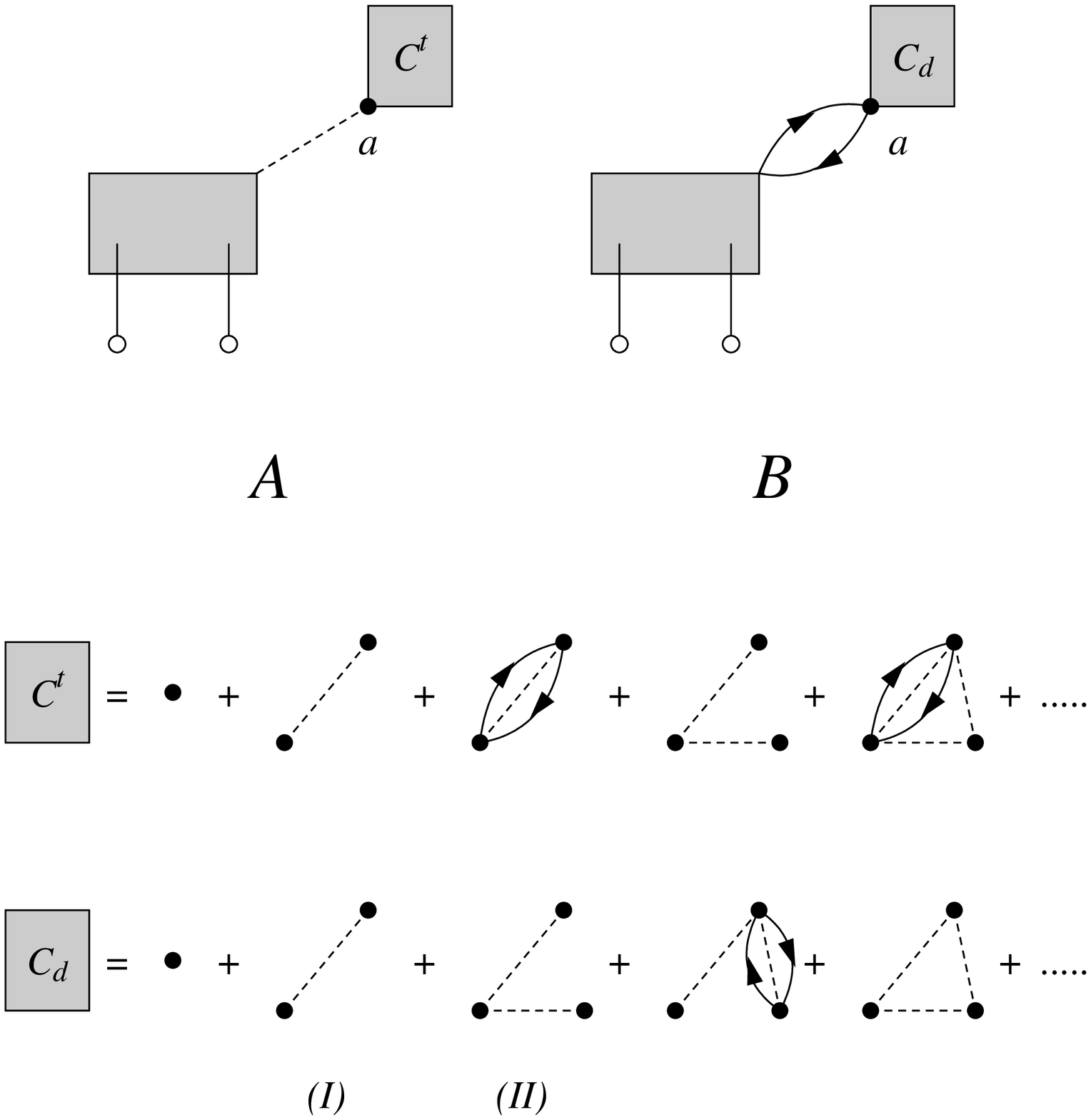}
\vskip 0.1 cm
\includegraphics[scale=0.5]{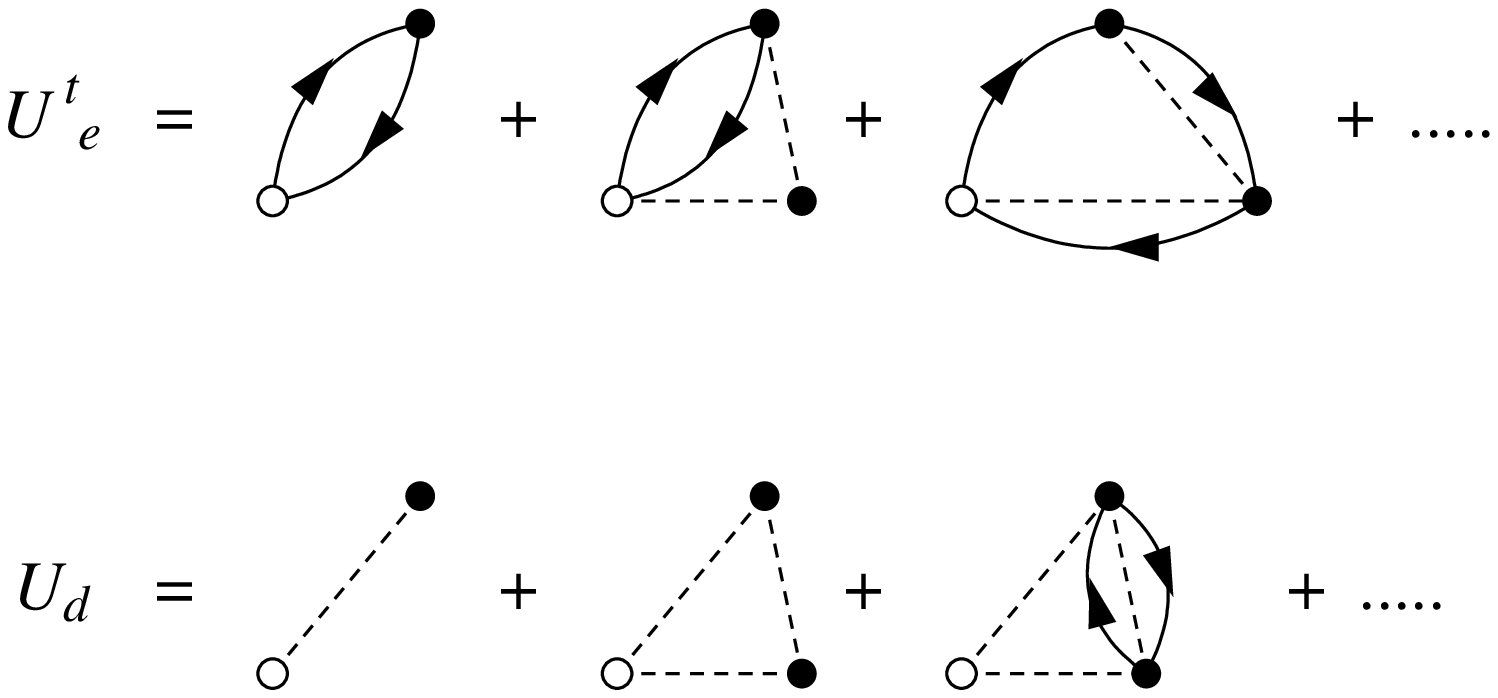}
\end{center}
\caption{ \small 
Graphical illustration of the vertex corrections. Since the Pauli 
exclusion principle allows each point to be reached by no more than
two exchange lines, we have to distinguish the reducible parts of the
diagrams. In $C_d$ the articulation point $a$ can be reached only by
dynamical correlations, while in $C^t$ also by statistical correlations.
  }
\label{fig:fin-vc}
\end{figure}

However, even in finite systems it is possible to recover the
irreducibility of the expansion by introducing the so-called vertex
corrections \cite{fan78a,fan79a}.  A graphical representation of this
idea is given in Fig. \ref{fig:fin-vc}.  Every reducible diagram can
be thought as composed by two parts, as indicated by the diagrams A
and B of the figure.  A first part contains the external points and is
the irreducible part of the diagram.  A second, reducible, part
contains only internal points, and it is linked to the irreducible
part through the articulation point $a$.  The total contribution of
these connected, and reducible, diagrams to the TBDF can be written as
the folding integral of the irreducible part with a function taking
into account the contribution of all the diagrams connected to the
articulation point and it is directly related to the OBDF
(\ref{eq:fin-obd}).

%%%%%%%%%%%%%%%%%%%%%%%%%%%%%%%%
% Figure vertex correction
%%%%%%%%%%%%%%%%%%%%%%%%%%%%%%%%
%%%%%%%%%%%%%%%%%%%%%%%%%%%%%%%%

It is necessary to distinguish the case where the irreducible part is
linked to the articulation point only by dynamical correlations, as in
the A diagram of the figure, from the case when there are statistical
correlations joining the articulation point, as in the diagram B.  To
simplify the drawing, we show in the A diagram only a single
dynamical correlation line connecting the irreducible part to the
articulation point.  In reality, there are no limitations on the number
of dynamical correlations.  In the case of the B diagram we show only
the statistical lines connecting the articulation point, but also
dynamical correlations may be present.

The fact to be considered is that the Pauli principle allows each
point to be reached by no more than two exchange lines.  When the
articulation point is of type $d$, i.e. linked to the irreducible part
of the diagram only by dynamical correlations, the Pauli principle is
not active.  In this case, the reducible part of the diagram can reach
the articulation point with both dynamical and statistical
correlations. We call $C^t(a) \equiv \rho^t(a)$ the sum of all the
possible linked diagrams containing the articulation point $a$ which
has isospin third component $t$.  This is really the OBDF
(\ref{eq:fin-obd}) of the nucleons with isospin third component $t$.

The situation changes when the articulation point is of type $e$, i.e.
linked to the irreducible part of the diagram also by statistical
correlations.  In this case, because of the Pauli principle, the
reducible part of the diagram can reach the articulation point
exclusively with dynamical correlations. We call $C_{d}(a)$ the sum of
the diagrams reaching the articulation point $a$ with dynamical
correlations only.

The evaluation of $C_{d}(a)$ can be performed 
by extending the diagrams
classification done in Sect.  \ref{sec:inf-boson} to the case of a
single external point.  All the linked diagrams, both simple and
composite ones, contribute to $C_{d}(a)$.  As an example, the
contribution of the $(II)$ diagram of Fig. \ref{fig:fin-vc} is
obtained by squaring the contribution of the $(I)$ diagrams and
dividing by two, in order to avoid double counting. The procedure used
in Sect. \ref{sec:inf-boson} to calculate the contribution of the
composite diagrams can be applied also in this case. If we call
$U_{d}(a)$ the sum of all the simple irreducible diagrams
connected to the point $a$ by dynamical correlations only, we can
write:
\beq
\nonumber
C_{d}(a) = 1+U_{d}(a)+\frac{1}{2!}U^{2}_{d}(a)+
\frac{1}{3!}U^{3}_{d}(a)+\cdot\cdot\cdot = \exp[U_{d}(a)]
\label{eq:fin-ce}
\eeq
It is understood that all the diagrams contained in $U_d(a)$ are
renormalized by the vertex corrections, therefore they must be
irreducible in each internal point.

For the calculation of $C^t(a)$ we have to consider also the diagrams
linked to the articulation point $a$ with statistical correlations.
We call $U_e^t(a)$ the sum of all these simple irreducible diagrams.
Because of the Pauli principle, one can construct composite diagrams
with $U_e^t(a)$ combining it only with any number of $U_d(a)$ that
produces $C_d(a)$.  By definition, $C^t(a)$ is given by all the
diagrams contributing to $C_{d}(a)$, i.e. all those reaching $a$ with
dynamical correlations, plus the diagrams constructed by associating
those with $U_e^t(a)$:
\beq
C^t(a)=
C_{d}(a)\left[\rho_0^t(a)+U_{e}^t(a)\right]=\rho^t(a) ,
\label{eq:fin-cd}
\eeq
where $\rho_0^t(a)$ indicates the uncorrelated one-body density for
nucleons with isospin $t$, and $\rho^t(a)$ is the corresponding OBDF.
In the absence of correlations, $U_d$ and $U_e^t$ are zero, therefore
$C^t$ is equal to the uncorrelated density, as is expected.  The
construction of the functions $U_{d,e}(1)$ is done by integrating the
composite daigrams over the coordinate $2$. This procedure requires a
careful attention to the avoid the possible overcounting problems
\cite{fan78a}.  The explicit expressions of $U_{d,e}(1)$ are given in
Appendix \ref{sec:app-fhncsoc}.

By introducing the vertex corrections idea, we recover the irreducibility of
the cluster expansion. Obviously the evaluation of the nodal diagrams is more
involved than in the case of infinite systems.  However, the basic ideas used
to calculate the nodal diagrams in the infinite system, are still valid in the
present case, and the expressions of the $N_{dd}(\br_1,\br_2)$,
$N_{de}^{\phantom{\ton} \ttw}(\br_1,\br_2)$, $N_{ee}^{\ton \ttw}(\br_1,\br_2)$
diagrams, given in Appendix \ref{sec:app-fhncsoc}, are a rather
straightforward extension of those presented for the infinite systems.  We
have shown the dependence on isospin associated to the exchange.  The only
relevant differences are related to the cyclic-cyclic
$N_{cc}^{\ton}(\br_1,\br_2)$ diagrams and they are worthy of a short
discussion.

\begin{figure}[htb]
\begin{center}
\includegraphics[scale=1.0]{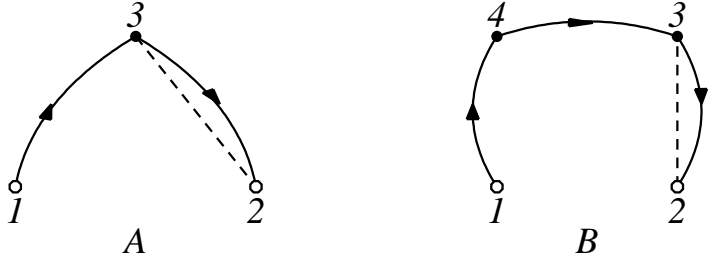}
\end{center}
\caption{ \small 
  Two $N_{cc}(\br_1,\br_2)$ diagrams. In infinite systems they give opposite 
  contributions since the integration over the point $4$ in the
  diagram B recovers the diagram A, except for a minus sign.  
  In finite systems, because of the
  presence of vertex corrections in the $3$ and $4$ points, the two
  diagrams give different contributions.
  }
\label{fig:fin-ncc}
\end{figure}

In analogy to the infinite system case, the nodal $N_{cc}^{\ton}(\br_1,\br_2)$
diagrams are generated by the folding products of $X_{cc}^{\ton}(\br_1,\br_2)$
or of $\rho_0^{\ton}(\br_1,\br_2)$. In the finite system case, the presence of
the vertex corrections generates the possibility of having nodal diagrams
where there are two consecutive statistical correlations
$\rho_0^{\ton}(\br_i,\br_j)$.  We show in Fig. \ref{fig:fin-ncc} an example of
this situation.  In the diagram A the point 3 is reached by a statistical
correlation on the left hand side, and by a dynamical correlation on the right
hand side.  In the diagram B, the point 4 is reached on both sides only by
statistical correlations.  In infinite systems, i. e. in the absence of vertex
corrections, because of the property (\ref{eq:inf-contrho}) of the
uncorrelated OBDM, the two diagrams give the same contribution, except for a
minus sign.  For this reason, in order to avoid overcounting, we did not
consider diagrams of the B type in the evaluation of the $N_{cc}^{\ton}$
contribution to the TBDF.  In finite systems each point is vertex corrected,
therefore Eq.  (\ref{eq:inf-contrho}) cannot be applied to describe the
integration over the point 4 in the B diagram, and, consequently, the
contribution of the diagram B is different from that of the diagram A.

%%%%%%%%%%%%%%%%%%%%%%%%%%%%%%%%%%
% Figure of Ncc diagrams
%%%%%%%%%%%%%%%%%%%%%%%%%%%%%%%%%%
%%%%%%%%%%%%%%%%%%%%%%%%%%%%%%%%%%

We take care of this fact by separating the contribution of
$N_{cc}^{\ton}$ in four different terms:
\begin{eqnarray}
\nonumber 
N_{cc}^{\ton}(\br_1,\br_2) & = &N_{cc}^{xx \ton}(\br_1,\br_2)+
N_{cc}^{x\rho \ton}(\br_1,\br_2) \\
&+& N_{cc}^{\rho x \ton}(\br_1,\br_2)+N_{cc}^{\rho\rho \ton}(\br_1,\br_2)
\,\,.
\label{eq:fin-ncc}
\end{eqnarray}
The superscripts $x$ and $\rho$ refer to the type of correlation
reaching the external points.  With the label $\rho$ we indicate that
the point is reached by a statistical correlation only, like the point
1 in Fig. \ref{fig:fin-ncc}, while with the label $x$ we indicate that
also dynamical correlations are present, as in the point 2 of the
figure.

The full set of FHNC equations containing the vertex corrections,
called Renormalized FHNC (RFHNC) equations, is given in Appendix
\ref{sec:app-fhncsoc}.

\subsection{Treatment of the operator dependent correlations}
\label{sec:fin-stdc}

In the discussion done so far for finite nuclei, we have considered
scalar correlations only.  As in the infinite system case, the use of
the operator dependent correlations is treated in the SOC
approximation.  However, the treatment of these correlations presented
in Sect.  \ref{sec:inf-stdc} cannot be straightforwardly extended to
finite nuclear systems. First, the systems we want to describe are not
saturated in isospin, and this changes the treatment of the isospin
dependent terms.  Second, the $jj$ coupling of the single-particle
wave functions modifies the calculation of the spin traces.  In any
case, we tackle the problem by using the strategy outlined in Sect.
\ref{sec:inf-stdc}.  We first reduce the product of operators to a sum
of single operators, and then we calculate the appropriated traces.
We find it convenient to treat separately the spin and isospin
operators. For this reason, we use Eq. (\ref{eq:inf-separ}), to
express the operator dependent correlation (\ref{eq:inf-stcorr}) as:
\beq
F_{ij}=\sum_{p=1}^{6}f_{p}(r_{ij})O^{p}_{ij}=
\sum_{l=0}^{1}(\btau_{i}\cdot\btau_{j})^{l}
\sum_{k=1}^{3}f_{2k-1+l}(r_{ij})P_{ij}^{k}
\,\,.
\label{eq:fin-stcorr}
\eeq

As we have already pointed out in the case of infinite systems, we
remember that, in addition to the product of operators coming from the
dynamical correlations and from the interaction, we have also to deal
with the spin and isospin exchange operator Eq. (\ref{eq:inf-siexch}),
coming from the statistical correlation.  Each exchange loop formed by
$n$ statistical correlations $\rho_0(x_1,x_2)$ carries $n-1$
spin-isospin exchange operators. By using the symbols defined in the
above equations, we express the operators generated by a single
exchange loop as:
\begin{equation}
\frac 1 4\sum_{p=1}^{6}\Gamma^p O^{p}_{12}=\frac 1 4\sum_{l=0}^{1}
(\btau_{1}\cdot\btau_{2})^{l}
\sum_{k=1}^{3}\Delta^{k}P_{ij}^{k}
\,\,,
\end{equation}
where the values of $\Gamma^p$ are those given in Tab.
\ref{tab:inf-delta} of Appendix \ref{sec:app-matrix}, from which 
we see that $\Delta^{k}=1-\delta_{k,3}$.

\subsubsection{Spin traces}
\label{sec:fin-straces}
We consider a closed loop of statistical correlations
(\ref{eq:fin-obdm}), involving only two-particles:
\begin{eqnarray}
\nonumber &~&
\rho_0 (x_i,x_j) \rho_0 (x_j,x_i) =
\\\nonumber &~& \sum_{s_1,s'_1,t_1}\rho_{0}^{s_1s'_1t_1}({\bf r}_i,{\bf r}_j)
              \chi_{s_1}^+(i)\chi_{s'_1}(j)
              \chi_{t_1}^+(i)\chi_{t_1}(j) \times
\\\nonumber &~& 
\sum_{s_2,s'_2t_2}\rho_{0}^{s_2s'_2t_2}({\bf r}_j,{\bf r}_i)
              \chi_{s_2}^+(j)\chi_{s'_2}(i)
              \chi_{t_2}^+(j)\chi_{t_2}(i)
\\\nonumber &~&
=
\sum_{s_1,s'_1,t_1} \sum_{s_2,s'_2 t_2}
\rho_{0}^{s_1s'_1t_1}({\bf r}_i,{\bf r}_j) 
\rho_{0}^{s_2s'_2t_2}({\bf r}_j,{\bf r}_i) \times
\\\nonumber &~&
 \chi_{s_1}^+(j) \chi_{s_2}^+(i) \chi_{t_1}^+(j) \chi_{t_2}^+(i)
\Pi^{\sigma,\tau}(i,j) 
\, \chi_{s'_1}(j)\chi_{s'_2}(i)\chi_{t_1}(j) \chi_{t_2}(i)
\,\,,
\end{eqnarray}
where we have used the operator (\ref{eq:inf-siexch}), 
to exchange the spins and isospin of the bra. Fixing the isospins,
which will be treated appart, the kernel of the above equation is:
\begin{eqnarray}
&~&
\nonumber
\sum_{s,s'} \rho_0^{s,s',t}(\br_i,\br_j) 
\chi_{s}^+ (j) \chi_{s'}(j) 
\\ \nonumber 
&~& =  \rho_{0}^{t}(\br_i,\br_j) 
\sum_s \chi^+_{s} (j) \chi_{s}(j) 
\\
&~&  
 +  \rho_{0j}^{t}(\br_i,\br_j) 
\sum_s (-1)^{s-1/2} \chi_{s}^+ (j) \chi_{-s}(j)
\,\,,
\label{eq:fin-tracerho}
\end{eqnarray}
where we used the uncorrelated OBDMs for parallel and antiparallel
spins defined in Eqs. (\ref{eq:fin-ropar},\ref{eq:fin-roantipar}).
The treatment of the parallel spin term is similar to that of the
infinite system case. The antiparallel term, which appears only in the
$jj$ coupling scheme, should be treated differently.

Following the scheme presented in Sect. \ref{sec:inf-stdc} we evaluate
the spin traces by considering three cases: the product of operators
acting on the same nucleonic pair, the product of operator forming a
ring (SOR) and the product of more than one operator acting on an
internal point. Consistently with the definitions (\ref{eq:inf-separ})
of the operators, in the following expressions the upper indexes
$i,j,k$ can assume the values 1,2 and 3, only.

%%%%%%%%%%%%%%
{\em \ref{sec:fin-straces}.a Parallel spin traces}\\
In analogy to the infinite system we find that the trace of the 
product of two operators $P^i$, acting on the same pair of nucleons is:
\beq
C(P_{12}^{k_1}P_{12}^{k_2})=
B^{2k_1-1}\delta_{k_1 k_2}=A^{k_1}\delta_{k_1 k_2}
\,\,,
\label{eq:fin-prodp2}
\eeq
where the values of $B^p$ are given in Tab. \ref{tab:inf-ap}. 
From this table we see that $A^k=k(k+1)/2$. 

We find that the product of two operators can be reduced to the
following sum of single operators:
\beq
P_{12}^{i}P_{12}^{j}=\sum_{k=1}^3I^{ijk} P_{12}^{k}
\,\,,
\label{eq:fin-prodp3}
\eeq
where the matrix $I^{ijk}$ is constructed by selecting only the values
of the odd indexes of the matrix $K^{pqr}$ given in Tab.
\ref{tab:inf-kpqr} of Appendix \ref{sec:app-matrix}:
\begin{equation}
I^{ij1} = \left( 
\begin{array}{rrr} 1 & \phantom{-}0 &\phantom{-}0 \\ 0 & 3 & 0 \\ 0 & 0 & 6
\end{array} \right) \,\,,
\  I^{ij2} = \left( 
\begin{array}{rrr} 0 & 1 & 0 \\ 1 & -2 & \phantom{-}0 \\ 0 & 0 & 2
\end{array} \right) \,\,,
\  I^{ij3} = \left( 
\begin{array}{rrr} 0 & \phantom{-}0 & 1 \\ 0 & 0 & 1 \\ 1 & 1 & -2
\end{array} \right) \,\,.
\label{eq:fin-iijk}
\end{equation}

In analogy to Sec. \ref{sec:inf-traces} we can use recursively Eqs.
(\ref{eq:fin-prodp2}) and (\ref{eq:fin-prodp3}).  For example, we
find for the trace of the product of three operators the expression:
\beq
C\Big(P^{i}_{12}P^{j}_{12}P^{k}_{12}\Big)
= \sum_{k_1=1}^3I^{ijk_1}C
\Big(P^{k_1}_{12}P^{k}_{12} \Big) 
= I^{ijk}A^{k}
\,\,.
\eeq

The evaluation of the product of operators forming a closed loop 
(SOR) follows the steps outlined for the infinite system case. 
In analogy to Eq. (\ref{eq:inf-xidef}), we find that also in the
present case this product of operators can be written as a sum of
single operators as: 
\beq
\sum_{\sigma_{2}}\int d \phi_{2}P^{i}_{12}P^{j}_{23}=
\sum_{k=1}^3
\int d\phi_{2}\,\zeta^{i j k}_{123}\,P^k_{13}
\,\,,
\label{eq:fin-prodp4}
\eeq
where the values $\zeta^{ijk}_{123}$ factors are given in Eqs.
(\ref{eq:inf-xiprop}).  By using the above equation we find that the
trace of multipole operator diagrams, such as those of Fig.
\ref{fig:inf-mod}, can be calculated as:
\begin{equation}
\int d\phi_{1}C(P_{mn}^{i}P_{m1}^{j}P_{mn}^{k}P_{n1}^{k'})=
      \sum_{k_1=1}^3 J^{ikk_1}  \int d\phi_{1}\zeta^{jk'k_1}_{m1n}
\,\,.
\end{equation}
The matrix $J^{ijl}$ is built by using the odd index values of the
matrix $L$ given in Tab. \ref{tab:inf-lpqr} of Appendix
\ref{sec:app-matrix}: 
\begin{equation}
J^{ij1} = \left( 
\begin{array}{rrr} 1 & \phantom{-}0 & \phantom{-}0 \\ 0 & 3 & 0 \\ 0 & 0 & 6
\end{array} \right) \,\,,
\ J^{ij2} = \left( 
\begin{array}{rrr} 0 & \phantom{-}3 & 0 \\ 3 & 6 & 0 \\ 0 & 0 & -6
\end{array} \right) \,\,,
\ J^{ij3} = \left( 
\begin{array}{rrr} 0 & 0 & 6 \\ 0 & 0 & -6 \\ 6 & -6 & 12
\end{array} \right) \,\,.
\label{eq:fin-jijk}
\end{equation}
%

%%%%%%%%%%%%%%%%%%
{\em \ref{sec:fin-straces}.b Antiparallel statistical function.}\\
Since the Eqs. (\ref{eq:fin-prodp3}) and (\ref{eq:fin-prodp4}) involve
only  
operators, they do not depend upon the spin structure of the wave
function, therefore they are valid also in the antiparallel spin case.
The change with respect to the parallel case, is in the basic trace
value (\ref{eq:fin-prodp2}) which is no longer valid.  For the
antiparallel statistical function of Eq. (\ref{eq:fin-tracerho}) we
write the trace of a single operator as:
\beq
C_j(P^k_{12}) = \frac 1 4 \sum_{s_1,s_2}
(-1)^{s_1+s_2} 
\chi_{s_1}^+ (1) \chi_{s_2}^+ (2) P^k_{12} \chi_{-s_1}(1)\chi_{-s_2}(2)
\,\,,
\eeq
and we find the values
\begin{equation}
C_j(P^k_{12}) = \delta_{k,2} \,\,.
\label{eq:fin-cj}
\end{equation}
We see that the contribution of the scalar operator is zero and that
of the spin operator is one, just the opposite results of those of the
parallel case. The result (\ref{eq:fin-cj}) for the tensor operator
has been obtained under the hypothesis of a spherical symmetry of the
system.

\subsubsection{Isospin expectation values}
\label{sec:fin-tiso}
In a system not saturated in isospin, we should not sum on the isospin
third components, since the various diagrams, nodal, elementary,
vertex corrections, etc., depend on these quantum numbers.  This means
that the values of these diagrams are different when they are
calculated for protons, neutrons, or mixed, clusters. We do not
calculate isospin traces, but the expectation values of products of
isospin operators.  In order to obtain these expectation values we
use the properties of the Pauli matrices which allows us to express
the product of $n$ isospin operator pairs acting on the same pair of
nucleons as:
\begin{equation}
\left(\btau_i \cdot \btau_j \right)^n = a_n + 
(1-a_n) \btau_i \cdot \btau_j 
\,\,,
\label{eq:fin-isorec1}
\end{equation}
with:
\beq
a_{n+1}=3(1-a_n) \,\,,
\hspace {1.0 cm} {\rm and} \hspace {1.0 cm} a_0=1  
\,\,.
\label{eq:fin-isorec2}
\eeq

\begin{figure}[ht]
\begin{center}
\includegraphics[scale=0.5,angle=90.]{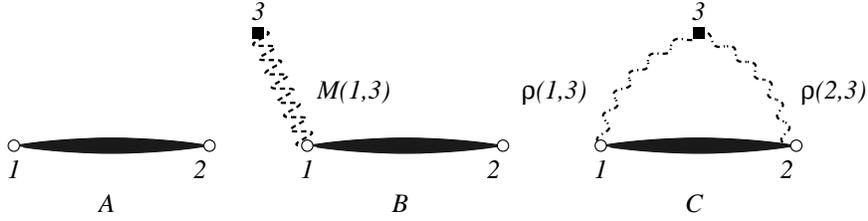}
\end{center}
\caption{ \small 
  Example of diagrams contributing to the energy
  expectation value. The black areas indicates the product of three
  operators $O_{12}^pO_{12}^qO_{12}^r$. The
  two operators $O^p$, $O^r$ come form the dynamical correlations,
  while  $O^q$ comes from the interaction. The dashed wiggly line of
  the diagram B indicate an operator dependent vertex correction
  $M$. This implies the presence of a single operator coming from the
  correlation, because of the SOC approximation, eventually associated
  with another operator coming from the spin-isospin exchange term.
  The wiggly dashed and dotted lines of diagram C indicate the
  correlated TBDF containing operators of the correlations and those
  related to the spin-isospin exchange terms. 
  }
\label{fig:fin-tiso}
\end{figure}

The recursive relation (\ref{eq:fin-isorec1}) expresses the product of
isospin operator pairs as a sum of a scalar term plus a term depending
from a single isospin operator pair.  The expectation value of the
operator sequence (\ref{eq:fin-isorec1}) is:
\beq
\chi_n^{\ton \ttw} = \chi_{\ton}^+ (1)\chi_{\ttw}^+ (2)
\left(\btau_1 \cdot \btau_2 \right)^n \chi_{\ton} (1)
\chi_{\ttw} (2) 
\,\,.
\label{eq:fin-itrace1}
\eeq
By using Eqs. (\ref{eq:fin-isorec1}) and (\ref{eq:fin-isorec2}) 
we obtain:
\beq
\chi_0^{\ton \ttw}= 1 \,\,,
\hspace {1.0 cm} {\rm and} \hspace {1.0 cm}
\chi_1^{\ton \ttw}= 2\delta_{\ton \ttw}-1
\,\,,
\eeq
and by applying the recursive relations, we have the more general
result:
\beq
\chi_n^{\ton \ttw} = 2a_n-1 +2(1-a_n) \delta_{\ton \ttw}
\,\,.
\label{eq:fin-chi}
\eeq
This result is the contribution to the cluster expansion of terms like
that represented by diagram A of Fig. \ref{fig:fin-tiso}. In this
figure we show the various types of diagrams which appear in the
calculation of the energy expectation value, and more precisely in the
calculation of the expectation value of the two-body interaction:
\beq
\frac {<\Psi|V_2|\Psi>} {<\Psi|\Psi>} 
= 
\frac
{ <\Phi|{\cal F}^\dagger V_2 {\cal F}|\Phi>}
{<\Psi|\Psi>} 
\label{eq:fin-expv2}
\eeq
In the figure we indicate with the black area joining the external
points 1 and 2, the product of the three operators $O_{12}^p O_{12}^q
O_{12}^r$, where $O^p$ and $O^r$ come from the dynamical correlations
${\cal F}$, while $O^q$, always in the middle, comes from the two-body
interaction $V_2$.  All the isospin expectation values necessary for
the calculation of the energy expectation value are taken into account
by Eq.  (\ref{eq:fin-chi}).

%%%%%%%%%%%%%%%%%%%%%%%%%%%%%%%%%%
% Figure isospin diagrams
%%%%%%%%%%%%%%%%%%%%%%%%%%%%%%%%%
\vskip 0.5 cm
\vskip 0.5 cm 
%%%%%%%%%%%%%%%%%%%%%%%%%%%%%%%%%%%%%%%%%%

The case of the A diagram of Fig. \ref{fig:fin-tiso} is an example of
our procedure. We find general recursive relationships connecting the
expectation values of isospin operators, and use them to obtain
expectation values. The case of the A diagram is the easiest one, but
we apply an analogous procedure to calculate the cases of the vertex
correction, represented by the B diagram, and the case of the nodal
diagram, represented by the C diagram of the figure.  The calculation
of the isospin expectation values for these two, more involved cases,
is presented in detail in Appendix \ref{sec:app-iso}.

The rules that 
we have presented in this 
section to calculate the spin traces
and the isospin matrix elements have been used to evaluate the OBDF and the
TBDF in the SOC approximation.  The expressions of the RFHNC equations which
allows us to calculate the cluster expansion of the TBDF in the SOC
approximation are given in Appendix \ref{sec:app-fhncsoc} by 
equations (\ref{eq:asoc-soc1}-\ref{eq:asoc-socend}).  The cluster
expansion of the TBDF allows us to calculate the energy expectation
value. The details of this calculation are given in next section.

\section{The evaluation of the ground state energy}
\label{sec:ene}

The discussion of the previous sections was devoted to the calculation
of the scalar TBDF defined by Eq. (\ref{eq:fin-tbdm}) when $q=1$.  The
TBDF contains all the many-body effects independent from the two-body
interaction, whose expectation value can be obtained as indicated by
Eq.  (\ref{eq:inf-tbmv}) for the infinite system.  For the calculation
of the TBDF we developed the cluster expansion techniques and we built
a set of hypernetted chain equations which allows us to evaluate in a
closed form the contribution of all those diagrams we have called
nodal and composite.  In this section, we apply these techniques to
the calculation of the energy expectation value (\ref{eq:inf-energy})
with a hamiltonian of the form:
\begin{equation}
H=-\sum_{i=1}^{A}\frac{\hbar^2}{2m}\nabla^{2}_{i}+\sum_{j>i=1}^{A}V_{ij}
+\sum_{k>j>i=1}^{A}V_{ijk}
\,\,.
\label{eq:ene-hami}
\end{equation}
The two-body interaction is expressed as:
\begin{equation}
V_{ij}=\sum_{p=1}^{8}v^{p}(r_{ij})O_{ij}^{p}
\,\,,
\label{eq:ene-v2}
\end{equation}
where $v^{p}(r_{ij})$ are scalar functions of the distance $r_{ij}$ of
the two interacting nucleons, and the $p=1,\ldots,6$ operators are
those defined in (\ref{eq:inf-operators}).  We consider in addition
the spin-orbit operators
\beq
O_{ij}^{p=7,8} = ({\bf L} \cdot {\bf S})_{ij}, 
({\bf L} \cdot {\bf S})_{ij} (\btau_i\cdot\btau_j)
\,\,,
\label{eq:ene-op78}
\eeq
where ${\bf L}$ is the relative angular momentum of the two
interacting nucleons, and ${\bf S}$ is the sum of their spins.  We
give in Sect. \ref{sec:ene-v3} the explicit expression of the
three-body interaction $V_{ijk}$.

The calculation of the energy expectation values is done by using the
trial wave function:
\begin{equation}
\Psi(1,\ldots,A)={\cal F}(1,\ldots,A)\Phi(1,\ldots,A)
      ={\cal S}\Big(\prod_{i<j}F_{ij} \Big)\Phi(1,\ldots,A)
\,\,,
\label{eq:ene-psi}
\end{equation}
where the two-body correlation function consider only the first six
operators as indicated by Eqs. (\ref{eq:inf-stcorr}) and
(\ref{eq:fin-stcorr}).  The uncorrelated state, $\Phi$, is a Slater
determinant composed by all the single particle wave functions
(\ref{eq:fin-spwf}) lying below Fermi surface.

In the following, we treat together the kinetic energy and the
two-body interaction up to the tensor channels. All the other parts of
the hamiltonian, two-body spin orbit interactions, three-body and
Coulomb interactions, are treated separately.  In the following
expressions the $t$ indexes run on protons and neutrons, the $p,q,r,s$
labels may assume values from 1 up to 6, and are used to generically
identify the different operator channels as in Eq.
(\ref{eq:inf-operators}). When we separate the spin and isospin
dependence as in Eq.  (\ref{eq:inf-separ}) we shall use the indexes
$l$=0,1, and $k=1,2,3$.

\subsection{Kinetic energy and $V^{6}_{12}$ part}
\label{sec:ene-kinv6}

We evaluate the expectation value of the kinetic energy, by using the
Jackson-Feenberg separation scheme \cite{jac61,jac62} as suggested in
Ref. \cite{fan79}. We obtain:
\begin{equation}
\langle T \rangle \equiv T_\phi + T_F -T_{c.m.}
\,\,,
\label{eq:ene-TJF}
\end{equation}
where we have defined 
\begin{equation}
T_{\phi} \equiv -\frac{\hbar^2}{4m}\Bigg(<\Phi^{*} 
{\cal F}^2\sum_{i=1}^A
\nabla_i^2\Phi >-<\sum_{i=1}^A(\nabla_i\Phi^*)\cdot ({\cal F}^2
\nabla_i\Phi)>
\Bigg)
\,\,,
\label{eq:ene-Tphi}
\end{equation}
\begin{equation}
T_{F} \equiv -\frac{\hbar^2}{4m}<\Phi^{*} 
\Big[ {\cal F} \Big(\sum_{i=1}^A
\nabla_{i}^{2} {\cal F}\Big)-\sum_{i=1}^A(\nabla_i {\cal F})^2\Big]\Phi>
\,\,,
\label{eq:ene-TF}
\end{equation}
and for the contribution of the center of mass term we have:
\beq
T_{c.m.} = -\frac{\hbar^2}{2mA}<\Psi^{*}
\Big(\sum_{i=1}^A\nabla_{i}\Big)^2\Psi>
\,\,.
\label{eq:ene-cm}
\eeq

In the above equations we have used the symbol $<>$, which has been
defined as \cite{co92}:
\beq
<X> = \frac 
 {\int d x_1 \ldots d x_A  X(x_1, \ldots, x_A)  }
 {<\Psi|\Psi>}
\,\,.
\label{eq:ene-simbol}
\eeq

Before attacking the problem of
calculating $\langle T \rangle$ we define some useful quantities:
\begin{equation}
\rho_{T1}^{\ton}(\br_1)=
\sum_{nljm}\Big[\phi^{\ton*}_{nljm}(\br_1)
\nabla_{1}^{2}\phi_{nljm}^{\ton}(\br_1)
-\nabla_{1}\phi^{\ton*}_{nljm}(\br_1)\cdot\nabla_{1}
\phi_{nljm}^{\ton}(\br_1) \Big]
\,\,,
\label{eq:ene-rhot1}
\end{equation}
\begin{equation}
\rho_{T2}^{\ton\ttw}(\br_1,\br_2)=
\rho_{0}^{\ton}(\br_1,\br_2)\nabla_{1}^{2}
\rho_{0}^{\ttw}(\br_1,\br_2)-\nabla_{1}\rho_{0}^{\ton}
(\br_1,\br_2)\cdot
\nabla_{1}\rho_{0}^{\ttw}(\br_1,\br_2)
\,\,,
\label{eq:ene-rhot2}
\end{equation}
\begin{equation}
\rho_{T3}^{\ton}(\br_1,\br_2)=2\nabla_{1}^{2}\rho_{0}^{\ton}(\br_1,\br_2)
\,\,,
\label{eq:ene-rhot3}
\end{equation}
\begin{eqnarray}
\nonumber
\rho_{T4}^{\ton}(\br_1,\br_2)
&=&\nabla_{1}\rho^{\ton}_{0}(\br_1,\br_2)
\cdot\nabla_{2}\rho_{0}^{\ton}(\br_1,\br_2) \\
\nonumber
&-&\rho_{0}^{\ton}(\br_1,\br_2)
\nabla_{1}\cdot\nabla_{2}
\rho_{0}^{\ton}(\br_1,\br_2) \\
\nonumber 
&+&\nabla_{1}\rho^{\ton}_{0j}(\br_1,\br_2)
\cdot\nabla_{2}\rho_{0j}^{\ton}(\br_1,\br_2) \\
&-&\rho_{0j}^{\ton}(\br_1,\br_2)
\nabla_{1}\cdot\nabla_{2}
\rho_{0j}^{\ton}(\br_1,\br_2) 
\,\,.
\label{eq:ene-rhot4}
\end{eqnarray}

The above expressions involve neither the interaction nor the
correlations, therefore they depend only on the uncorrelated
many-body state, $\Phi$. The expressions of these quantities in terms
of single particle wave functions (\ref{eq:fin-spwf}) are given in
Appendix \ref{sec:app-obdt}. In terms of these quantities the center
of mass contribution can be expressed as:
\beq
T_{cm}=-\frac{\hbar^2}{4mA}\sum_{\ton}  \Big(\int d\br_1
\rho_{T1}^{\ton}(\br_1)-
\int d\br_1d\br_2\rho_{T4}^{\ton}(\br_1,\br_2)\Big)
\,\,.
\eeq

The operator structure of $T_F$ can be easily associated to that
required by the calculation of the interaction expectation value.  For
this reason, we calculate together $T_F$ and 
$V_2 \equiv {\displaystyle\sum_{p=1,6} <v^p O^p>} $.  
The contribution of $T_F+V_2 \equiv W$ is also called
{\em interaction energy} \cite{fab98}.

The structure of the SOC approximation generate various cases that we
separate in four parts, i.e. :
\begin{equation}
W=W_0+W_s+W_c+W_{cs}
\,\,.
\label{eq:ene-W}
\end{equation}

\begin{figure}[htb]
\begin{center}
\includegraphics[scale=0.5] {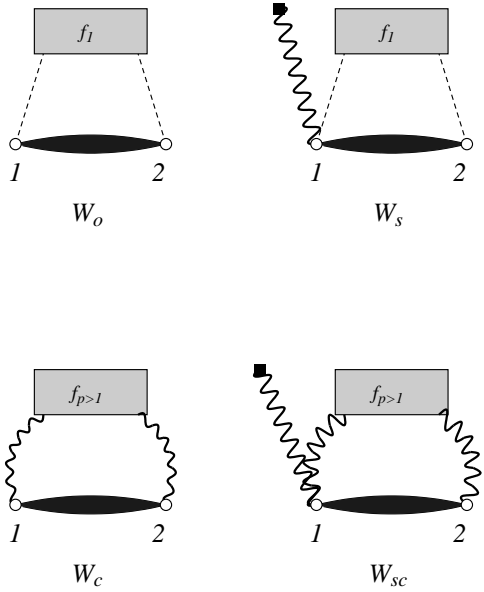}
\end{center}
\caption{\small Graphical representation of the W terms 
of Eq. (\protect\ref{eq:ene-W}). }
\label{fig:ene-W}
\end{figure}

We sketch in Fig. \ref{fig:ene-W} the characteristics of the four
terms. The black bands between the interacting points 1 and 2,
indicate $O^p O^q O^r$, operators $O^p$ and $O^r$ coming from the
correlation and $O^q$ from the interaction. 

%%%%%%%%%%%%%%%%%%%%%%%%%%%%%%%
% 
% Figure of the W terms
%
%%%%%%%%%%%%%%%%%%%%%%%%%%%%%%
%\vskip 0.5 cm 
%\vskip 0.5 cm 
%%%%%%%%%%%%%%%%%%%%%%%%%%%%%%

In the term $W_0$ we calculate the sum of all the diagrams connected
to the interaction points by scalar correlation functions only. The
interaction point are vertex corrected. If the product $O^p O^q O^r$
produces a scalar term, the vertex correction contains all the
operators types.  If, instead, the product $O^p O^q O^r$ generates
operator terms, the vertex correction contains scalar terms only. The
explicit expression of $W_0$ is given in Appendix \ref{sec:app-ene}.

With $W_s$ we consider the sum of operator rings touching a single
interaction point. These diagrams may include also the presence of
scalar operator chains such as those considered in $W_0$.  The $W_c$
term is the sum of all the diagrams forming a SOC between the two
interacting points.  We present in Appendix \ref{sec:app-ene} the
derivation of the explicit expressions of the $W_s$ and $W_c$
diagrams.

The contribution of the $W_{cs}$ term is given by the sum of all the
diagrams with operator rings reaching a single interaction point, and,
in addition, the SOC between the two interacting points. The 
$W_{cs}$ term is obtained by the combination of the topologies of the 
$W_{s}$ and $W_{c}$ terms. Because of the large number of operators
present in $W_{cs}$, we do not calculate explicitly its contribution,
but we rather estimate it by using the 
prescription proposed in \cite{fab98}:
\beq
W_{cs}\sim \frac{W_{s}W_{c}}{W_{0}}
\,\,.
\label{eq:ene-Wcs}
\eeq
Nuclear matter calculations \cite{pan79} where, more refined
computational schemes are used, indicate that the largest contribution
of $W_{cs}$ is two orders of magnitude smaller than those of the other
$W$ terms. We have compared nuclear matter estimations of the values
of $W_{cs}$ obtained with our prescription with those calculated more
accurately in \cite{pan79}, and we have found agreement up to the
second significant figure.

\begin{figure}[htb]
\begin{center}
\includegraphics[scale=0.5]{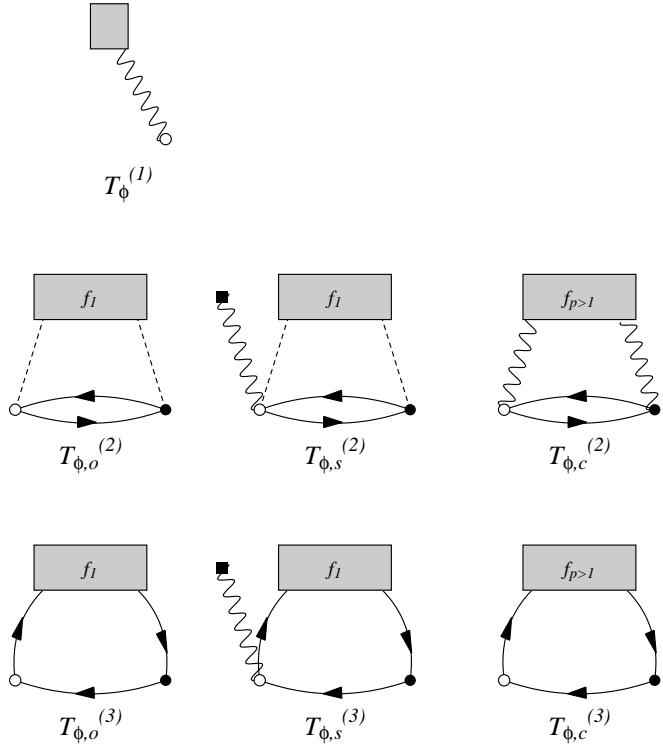}
\end{center}
\caption{\small  Graphical representation of the  $T_\phi$ terms 
of Eq. (\protect\ref{eq:ene-Tphi}).}
\label{fig:ene-Tphi}
\end{figure}

In the evaluation of the {\sl interaction energy} $W$, the $T_F$ part
of the kinetic energy is included. We describe now how we calculate
the contribution of the first term of the kinetic energy
(\ref{eq:ene-TJF}), the $T_{\phi}$ term, where the $\nabla_1$ operator
acts on the mean-field wave functions.  In Ref. \cite{co92} we found
it convenient to separate the contribution of $T_\phi$ in three parts:
\begin{equation}
T_\phi = T_\phi^{(1)} + T_\phi^{(2)} +  T_\phi^{(3)} 
\,\,.
\label{eq:ene-Tphi2}
\end{equation}
where each part is characterized by the type of statistical
correlations reaching the interacting point $1$.  In Fig.
\ref{fig:ene-Tphi} we show some diagrams which identify each term. We
have denoted the interacting point $1$, by an open circle. The
nomenclature {\sl interacting point} is due to the fact that this is
the point on which the differential operator is acting.  In the
$T_\phi^{(1)}$ type of diagrams, the interacting point is connected to
the other points by means of dynamical correlations only.  In
$T_\phi^{(2)}$, the statistical correlations reaching the point $1$,
form a closed loop involving only another single point. In the
diagrams of Fig. \ref{fig:ene-Tphi} we call $2$ this other point.
Finally, in the $T_\phi^{(3)}$ type of diagrams the statistical
correlations reaching the point $1$, form close loops which involve at
least two other internal points.

%%%%%%%%%%%%%%%%%%%%%%%%
% Figure T_\phi
%%%%%%%%%%%%%%%%%%%%%%%
%%%%%%%%%%%%%%%%%%%%%%%%

The expression of the $T_\phi^{(1)}$ is the expectation value of a
one-body operator dressed by the vertex correction, and is given in
Appendix \ref{sec:app-ene}.

The calculation of the other two terms is more involved, since we have
to calculate the expectation value of a two-body
operator. Nevertheless, the structure of these operators is the same
as the ones in the exchange terms of $W$. So we further classify the
$T_\phi^{(2)}$ and $T_\phi^{(3)}$ terms in the same way that we did
for the $W$ terms.

We consider $T_{\phi,0}^{(2,3)}$ terms where the points 1 and 2 are
connected to the other nucleons by a scalar operator chain.  As
indicated in Fig. \ref{fig:ene-Tphi} this means that these points are
reached by scalar correlations only, whose contribution can be
calculated in terms of RFHNC diagrams.  This is indicated by the $f_1$
label in the gray box. A second class of diagrams is that we labeled
as $T_{\phi,s}^{(2,3)}$. In this case, we consider that, in addition
to the scalar chains of $T_{\phi,0}^{(2,3)}$, the interaction point is
reached by a ring of $p>1$ operator dependent dynamical correlations.
Finally, with $T_{\phi,c}^{(2,3)}$ we indicate the sum of the diagrams
where the points 1 and 2 form a SOC. In the figure, this is indicated
by the wiggly lines and by the $f_{p>1}$ label in the gray box. As in
the case of $W$ terms, we may have also diagrams which are the
topological combination of $T_{\phi,s}^{(2,3)}$ and
$T_{\phi,c}^{(2,3)}$. We have estimated that their contribution is
negligible.

The separation of the $T_{\phi}$ contribution in three parts, was
proposed, and used, in \cite{co92}, where the finite nucleon systems
treated were saturated in both spin and isospin, and only scalar
correlations were considered.  The presence of operator dependent
correlations, requires a further classification of the various terms.
Clearly, $T_{\phi}$ depends on the isospin third components of the
particles $1$ and $2$.  Furthermore, since we work in a $jj$ coupling
scheme, we have to distinguish in the calculation of the
$T_\phi^{(2,3)}$ terms, the cases when the statistical correlations
have parallel and antiparallel spin components. The complete list of
expressions of the various terms composing $T_{\phi}$ is given in
Appendix \ref{sec:app-ene}.

\subsection{Spin-orbit and Coulomb terms}
\label{sec:ene-socou}
The contribution of the Coulomb interaction is:
\begin{equation}
<v_{Coul}>=
<\Psi^*\sum_{j>i=1}^A P^{1/2}_i P^{1/2}_j \frac{e^2}{r_{ij}}
\Psi>
\,\,,
\end{equation}
where  the 
$P^{1/2}_i$ projection operator selects the protons.
The Coulomb interaction is added to the scalar part of the $V_{ij}$
interaction when two protons interact. This means that its
contribution is consistently calculated following the methodology
described in Sect. \ref{sec:ene-kinv6} for all the proton-proton $W$
terms. 

We calculate the contribution of the spin-orbit terms of the
potential, i.e. the $p=7,8$ channels in Eq. (\ref{eq:ene-v2}), by
considering only diagrams containing scalar chains between the
interacting points.  In other words, for the spin-orbit interaction,
we calculate only the $W_0$ term of Eq. (\ref{eq:ene-W}). The explicit
expression of the spin-orbit contribution is given by Eq.
(\ref{eq:aene-ls}) in Appendix \ref{sec:app-ene}.

%%%%%%%%%%%%%%%%%%%%%%%%
% Figure Fujita Miyazawa
%%%%%%%%%%%%%%%%%%%%%%%%
\vskip 0.5 cm 
\begin{figure}[htb]
\begin{center}
\includegraphics[scale=0.8]{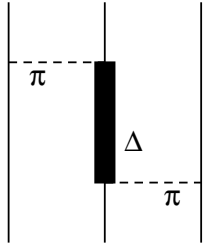}
\end{center}
\caption{\small Fujita-Miyazawa term of the three-body force. }
\label{fig:ene-fujita}
\end{figure}
%%%%%%%%%%%%%%%%%%%%%%%%

\subsection{The three-body potential}
\label{sec:ene-v3}

In our calculation we use three-nucleon potentials of Urbana type
\cite{car83}. The explicit expression of this potential is:
\beq
V_{ijk}=v^{2\pi}_{ijk}+v^{R}_{ijk}
\,\,.
\label{eq:ene-v3}
\eeq

The $v^{2\pi}_{ijk}$ term, describes a three--nucleon force produced
when one of the interacting nucleons is transformed into a $\Delta$ by
a first interaction with another nucleon, and it turns back to a
nucleonic state after interacting with a third nucleon (see fig.
\ref{fig:ene-fujita}). This term called Fujita-Miyazawa \cite{fuj57}
describes the long range part of the three--body interaction and
produces an attractive contribution. The second term of Eq.
(\ref{eq:ene-v3}), $v^R_{ijk}$, has a shorter range and a
phenomenological origin.  The explicit expressions of the two terms of
Eq. (\ref{eq:ene-v3}) are:
\begin{eqnarray}
\nonumber
v_{123}^{2\pi}&=&A_{2\pi}\sum_{cycl}
\Big(\{X_{31},X_{32}\}\{\btau_{3}
\cdot\btau_{1},\btau_{3}\cdot\btau_{2}\}
\\
&&+\frac{1}{4}[X_{31},X_{32}][\btau_{3}
\cdot\btau_{1},\btau_{3}\cdot\btau_{2}]\Big)
\label{eq:ene-v3v2pi}
\,\,,
\\
v^{R}_{123}&=&U_{0}\sum_{cycl}T^{2}_{\pi}(r_{31})T^{2}_{\pi}(r_{32})
\,\,.
\label{eq:ene-v3vr}
\end{eqnarray}
where the sums run on all the possible cyclic combinations of the 1,2
and 3 indexes. In the above equations we have used the terms
$T$ and $X$ defined by:
\begin{eqnarray}
T_{\pi}(r)&=&\frac{e^{-\mu r}}{\mu r}
\left[1+\frac{3}{\mu r}+\frac{3}{(\mu r)^2}\right]
(1-e^{-cr^{2}})^2
\,\,,
\label{eq:ene-tpi} 
\\
X_{ij}&=&Y_{\pi}(r_{ij})\bsigma_{i}\cdot\bsigma_{j}+T_{\pi}(r_{ij})S_{ij}
= \sum_{k=2}^3 X^k (r_{ij}) P^k_{ij}
\,\,,
\label{eq:ene-x}
\end{eqnarray}
where:
\beq
Y_{\pi}(r)=\frac{e^{-\mu r}}{\mu r}(1-e^{-cr^{2}})
\,\,,
\label{eq:ene-ypi} 
\eeq
where, as usual, $S_{ij}$ indicates the tensor operator
(\ref{eq:inf-tensor}), and the symbols $\{,\}$ and $[,]$ indicate the
anticommutator and commutator operators, respectively.  The values of
the constants of the $v_{123}^{2\pi}$ term are: $\mu=0.7 fm^{-1}$,
and $c=2 fm^{-2}$. The parameters $A_{2\pi}$, and $U_{0}$ of the
$v^{R}_{123}$ term are fixed to reproduce the $^3$H binding energy
\cite{car83}.

%%%%%%%%%%%%%%%%%%%%%%%%%%%
% Figure Three body diagrams
%%%%%%%%%%%%%%%%%%%%%%%%%%%
\begin{figure}[htb]
\begin{center}
\includegraphics[scale=0.5]{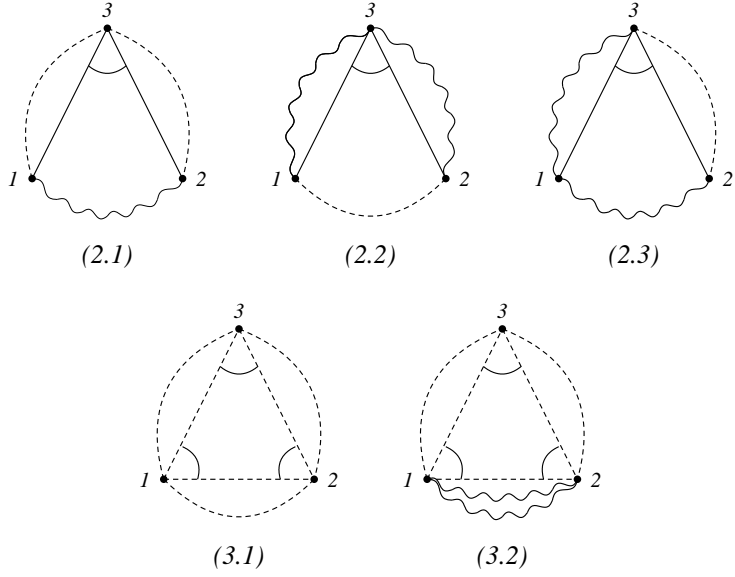}
\end{center}
\caption{\small
Cluster diagrams considered for the three-body force
expectation value. The 2.1, 2.2, 2.3 diagrams are related to the
$<v^{2\pi}_{ijk}>$ part of the force, and the 3.1 and 3.2 diagrams are
related to $<v^{R}_{ijk}>.$ The points denote the particle
coordinates. The dashed, wavy, and double-wavy lines represent
generalized scalar, operator and single-operator ring correlation
bonds, respectively.}
\label{fig:ene-3b}
\end{figure}
%%%%%%%%%%%%%%%%%%%%%%%%%%%
In analogy to the calculations in nuclear and neutron matter 
\cite{pan79,car83,wir88,akm97,akm98}, we evaluate the contribution of
the three-body interaction by considering only the sum of the five
diagrams presented in Fig. \ref{fig:ene-3b}. In Ref. \cite{car83} it
was shown that the diagrams (2.1), (2.2) and (2.3),
provide the relevant contribution to $<v_{123}^{2\pi}>$. The other two
diagrams are those important for $<v^{R}_{123}>$.

In the diagram (2.1) the pairs of nucleon connected by operators
$X_{ij}$ (pairs 31 and 32) are dressed by scalar
correlations, whereas the remaining pair (12) is also dressed by all the
other operator dependent correlations $p>1$, in the SOC approximation.
In the (2.2) diagram, the pairs 31 and 32 are linked by operator
dependent correlations in the SOC approximation, while the pair 12 is
dressed by the scalar correlation. In the (2.3) diagram
there is a cyclic permutation of the operator dependent correlations.  

The $<v^{R}_{123}>$ term is calculated by evaluating the (3.1) and
(3.2) diagrams. We calculate the case where all the pairs are dressed
by scalar correlations at all orders, diagram (3.1) and the case when
there is in addition a SOC correction for a single pair of nucleons,
diagram (3.2).

The detailed derivation of the expressions of the three-body potential
is given in Appendix \ref{sec:app-ene}.

\section{Specific applications}
\label{sec:results}

We have applied the formalism presented in the previous sections to
the description of the \car, \oxy, \caI, \caII and \lead nuclei.  The
only inputs required by our calculations are the two- and three-body
nuclear interactions. In Sect. \ref{sec:res-int}, we present those
chosen for our studies. The single particle wave functions, and the
correlation functions, fixed by the minimization procedure
(\ref{eq:in-varp}), are presented in Sects. \ref{sec:res-spwf} and
\ref{sec:res-corr}, respectively.  The theoretical and computational
reliability of our calculations has been tested by verifying the
exhaustion of some sum rules. This discussion is developed in Sect.
\ref{sec:res-sumrules}. After that, we discuss in Sect.
\ref{sec:res-ene} the results regarding the most important observable
of our calculations: the binding energy.  We compare results obtained
by using two different interactions.  We continue our discussion by
presenting a set of quantities which we have chosen to investigate the
effects, and the relevance, of the short-range correlations. These
quantities are: matter and charge density distributions, Sect.
\ref{sec:res-obd}, momentum distributions, Sect. \ref{sec:res-obdm},
natural orbits, Sect. \ref{sec:res-no}, two-body density
distributions, Sect. \ref{sec:res-tbdm}, and, finally, quasi-hole wave
functions and spectroscopic factors, Sect. \ref{sec:res-specf}.

\subsection{The nuclear interaction}
\label{sec:res-int}

The definition of 
the hamiltonian (\ref{eq:ene-hami}) requires the definition of 
both two- and three-body forces. We use two-body forces constructed
to reproduce the data of the phase-shifts analysis of the large body
of nucleon-nucleon scattering \cite{sto93,arn92}.

We have used nucleon-nucleon interactions of the Argonne-Urbana
family.  These interactions are local and non-relativistic, and are
expressed as a sum of operator dependent terms as indicated in Eq.
(\ref{eq:ene-v2}).  The most recent interaction of this type, fitting
the phase-shifts of Refs.  \cite{sto93,arn92}, is the Argonne $v_{18}$
interaction (AV18) \cite{wir95}, containing 18 operator terms, some of
them breaking the charge symmetry.

In our calculations we have considered interactions containing up to
eight operators channels, see Eqs.  (\ref{eq:inf-operators}) and
(\ref{eq:ene-op78}). For this reason, we used a truncated version of
the AV18 potential, called Argonne $v'_{8}$ (AV8'), and introduced in
Ref.  \cite{pud97} because its simpler parameterization allowed a
simplification of the numerically involved quantum Monte Carlo
calculations. This interaction is not a simple truncation of the full
AV18 interaction, but its parameters have been slightly modified to
simulate the effects of the missing channels. The AV8' interaction
reproduces the results of the full interaction for the $S$ and $P$
scattering waves and also the $^3D_1$ wave. The details of the
construction of the AV8' interaction are given in Ref. \cite{pud97}.

%%%%%%%%%%%%%%%%%%%%%%%%%%%%%%%
% 
% Figure of the interaction
%
%%%%%%%%%%%%%%%%%%%%%%%%%%%%%%
%\vskip 0.5 cm 
\begin{figure}[]
\begin{center}
\includegraphics[scale=0.5] {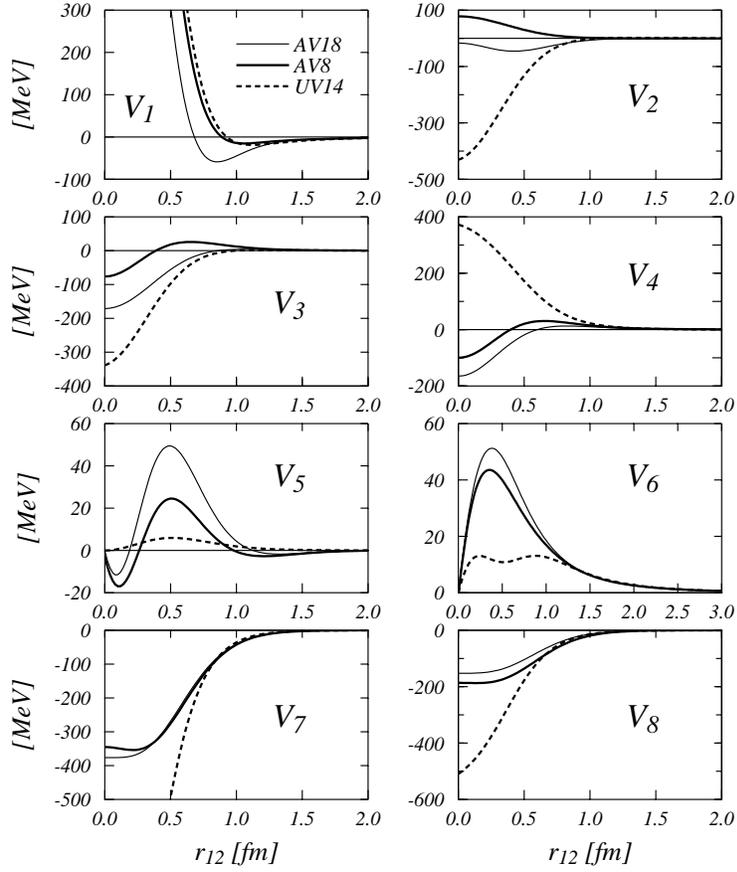}
\end{center}
\caption{\small The AV8' and UV14 
  nucleon-nucleon interactions as a function of the relative distance,
  in the channels used in our calculations. The continuous thin line,
  show the behavior of the Argonne $v_{18}$ potential.
  }
\label{fig:res-force}
\end{figure}
\vskip 0.5 cm 
%%%%%%%%%%%%%%%%%%%%%%%%%%%%%%

The major part of our calculations have been done with the AV8'
interaction.  However, in order to test the sensitivity of our results
to the nuclear interaction, we have also used the Urbana $v_{14}$
interaction, fixed in Ref.  \cite{lag81a}, to reproduce the set of
phase shifts data available at the beginning of the '80s.  In reality
we have used only the first eight channels of the interaction, without
any readjustment of the parameters values. For this reason, we shall
refer to this interaction as UV14, understanding that we used the
operator channels only up to the spin-orbit ones.

The two-nucleon interactions are implemented with the three-nucleon
interactions fixed to reproduce the $^3$H binding energy. This means
that associated to each two-nucleon potential there is a three-body
force. Even though more elaborated versions of the three-nucleon
forces have been recently proposed \cite{pie01a}, in our calculations,
we use the original expression \cite{car83}, as presented
in Sect. \ref{sec:ene-v3}.

We use the Urbana UIX three-nucleon interaction \cite{pud97} together
with the AV8' potential, and the Urbana UVII interaction \cite{sch86}
together with the UV14 potential. The values of the parameters of Eqs.
(\ref{eq:ene-v3v2pi},$\ldots$,\ref{eq:ene-x}), fixing these two forces are
given in Tab. \ref{tab:res-3body}.

%%%%%%%%%%%%%%%%%%%%%%%%%%%%%%%%%%%%%%%%%%%%
% Table three-nucleon force
%%%%%%%%%%%%%%%%%%%%%%%%%%%%%%%%%%%%%%%%%%%%
\vskip 0.5 cm 
\begin{table}[ht]
\begin{center}
\begin{tabular}{|c|c|c|}
\hline
           & UVII        &  UIX     \\ \hline
$A_{2\pi}$ &  $-0.03330$   & $-0.02930$   \\ \hline
$U_{0}$    &   0.003700  & 0.004800   \\ \hline
\end{tabular}
\end{center}
\caption{\small 
Parameters used in  Eqs. 
(\protect\ref{eq:ene-v3v2pi},$\ldots$, 
\protect\ref{eq:ene-x}) 
to fix the three-body UVII and UIX interactions.
In both cases $\mu=0.7 fm^{-1}$ and $c=2 fm^{-2}$.
}
\label{tab:res-3body}
\end{table}
%%%%%%%%%%%%%%%%%%%%%%%%%%%%%%%%%%%%%%%%%%%%

\subsection{The single particle wave functions}
\label{sec:res-spwf}

%%%%%%%%%%%%%%%%%%%%%%%%%%%%%%%%%%%%%%%%%%%%
% Table Woods-Saxon
%%%%%%%%%%%%%%%%%%%%%%%%%%%%%%%%%%%%%%%%%%%%
\begin{table}[htb]
\begin{center}
\begin{tabular}{|c|c|c|c|c|c|}
\hline
             &\car&  \oxy   & \caI & \caII & \lead  \\ 
\hline
$V_0^p$      &-62.00   & -52.50      &-57.5      & -59.50    & -60.40 \\ 
\hline 
$V_{ls}^p$   &-3.20    & -7.00       & -11.11    & -8.55     & -6.75  \\ 
\hline
$a_{0}^p$    & 0.57    &  0.53       & 0.53      & 0.53      &  0.79  \\ 
\hline
$a_{ls}^p$   & 0.57    &  0.53       & 0.53      & 0.53      &  0.79 \\
 \hline
$R_{0}^p$    & 2.86    &  3.20       & 4.10      & 4.36      &  7.46 \\ 
\hline
$R_{ls}^p$   & 2.86    &  3.20       & 4.10      & 4.36      &  7.46 
        \\ \hline
$R_{Coul}$   & 2.86    &  3.20       & 4.10      & 4.36      &  7.46 
          \\ \hline
$ V_0^n$     &-62.00   & -52.50      & -55.00    &-50.00     & -44.32
         \\ \hline
$V_{ls}^n$   & -3.15   & -6.54       & -8.50     & -7.74     & -6.08
         \\ \hline
$a_{0}^n$    & 0.57    &  0.53       & 0.53      & 0.53      &  0.66 
        \\ \hline
$a_{ls}^n$   &  0.57   &  0.53       & 0.53      & 0.53      &  0.66
         \\ \hline
$R_{0}^n$    & 2.86    &  3.20       & 4.10      & 4.36      &  7.46 
        \\ \hline
$R_{ls}^n$   & 2.86    &  3.20       & 4.10      & 4.36      &  7.46 
        \\ \hline
\end{tabular}
\caption{\small The values of the parameters of the Woods-Saxon
  potential well, Eqs. (\ref{eq:fin-wspot}) and
  (\ref{eq:fin-coulpot}).  The superscripts $p$ and $n$ indicate
  protons and neutrons respectively. The values of $V_{0}$ and
  $V_{ls}$ are expressed in MeV, those of all the other parameters in
  fm.  }
\label{tab:res-ws}
\end{center} 
\end{table}
\vskip 0.5 cm 
%%%%%%%%%%%%%%%%%%%%%%%%%%%%%%%%%%%%%%%%%%%%

In our calculations, the search of the minimum is done by making
variations on the correlation function and on the mean-field potential
generating the set of single particle states. We have already said in
Sect. \ref{sec:fin-jj} that we used a mean-field basis generated by
two Woods-Saxon wells, one for protons and another one for neutrons,
both containing spin-orbit terms. The analytical expressions of the
Woods-Saxon wells, Eqs. (\ref{eq:fin-wspot}) and
(\ref{eq:fin-coulpot}), involve thirteen parameters. Variational
calculations done by changing all these parameters would be extremely
heavy, from the computational point of view. In reality, we found
\cite{fab00}, that the energy minimum is more sensitive to the
correlation function, than to the single particle basis. To be more
precise, we found that when the correlation functions provides an
energy minimum, changes of the potential do not produce large
differences of this value. In Ref. \cite{fab00} it has been shown
that, in \oxy, variations of 47\% of the central well changed by only 
1.2 \% the energy value. In calculations of the \caII nucleus we found
a change of 9\% in the energy value by doubling the depths of the
neutrons and protons wells.

These findings induced us to make calculations by using, for each
nucleus, a fixed set of Woods-Saxon parameters. The values of these
parameters, given in Tab. \ref{tab:res-ws}, are taken from the
literature \cite{ari96}. They have been fixed to reproduce the charge
root mean square radii and the single particle energies around the
Fermi surface. One could consider the requirement of reproducing these
data, a further variational constraint. In any case, we have further
verified that our energy minima are only slightly modified by large
changes of the potential.

\subsection{The correlation functions}
\label{sec:res-corr}
The correlation function is fixed by the minimization procedure
(\ref{eq:in-varp}), and the result is independent of the starting
expression of the correlation function. In practice, however, in order
to minimize the computational effort, it is convenient to choose
expressions of the correlation functions containing a limited number
of parameters, and behaving at large interparticle distances as
intuitively expected.  This means that asymptotically $f_1$ should
reach the value 1, while the other correlation functions, should be
zero. A commonly used expression for the scalar term of the
correlation is the gaussian form:
\[
f_p(r) = \delta_{p,1} + a_p\,e^{-b_pr^2}
\]
where $a_p$ and $b_p$ are the free parameters to be changed in the
variation procedure. For example in Refs. \cite{co92,co94,ari96}
correlations of this type have been used. 

%%%%%%%%%%%%%%%%%%%%%%%%%%%%%%%
% 
% Figure of the correlation
%
%%%%%%%%%%%%%%%%%%%%%%%%%%%%%%
%\vskip 0.5 cm 
\begin{figure}[]
\begin{center}
\includegraphics[scale=0.6] {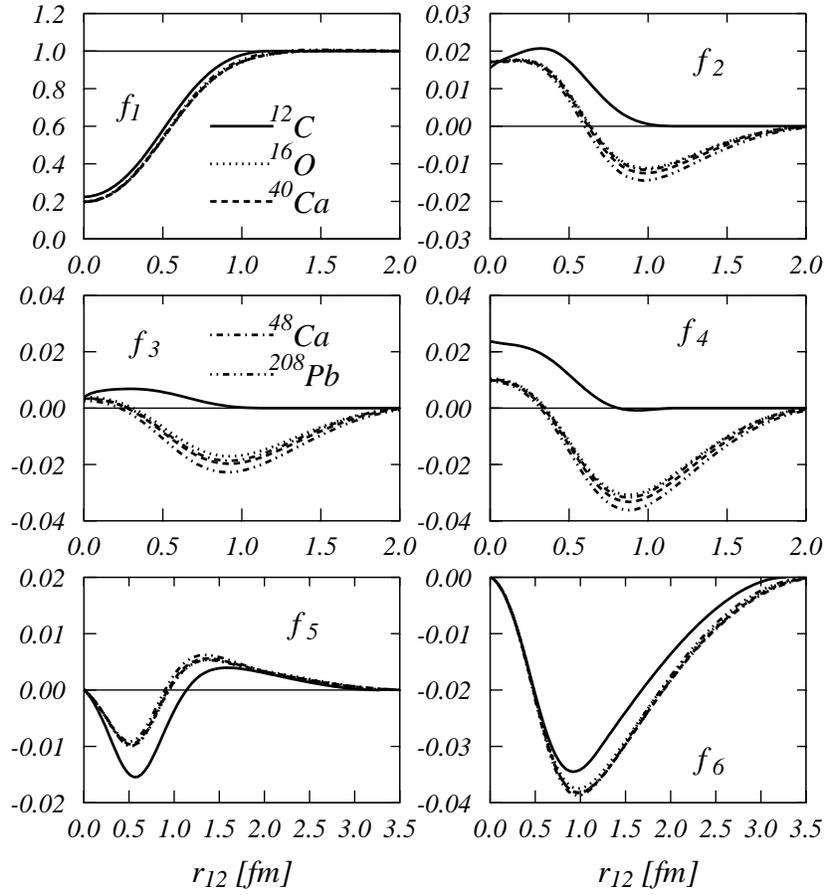}
\end{center}
\caption{\small 
  Two-body correlation functions $f_p$, obtained with the Euler
  procedure, as a function of the two-nucleon distance. In this
  calculations the AV8'+UIX interaction has been used.
  }
\label{fig:res-corr}
\end{figure}
\vskip 0.5 cm 
%%%%%%%%%%%%%%%%%%%%%%%%%%%%%%

In our calculations, we found more convenient, from the physical point
of view, and also in terms of number of variational parameters, to fix
the correlation functions by using what we called the {\em Euler
  procedure}.  The basic idea is to use as variational parameters the
distances where the correlation functions $f_p(r)$ reach their
asymptotic values.  For fixed values $d_p$ of these distances, that we
call {\sl healing distances}, the functions $f_p(r)$ are obtained by
doing a minimization of the energy calculated up to the second order
cluster expansion. This means that we solve the FHNC/SOC equations
when all the nodal diagrams are zero. We give a detailed description
of the Euler procedure in Appendix \ref{sec:app-eul}.

The application of the Euler procedure involves a single variational
parameter, the healing distance $d_p$, for each operator channel of
the correlation, therefore in our calculations we have to deal with
six variational parameters. On the other hand, it is know from nuclear
and neutron matter calculations \cite{pan79,wir88,akm98} that the
healing distances of the four central channels $p\leq 4$ are rather
similar, as are those of the two tensor channels $p=5,6$.  For this
reason, we performed our calculations by using only two variational
parameters, a healing distance $d_c$ for the four central channels
$p\leq 4$, and another one $d_t$, for the two tensor channels.

The values of the healing distances providing the energy minima for
the five nuclei considered, are given in Tab. \ref{tab:res-heal}.  In
this table, we compare the values obtained by using both interactions.
In Fig.  \ref{fig:res-corr} we show the two-body correlation
functions obtained for the AV8'+UIX interaction as a function of
the two-nucleon distance.

%%%%%%%%%%%%%%%%%%%%%%%%%%%%%%%%%%%%%%%%%%%%
% Table healing distances
%%%%%%%%%%%%%%%%%%%%%%%%%%%%%%%%%%%%%%%%%%%%
\vskip 0.5 cm 
\begin{table}[ht]
\begin{center}
\begin{tabular}{|l|r|rrrrr|}
\hline
  &       & \car & \oxy & \caI & \caII & \lead \\
\hline
AV8' +   & $d_c$   & 1.20 & 2.10 & 2.15 & 2.10 & 2.20 \\
UIX      & $d_t$   & 3.30 & 3.70 & 3.66 & 3.70 & 3.60 \\
\hline
UV14 + & $d_c$   & 1.40 & 2.10 & 2.15 & 2.10 & 2.20 \\
UVII     & $d_t$   & 3.30 & 3.80 & 3.86 & 3.90 & 3.80 \\
\hline
\end{tabular}
\end{center} 
\caption{
\small Values, in fm, of the {\sl healing distances}, 
which minimize the energy functional with the two
different interactions we have adopted.
}
\label{tab:res-heal} 
\end{table}
\vskip 0.5 cm 
%%%%%%%%%%%%%%%%%%%%%%%%%%%%%%%%%%%%%%%%%%%%

Various remarks are in order after observing these results. The most
evident one is the scarce dependence on the type of nucleus, with the
only exception of \car. We shall see that, also in the calculations
of other quantities, this nucleus always produces anomalies in the
general trend of our results. We think that this is due to the fact
that \car is not really a good doubly closed shell nucleus,
therefore, open shell effects, not included in our calculations, are
relevant. The following observations are done by excluding the \car
results.

The results of Tab. \ref{tab:res-heal} show the scarce dependence of
the healing distances on the interaction model.  The values of $d_c$
are identical for both the interaction models we have used. There are
small differences in the $d_t$ values, those of the UV14+AVII, are
slightly larger than those obtained with AV8'+UIX.

Also the dependence on the nucleus is rather weak. The variations of
the healing distances are very small, and, the curves of Fig.
\ref{fig:res-corr} relative to each nucleus are rather similar for
each considered channel.

The values of the scalar correlation functions $f_1$, are one order of
magnitude larger than those of the other correlations. The behavior
of the $f_1$ functions reflects the presence of the repulsive core in
the scalar channel of the nucleon-nucleon force.  The correlation
function hinders two nucleons from approaching each other too much.

The healing distances of the tensor correlations are larger than those
of the central correlations. Also this effect reflects a
characteristic of the interaction where the tensor channels have
slightly larger interaction range than the central ones (see Fig.
\ref{fig:res-force}).

\subsection{The sum rules}
\label{sec:res-sumrules}
The numerical solution of the FHNC/SOC equations is not trivial at
all. We have to deal with a set of interrelated, hypernetted, integral
equations. The numerical technique used to solve this set of equation
is based on an iterative procedure. In analogy to what we have
discussed in Sect. \ref{sec:inf-boson} for the HNC equations
(\ref{eq:inf-hnc1},\ref{eq:inf-hnc2},\ref{eq:inf-hnc3}), we started
the calculation of the FHNC/SOC equations by setting the nodal
diagrams to zero. The various integrals equations are calculated and
they provide new values of the nodal diagrams that are used again to
solve the FHNC/SOC equations. The convergence test is done on the energy,
and we stop the iterative procedure when the energy calculated in two
different iteration loops differ by less than 1 keV.  Every calculation
is done by using a fixed correlation function. The numerical
convergence of the solution does not ensure that this solution is
acceptable from the physical point of view.  For example we found
numerically convergent solutions which provided a wrong number of
nucleons.

In addition to these computational problems, we should remember that
our calculations do not solve exactly the many-body Schr\"odinger
equation.  The solution of the FHNC/SOC set of equations does not
include the elementary diagrams.  Furthermore, we do not consider the
contribution of those operator dependent terms which are beyond the
SOC approximation.

An important tool used to verify the numerical, and theoretical,
accuracy of the calculations is, the test of the sum rules exhaustion.
For this purpose, we have evaluated the following sum rules:
\begin{eqnarray}
S_{1}^{\ton}&\equiv&\frac{1}{{\cal N}_{\ton}}\int d\br_{1}
\rho^{\ton}(\br_1)=1
\,\,,
\label{eq:res-s1}
\\
\nonumber 
S_{2}^{\ton \ttw}& \equiv &\frac{1}{{\cal N}_{\ton}({\cal N}_{\ttw}
-\delta_{\ton \ttw})}\int d\br_{1}d\br_{2}
\rho_{2}^{1,\ton \ttw}(\br_1,\br_2)\\
\nonumber 
&=&\frac{1}{{\cal N}_{\ton}({\cal N}_{\ttw}-\delta_{\ton \ttw})}
\int d\br_1d\br_2\frac{f_{2k_1-1+l_1}(r_{12})f_{2k_2-1+l_2}(r_{12})}
{f_{1}^2(r_{12})}\\ \nonumber 
&&\Big\{\rho^{\ton \ttw}_{2,dir}(\br_1,\br_2)A^{k_1}\delta_{k_1k_2}
\chi^{\ton \ttw}_{l_1+l_2}\\
\nonumber
&&+\left[ \rho_{2,exc}^{\ton \ttw}(\br_1,\br_2)A^{k_3}I^{k_1k_2k_3}+
\rho_{2,excj}^{\ton \ttw}(\br_1,\br_2)I^{k_1k_2k_4}I^{k_4k_32} \right] \\
& & \Delta^{k_3}\sum_{l_3=0}^1\chi^{\ton \ttw}_{l_1+l_2+l_3} \Big\}=1
\,\,,
\label{eq:res-s2}
\\
\nonumber
S_{2,\sigma}
& \equiv &\frac{1}{3A}\sum_{\ton\ttw=p,n}
\int d\br_{1}d\br_{2} \rho_{2}^{3,\ton \ttw}(\br_1,\br_2)
\\ 
\nonumber
&=&\frac{1}{3A}\sum_{\ton\ttw=p,n}
\int d\br_1d\br_2\frac{f_{2k_1-1+l_1}(r_{12})f_{2k_2-1+l_2}(r_{12})}
{f_{1}^2(r_{12})}\\ \nonumber 
&&\Big\{\rho^{\ton\ttw}_{2,dir}(\br_1,\br_2)
I^{k_12k_2}\chi^{\ton\ttw}_{l_1+l_2}\\
\nonumber
&&+\left[ \rho_{2,exc}^{\ton\ttw}(\br_1,\br_2)
A^{k_3}I^{k_12k_4}I^{k_4k_2k_3}
+\rho_{2,excj}^{\ton\ttw}(\br_1,\br_2)
I^{k_12k_4}I^{k_4k_2k_5}I^{k_5k_32} \right] \\
& & \Delta^{k_3} \sum_{l_3=0}^1 \chi^{\ton\ttw}_{l_1+l_2+l_3}\Big\}=-1
\,\,.
\label{eq:res-s2sigma}
\end{eqnarray}
In the above equations, a sum on the repeated indexes is understood.
The operator dependent TBDFs $\rho_2^q$ are defined in Eq.
(\ref{eq:fin-tbdm}).  The sum rule $S_{2,\sigma}$ is related to the
two-body density spin function.  This sum rule is valid only for spin
saturated systems and for correlations not containing tensor operator
terms. For this reason, we expect $S_{2,\sigma}$ to be exhausted in
\oxy and \caI nuclei only, and in the absence of the $O^{p=5,6}$ operator
terms in the correlation.

%%%%%%%%%%%%%%%%%%%%%%%%%%%%%%%%%%%%%%%%%%%%%%
% Table sum rule
%%%%%%%%%%%%%%%%%%%%%%%%%%%%%%%%%%%%%%%%%%%%%%
\begin{table}[htb]
\begin{center}
\begin{tabular}{|c|c|c|c|c|c|}
\hline
                 & \car & \oxy & \caI & \caII & \lead \\ \hline
$S_{1}^p(f_1)$   & 1.000    & 1.000   &1.000  & 1.000   & 0.999  \\ \hline
$S_{1}^n(f_1)$   & 1.000    & 1.000   &1.000  & 0.999   & 0.999  \\ \hline
$S_{1}^p(f_4)$   & 0.998    & 0.996   &0.993  & 1.005   & 1.008 \\ \hline
$S_{1}^n(f_4)$   & 0.998    & 0.996   &0.993  & 1.004   & 1.004 \\ \hline
$S_{1}^p(f_6)$   & 0.997    & 1.006   &1.008  & 0.994   & 1.002   \\ \hline
$S_{1}^n(f_6)$   & 0.997    & 1.006   &1.008  & 0.996   & 1.000   \\ \hline
$S_{2}(f_1)$     & 1.004    & 1.003   &1.001  & 1.000   & 0.998 \\\hline
$S_{2}(f_4)$      & 0.995    & 0.999   &0.989  & 1.012   & 1.014 \\\hline
$S_{2}(f_6)$     & 0.996    & 0.998   &0.978  & 0.994   & 1.003  \\ \hline
$S_{2\sigma}(f_1)$ &      & -0.95   &-0.929 &      &  \\\hline
$S_{2\sigma}(f_4)$&       & -1.080  &-1.101 &      &   \\\hline
\end{tabular}
\end{center}
\caption{\small Sum rules exhaustion calculated for the AV8'+UIX
                 interaction. The indexes $f_p$ indicate the number of
                 operator terms of the correlation. 
}
\label{res:tab-sr}
\end{table}
%%%%%%%%%%%%%%%%%%%%%%%%%%%%%%%%%%%%%%%%%%%%%%%%%%%

To obtain the expressions
(\ref{eq:res-s1},$\ldots$,\ref{eq:res-s2sigma}) of the sum rules, we
have considered all the possible types of correlation operator between
the 1 and 2 coordinates, which have been vertex corrected by the
scalar correlations. This is the same approach used to calculate the
$W_{0}$ term of the interaction energy Eq. (\ref{eq:ene-W}).

In Tab. \ref{res:tab-sr}, we show the sum rule values calculated with
the AV8'+UIX interaction, by using different types of correlations:
purely scalar correlations, $f_1$, central correlation, $f_4$, and
correlations containing also tensor terms, $f_6$.  To simplify the
presentation of the results, we indicate with $S_2$ the quantity:
\begin{equation}
S_2= \frac 1 {A(A-1)}\sum_{\ton \ttw} {\cal N}_{\ton}({\cal N}_{\ttw}
-\delta_{\ton \ttw}) S_{2}^{\ton \ttw} \,\,,
\end{equation}
which must be equal to one. The $f_1$ results give an indication of
the error made by neglecting the elementary diagrams. The differences 
with the other sum rules, is a measure of the validity of the SOC
approximation. 

Apart from the $S_{2\sigma}$ values, which we shall discuss
separately, the various sum rules are satisfied at the level of few
parts per thousand. The $f_1$ sum rules, give the best results which
are only slightly spoiled by the other correlations.

%%%%%%%%%%%%%%%%%%%%%%%%%%%%%%%%%%%%
%
% Figure elementary fhnc diagram
%
%%%%%%%%%%%%%%%%%%%%%%%%%%%%%%%%%%%%
\vskip 0.5 cm
\begin{figure}[]
\begin{center}
\includegraphics[scale=0.5]{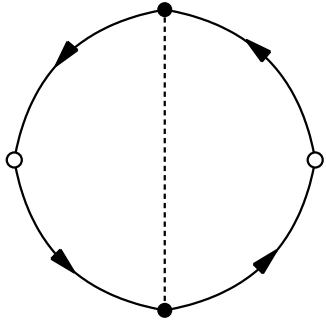}
\end{center}
\caption{\small  
Elementary diagram included in the RFHNC-1 calculations of
Ref. \protect\cite{co92}. 
}
\label{fig:fin-elementary}
\end{figure}
\vskip 0.5 cm
%%%%%%%%%%%%%%%%%%%%%%%%%%%%%%%%%%%%

Different remarks should be made for the spin sum rule $S_{2\sigma}$.
As already pointed out, in our calculations, these sum rules have to
be satisfied only for the \oxy and \caI nuclei, and for the $f_1$ and
$f_4$ cases. The values of Tab. \ref{res:tab-sr} indicate that the
$S_{2\sigma}$ sum rule is satisfied only at the 5-10\% level. In Ref.
\cite{co92}, we have verified that the inclusion of the elementary
diagram shown in Fig. \ref{fig:fin-elementary} improves the exhaustion
of the $S_{2\sigma}$ sum rule up to few parts per thousand. This is
the same level of accuracy obtained for the other sum rules.

\subsection{The ground state energies}
\label{sec:res-ene}

The most important results of our work are summarized in Tab.
\ref{tab:res-energy} where we give the values of the binding energies
per nucleon, for all the five nuclei considered. We show the results
obtained with the two interactions AV8'+UIX and U14+UVII, and we
compare them with the experimental energies \cite{aud93}.  We present
the various terms contributing to the total energy: the kinetic energy
$T$, the two-body interaction, where the contribution of the first six
channels $V^{6}_{2-body}$ and that of the spin-orbit interaction
$V_{LS}$ are separately given, the Coulomb interaction $V_{Coul}$ and
the three-body force $V_{3-body}$.  In the kinetic energy term the
spurious contribution of the center of mass motion,
Eq. (\ref{eq:ene-cm}), calculated as discussed in \cite{co92}, has
already been subtracted.

The various terms show some saturation properties.  For example, the
values of the kinetic energies per nucleon, $T$, increase up to \caI
and then they remain almost stable.  An analogous behavior is shown
by the $V^{6}_{2-body}$ terms whose contribution per nucleon increases
with increasing number of nucleons up to \caI, and afterword it
remains almost constant.

We have mentioned the fact that the spin-orbit terms are not treated
consistently in the FHNC/SOC computational scheme, but they are
evaluated by using some approximation. In any case, in all the nuclei
considered, their contributions are of the order of a few percent with
respect to the $V^{6}_{2-body}$ contributions. We have done
calculations in \oxy and \caI after switching off the spin-orbit terms
in the mean field potential. In this case the spin-orbit partner
single particle wave functions are identical. The differences in the
total spin-orbit contributions, with respect to the values given in
Tab. \ref{tab:res-energy} are within the numerical uncertainty.

%%%%%%%%%%%%%%%%%%%%%%%%%%%%%%%%%%%%%%%%%%%%%%
% energies per nucleon
%%%%%%%%%%%%%%%%%%%%%%%%%%%%%%%%%%%%%%%%%%%%%%
\vskip 0.5 cm 
\begin{table}[ht]
\begin{center}
\begin{tabular}{|c|r|r|r|r|r|}
\hline
AV8'+UIX         & \car & \oxy  & \caI & \caII & \lead \\ 
\hline
$ T$       & 27.13   &  32.33     & 41.06     & 39.64     &  39.56
\\ 
\hline
$V^{6}_{2-body}$ &-29.13   & -38.15     & -48.97    & -46.60    &
-48.43    \\ 
\hline
$V_{Coul}$       & 0.67    &  0.86      & 1.97      & 1.57      & 3.97
\\
 \hline
$V_{LS}$         & -0.25   & -0.38      & -0.39     & -0.35     &
 -0.45    \\
 \hline
$T+V_{2}$& -1.58   & -5.34      & -6.34     & -5.74     &
 -5.35    \\ 
\hline
$V_{3-body}$     & 0.67    &  0.86      & 1.76      & 1.61      &
1.91     \\
 \hline
$E$       & -0.91   & -4.48      & -4.58     & -4.14     &
 -3.43    \\ 
\hline
\hline
U14+UVII         & \car & \oxy  & \caI & \caII & \lead \\ 
\hline
$ T$             & 24.63   &  29.25     & 37.32     & 36.12     &
36.07    \\
 \hline
$V^{6}_{2-body}$ &-27.08   & -35.84     & -46.65    & -44.40    &
 -46.28    \\
 \hline
$V_{Coul}$       & 0.68    &  0.88      & 2.01      & 1.59      & 4.00      
 \\ \hline
$V_{LS}$         & 0.05   &   0.03      & 0.08     &  0.09     &  0.04
 \\ 
\hline
$T+V_{2}$& -1.72   & -5.68      & -7.24     & -6.71     &
-6.17    \\
 \hline
$V_{3-body}$     & 0.54    &  0.69      & 1.46      & 1.32      &
 1.61     \\ 
\hline
$E$       & -1.18   & -4.99      & -5.77     & -5.27     &
-4.55    \\ \hline \hline
$E_{exp}$       & -7.68   & -7.97      & -8.55     & -8.66     &
-7.86    \\ 
\hline
\end{tabular}
\caption{\small Energies per nucleon in MeV, 
obtained by using the $AV8'+UIX$ and $U14+UVII$ interactions.
We have indicated with $T$ the kinetic energy, with $V^6_{2-body}$
the contribution of the first six channels of the two-body
interaction, with $V_{LS}$ the spin-orbit contribution, with 
$V_{Coul}$ the contribution of the Coulomb interaction and with 
$V_{3-body}$ the total contribution of the three-body force. The rows
labeled $T+V_{2}$ show the energies obtained by considering the
two-body interactions only. The experimental energies are from Ref. 
\protect\cite{aud93}.
}
\label{tab:res-energy}
\end{center} 
\end{table}
\vskip 0.5 cm 
%%%%%%%%%%%%%%%%%%%%%%%%%%%%%%%%%%%%%%%%%%%%%%%%%%%%%%%%%%%%%%%%%

%%%%%%%%%%%%%%%%%%%%%%%%%%%%%%%%%%%%%%%%%%%%%%%%%%%%%%%%%%%%%%%
% Table 208Pb energy 
%%%%%%%%%%%%%%%%%%%%%%%%%%%%%%%%%%%%%%%%%%%%%%%%%%%%%%%%%%%%%%%
\begin{table}[htb]
\begin{center}
\begin{tabular}{|c|c|c|c|c|}
\hline
                   &$(0)$ &   $(s)$     &  $(c)$       &$(0)+(s)+(c)$ \\ \hline
$T^{(1)}_{\Phi}$   &16.05 &             &              & 16.05      \\ \hline
$T^{(2)}_{\Phi}$   &4.16  & -0.35       &  -0.13       & 3.68       \\ \hline
$T^{(3)}_{\Phi}$   &0.42  & -0.08       &              & 0.34       \\ \hline
$T_{F}$            &17.34 &  1.62       &  0.55        & 19.51      \\ \hline
$T^{(2)}_{\Phi j}$ &0.17  &             &              & 0.17       \\ \hline
$T^{(3)}_{\Phi j}$ &-0.012&             &              & -0.012      \\ \hline
$T_{Fj}$           & 0.0  &             &              & 0.0        \\ \hline
$T_{c.m.}$         & 0.02 &             &              & 0.02       \\ \hline
$v^{1}$            &-0.03 &  -0.10      &  -0.47       & -0.6       \\ \hline
$v^{2}$            &-1.29 &  0.04       &  -0.19       & -1.44      \\ \hline
$v^{3}$            &-3.87 &  0.18       &  -0.67       & -4.36      \\ \hline
$v^{4}$            &-17.57&  -0.41      &  -1.49       & -19.47     \\ \hline
$v^{5}$            &0.11  &  -0.02      &   0.04       & 0.13       \\ \hline
$v^{6}$            &-21.42&  -1.37      &   0.10       & -22.69      \\ \hline
$v^{1}_j$          &0.006 &             &              & 0.006       \\ \hline
$v^{2}_j$          &0.0   &             &              & 0.0      \\ \hline
$v^{3}_j$          &0.004 &             &              & 0.004       \\ \hline
$v^{4}_j$          &0.012 &             &              & 0.012       \\ \hline
$v^{5}_j$          &0.0   &             &              & 0.0      \\ \hline
$v^{6}_j$          &-0.005&             &              & -0.005      \\ \hline
$v^R_{123}$        &      &             &              &  3.282      \\ \hline
$v^{2\pi}_{123}$   &      &             &              & -1.368      \\ \hline
\end{tabular}
\caption{\small Contributions, in MeV, of the various terms forming
the \lead energy, calculated with the AV8'+UIX interaction.  The
various terms are defined in Sect. \protect\ref{sec:ene}.  The terms
T of the kinetic energy are defined by Eqs. (\ref{eq:ene-TJF},
\ref{eq:ene-Tphi}, \ref{eq:ene-TF}, \ref{eq:ene-Tphi2}).  
The $v^p$ terms indicate the
six-channels of the two-body interaction.  The three-body terms are
defined in Eqs. (\ref{eq:ene-v3v2pi},\ref{eq:ene-v3vr}).  The
subscript $j$ indicate the contribution produced by antiparallel spin
densities.  The labels $(0)$, $(s)$ and $(c)$ indicate the various
approximation of the energy related to the pieces $W_{0}$, $W_{s}$,
and $W_{c}$ as defined in Eq. (\ref{eq:ene-W}).  }
\label{tab:res-e208pb}
\end{center} 
\end{table}
%%%%%%%%%%%%%%%%%%%%%%%%%%%%%%%%%%%%%%%%%%%%%%%%%%%%%%%%%%%%%%%

As expected, the results of Tab. \ref{tab:res-energy} show that the
binding is obtained by a subtle subtraction between the repulsive
kinetic energy term and the attractive contribution of the two-body
potential.  The sum of only these contributions for the AV8'
interaction, provide -2.25, -6.20, -8.30, -7.31 and -9.32 MeV for the
\car, \oxy, \caI, \caII and \lead nuclei respectively. The sum in the
U14 model provide -2.40, -6.56, -9.25, -8.19 and -10.17 MeV.  It is
evident that the U14 interaction is more attractive than the
AV8'. This depends on the intrinsic structure of the interaction and
its parametrization.

The contribution of the Coulomb term $V_{Coul}$ is evaluated within
the complete FHNC/SOC computational scheme.  As expected, the
behavior with increasing size of the nucleus does not show saturation
because of the long range nature of the interaction. The Coulomb terms
behave as expected, their contributions increase with increasing
number of protons.  The apparent inversion of this trend from \caI to
\caII is due to the representation in terms of energy per nucleon,
which in this case is misleading, since the proton number is the same
for the two nuclei. In this case it is better to compare the total
values of the Coulomb energies, 78.80 MeV for \caI and 75.36 for
\caII. The 4.4\% difference between these two values is due to the
different structure of the two nuclei. The inclusion of the Coulomb
repulsion reduces the nuclear binding energies.

In addition, there is the contribution of the three-body force.  As
discussed in sect. \ref{sec:ene-v3} the two terms composing this
interactions provide contributions of different sign; the
Fujita-Miyazawa term $v^{2\pi}_{123}$ is attractive, while the other 
term $v^{R}_{123}$ is repulsive. In our calculations, the total
contribution of the UVII and UIX three-body interactions is always
globally repulsive. This feature is common to the FHNC/SOC nuclear
matter results \cite{wir88,akm98}.

The comparison with the experimental energies indicates a general
underbinding of about 4.0 MeV per nucleon. This is roughly the same
underbinding obtained, at the saturation density, by the most recent
FHNC/SOC nuclear matter calculations \cite{akm98}.

The behavior of the \car nucleus is anomalous in this general trend.
This nucleus is barely bound in our calculations. Some crucial physics
ingredient, relevant in \car, but negligible for the other nuclei, is
missing in our approach. Probably, this has to do with soft
deformations of the \car nucleus, effects which we are unable to treat.

The comparison between the two interactions indicates that the
UV14+UVII interaction is more attractive than the AV8'+UIX force.
This fact is already present when only the two-body interactions are
considered, and it is enhanced by the inclusion of the three-body
force.  The contributions of the spin-orbit term in the two cases have
different sign, they are attractive for AV8' and slightly repulsive
for UV14.  Globally, the differences in the total energies, calculated
with the two interactions, vary from a minimum of 5\% (\oxy) to a
maximum of 18\% (\lead).

We have done a detailed study of the relevance of the various terms
contributing to the energy, as they have been presented in Sect.
\ref{sec:ene}. As an example, we show in Tab. \ref{tab:res-e208pb} the
various contributions obtained in the calculation of the \lead energy
with the AV8'+UIX interaction.  We have obtained analogous results for
all the other nuclei investigated and also for the other interaction.

The larger contributions to the energy come from the terms
calculated with what we have called the (0) approximation in Eq.
(\ref{eq:ene-W}). This is the contribution of those diagrams
containing all the scalar dressings of the $(1,2)$ pair. The (s)
and (c) terms, more difficult to calculate, give much smaller 
contributions.

It is interesting to observe that, the contributions given by the
terms depending on one-body densities with antiparallel spins, i. e.
the terms labeled with the $j$ subscript in the table, are very small.
In Tab. \ref{tab:res-e208pb} we show their largest values, since we
found their contributions to be even smaller for nuclei with saturated $l$
shells such as \oxy and \caI.

The study of the contributions of the various channels of the two-body
interaction, indicates that the spin-isospin ($p=4$) and
tensor-isospin ($p=6$) terms are the main source for the binding. This
is a common feature for all the nuclei we have considered
\cite{bis06t}.

The results for the UV14+UVII interaction are analogous to those shown
in Tab. \ref{tab:res-e208pb} for the AV8'+UIX interaction.  In the
remaining part of Sect. \ref{sec:results}, we shall present some
quantities with the aim of studying the effects of the Short-Range
Correlations (SRC). We have found that the differences between the
results obtained with the two interaction models are smaller than the
effects we are looking for. For this reason, henceforth, we shall
present only the results obtained by using the AV8'+UIX
interaction.

\subsection{The one-body distribution functions}
\label{sec:res-obd}

The OBDF, $\rho^t(\br)$, has been defined by Eq. (\ref{eq:fin-obd}).
The physical meaning of this quantity is the probability density of
finding a nucleon of type $t$, in the position $\br$ with respect to
the nuclear center. Since we have assumed that our systems are
spherical, this probability depends only on the distance from the
center of the nucleus. The expression of the density distribution in
terms of the FHNC/SOC quantities is given by Eqs. (\ref{eq:fin-cd}) and
(\ref{eq:asoc-cd}).

%%%%%%%%%%%%%%%%%%%%%%%%%%%%%%%
% 
% Figure of the neutron density
%
%%%%%%%%%%%%%%%%%%%%%%%%%%%%%%
%\vskip 0.5 cm 
\begin{figure}[]
\begin{center}
\includegraphics[scale=0.5] {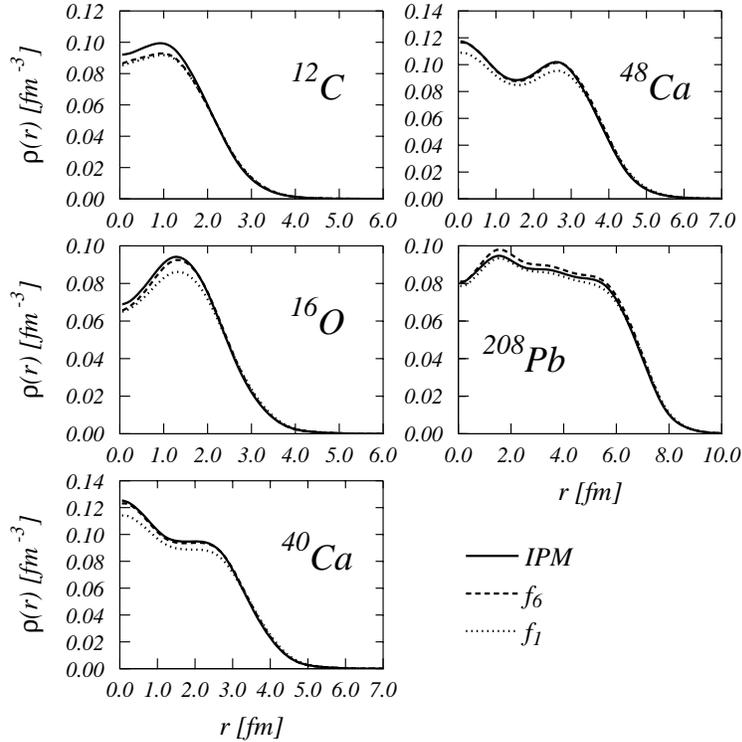}
\end{center}
\caption{\small Neutron density distributions. The full lines are the
  IPM distributions, the dotted ones have been obtained by using
  scalar correlations only, and the dashed lines show the results of
  the complete calculation. 
  }
\label{fig:res-ndens}
\end{figure}
%%%%%%%%%%%%%%%%%%%%%%%%%%%%%%
%%%%%%%%%%%%%%%%%%%%%%%%%%%%%%%
% 
% Figure of the charge density
%
%%%%%%%%%%%%%%%%%%%%%%%%%%%%%%
%\vskip 0.5 cm 
\begin{figure}[]
\begin{center}
\includegraphics[scale=0.5] {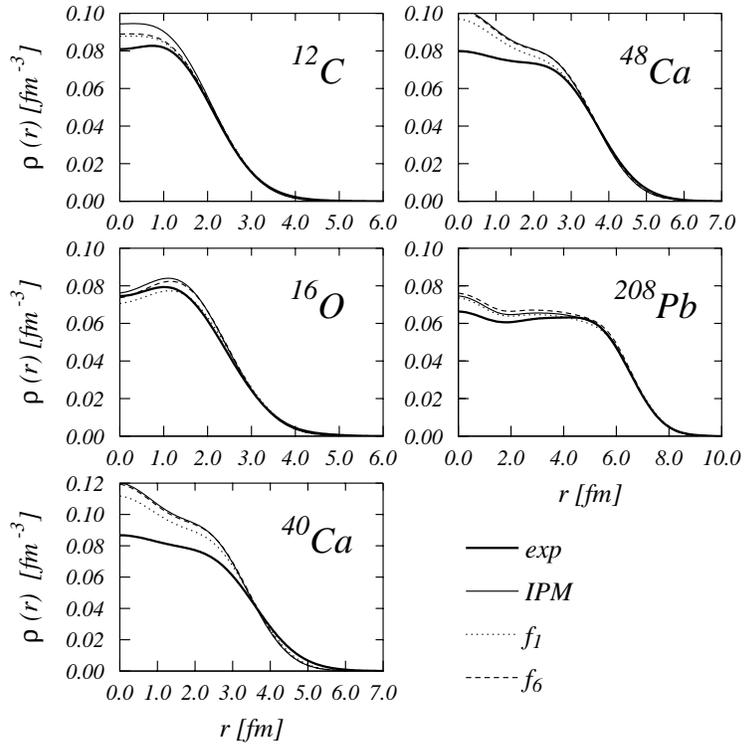}
\end{center}
\caption{\small 
  Charge density distributions, obtained by folding the
  proton distributions with the electromagnetic form factor. The 
  full thick lines show the empirical distributions
  \protect\cite{dej87}. The other curves have the same meaning as in
  Fig. \protect\ref{fig:res-ndens}.
  }
\label{fig:res-pdens}
\end{figure}
%%%%%%%%%%%%%%%%%%%%%%%%%%%%%%

We show in Fig. \ref{fig:res-ndens} the neutron density distributions
for the five nuclei we have considered. The full lines show the
IPM results, the dotted lines those obtained by using scalar, $f_1$, 
correlations only, and the dashed lines show the results of the full
FHNC/SOC calculations. In an analogous way we show in Fig. 
\ref{fig:res-pdens} the charge distributions, obtained by folding
the proton distributions with the proton electromagnetic form
factor. We have used a dipole form for this form factor. In this 
figure the continous thick lines indicate the empirical charge 
distributions extracted from elastic electron scattering experiments
\cite{dej87}.

The results obtained with scalar interactions produce distributions
which are smaller at the center of the nucleus with respect to the
mean-field distributions.  This effect is reduced when all the
correlations are included in the calculation.  These findings are in
agreement with the results of Ref. \cite{ari97} where a first-order
cluster expansion was used.

%%%%%%%%%%%%%%%%%%%%%%%%%%%%%%%%%%%%%%%%%%%%%%%%%
% 
% Figure of the elastic cross sections
%
%%%%%%%%%%%%%%%%%%%%%%%%%%%%%%%%%%%%%%%%%%%%%%%%%%
%\vskip 0.5 cm 
\begin{figure}[]
\begin{center}
\includegraphics[scale=0.5, angle=90] {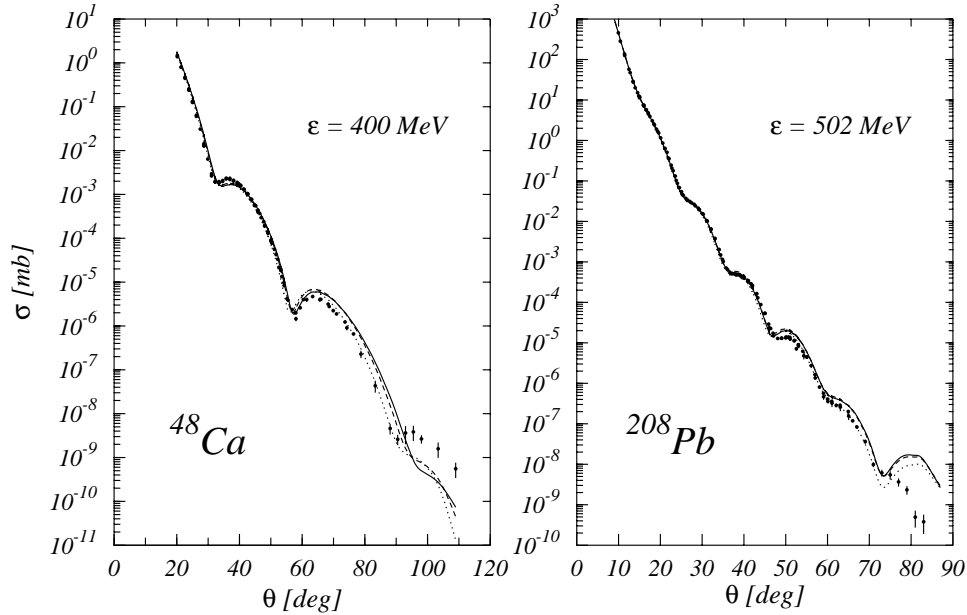}
\end{center}
\caption{\small Elastic electron scattering cross sections on \caII
  and \lead target nuclei, as a function of the scattering angle. The
  theoretical cross sections have been calculated in distorted wave
  Born approximation \protect\cite{ann95a,ann95b}. The full lines show
  the IPM results, the dotted lines the results obtained with the
  $f_1$ correlations, the dashed lines with the  $f_6$ correlations.
  We have indicated with $\epsilon$ the electron energy.
  }
\label{fig:res-xsec}
\end{figure}
\vskip 0.5 cm 
%%%%%%%%%%%%%%%%%%%%%%%%%%%%%%%%%%%%%%%%%%%%%%%%%%%%

We used the charge density distributions to calculate the elastic
electron scattering cross sections within a distorted wave Born
approximation \cite{ann95a,ann95b}. We compare in Fig.
\ref{fig:res-xsec} our results with the experimental data of the \caII
and \lead nuclei \cite{cav80t,sic}.  In Ref. \cite{fab00} a similar
figure for the \oxy and \caI nuclei is shown.  It is evident that the
main discrepancies with the data appear at large scattering angles.
The use of the correlations slightly improves tha agreement with the
experiment. The effects of the $f_6$ correlations go in the opposite
direction of those of the $f_1$ correlation (as has already been
remarked) in the neutron density, and in the charge distributions
cases.

We would like to point out here that the density distributions are the
only quantities amongst those we have investigated where the use of
operator dependent correlations reduces the effects of purely scalar
correlations. In all the other cases, as we are going to show, the
effects of the $f_6$ correlations are larger than those of the $f_1$
correlation.

\subsection{Momentum distributions} 
\label{sec:res-obdm}

\begin{figure}[]
\begin{center}
\includegraphics[scale=0.7] {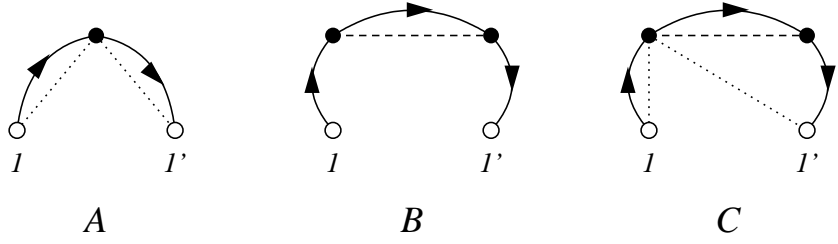}
\end{center}
\caption{\small 
 Example of diagrams used to calculate the OBDM 
 (\protect\ref{eq:res-obdm}). The dotted lines indicate the 
  new dynamical  correlation function $h_w=f_1-1$, while the 
  dashed line indicate the usual dynamical  correlation function 
  $h=f_1^2-1$.
  }
\label{fig:res-diank}
\end{figure}

The momentum distribution is related to the probability of finding a
nucleon with a certain value of the momentum. This quantity is related
to the Fourier transform of the OBDM, which is defined in analogy to
the OBDF (\ref{eq:fin-obd}) but for different values of the non
integrated variable:
\begin{eqnarray}
\nonumber
&~&
\rho(x_1,x'_1)  =  \sum_{s,s',t}
\rho^{s,s';t}(\br_1,\br'_{1}) \chi^+_s(1)\chi^+_t(1)\chi_{s'}(1')\chi_t(1')
 \\
&~& = \frac{A}{<\Psi|\Psi>} \int dx_2  \ldots dx_A \,\,
\Psi^{\dagger}(x_1,x_2,\ldots,x_A) \, 
\Psi(x'_1,x_2,\ldots,x_A) \,\,\,.
\label{eq:res-obdm}
\end{eqnarray}
The evaluation of this quantity merits some comments.  For simplicity,
we consider in this discussion only scalar correlations. In order to
perform the cluster expansion of the OBDM, it is necessary to
define a new dynamical correlation function:
\begin{equation}
h_w(r_{ij})=f_1(r_{ij})-1  \,\,,
\label{eq:res-hw}
\end{equation}
where $i$ can be either $1$ or $1'$, and $j=2,\ldots,A$.  This takes
into account the fact that the coordinate $x_1$ is present only in the
bra, while the coordinate $x_1'$ is present only in the ket.  The
coordinates describing the other particles can appear in both bra and
ket states, and generate the usual $h=f_1^2-1$ dynamical correlation
function. For the calculation of Eq. (\ref{eq:res-obdm}), we found it 
convenient to define a new type of sub-determinant, in analogy to what
has been done in Eq. (\ref{eq:inf-subdet}):
\begin{equation}
\Delta'_{p}(1,1',2,...,p)=
\begin{tabular}{|cccc|}
$\rho_0(x_1,x_{1'})$&$\rho_0(x_1,x_2)$&\ldots&$\rho_0(x_1,x_p)$\\
$\rho_0(x_2,x_{1'})$&$\rho_0(x_2,x_2)$&\ldots&$\rho_0(x_2,x_p)$\\
\vdots & \vdots & $\ddots$ & \vdots \\
$\rho_0(x_p,x_{1'})$&$\rho_0(x_p,x_2)$&\ldots&$\rho_0(x_p,x_p)$
\end{tabular}
\,\,.
\end{equation}
Also for this new sub-determinant the properties
(\ref{eq:inf-subdetp1}$\ldots$ \ref{eq:inf-subdetp3}) remain valid,
therefore we can apply the usual cluster expansion techniques,
developed for the calculation of the TBDF. However, we have to
consider that the separation of the $x_1$ and $x_{1'}$ coordinates
which refer to the same particle, implies the absence of dynamical
correlations between these two coordinates, as is shown in the
diagrams of Fig. \ref{fig:res-diank}.  In addition, we should take
care of the fact that the statistical loops containing the coordinate
$x_1$ must contain also $x_{1'}$ and must be open in these two points,
see again the diagrams of Fig. \ref{fig:res-diank}.

%%%%%%%%%%%%%%%%%%%%%%%%%%%%%%%
% 
% Figure diagrams of momentum distribution 
%
%%%%%%%%%%%%%%%%%%%%%%%%%%%%%%
%\vskip 0.5 cm 
\vskip 0.5 cm 
%%%%%%%%%%%%%%%%%%%%%%%%%%%%%%

The calculation of the OBDM proceeds in analogy to that of the TBDF.
The denominator of Eq. (\ref{eq:res-obdm}) simplifies the contribution
of the unlinked diagrams of the numerator.  At this point, we
encounter a difficulty, since it is not possible to cancel those
reducible diagrams where the articulation point is one of the points
in the open statistical loop containing to $1$ and
$1'$. This is not a specific problem of the
finite systems. Its solution is based on the use of the vertex
corrections which have been introduced for the first time to calculate
the momentum distribution of an infinite system of particles
\cite{fan78a}.

Coming back to our specific case, we found that the calculation of the
OBDM requires the use of a new vertex correction, which we call
$C_w(\br_i)$, understanding that $i$ can be $1$ or $1'$. New types of
diagrams appear. We call them $wd$, $we$, $ww$, $w_cc$, and $w_cw_c$.
The label $w$ is associated to the new dynamical correlation $h_w$
and, as $d$, denotes that no statistical line arrive at the corresponding 
point. With the label $w_c$ we indicate diagrams analogous to those we have
called $c$ in the calculation of the TBDF. These diagrams have an
open statistical loop with the $h_w$ dynamical correlations, as the
diagrams of Fig. \ref{fig:res-diank}. The diagrams contributing to 
the OBDM are of $w_c w_c$ type and the rest of classes diagrams are
auxiliary quantities needed to calculate them. The RFHNC/SOC
expressions of the different quantities involved in the calculation of
the OBDM, are presented in Appendix \ref{sec:app-fhncsoc} where the
parallelisms with the TBDF expressions are also pointed out.

Because of the spherical symmetry of the systems we are describing,
the quantity of interest in our calculations is:
\begin{equation}
\rho^t(\br_1,\br'_{1}) =
\sum_{s=\pm 1/2} \left[
\rho^{s,s;t}(\br_1,\br'_{1}) + \rho^{s,-s;t}(\br_1,\br'_{1})
\right] \,\,,
\label{eq:res-obdm1}
\end{equation}
whose diagonal part, $\br_{1'}=\br_1$, is the OBDF. We obtain the
momentum distributions  
of protons or neutrons as:
\begin{equation}
n^t(k)= \frac {1}{(2\pi)^3} \frac {1} {{\cal N}_t} \int d\br_1 d\br'_1 \, 
e^{i {\bf k} \cdot (\br_1-\br'_1)} 
\rho^t(\br_1,\br'_1) \,\,,
\label{eq:res-md}
\end{equation}
which is normalized:
\begin{equation}
\int d{\bf k} \,n^t(k)= 1 \,\,.
\label{eq:res-nor}
\end{equation}

The uncorrelated OBDMs are obtained by inserting in Eq.
(\ref{eq:fin-obdm}) the Slater determinant, $\Phi$, formed by the
single particle wave functions (\ref{eq:fin-spwf}). We obtain the
expressions (\ref{eq:fin-ropar}) for $s'=s$ and (\ref{eq:fin-roantipar})
for $s'\ne s$.

For the correlated OBDM we obtain the expression:
\begin{eqnarray}
\nonumber
\rho^{t}(\br_1,\br'_{1})&=&-2C_{w,11}^{t}(\br_1)
C_{w,11}^{t}(\br'_{1}) g_{w_cw_c}^{t}(\br_1,\br'_{1})
\\
&&
- 2C_{w,22}^{t}(\br_1)C_{w,22}^{t}(\br_{1'})
\sum_{p>1}  
A^k\Delta^{k}g_{w_cw_c,p}^{t}(\br_1,\br'_{1}) \,\,.
\label{eq:fin-obdm2}
\end{eqnarray}
with $p=2k-1+l$. All the FHNC/SOC quantities have been defined in
Appendix \ref{sec:app-fhncsoc}.

%%%%%%%%%%%%%%%%%%%%%%%%%%%%%%%
% 
% Figure of momentum distribution 
%
%%%%%%%%%%%%%%%%%%%%%%%%%%%%%%
%\vskip 0.5 cm 
\begin{figure}[]
\begin{center}
\includegraphics[scale=0.5] {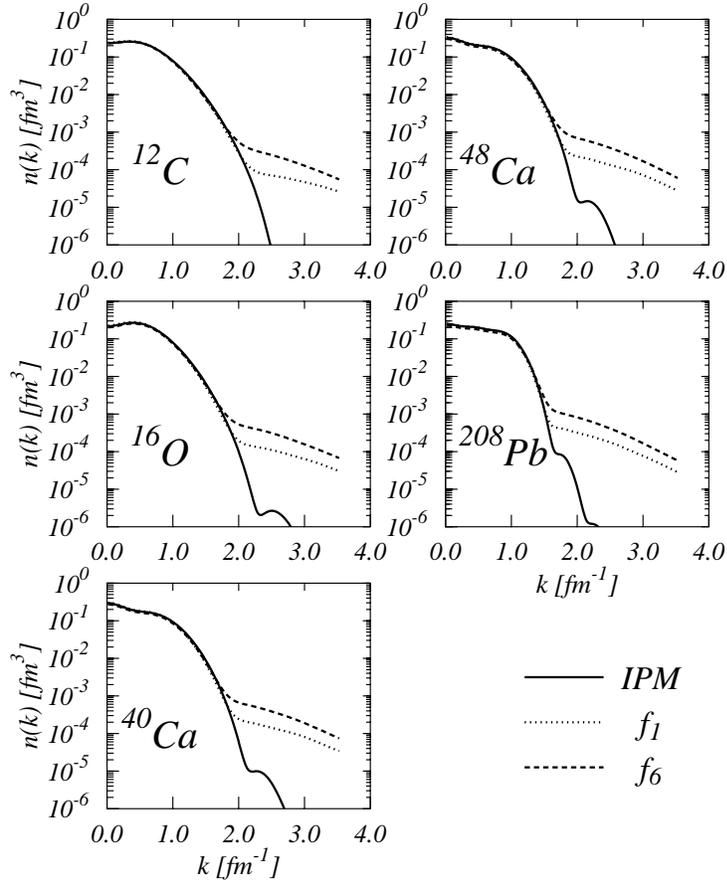}
\end{center}
\caption{\small The proton momentum distributions 
  of the \car, \oxy, \caI, \caII and \lead nuclei calculated in the
  IPM model (full lines), by using the scalar correlation only
  (dotted lines) and the full operator dependent correlations 
  (dashed lines).  }
\label{fig:res-nk}
\end{figure}
\vskip 0.5 cm 
%%%%%%%%%%%%%%%%%%%%%%%%%%%%%%

In Fig. \ref{fig:res-nk}
we compare the \car, \oxy, \caI, \caII and \lead momentum
distributions calculated in the IPM model, with those obtained by
using $f_1$ and $f_6$ correlations.

%%%%%%%%%%%%%%%%%%%%%%%%%%%%%%%
% 
% Figure of proton and neutron momentum distribution 
%
%%%%%%%%%%%%%%%%%%%%%%%%%%%%%%
%
\begin{figure}[]
\begin{center}
\includegraphics [scale=0.5, angle=90]{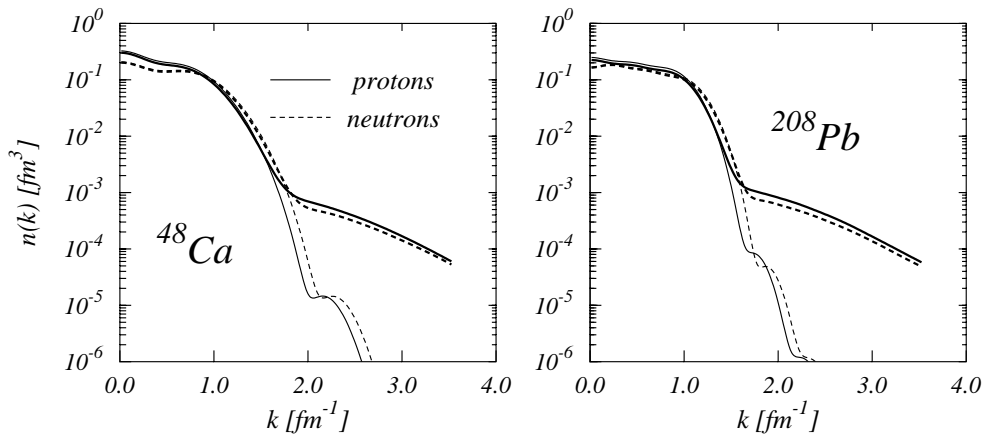}
\caption{\small Protons (full lines) and neutrons (dashed lines)
                 momentum distributions of the \caII and \lead. The
                 thick lines show the results of our calculations, the
                 thin lines the IPM results. 
}
\label{fig:res-nkpn} 
\end{center}
\end{figure}
%%%%%%%%%%%%%%%%%%%%%%%%%%%%%%%
%

The general behavior of the momentum distributions, is very similar
for all the nuclei we have considered. Correlated and IPM
distributions almost coincide in the low momentum region up to a
precise value, when they start to deviate. The correlated
distributions show high momentum tails, which are orders of magnitudes
larger than the IPM results.  The value of $k$ of which 
uncorrelated and
correlated momentum distributions start to deviate is smaller the
heavier is the nucleus.  It is about 1.9 fm$^{-1}$ for \car, and 1.5
fm$^{-1}$ for \lead. We recall that the value of the Fermi momentum of
symmetric nuclear matter at the saturation point is 1.36 fm$^{-1}$.
The results presented in Fig. \ref{fig:res-nk} clearly show that the
effects of the scalar correlations are smaller than those obtained by
including the operator dependent terms.

In our calculations, we have found that the proton and neutron momentum
distributions for nuclei with $N=Z$ are very similar.  For this
reason, we show in Fig. \ref{fig:res-nkpn} the proton and neutron
momentum distributions of the two nuclei we have investigated with $N
\ne Z$: the \caII and \lead nuclei.  The thicker lines show the
results of our RFHNC/SOC calculations, while the thinner ones the IPM
distributions.  The main differences between the two distributions are
in the zone where the $n(k)$ values drop by orders of magnitudes.
This zone, corresponds, in the infinite systems, to the discontinuity
region of the momentum distribution, related to the Fermi momentum. In
a finite system, the larger number of neutrons implies that the
neutron Fermi energy and, consequently, the effective Fermi momentum,
is larger than that of the protons.  For this reason, the
discontinuity regions of the neutron momentum distributions are
located at larger values of $k$ with respect to the protons momentum
distributions.

After the discontinuity region, the behaviors of the distributions
are dominated by the SRC effects, and the protons and neutrons results
are very close. In terms of relative difference, the SRC effects are
essentially the same for protons and neutrons \cite{bis07}.  A
discussion about the role of SRC effects on the proton and neutron
momentum distributions in asymmetric nuclear matter is open, and our
results are in agreement with the findings of Ref.  \cite{boz04}, but
disagree with those of Ref. \cite{fri05}.

The increase of the momentum distribution at large $k$ values, induced
by the SRC, is a well known result in the literature, see for example
the review of Ref. \cite{ant88}. The momentum distributions of
medium-heavy nuclei, have been usually obtained by using approximated
descriptions of the cluster expansion, which is instead considered at
all orders in our treatment.  We found in \cite{bis07} that the
results of the approximate treatment provide only a qualitative
description of the correlation effects. They produce a high-momentum
enhancement, but they underestimate the correct results by orders of
magnitude.

\subsection{Natural orbits}
\label{sec:res-no}
We have studied the effects of the SRC on the natural orbits which are
defined as those single particle wave functions forming the basis
where the OBDM is diagonal:
\begin{equation}
\rho^t(\br_1,\br'_{1})=\sum_{nlj}c_{nlj}^t\phi_{nlj}^{*\,t,NO}(\br_1)
\phi_{nlj}^{t,NO}(\br'_{1}) \,\,.
\label{eq:res-no}
\end{equation}

In the above equation the $c^t_{nlj}$ coefficients, called occupation
numbers, are real numbers.  In the IPM the natural orbits correspond
to the mean-field wave functions of Eq. (\ref{eq:fin-spwf}), and the
$c^t_{nlj}$ numbers are 1, for the states below the Fermi surface and
0 for those above it.

%%%%%%%%%%%%%%%%%%%%%%%%%%%%%%%%%%%%%%%%%%%%%%
% Figure occupation numbers protons
%%%%%%%%%%%%%%%%%%%%%%%%%%%%%%%%%%%%%%%%%%%%%%
\begin{figure}[]
\begin{center}
\includegraphics [scale=0.68]{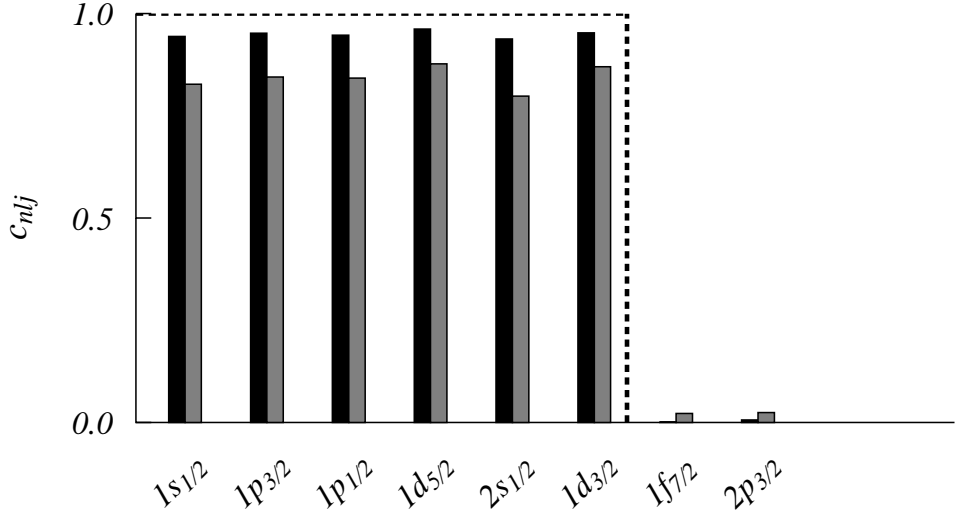}
\caption{\small Occupation numbers of the proton natural 
orbits of the \caII nucleus. The dashed line indicates the IPM values. 
The black bars show the values obtained with the scalar corrrelation
and the gray bars those values obtained with the full correlation. 
}
\label{fig:res-natorb1} 
\end{center}
\end{figure}
%%%%%%%%%%%%%%%%%%%%%%%%%%%%%%%%%%%%%%%%%%%%

%%%%%%%%%%%%%%%%%%%%%%%%%%%%%%%%%%%
% Figure occupation numbers neutrons
%%%%%%%%%%%%%%%%%%%%%%%%%%%%%%%%%%%%
\begin{figure}[]
\begin{center}
\includegraphics [scale=0.68]{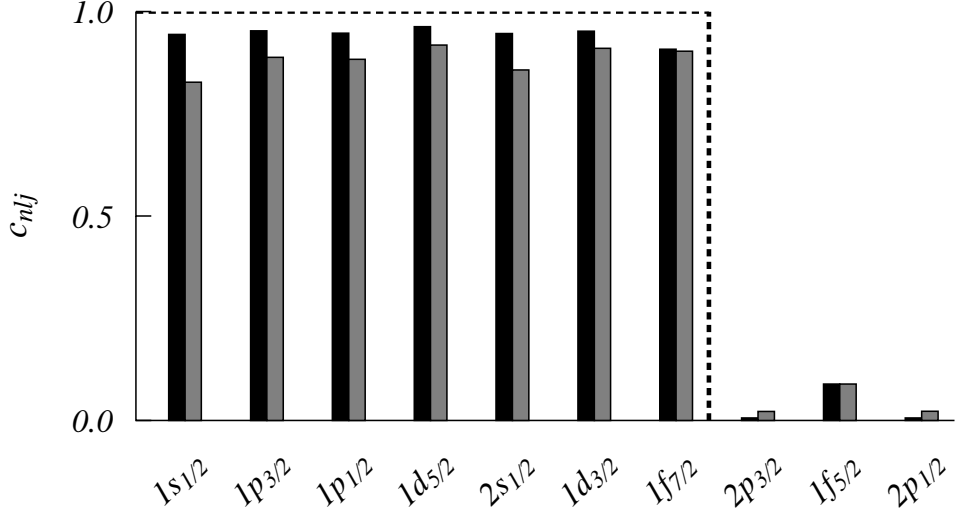}
\caption{\small The same as Fig. \protect\ref{fig:res-natorb1} for the 
 neutron natural orbits of the \caII nucleus. 
}
\label{fig:res-natorb2} 
\end{center}
\end{figure}
%%%%%%%%%%%%%%%%%%%%%%%%%%%%%%%%%%%%%%%%%%%%

%%%%%%%%%%%%%%%%%%%%%%%%%%%%%%%%%%%%%%%%%%%%%
% figure natural orbits 
%%%%%%%%%%%%%%%%%%%%%%%%%%%%%%%%%%%%%%%%%%%%%
\begin{figure}[]
\begin{center}
\includegraphics [scale=0.50]{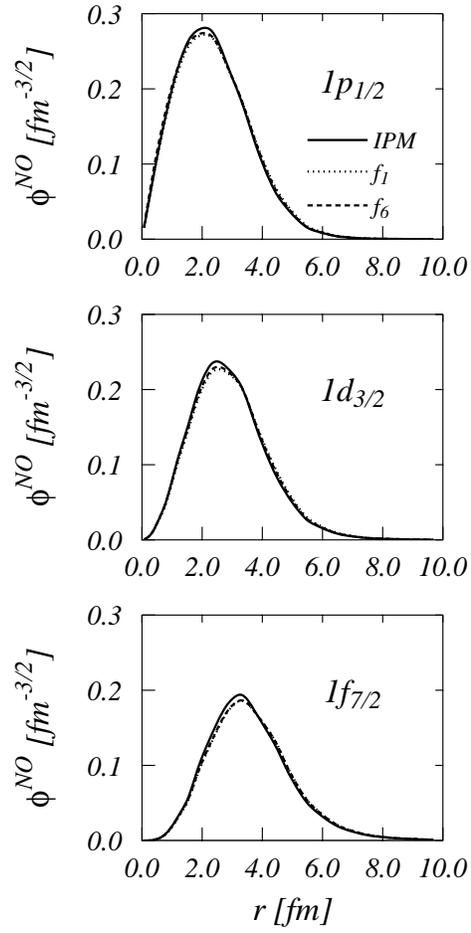}
\caption{\small Natural orbits for some neutron states in \caII.
The full lines indicate the IPM orbits, the dotted lines those
 obtained with scalar correlations only and the dashed lines those
 obtained with the complete operator dependent correlation.
 }
\label{fig:res-natorb3} 
\end{center}
\end{figure}
%%%%%%%%%%%%%%%%%%%%%%%%%%%%%%%%%%%%%%%%%%%%%

In order to obtain the natural orbits we found it convenient to express
the OBDM of Eq. (\ref{eq:res-no}) as:
\begin{equation}
\rho^t(\br_1,\br_{1'})=A^t(\br_1,\br_{1'})
\rho_0^t(\br_1,\br_{1'})+B^t(\br_1,\br_{1'}) \,\,,
\label{eq:res-rhosplit}
\end{equation}
where 
$\rho^t_{0}(\br_1,\br_{1'})$ is the uncorrelated OBDM of Eq. 
(\ref{eq:fin-ropar}), and the other two quantities are defined as:
\begin{eqnarray}
\nonumber
A^{t}(\br_1,\br'_1)&=&2C_{w,11}^{t}(\br_1)
C_{w,11}^{t}(\br'_1)
g_{ww}^{tt}(\br_1,\br'_1)
+\\ &&
2C_{w,22}^{t}(\br_1)C_{w,22}^{t}(\br'_1)\sum_{p>1}  
A^k\Delta^{k}g_{ww,p}^{tt}(\br_1,\br'_1) \,\,,
\label{eq:res-arho}\\
\nonumber
B^{t}(\br_1,\br'_1)&=&-2C_{w,11}^{t}(\br_1)
C_{w,11}^{t}(\br'_1)
g_{ww}^{tt}(\br_1,\br'_1)
N_{w_cw_c}^{t}(\br_1,\br'_1)
- \\
\nonumber &&
2C_{w,22}^{t}(\br_1)C_{w,22}^{t}(\br'_1)\sum_{p>1}  
A^k\Delta^{k}\Big[g_{ww,p}^{tt}(\br_1,\br'_1)
N_{w_cw_c}^{t}(\br_1,\br'_1)+
\\ \nonumber & &
g_{ww}^{tt}(\br_1,\br'_1)
N_{w_cw_c,p}^{t}(\br_1,\br'_1)\Big] \,\,.
\label{eq:res-brho}
\end{eqnarray}
The expressions of the various FHNC/SOC quantities used in the 
above equations are given in Appendix \ref{sec:app-fhncsoc}. 

We expand the OBDM on a basis of spin spherical harmonics
${\bf Y}^m_{lj}$ defined in Eq. (\ref{eq:fin-spwf}),
\begin{equation}
\rho^t(\br_1,\br'_{1}) = \sum_{ljm} \frac {1}{2j+1}
\left[{\cal A}_{lj}^t(r_1,r'_1) + {\cal B}_{lj}^t(r_1,r'_1) 
\right]
{\bf Y}^{* m}_{lj}(\Omega)  {\bf Y}^m_{lj}(\Omega')
\label{eq:res-rhoexp}
\end{equation}
where $\Omega$ and $\Omega'$ indicate the polar angles identifying 
$\br_1$ and $\br'_1$.
The explicit expressions of the ${\cal A}$ and ${\cal B}$ coefficients
are:
\begin{eqnarray}
\nonumber
{\cal A}_{lj}^t(r_1,r'_1) &=&
(2l+1)\sum_{nl_1l_2j_2}(2l_2+1)(2j_2+1)
\left( \begin{array}{ccc}
       l & l_1 & l_2\\
       0 & 0 & 0 
      \end{array}\right)^2
\left\{ \begin{array}{ccc}
       j_2 & l_1 & j\\
       l & \half & l_2 
      \end{array}\right\}^2 
\nonumber
\\
&~&R^t_{nl_2j_2}(r_1)R^t_{nl_2j_2}(r_2) A^t_{l_1}(r_1,r'_1)
\label{eq:res-cala}
\end{eqnarray}
with
\begin{equation}
A_l^t(r_1,r'_1)=\frac 2 {2l+1}\int d\Omega A^t(\br_1,\br'_1)
P_{l}(\cos\theta_{11'})
\end{equation}
and 
\begin{equation}
{\cal B}_{lj}^t(r_1,r'_1) 
=\frac{4\pi}{2l+1} \int d (\cos\theta_{11'}) B^t(\br_1,\br'_1)
P_{l}(\cos\theta_{11'})
\label{eq:res-calb}
\end{equation}
In the above equations we have used the 3j and 6j Wigner symbols
\cite{edm57} and we have indicated with $\theta_{11'}$ the angle
between $\br_1$ and $\br'_{1'}$. The term ${\cal A}$ depends on both
orbital and total angular momenta of the single particle, $l$ and $j$
respectively, and ${\cal B}$ depends only on the orbital angular
momentum $l$.

As has been done in Refs.  \cite{lew88} and \cite{pol95} we
identify the various natural orbits with a number, $\alpha$, ordering
them with respect to the decreasing value of the occupation
probability. The general behavior of our results, is analogous to
that described in Ref. \cite{lew88} where a system of $^3$He drops,
composed by 70 atoms, have been studied. The orbits corresponding to
states below the Fermi level in the IPM picture, have occupation
numbers very close to unity for $\alpha=1$, and very small in all the
other cases.

As example of our results, we show in Figs. \ref{fig:res-natorb1} and
\ref{fig:res-natorb2} the protons and neutrons occupation numbers for
the natural orbits with $\alpha=1$ of the \caII nucleus.  In the
figures, the IPM results are indicated by the dashed lines.  The black
bars show the values obtained by using scalar correlations only, the
gray bars those obtained with the complete operator dependent
correlations.

The correlated occupation numbers are smaller than one for orbits
below the Fermi surface, and larger than zero for those orbits above
the Fermi surface. This effect is enhanced by the operator dependent
correlations. We observe that for the states above the Fermi surface
the gray bars are larger than the black ones, indicating that also
for these states the operator dependent correlations, produce larger
effects than the scalar ones.

We show in Fig. \ref{fig:res-natorb3} some $\alpha=1$ natural orbits
for three neutron states in \caII. In this figure, we compare the IPM
results (full lines) with those obtained with scalar correlation only
(dotted lines), and with the full operator dependent correlation
(dashed lines). The effect of the correlations is a lowering of the
peak and a small widening of the function. Despite the small effect,
it is interesting to notice the inclusion of operator dependent terms
diminishes the correlation effect. This fact is consistent with the
results on the density distributions we have presented in Sect.
\ref{sec:res-obd}.

%
% Table Occupation numbers
%
\begin{table}[pt]
\begin{center}
\begin{tabular}{r|ccc|ccc|ccc}
\hline
 State       & & $\alpha=1$ & & &  $\alpha=2$ & & & $\alpha=3$ & \\  
\hline
             & $f_1$ & $f_6$ & PMD & $f_1$ & $f_6$ & PMD & 
             $f_1$ & $f_6$ & PMD \\ 
$1s_{1/2}$ (p)  & 0.956 & 0.873 & 0.921 & 0.011 & 0.038 & 0.013 
                & 0.002 & 0.007 & 0.002 \\
           (n)  & 0.957 & 0.873 &       & 0.012 & 0.039 & 
                & 0.003 & 0.008 &     \\
$1p_{3/2}$ (p)  & 0.973 & 0.921 & 0.947 & 0.004 & 0.013 & 0.007 
                & 0.001 & 0.003 & 0.001 \\
           (n)  & 0.973 & 0.924 &       & 0.004 & 0.014 & 
                & 0.002 & 0.004 &       \\
$1p_{1/2}$ (p)  & 0.970 & 0.923 & 0.930 & 0.003 & 0.012 & 0.008 
                & 0.001 & 0.003 & 0.002 \\
           (n)  & 0.970 & 0.922 &       & 0.004 & 0.013 & 
                & 0.002 & 0.003 &       \\
$1d_{5/2}$ (p)  & 0.001 & 0.005 & 0.016 & 0.013 & 0.003 & 0.003 
                & 0.000 & 0.000 & 0.000 \\
           (n)  & 0.001 & 0.005 &       & 0.001 & 0.003 & 
                & 0.000 & 0.000 &       \\
$1d_{3/2}$ (p)  & 0.002 & 0.005 & 0.019 & 0.001 & 0.003 & 0.005 
                & 0.000 & 0.000 & 0.001 \\
           (n)  & 0.001 & 0.005 &       & 0.001 & 0.003 & 
                & 0.000 & 0.000 &       \\
\hline
\end{tabular}
\end{center}
\caption{\small Protons (p) and neutrons (n) natural orbits occupation 
numbers for \oxy. The PMD values are those of Ref. \cite{pol95}.
The $f_1$ values have been obtained with scalar correlations only, and
the $f_6$ values with the complete state dependent corrrelation
function. 
}
\label{tab:res-ocn}
\end{table}
%%%%%%%%%%%%%%%%%%%%%%%%%%%%%%%%%%%%%555

In Tab. \ref{tab:res-ocn} we show the occupation numbers of the \oxy
protons and neutrons natural orbits also for $\alpha > 1$, and we make
a direct comparison with the results of Ref. \cite{pol95}. As already
said in the discussion of the \caII results, the inclusion of the
state dependent correlations increases the differences with respect to
the IPM. The occupation numbers of the orbits below the Fermi surface
are smaller than those obtained with scalar correlations only. The
situation is reversed for the orbits with $\alpha > 1$ or above the
Fermi level. For the states below the Fermi surface, our full
calculations produce correlation effects slightly larger than those
found in \cite{pol95}, whose results are closer to those we obtain
with scalar correlations only.  For orbits above the IPM Fermi
surface, our occupation numbers are always smaller than those of Ref.
\cite{pol95}.

\subsection{The two-body distribution functions}
\label{sec:res-tbdm}

We have already defined the state dependent 
TBDF in Eq. (\ref{eq:fin-tbdm}).  
In our FHNC/SOC computational scheme, we calculate the
TBDF as:
\begin{eqnarray}
\nonumber
\rho^{2k_3-1+l_3,\ton\ttw}_2(\br_1,\br_2) & = & 
\frac{f_{2k_1-1+l_1}(r_{12})f_{2k_2-1+l_2}(r_{12})}
{f_{1}^2(r_{12})}\\ \nonumber 
&&\Bigg\{
I^{k_1k_3k_2}A^{k_2}\chi^{\ton \ttw}_{l_1+l_2+l_3}
\rho^{\ton \ttw}_{2,dir}(\br_1,\br_2)\\
\nonumber
&&+\Big[I^{k_4k_1k_5}I^{k_3k_2k_5} 
A^{k_5} \rho_{2,exc}^{\ton \ttw}(\br_1,\br_2)\\ \nonumber & & 
+I^{k_4k_1k_5}I^{k_3k_2k_6} I^{k_5 k_6 2} 
\rho_{2,excj}^{\ton \ttw}(\br_1,\br_2)  \Big]
\\ 
& & \Delta^{k_4} \sum_{l_4=0}^1\chi^{\ton \ttw}_{l_1+l_2+l_3+l_4}\Bigg\}
\end{eqnarray}
where a sum is understood on every repeated index. The FHNC/SOC 
quantities are defined in Appendix \ref{sec:app-fhncsoc}.

%%%%%%%%%%%%%%%%%%%%%%%%%%%%%%%%%%%%%%%%%%%%%
% figure tbdm scalar for all the nuclei
%%%%%%%%%%%%%%%%%%%%%%%%%%%%%%%%%%%%%%%%%%%%%
\begin{figure}[]
\begin{center}
\includegraphics [scale=0.50]{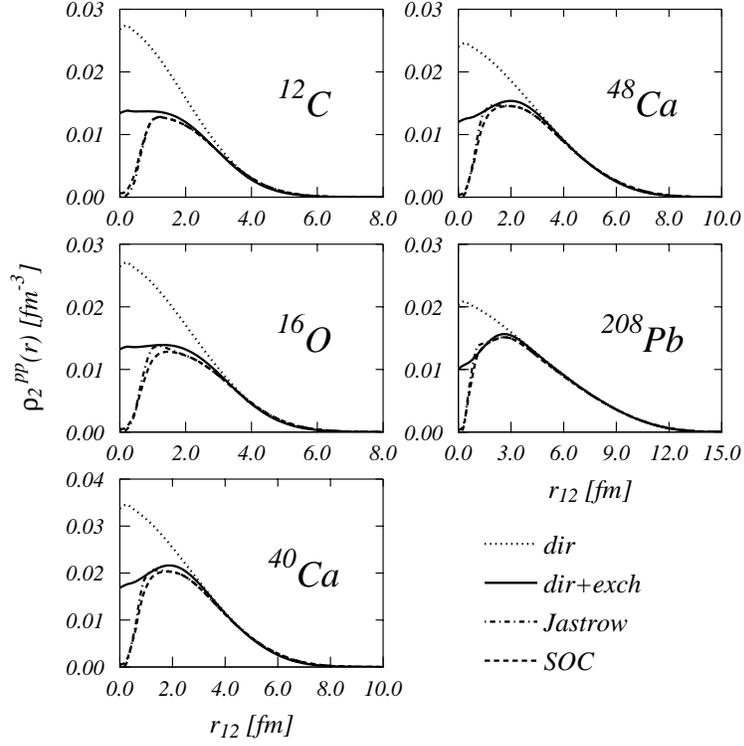}
\caption{\small Proton-proton scalar (q=1) TBDFs
(\protect\ref{eq:fin-tbdm}) as a function of the distance between two
nucleons The dotted lines show the uncorrelated TBDF obtained as a
product of two uncorrelated OBDMs. The full lines show the
uncorrelated TBDF obtained by using in Eq. (\protect\ref{eq:fin-tbdm})
Slater determinants. The other lines have been obtained with the
scalar ($f_1$) and full ($f_6$) correlations. 
 }
\label{fig:res-tbdmpp} 
\end{center}
\end{figure}
%%%%%%%%%%%%%%%%%%%%%%%%%%%%%%%%%%%%%%%%%%%%%

We first discuss the case $q=1$ which, apart from some
constant factors, gives the probability of finding a nucleon in
$\br_1$, and at the same time, another nucleon in $\br_2$. 
The IPM two-body densities are obtained by inserting a Slater
determinant in Eq. (\ref{eq:res-rtbdm}).  For $q=1$ TBDF
we obtain the expression:
\beq
\rho^{\ton \ttw}_{2,0}(\br_1,\br_2) =
\rho^{\ton}_0(\br_1) \rho^{\ttw}_0(\br_2)
- 2 \delta_{\ton \ttw} \Big\{
\Big[\rho^{\ton}_0(\br_1,\br_2)\Big]^2
+\Big[\rho^{\ton}_{0j}(\br_1,\br_2)\Big]^2 \Big\}
\label{eq:res-uobdm}
\,\,,
\eeq
where the $\rho^{\ton}_{0(j)}(\br_1,\br_2)$ are the uncorrelated OBDMs
defined in Eqs. (\ref{eq:fin-ropar}) and (\ref{eq:fin-roantipar}).

The relevant information about the TBDF is contained in the
function:
\beq
\rho^{q,\ton \ttw}_{2}(r_{12}) =
\int d {\bf R}_{12}  \rho^{q,\ton \ttw}_{2}(\br_1,\br_2) \,\,,
\label{eq:res-rtbdm}
\eeq
where $r_{12}=|\br_1 - \br_2|$ is the relative distance and 
${\bf R}_{12} = (\br_1 + \br_2)/2$ the center-of-mass of the nucleonic
pair.

We show in Fig. \ref{fig:res-tbdmpp} the proton-proton scalar, q=1,
TBDFs, Eq. (\ref{eq:res-rtbdm}), for all the nuclei considered as a
function of the relative distance of the pair. The dotted lines
represent the uncorrelated joint probability density of finding the
two nucleons at a certain distance, as often given in the
literature (see e.g.  Ref. \cite{ben93}). These lines have been
obtained as a product of the uncorrelated one-body densities. This
definition of the uncorrelated two-body densities can be meaningful
from the probabilistic point of view, but it is misleading in our
framework, since it corresponds to using only the first, direct, term
of Eq. (\ref{eq:res-uobdm}). In fermionic systems the uncorrelated
two-body density is given by the full expression (\ref{eq:res-uobdm})
which contains also the exchange term. These complete uncorrelated
TBDF are shown by the full lines.  The effects of the SRC can be
deduced by comparing these lines with the dashed-dotted lines obtained
with the $f_1$ scalar correlations only, and with the dashed lines
showing the full FHNC/SOC results.

In all our results the correlations reduce the values of the TBDF at
short internucleon distances. The exchange term of the uncorrelated
density already contributes to this reduction, but the major effect is
produced by the SRC, and mainly by the scalar correlations. We found
similar results for the neutron-neutron TBDF. When the TBDF are
composed of different particles, the results change only slightly.
Beside a strong reduction at small distances, the correlations produce
enhancements, with respect to the IPM results, around 2 fm, in all the
nuclei considered \cite{bis06}.

%%%%%%%%%%%%%%%%%%%%%%%%%%%%%%%%%%%%%%%%%%%%%
% figure tbdm
%%%%%%%%%%%%%%%%%%%%%%%%%%%%%%%%%%%%%%%%%%%%%
\begin{figure}[]
\begin{center}
\includegraphics [scale=0.50]{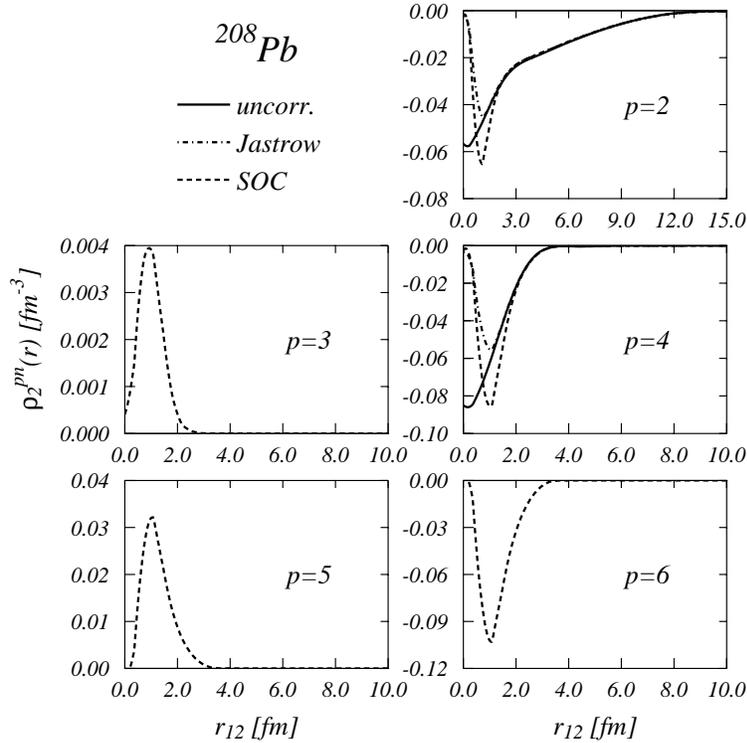}
\caption{\small 
Proton-neutron operator dependent (q$>$1) TBDFs
(\protect\ref{eq:fin-tbdm}) as a function of the distance between two
nucleons for the \lead nucleus. 
 }
\label{fig:res-tbdmop} 
\end{center}
\end{figure}
%%%%%%%%%%%%%%%%%%%%%%%%%%%%%%%%%%%%%%%%%%%%%

To discuss the effects of the correlations on the other operator
dependent TBDFs (\ref{eq:fin-tbdm}) we show in Fig.
\ref{fig:res-tbdmop} the TBDFs for the \lead nucleus, for the various
operators. We first notice that the tensor TBDFs, $p=5,6$, are
different form zero only when the $f_6$ correlation is used. This
occurs because, in order to get a spin trace different from zero in
Eq. (\ref{eq:fin-tbdm}), at least two-tensor operatosr are needed.
Also the spin term $p=3$ differs from zero in the $f_6$ case only.
Here there are different reasons why both terms of Eq.
(\ref{eq:res-uobdm}) are zero.  The first term is zero because of the
spin trace, while the second term vanishes because the two nucleons
have different isospin.

The most remarkable feature is the range of the various TBDFs. The
scalar TBDFs of Fig. \ref{fig:res-tbdmpp} and the isospin TBDF of Fig.
\ref{fig:res-tbdmop} extend up to relative distances comparable to the
dimensions of the nucleus, 15 fm. In contrast, all the other density
functions have much smaller ranges, of the order of 3-4 fm. We found
similar results for all the nuclei considered \cite{bis06,bis06t}.

\subsection{Quasi-hole wave functions and spectroscopic factors}
\label{sec:res-specf}

The quasi-hole wave function is defined as:
\begin{equation}
\psi_{nljm}^{t}(x)=
\frac{\sqrt{A}<\Psi_{nljm}^t(1,...,A-1)|\delta(x-x_A)P^{t}_{A}|\Psi(1,...,A)>}
{<\Psi_{nljm}^t|\Psi_{nljm}^t>^{1/2}
<\Psi|\Psi>^{1/2}} \,\,,
\label{eq:res-qhfun}
\end{equation}
where $\Psi_{nljm}^t(1,...,A-1)$ and $\Psi(1,...,A)$ are the states
of the nuclei formed by $A-1$ and $A$ nucleons respectively, and
$P^t_A$ is the isospin projector.  In analogy to the ansatz
(\ref{eq:in-trial}), we assume that the state of the nucleus with $A-1$
nucleons can be described
as:
\begin{equation}
\Psi_{nljm}^t(1,...,A-1)={\cal F}(1,...,A-1)\Phi_{nljm}^t(1,...,A-1) \,\,,
\label{eq:res-psiam1} 
\end{equation}
where $\Phi_{nljm}^t(1,...,A-1)$ is the Slater determinant obtained by
removing from $\Phi(1,...,A)$ a single state characterized by the quantum
numbers $nljmt$.  For the system of $A-1$ nucleons we use the same
correlation function fixed for the system of $A$ nucleons.  In an
uncorrelated system the quasi-hole wave functions coincide with the
hole mean-field wave functions (\ref{eq:fin-spwf}).

We are interested in the radial part of the quasi-hole wave function,
which we obtain by multiplying equation (\ref{eq:res-qhfun}) by the
vector spherical harmonics ${\bf Y}^{*m}_{lj}(\Omega)$ and, then, by
integrating over the angular coordinates $\Omega$, and summing over
$m$.  It is useful to rewrite the radial part of the quasi-hole wave
function as \cite{fab01}:
\begin{eqnarray}
\nonumber 
\psi_{nlj}^{t}(r)& = &\frac 1 {2j+1}
\sum_m \int d\Omega \,
{\bf Y}^{*m}_{lj}(\Omega)
\, \psi_{nljm}^{t}(x) \\ 
& = & \frac 1 {2j+1}\sum_m {\cal X}_{nljmt}^t(r)[{\cal N}_{nlj}^t]^{1/2} \,\,,
\label{eq:res-qhprod}
\end{eqnarray}
where we have defined:
\begin{equation}
{\cal{X}}_{nljmt}^t(r)= 
\frac{\sqrt{A}<\Psi_{nljm}^t(A-1)|{\bf Y}^{*m}_{lj}(\Omega)  
\delta(\br-\br_{A})P^{t}_{A}|\Psi(A)>}{<\Psi_{nljm}^t|
\Psi_{nljm}^t>} \,\, ,
\label{eq:res-calx}
\end{equation}
and 
\begin{equation}
{\cal N}_{nljmt}^t=\frac{<\Psi_{nljm}^t(1,...,A-1)|\Psi_{nljm}^t(1,...,A-1)>}
{<\Psi(1,...,A)|\Psi(1,...,A)>} \,\, .
\label{eq:res-caln}
\end{equation}

Following the procedure outlined in Ref.  \cite{fab01}, we consider
separately the cluster expansions of the two terms ${\cal N}^t$ and
${\cal X}^t$. For ${\cal X}^t$ we have:
\begin{eqnarray}
\nonumber
{\cal X}_{\alpha}^{t}(r)&=&
C^{t,\alpha}_{w,11}(\br)
\Bigg\{
R^{t}_{nlj}(r)+
\int d\br_1R^{t}_{nlj}(r_1)P_{l}(\cos{\theta}) 
\Bigg[
g_{w_c c}^{t,\alpha}(\br,\br_1)C_{d,11}^{t,\alpha}(\br_1)
 \\ 
\nonumber 
&~&
+\rho_{0}^{t,\alpha}(\br,\br_1)
-N^{(\rho)t,\alpha}_{c w_c}(\br_1,\br)
\Bigg]
\Bigg\}
\\ &&
+ C^{t,\alpha}_{w,22}(\br)
\int d\br_1R^{t}_{nlj}(r_1)P_{l}(\cos{\theta})
{\cal X}_{SOC}^{t}(\br,\br_1)
 \,\,,
\end{eqnarray}
and for ${\cal N}^t$ we obtain: 
\begin{eqnarray}
\nonumber
\Big[ {\cal N}_{\alpha}^{t} \Big]^{-1}&=&\int
d\br C_{d,11}^{t,\alpha}(\br)
\Bigg\{
|\phi^{t}_{\alpha}(\br)|^2+
\int d\br_1\phi^{t *}_{\alpha}(\br)\phi^{t}_{\alpha}(\br_1)
2
\Bigg[
g_{cc}^{t,\alpha}(\br,\br_1)C_{d,11}^{t,\alpha}(\br_1)
\\
\nonumber
&& +\rho_{0}^{t,\alpha}(\br,\br_1)-N_{cc}^{(\rho)t,\alpha}(\br_1,\br)
\Bigg] \Bigg\} 
\\ &&
+\int d\br \phi^{t *}_{\alpha}(\br)C_{d,22}^{t,\alpha}(\br)
\int d\br_1\phi^{t}_{\alpha}(\br_1) {\cal N}_{SOC}^{t}(\br,\br_1) 
\,\,,
\end{eqnarray} 
where we have indicated with $\alpha$ the set of the $nljm$ quantum
numbers. All the FHNC quantities have a superscript $\alpha$ since these
equations must be built by using:
\begin{equation}
\rho_{0}^{t,\alpha}(\br,\br_1)= \rho_{0}^{t}(\br,\br_1)-
\phi^{t *}_{\alpha}(\br)\phi^{t}_{\alpha}(\br_1) \,\,,
\end{equation}
instead of $\rho_{0}^{t}(\br,\br_1)$. The expressions of  
${\cal N}_{SOC}^{t}(\br,\br_1)$ and
${\cal X}_{SOC}^{t}(\br,\br_1)$, are:
\begin{eqnarray}
\nonumber
{\cal X}_{SOC}^{t_{1}}(\br,\br_1)&=&
\sum_{k=1}^{3}A^k\sum_{t_{2}=p,n}
\Big\{ (1-\delta_{k,1}){\cal X}_{2k-1,2k-1}^{t_{1}t_{2}}(\br,\br_1)
\\
\nonumber
&&+\chi_{1}^{t_{1}t_{2}}\Big[{\cal X}_{2k-1,2k}^{t_{1}t_{2}}(\br,\br_1)
+{\cal X}_{2k,2k-1}^{t_{1}t_{2}}(\br,\br_1)\Big]
\\
& &+
\chi_{2}^{t_{1}t_{2}}{\cal X}_{2k,2k}^{t_{1}t_{2}}(\br,\br_1)\Big\}
\,\,,\\
\nonumber
{\cal N}_{SOC}^{t_{1}}(\br,\br_1)&=&
\sum_{k=1}^{3}A^k\sum_{t_{2}=p,n}
\Big\{
(1-\delta_{k,1}){\cal N}_{2k-1,2k-1}^{t_{1}t_{2}}(\br,\br_1)
\\
\nonumber
&~&+\chi_{1}^{t_{1}t_{2}}\Big[{\cal
  N}_{2k-1,2k}^{t_{1}t_{2}}(\br,\br_1)
\\
&~&+{\cal X}_{2k,2k-1}^{t_{1}t_{2}}(\br,\br_1)\Big]+
\chi_{2}^{t_{1}t_{2}}{\cal
N}_{2k,2k}^{t_{1}t_{2}}(\br,\br_1)\Big\}
\,\,.
\end{eqnarray}
where the indexes $t$ refer
to the isospin and the functions and we have defined:
\begin{eqnarray}
\nonumber
{\cal X}_{pq}^{t_{1}t_{2}}(\br,\br_1)&=&\frac{1}{2}
\Big\{ h_{w,p}^{t_{1}t_{2} ,\alpha}(\br,\br_1)
g_{w d}^{t_{1}t_{2} ,\alpha}(\br,\br_1)
C_{d,pq}^{t_{2},\alpha}(\br_1) \\
\nonumber &~& 
\Big[-\rho_{0}^{t_{2},\alpha}(\br,\br_1)+
N_{w_cc}^{t_{2},\alpha}(\br,\br_1)\Big]\\
\nonumber &~&
+g_{w d}^{t_{1}t_{2},\alpha}(\br,\br_1)C^{t_{2},\alpha}_{d,pq}(\br_1)
N_{w_cc,p}^{t_{2},\alpha}(\br,\br_1)- \\
&~&
N_{cw_c,p}^{(\rho)t_{2},\alpha}(\br_1,\br)\Big\}\Delta^{k_2}
\,\,, \\
\nonumber
{\cal N}_{pq}^{t_{1}t_{2}}(\br,\br_1)&=&
\Big\{ h^{t_{1}t_{2} ,\alpha}_{p}(\br,\br_1)
g_{dd}^{t_{1}t_{2},\alpha}(\br,\br_1)
C_{d,pq}^{t_{2},\alpha}(\br_1) \\
\nonumber
&~& \Big[-\rho_{0}^{t_{2},\alpha}(\br,\br_1)+
N_{cc}^{t_{2},\alpha}(\br,\br_1)\Big]\\
&&+g_{dd}^{t_{1}t_{2}}(\br,\br_1)C_{d,pq}^{t_{2},\alpha}(\br_1)
N^{t_{2},\alpha}_{cc,p}(\br,\br_1)-
N_{cc,p}^{(\rho)t_{2},\alpha}(\br_1,\br) \Big\}
\Delta^{k_2}
\,\,,
\end{eqnarray}
with $q=2k_2-1+l$. The other terms are defined in 
Appendix \ref{sec:app-fhncsoc}.

The knowledge of the quasi-hole functions allows us to calculate the
spectroscopic factors:
\begin{equation}
S_{nlj}^t=\int dr\,r^2\,|\psi_{nlj}^t(r)|^2
\,\,.
\label{eq:res-sf}
\end{equation}
%

%%%%%%%%%%%%%%%%%%%%%%%%%%%%%%%%%
% Figure spectroscopic factors
%%%%%%%%%%%%%%%%%%%%%%%%%%%%%%%%%
\begin{figure}[]
\begin{center}
\includegraphics [scale=0.50]{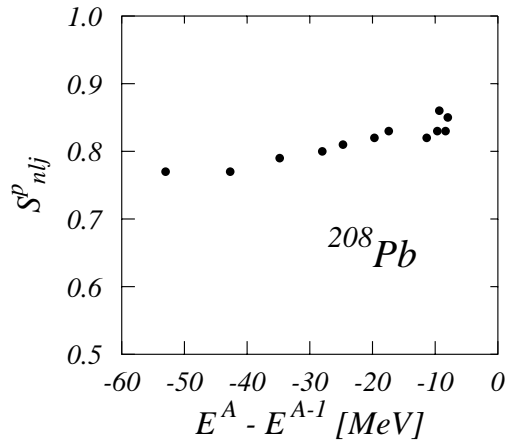}
\caption{\small 
 Protons spectroscopic factors of the \lead nucleus as a function of
 the separation energies. 
}
\label{fig:res-sf} 
\end{center}
\end{figure}
%%%%%%%%%%%%%%%%%%%%%%%%%%%%%%%%%%

The inclusion of the correlations produce spectroscopic factors
smaller than one, the mean-field value.  In general, this effect
increases together with the complexity of the correlation.  The $f_6$
results are smaller than those of $f_4$, which are smaller than those
obtained with $f_1$.

%%%%%%%%%%%%%%%%%%%%%%%%%%%%%%%%%%%%%%%%%%%%
% figure quasi-hole 
%%%%%%%%%%%%%%%%%%%%%%%%%%%%%%%%%%%%%%%%%%%%
\begin{figure}[]
\begin{center} 
\includegraphics[scale=0.50]{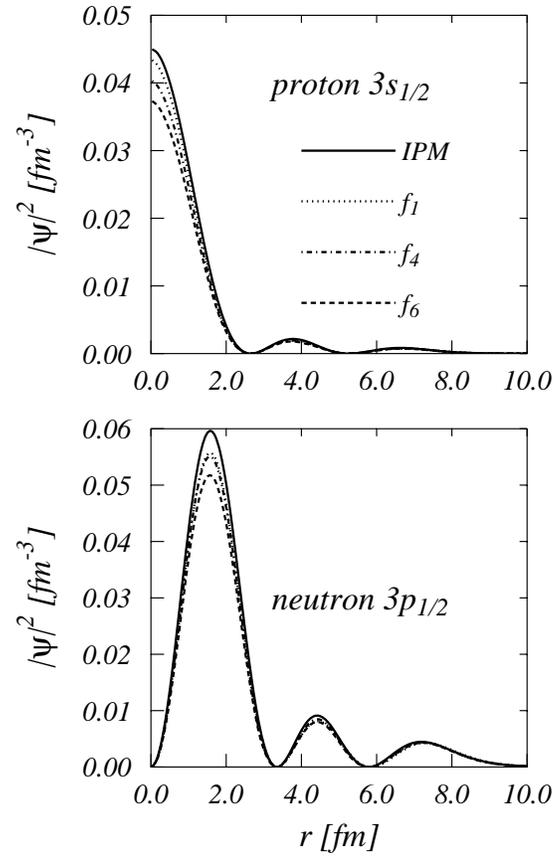} 
\caption{\small Square of the \lead proton $3s_{1/2}$ 
and neutron  $3p_{1/2}$ quasi-hole functions.
The various lines show the results obtained by using different type of
correlations. 
}
\label{fig:res-qh_fun}
\end{center} 
\end{figure}
%%%%%%%%%%%%%%%%%%%%%%%%%%%%%%%%%%%%%%%%%%%%

We found that in our calculations the effects of the correlations are
larger on the internal shells \cite{bis07}.  This fact emerges by
observing that for a fixed set of $lj$ quantum numbers, the
spectroscopic factors increases with $n$ and at the same time, that
values of the spectroscopic factors become larger when $n$ and 
$lj$ values increase.  This effect is well represented in Fig.
\ref{fig:res-sf} where we show with the black points the proton
spectroscopic factors of the \lead nucleus as a function of the
separation energies, defined as the difference between the energy of
the A-nucleons system and that of the correspondent A-1-nucleons
system.  We have associated the spectroscopic factor of each level to
its empirical separation energy.  The behavior of the black points of
the figure indicates that, in our calculations, correlation effects
are larger on the more bound levels.

In Fig. \ref{fig:res-qh_fun}, as example of the correlations effects on
the quasi-hole wave functions, we show the squares of the proton
$3s_{1/2}$ and neutron $3p_{1/2}$ quasi-hole wave functions.  The
global effect is a lowering of the wave function in the nuclear
interior, and this effect increases with increasing complexity of the
correlation.

%%%%%%%%%%%%%%%%%%%%%%%%%%%%%%%%%%%%%%%%%%%%%%%%%%%%
% Figure 205Tl
%%%%%%%%%%%%%%%%%%%%%%%%%%%%%%%%%%%%%%%%%%%%%%%%%%%%
\begin{figure}[]
\begin{center}
\includegraphics[scale=0.60] {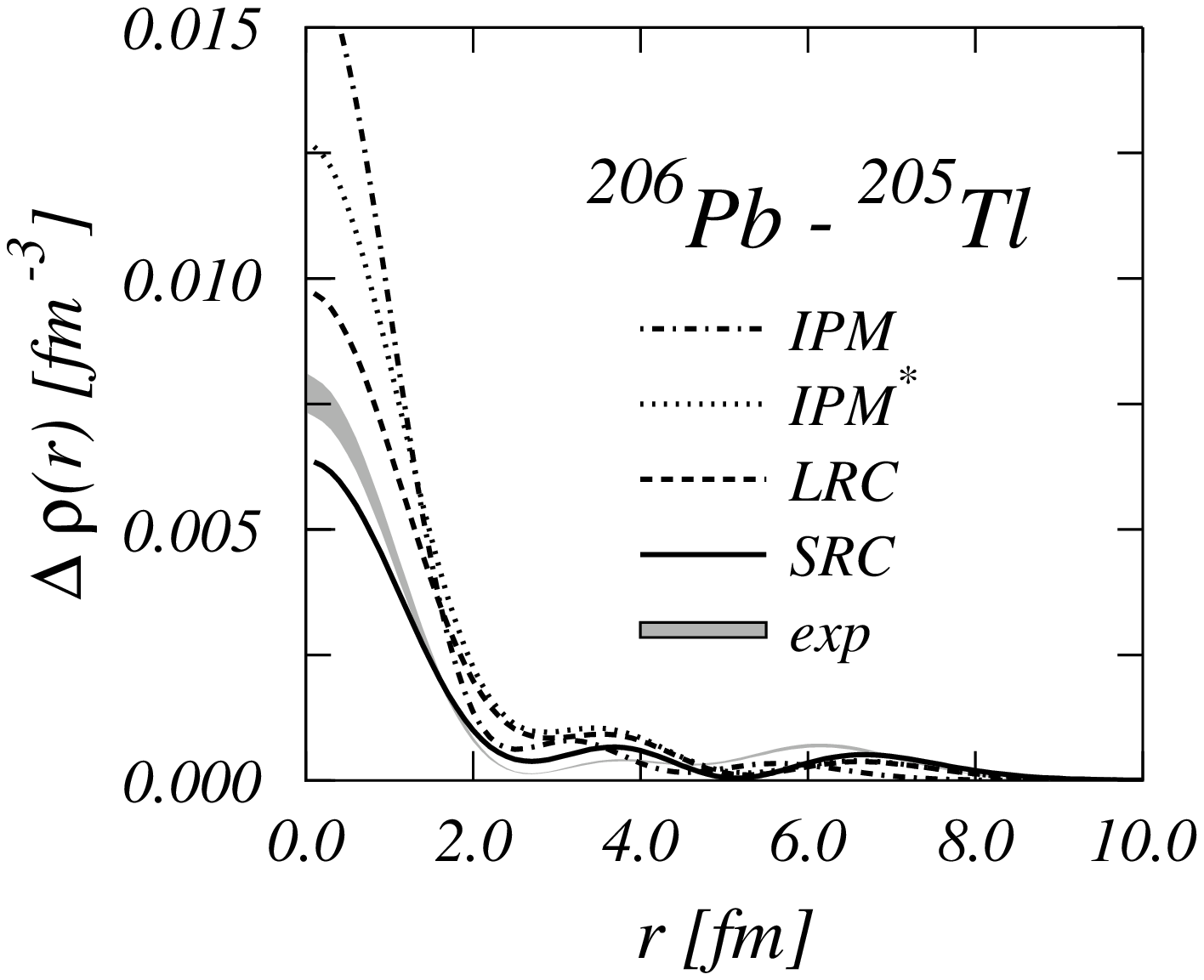} 
\caption{\small Differences between charge density distributions of 
$^{206}$Pb and $^{205}$Tl. See the text for the explanation of the
various lines. 
}
\label{fig:res-205tl}
\end{center}
\end{figure}
%%%%%%%%%%%%%%%%%%%%%%%%%%%%%%%%%%%%%%%%%%%%%%%%%%%%

In Fig. \ref{fig:res-205tl} we show with a gray band the difference
between the empirical charge distributions of $^{206}$Pb and
$^{205}$Tl \cite{cav82}. The dashed dotted line, labeled as IPM, has
been obtained by using the fact that the difference between the two
charge distributions can be described as a single $3s_{1/2}$ proton
hole in the core of the lead nucleus.  This curve has been obtained by
folding the IPM line of Fig. \ref{fig:res-qh_fun} with the electric proton
form factor in its dipole form.  In a slightly more elaborated
picture, the ground state of the $^{205}$Tl is composed of the
$3s_{1/2}$ proton hole in the $^{206}$Pb ground state, plus the
coupling of the $2d_{5/2}$ and $2d_{3/2}$ protons levels with the
first $2^+$ excited state of $^{206}$Pb \cite{zam75,kle76}.  This
description of the $^{205}$Tl, charge distribution, labeled IPM$^*$,
and shown by the dotted line, is still in a IPM framework. The dashed
line has been obtained by adding to the dotted line the core
polarization effects produced by long-range correlations. These
effects have been calculated by following the Random Phase
Approximation approach of Refs. \cite{co87c,ang01}. The full line has
been obtained when our SRC effects are also included.

The various effects presented in Fig. \ref{fig:res-205tl} have been
obtained in different theoretical frameworks, and the final result
does not have any pretense of being a well grounded and coherent
description of the empirical charge differences.  The point we want to
make by showing this figure is that the effects of the SRC are of the
same order of magnitude as those commonly considered in nuclear
structure calculations.

\section{Perspectives}
\label{sec:per}

In this section we give a short review of the possible developments of
the theory. We present only those topics which have been already
formulated.  Some of these subjects have been already well studied in
nuclear matter, and the work to be done consists in adapting the
formulation to the case of finite nuclei.  Other topics are still at
the level of a very abstract general formulation, valid for any kind
of many-body systems.  We first present some possible applications of
the theory, which do not require changes in its basic hypotheses.
Then, we discuss some extensions of the theory.

The main goal of the RFHNC/SOC equations is the evaluation of the
hamiltonian expectation value, in order to apply the variational
principle. Once the parameters of the correlation function and of the
mean-field potential which minimize the energy functional, Eq.
(\ref{eq:in-varp}), have been found, it is relatively straightforward
to apply the RFHNC/SOC equations to evaluate expectation values of
other operators. This is the strategy used in Sect. \ref{sec:results},
to calculate all the quantities other than the energy.

So far our approach is aimed at the description of the nuclear ground
state, therefore we can only obtain expectation values of ground state
observables.  On the other hand, a clever use of completeness
relations allows us to get information on excited states by
calculating expectation values of operators between ground states.
This is, for example, the case of the sum rules. The enhancement
factor of the electric-dipole sum rule has been calculated in nuclear
matter \cite{fab85}, and the same approach can be applied to finite
nuclei.

Dynamical response functions and hole spectral functions have been
calculated in nuclear matter by using the FHNC/SOC formalism
\cite{fan87,ben89,fab89,fab97}.  The responses of the system for a
momentum transfer $\bqu$ and an excitation energy $\omega$ have been
evaluated by using the expression \cite{fet71}:
\beq
S(\bqu,\omega) =  \frac{1}{\pi}
{\rm Im} \Bigg( 
<\Psi_0 | 
{\displaystyle \frac{ \rho^\dagger(\bqu) \rho(\bqu)} 
{H - E_0 - \omega -i \eta} }
| \Psi_0> <\Psi_0 | \Psi_0 >^{-1} \Bigg)
\,\,,
\label{eq:per-response}
\eeq
where $\Psi_0$ and $E_0$ are the ground state wave function and energy
respectively, $H$ the nuclear hamiltonian and $\rho$ the external
operator exciting the system. Analogous expressions have been used to
calculate the spectral functions. Also in this case the completeness
of the excited states has been used, and the response is expressed as
the ground state expectation value of an operator.  Formally, the cluster
expansion and the RFHNC/SOC resummation techniques can be applied
without any major problems to evaluate these expectation values.
However, from the pragmatical point of view, we have to consider that
the expression of the global operator is extremely involved, as is 
shown by Eq. (\ref{eq:per-response}).  This global operator combines
the operator describing the effect of the external probe, usually a
relatively simple one-body operator, with the hamiltonian
(\ref{eq:ene-hami}) composed by one- two- and three-body terms. The
effort will be rewarded by the results, since the evaluation of the
response functions gives direct access to the calculation of the cross
sections, and of other observables.

The RFHNC/SOC theory can be applied to describe hypernuclei by
considering the hyperon as an impurity in the nucleonic fluid. The
FHNC equations for an impurity in homogeneous matter have been derived
in Refs. \cite{fab82,bor94} to describe the presence of atomic $^4$He
in liquid $^3$He. A first application of this theory to the case of
single $\Lambda$ hypernuclei has been done in Ref. \cite{ari01}. In
these calculations we used simple interactions and correlations. The
nucleon-nucleon interaction contained only the first four central
channels, and the $\Lambda$-nucleon interaction the scalar and spin
channels only. All the correlations were purely scalar functions.

%%%%%%%%%%%%%%%%%%%%%%%%%%%%%%%
% 
% Figure of the hyperon binding energy 
%
%%%%%%%%%%%%%%%%%%%%%%%%%%%%%%
\vskip 0.5 cm 
\begin{figure}[htb]
\begin{center}
\includegraphics[scale=0.5] {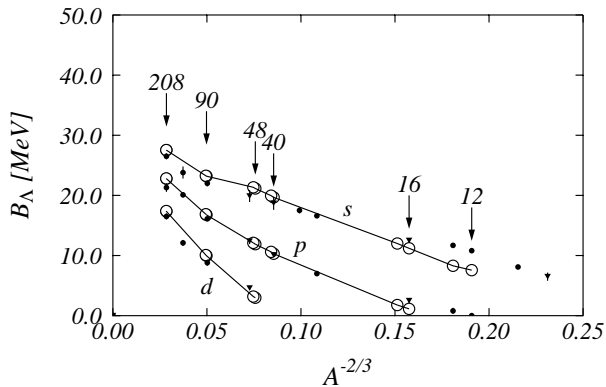}
\end{center}
\caption{\small 
  Binding energy of the $\Lambda$ hyperon for the 1$s$, 1$p$ and 1$d$
  states as a function of $A^{-2/3}$. The white circles are the
  energies calculated as indicated in Ref. \cite {ari01} by using a
  Woods-Saxon mean-field well for the hyperon. The experimental
  energies are from Ref. \cite{pil91} (triangles) and \cite{has96}
  (squares).  The full lines connecting the theoretical values have
  been drawn to guide the eye.
  }
\label{fig:per-hyperon}
\end{figure}
\vskip 0.5 cm 
%%%%%%%%%%%%%%%%%%%%%%%%%%%%%%

Despite the simple ingredients used in the calculations, the agreement
with the experimental $\Lambda$ binding energies of Refs.
\cite{pil91,has96} is rather good, as is seen in Fig.
\ref{fig:per-hyperon}. The extension of the formalism of Ref.
\cite{ari01} to the case of operator dependent correlations, which
allows us to deal with realistic hamiltonians, is technically rather
involved. On the other hand, the results of Fig. \ref{fig:per-hyperon}
are very encouraging and the potentialities of our technology to make
predictions are wide, and rather unique.

In the introduction, we have mentioned that our approach is the lowest
order approximation of the CBF theory formulated at the beginning of
the '60s \cite{cla66a,cla66b,fee69}.  The starting point of the CBF
theory is the construction of a basis of normalized, but in general
non-orthogonal, state vectors of the type:
\beq
|\Psi_m(1,...,A)>= \frac 
{F(1,...,A)|\Phi_m(1,....,A)>} 
{<\Phi_m| F^\dagger F |\Phi_m>^{1/2}}
\label{eq:per-cbfstate} 
\,\,,
\eeq
where $F$ is the many-body correlation function acting on a complete
orthonormal set of model states $|\Phi_m>$. The CBF theory constructs
a coherent perturbation theory on the correlated basis formed by the
states given by Eq. (\ref{eq:per-cbfstate}), \cite{kro02}.  In the
limit $F \rightarrow 1$ the application of the variational
principle would provide the Hartree-Fock equations, and the excited
states (\ref{eq:per-cbfstate}) would be constructed as particle-hole
excitations of the single particle basis. The so-called residual
interaction would mix these single particle excitations in a
perturbative expansion.  It appears evident how our calculations,
based on the application of the variational principle
(\ref{eq:in-varp}), can be considered, in the CBF framework, as the
first step of a perturbation expansion.  This first step is necessary
to fix the correlation function $F$ and the wave functions basis.

An improvement of our work consists in the inclusion of
higher-order perturbative terms in the CBF expansion. This has already
been done for nuclear matter, where perturbative corrections to the
binding energy \cite{fan83}, momentum distributions \cite{fan84},
responses \cite{fan87,fab89,fab97}, and spectral functions
\cite{ben89,ben90}, have been evaluated. With the help of the
perturbative expansion, the nucleon-self energy in nuclear matter
has also been calculated and the optical potential has been evaluated
\cite{fan82,fan83}. 

The use of the CBF theory allows the description of nuclear excited
states. In this respect, they can be treated within  the
correlated Random Phase Approximation theory, whose basic equations
have been obtained by using a time-dependent Hartee-Fock approach
\cite{cla81,kro82,def98t,kro02}. 

Another line of development of our theory consists in modifying the
Jastrow anstaz (\ref{eq:in-jastrow}). For example, correlation
functions composed of scalar Jastrow function and linear
state-dependent correlations have been proposed in Ref. \cite{pie92}.
These correlations have been applied to the description of light
nuclei using Variational Monte Carlo techniques with promising results
\cite{bue02,pra03}.  With linear state-dependent correlations, the
structure of our RFHNC equations becomes simpler. However, this
approach has serious drawbacks, since it requires the use of up to
six-body distribution functions.

The developments described so far are thought to be applied to doubly
closed shell nuclei. The description of other nuclei is related to a
change of the single particle basis, which should consider open shell
and deformations.  Also in this case, there are no problems in
principal in applying our theory, but the technicalities are rather
involved.

\section{Closing remarks}
\label{sec:conclusions}

The theory we have presented aims to describe the properties of
medium-heavy nuclei by solving the many-body Schr\"odinger equation
with microscopic nucleon-nucleon interactions.  This is an {\sl ab
  initio} calculation, since there are no free parameters to be
adjusted. However, the technique of solving the many-body
Schr\"odinger equation requires some approximations.

The most basic approximation is related to the use of the variational
principle. The search for the minimum of the energy functional, Eq.
(\ref{eq:in-varp}), is done by spanning the space of many-body wave
functions which can be expressed as a product of a correlation
function times an independent particle model wave function, see Eq.
(\ref{eq:in-trial}). The many-body correlation function is expressed
as a product of two-body correlation functions which hinder two
nucleons from approaching each other for distances smaller than the
range of the strongly repulsive core of the nucleon-nucleon
interaction. The use of the ansatz (\ref{eq:in-trial}) allows an
explicit treatment of a great number of correlations. However, these
are not all the possible correlations, and they are specifically
related to the strongly repulsive core of the interaction. Other types
of correlations, such as those produced by collective motions of the
nucleons, are not so well described. Since this is the starting
hypothesis of the theory, we cannot test its validity within the
theory itself. The only possibility of making this test, is a
comparison of our results with those obtained by other microscopic
theories which are approximation free, such as Quantum Monte Carlo
calculations.  The validity of the variational ansatz has been tested
in the literature by comparing Variational Monte Carlo and Quantum
Monte Carlo results for light nuclei. This comparison, done for nuclei
with A$<$8, indicates a minimum difference of 0.1 MeV per nucleon in
$^4$He and a maximum one of about 1.0 MeV per nucleon in $^8$Li
\cite{wir00}. Hence, this difference seems to increase with the number
of nucleons.

The evaluation of the variational energy functional (\ref{eq:in-varp})
in Ref. \cite{wir00} is done with a Monte Carlo integration technique,
i. e. without approximation. In our approach we calculate the energy
functional by doing a cluster expansion. After a topological analysis
of the various diagrams, we sum two categories of diagrams, the nodal
and the composite ones, in closed form, by solving the RFHNC integral
equations. The procedure we have used does not consider a certain type
of diagrams, the elementary ones. This is another approximation of our
computational scheme. The role of the elementary diagrams has been
studied in quantum liquids and in nuclear matter, and they have been
found to be more important in the former systems than in the latter
ones. This is because, in many-body jargon, liquid helium is denser
than nuclear matter. This means that the number of particles in a
volume characterized by the range of the repulsive core of the
interaction, is larger in liquid helium than in nuclear matter.

We have calculated in Ref. \cite{co92} the contribution of the
elementary diagram of Fig. \ref{fig:inf-elementary}, the simplest one.
The calculation has been done for a simple interaction, the B1 force
of Brink and Boeker \cite{bri67}, with scalar correlations and in
model nuclei with isospin degeneracy and single particle wave
functions treated in the $ls$ coupling scheme. We obtained for \oxy
and \caI a repulsive contribution of about 1.1 MeV per nucleon.  This
is about 2\% of the contribution of the potential energy produced
by the interaction. However, the total energy is obtained as a
difference between kinetic and potential energies, and, in this case,
1 MeV is not negligible. The contribution of the elementary diagrams
should be further investigated.

The third approximation of our computational scheme is the SOC. It is
clear that the {\it ad hoc} SOC approximation breaks the formal
completeness of the FHNC theory which holds when only scalar correlations 
are used.  We have difficulties in making estimates of the
contribution to the energy of the diagrams excluded by the SOC
approximation. We can only say that the sum rules exhaustions are
slightly worsened when the SOC approximation is used. We are talking
in any case of few parts on a thousand. On the other hand, we cannot
exclude that the diagrams not considered are irrelevant for the sum
rules, but not totally negligible for the calculation of the energy.

The three approximations just discussed are somehow intrinsic to the
computational scheme. We now talk about the simplifications we have
done in the specific applications of the RFHNC/SOC formalism.  
In our calculation we have used a nucleon-nucleon interaction limited
to the first eight channels. Modern interactions, with isospin
symmetry breaking terms, have up to eighteen channels. Our correlation
functions are also limited to the first six channels. We have
estimated the contribution of these missing terms of the interaction
and of the correlation by doing a bold extrapolation of nuclear matter
results. The results of this estimate are shown in
Tab. \ref{tab:con-v910}. The contribution of the neglected channels is
small when compared with the contribution of the interaction term,
$v^6_{2-body}$ in Tab. \ref{tab:res-energy}. However, since the energy
is obtained by a subtracting potential energy to the kinetic one,
their contribution is not negligible on the final result. 

%%%%%%%%%%%%%%%%%%%%%%%%%%%%%%%%%%%%%%%%%%%%
% Table correction
%%%%%%%%%%%%%%%%%%%%%%%%%%%%%%%%%%%%%%%%%%%%
\vskip 0.5 cm 
\begin{table}[ht]
\begin{center}
\begin{tabular}{|l|ccccc|}
\hline
                  & \car & \oxy & \caI & \caII & \lead   \\ 
\hline
$<k_F>$           & 1.09 & 1.09 & 1.19 & 1.19  & 1.21    \\ 
$E$               &-0.91 &-4.48 &-4.58 &-4.14  & -3.43   \\ 
$\Delta E$        &-0.93 &-0.93 &-1.37 &-1.37  & -1.48   \\ 
$E$+$\Delta E$ &-1.84 &-5.41 &-5.95 &-5.51  & -4.91   \\ 
\hline
\end{tabular}
\end{center}
\caption{\small Average Fermi momenta $<k_F>$, in fm$^{-1}$, used to
estimate the corrections to the binding energies produced by the
interaction channels beyond the spin-orbit ones, from nuclear matter
calculations. All the energy values are expressed in MeV.
The $E$ rows gives the results of Tab.
\protect\ref{tab:res-energy} for the AV8'+UIX interaction. The $\Delta
E$ rows values have been obtained by doing a local density
approximation interpolation of the nuclear matter results given in
Tab. III of Ref. \protect\cite{fab00}.
}
\label{tab:con-v910}
\end{table}
%%%%%%%%%%%%%%%%%%%%%%%%%%%%%%%%%%%%%%%%%%%%

Our minimization procedure has been done on two parameters only, the
healing distance for all the central correlation channels, and that of
the two tensor correlations. In principle, our model could handle six
variational parameters for the correlation, plus other thirteen
describing the mean field Woods-Saxon potential generating protons
and neutrons single particle states. The choice of using only two
variational parameters, dictated mainly by computational reasons,
could seem rather limiting. In reality, we are confident that
our minima are rather close to the minima that a full minimization
with all the variational parameters could obtain.
Concerning the correlations, there is a nuclear matter experience
\cite{pan79} indicating that the four central correlations heal at
the same distance, noticeably different from the tensor healing
distance. We have done some calculations in \caI by changing the
healing distances of all the correlations, and we did not find
significant improvements of the energy value.  The case of the single
particle basis has been discussed with some detail in Sect.
\ref{sec:res-spwf}. Also in this case, we have tested that, when the
minimization on the correlation function has been done, even large
changes of the mean-field potential do not produce sensitive changes in
the energy minimum.

At present, the most evident problem of our calculations is related to
the behavior of the three-body force. In few body and light nuclei the
contribution of this part of the hamiltonian is attractive, while in
our calculations it is repulsive. This happens also in variational
nuclear matter calculations \cite{wir88,akm98}.  There are two
possible solution of the puzzle. The one which should be
explored first, is that our present treatment of the three-body force
is not accurate enough.  For example, we could find that the set of
diagrams we are now considering, see Sect. \ref{sec:ene-v3}, is not
sufficient, or that the inclusion of a three-nucleon correlation
function is necessary. If after improving the description of the
three-body force, its contribution remains repulsive, we have to
deduce that the three-body interactions we use, are tailored to
provide good descriptions of few body systems and light nuclei, but
they are not adequate for medium, heavy and infinite nuclear systems.

The three-body force puzzle, is an example of the potentiality of our
approach, and, in general, of all the microscopic calculations, in 
nuclear structure. These calculations allows us to investigate
phenomena that mean-field based effective theories cannot study. We
have pointed out a few examples, by showing the effects of the SRC on
momentum distributions, natural orbits, quasi-hole wave functions and
spectroscopic factors. But the relevance of microscopic calculations
goes beyond that, since these calculations have reached such an
accuracy to put constraints on the nuclear hamiltonian itself.
Microscopic nuclear structure calculations will become more and more
important in the near future, and, among them, the RFHNC/SOC
computational scheme will play a relevant role.

\vskip 1.0 cm 
\section*{Final note} 
We had just started to write this article, when Adelchi Fabrocini passed
away. The reader who had the privilege of knowing him is certainly
aware of the absence of his touch in the writing of this article.
Adelchi's contribution to the realization of the work presented here,
has been enormous and fundamental.  We miss his talent, his
leadership, his experience and, not least, his subtle sense of humor.
\vskip 1.0 cm
\section*{Acknowledgments}
We thank S. Fantoni who triggered the idea of the project and, with
us, started the work. We thank P. Folgarait and I. E. Lagaris for
their contributions in the early stages of the work. A special thank
to P. Rotelli whose careful, and critical, reading of the
manuscript helped us a lot in improving the presentation.  This work
has been partially supported by the agreement INFN-CICYT, by the
Spanish Ministerio de Educaci\'on y Ciencia (FIS2005-02145) and by
the MURST through the PRIN {\sl Teoria della struttura nucleare e
  della materia nucleare}.
\newpage
%{\bf Appendices}
\appendix
\section{Matrices of the spin and isospin traces}
\label{sec:app-matrix}
In this appendix we give the numerical values of the various matrices
used to calculate the spin and isospin traces, when state dependent
correlations are used. 

%%%%%%%%%%%%%%%%%%%%%%%%%%%%%%%%%%%%%%%%%%%%%%%
% Table Delta
%%%%%%%%%%%%%%%%%%%%%%%%%%%%%%%%%%%%%%%%%%%%%%%
%
%\vskip 0.5 cm
\begin{table}[ht]
\begin{center}
\begin{tabular}{|c|c|c|c|c|c|c|c}
\hline
$p$ & $1$ & $2$ & $3$ & $4$ & $5$ & $6$ \\ \hline
$\Gamma^p$ & $1$ & $1$ & $1$ & $1$ & $0$ & $0$ \\ \hline
\end{tabular}
\end{center}
\caption{\small The values of $\Gamma^p$ defined in Eq. 
(\protect\ref{eq:inf-siexch}). The values of $p$ indicate the
operator channel.
}
\label{tab:inf-delta}
\end{table}
%%%%%%%%%%%%%%%%%%%%%%%%%%%%%%%%%%%%%
% Table of A^p values traces 
%%%%%%%%%%%%%%%%%%%%%%%%%%%%%%%%%%%%%%
%\vskip 0.5 cm
\begin{table}[ht]
\begin{center}
\begin{tabular}{|c|c|c|c|c|c|c|c}
\hline
$p$ & $1$ & $2$ & $3$ & $4$ & $5$ & $6$ \\ \hline
$B^p$ & $1$ & $3$ & $3$ & $9$ & $6$ & $18$ \\ \hline
\end{tabular}
\end{center}
\caption{\small The values of $B^p$ defined in Eq. 
(\protect\ref{eq:inf-ctrace2}). 
}
\label{tab:inf-ap}
\end{table}
%%%%%%%%%%%%%%%%%%%%%%%%%%%%%%%%%%%%%%
% Table K^pqr
%%%%%%%%%%%%%%%%%%%%%%%%%%%%%%%%%%%%%%
\vskip 0.5 cm
\begin{table}[ht]
{\scriptsize
\begin{center}
\begin{tabular}{|cccccccccccccccccccc|} 
\hline
$K^{pqr}$   &   &  &   &q   & &   &   &  & &   &   &   &   &   &   &  &  & &  \\
   &   & p &  &    &1& 2 & 3 & 4&5& 6 &   &   &   & 1 & 2 &3 &4&5&6 \\
&&&&&&&&&&&&&&&&&&&\\
   \cline{6-11} \cline{15-20} 
r=1&   &1  &  &    &1& 0 & 0 & 0&0& 0 &   &r=2&   &0& 1 & 0 & 0&0& 0\\
   &   &2  &  &    &0& 3 & 0 & 0&0& 0 &   &   &   &1&-2 & 0 & 0&0& 0\\
   &   &3  &  &    &0& 0 & 3 & 0&0& 0 &   &   &   &0& 0 & 0 & 3&0& 0\\
   &   &4  &  &    &0& 0 & 0 & 9&0& 0 &   &   &   &0& 0 & 3 & -6&0& 0\\
   &   &5  &  &    &0& 0 & 0 & 0&6& 0 &   &   &   &0& 0 & 0 & 0&0& 6\\
   &   &6  &  &    &0& 0 & 0 & 0&0& 18 &   &   &  &0& 0 & 0 & 0&6& -12 \\
   \cline{6-11} \cline{15-20} 
&&&&&&&&&&&&&&&&&&&\\
   \cline{6-11} \cline{15-20} 
r=3&   &1  &  &    &0& 0 & 1 & 0&0 &0 &   &r=4&   & 0 & 0 &0 &1&0&0\\
   &   &2  &  &    &0& 0 & 0 & 3&0 &0 &   &   &   & 0 & 0 &1 &-2&0&0\\
   &   &3  &  &    &1& 0 &-2 & 0&0 &0 &   &   &   & 0 & 1 &0 &-2&0&0\\
   &   &4  &  &    &0& 3 & 0 &-6&0 &0 &   &   &   & 1 & -2 &-2 &4&0&0\\
   &   &5  &  &    &0& 0 & 0 & 0& 2&0 &   &   &   & 0 & 0 &0 &0&0&2\\
   &   &6  &  &    &0& 0 & 0 & 0& 0&6 &   &   &   & 0 & 0 &0 &0&2&-4\\
   \cline{6-11} \cline{15-20} 
&&&&&&&&&&&&&&&&&&&\\
   \cline{6-11} \cline{15-20} 
r=5&   &1  &  &    &0& 0 & 0 & 0&1& 0 &   &r=6&   & 0 & 0 &0 &0&0&1\\
   &   &2  &  &    &0& 0 & 0 & 0&0& 3 &   &   &   & 0 & 0 &0 &0&1&-2\\
   &   &3  &  &    &0& 0 & 0 & 0&1& 0 &   &   &   & 0 & 0 &0 &0&0&1\\
   &   &4  &  &    &0& 0 & 0 & 0&0& 3 &   &   &   & 0 & 0 &0 &0&1&-2\\
   &   &5  &  &    &1& 0 & 1 & 0&-2& 0 &   &   &   & 0 & 1 &0 &1&0&-2\\
   &   &6  &  &    &0& 3 & 0 & 3&0& -6 &   &   &   & 1 &-2 &1 &-2&-2&4\\
   \cline{6-11} \cline{15-20} 
&&&&&&&&&&&&&&&&&&&\\
\hline 
\end{tabular} 
\end{center}
}
\vskip 0.5 cm
\caption{\small The values of the matrix $K^{pqr}$ 
defined in Eq. 
(\protect\ref{eq:inf-kmatrix}). The values of $p$,$q$ and $r$ indicate the
operator channel.}
\label{tab:inf-kpqr}
\end{table}
%%%%%%%%%%%%%%%%%%%%%%%%%%%%%%%%%%%%%%%%
% Table Lpqr
%%%%%%%%%%%%%%%%%%%%%%%%%%%%%%%%%%%%%%%%
\vskip 0.5 cm 
\begin{table}[ht]
{\scriptsize
\begin{center}
\begin{tabular}{|cccccccccccccccccccc|} 
\hline
$L^{pqr}$   &   &  &   &q   & &   &   &  & &   &   &   &   &   &   &  &  & &  \\
   &   & p &  &    &1& 2 & 3 & 4&5& 6 &   &   &   & 1 & 2 &3 &4&5&6\\
&&&&&&&&&&&&&&&&&&&\\
   \cline{6-11} \cline{15-20} 
r=1&   &1  &  &    &1& 0 & 0 & 0&0& 0 &   &r=2&   &0& 3 & 0 & 0&0& 0 \\
   &   &2  &  &    &0& 3 & 0 & 0&0& 0 &   &   &   &3& 6 & 0 & 0&0& 0 \\
   &   &3  &  &    &0& 0 & 3 & 0&0& 0 &   &   &   &0& 0 & 0 & 9&0& 0 \\
   &   &4  &  &    &0& 0 & 0 & 9&0& 0 &   &   &   &0& 0 & 9 & 18&0& 0 \\
   &   &5  &  &    &0& 0 & 0 & 0&6& 0 &   &   &   &0& 0 & 0 & 0&0& 18  \\ 
   &   &6  &  &    &0& 0 & 0 & 0&0& 18 &   &   &  &0& 0 & 0 & 0&18& 36 \\
   \cline{6-11} \cline{15-20} 
&&&&&&&&&&&&&&&&&&&\\
   \cline{6-11} \cline{15-20} 
r=3&   &1  &  &    &0& 0 & 3 & 0&0& 0 &   &r=4&   & 0 & 0 &0 &9&0&0\\
   &   &2  &  &    &0& 0 & 0 & 9&0& 0 &   &   &   & 0 & 0 &9 &18&0&0\\
   &   &3  &  &    &3& 0 & 6 & 0&0& 0 &   &   &   & 0 & 9 &0 &18&0&0\\
   &   &4  &  &    &0& 9 & 0 &18&0& 0 &   &   &   & 9 & 18 &18 &36&0&0\\
   &   &5  &  &    &0& 0 & 0 & 0&-6&0 &   &   &   & 0 & 0 &0 &0&0&-18\\
   &   &6  &  &    &0& 0 & 0 & 0&0&-18&   &   &   & 0 & 0 &0 &0&-18&-36\\
   \cline{6-11} \cline{15-20} 
&&&&&&&&&&&&&&&&&&&\\
   \cline{6-11} \cline{15-20} 
r=5&   &1  &  &    &0& 0 & 0 & 0&6& 0 &   &r=6&   & 0 & 0 &0 &0&0&18\\
   &   &2  &  &    &0& 0 & 0 & 0&0& 18 &   &   &   & 0 & 0 &0 &0&18&36\\
   &   &3  &  &    &0& 0 & 0 & 0&-6& 0 &   &   &   & 0 & 0 &0 &0&0&-18\\
   &   &4  &  &    &0& 0 & 0 & 0&0& -18 &   &   &   & 0 & 0 &0 &0&-18&-36\\
   &   &5  &  &    &6& 0 &-6 & 0&12& 0 &   &   &   & 0 &18 &0 &-18&0&36\\
   &   &6  &  &    &0&18 & 0 & -18&0& 36 & & & & 18 & 36 &-18 &-36&36&72\\
   \cline{6-11} \cline{15-20} 
&&&&&&&&&&&&&&&&&&&\\
\hline
\end{tabular} 
\end{center}
}
\vskip 0.5 cm
\caption{\small The values of the matrix $L^{pqr}$ 
defined in Eq. 
(\protect\ref{eq:inf-lpqrdef1}). The values of $p$,$q$ and $r$ indicate the
operator channel.}
\label{tab:inf-lpqr}
\end{table}
%%%%%%%%%%%%%%%%%%%%%%%%%%%%%%%%%%%%%%%%%%%%%%%
% Table Dpq
%%%%%%%%%%%%%%%%%%%%%%%%%%%%%%%%%%%%%%%%%%%%%%%
\vskip 0.5 cm
\begin{table}[ht]
\begin{center}
\begin{tabular}{|cc|rrrrrr|}
\hline
   & p &   &   &   &   &   &    \\  
q  &   & 1 & 2 & 3 & 4 & 5 & 6  \\ 
\hline
1 &   & 0 & 0 & 0 & 0 & 0 & 0 \\
2 &   & 0 & $-\frac{4}{3}$ & 0 & $-\frac{4}{3}$ & 0 & $-\frac{4}{3}$ \\
3 &   & 0 & 0 & $-\frac{4}{3}$ & $-\frac{4}{3}$ & 
    $-\frac{4}{3}$ & $-\frac{4}{3}$ \\
4 &   & 0 & $-\frac{4}{3}$ & $-\frac{4}{3}$ & $-\frac{8}{9}$&
$ -\frac{4}{3}$ & $-\frac{8}{9}$\\
5 &   & 0 & 0 & $-\frac{4}{3}$ & $-\frac{4}{3}$ &
 $-\frac{4}{3}$ & $-\frac{4}{3}$\\
6 &   & 0 & $-\frac{4}{3}$ & $-\frac{4}{3}$ & $-\frac{8}{9}$ & 
$-\frac{4}{3}$ & $-\frac{8}{9}$ \\
\hline
\end{tabular}
\end{center}
\vskip 0.5 cm 
\caption{\small The values of the matrix $E_{pq}$ defined in Eq. 
(\protect\ref{eq:inf-dpqdef}).
}
\label{tab:inf-dpq}
\end{table}
%%%%%%%%%%%%%%%%%%%%%%%%%%%%%%%%%%%%%%%%%%%%%%%

\section{The RFHNC/SOC equations for nuclear finite systems}
\label{sec:app-fhncsoc}
We present here the set of RFHNC equations for finite nuclear systems
with different number of protons and neutrons, and in a $jj$ coupling
scheme.  The upper index $t$ distinguishes the contributions of
protons and neutrons. In this appendix, in addition to the terms
necessary to calculate the one- and two-body density functions, which
are used for the evaluation of the energy of the system, we present
also those terms which are required in the evaluation of the OBDM,
necessary ingredients for the calculation of the momentum
distribution. 

We show the expressions of the various parts necessary to
build the one- and two-body density functions used in the calculation
of the energy of the nucleus, and also those parts required to obtain
the OBDM necessary for the momentum distribution evaluation.

The various terms can be written in general as $Y_{mn}^{t_i
  t_j}(\br_i,\br_j)$. An important simplification in the writing of the
equations is the property:
\begin{equation}
Y_{mn}^{t_i t_j}(\br_i,\br_j)=Y_{nm}^{t_j t_i}(\br_j,\br_i) \,\,.
\end{equation}

We present first the set of RFHNC equations for the scalar part of the
correlation $f_1$, and, in a second step, the equations involving
operator dependent correlations, i.e. the RFHNC/SOC equations.

We start our presentation by considering the dynamical diagrams, i.e.
those where the external points are reached only by dynamical
correlations.  We used two types of dynamical correlations,
$h_d=f_1^2-1$, and $h_w=f_1-1$. This last correlation appears in the
calculation of the OBDM, where it connects only the external points.
The dynamical TBDFs can be written as:
\begin{eqnarray}
g_{dd}^{\ton\ttw}(\br_1,\br_2)&=&f_1^2(r_{12})\exp
\Big[N_{dd}^{\ton\ttw}(\br_1,\br_2)+
E^{\ton\ttw}_{dd}(\br_1,\br_2)\Big] 
\label{eq:asoc-gdd}
\,\,,\\
g_{wd}^{\ton\ttw}(\br_1,\br_2)&=&f_1(r_{12})\exp
\Big[N_{wd}^{\ton\ttw}(\br_1,\br_2)+
E^{\ton\ttw}_{wd}(\br_1,\br_2)\Big]  \,\,,\\
g_{ww}^{\ton\ttw}(\br_1,\br_2)&=&\exp\Big[N_{ww}^{\ton\ttw}(\br_1,\br_2)+
E^{\ton\ttw}_{ww}(\br_1,\br_2)\Big] \,\,. 
\end{eqnarray}
These distribution functions are related to the non-nodal diagrams 
$X^{\ton\ttw}$ by the equation:
\begin{equation}
g_{mn}^{\ton\ttw}(\br_1,\br_2)=1+N_{mn}^{\ton\ttw}(\br_1,\br_2)+
X^{\ton\ttw}_{mn}(\br_1,\br_2)
\,\,, 
\end{equation}
with $m,n=d,w$. In the presence of purely scalar correlation functions,
the dynamical functions do not depend on the isospin of the external
particles. However, in the above equations, we wrote the explicit
isospin dependence in view of the treatment with the state dependent
correlations.

The next step is to consider the case when the  dynamical correlations
link only one external point, let's say point $1$, while the other
point is reached by the statistical correlations, forming an exchange
loop. In this case, the functions to be considered are: 
\begin{eqnarray}
\nonumber g_{me}^{\ton\ttw}(\br_1,\br_2)& = & g_{md}^{\ton\ttw}(\br_1,\br_2)
\Big[N_{me}^{\ton\ttw}(\br_1,\br_2)+E_{me}^{\ton\ttw}(\br_1,\br_2)\Big] \\
&=& N_{me}^{\ton\ttw}(\br_1,\br_2)+X_{me}^{\ton\ttw}(\br_1,\br_2)\,\,,
\end{eqnarray}
with $m=d,w$. 

In the case when both the two external points are reached by exchange
loops, the equations are:
\begin{eqnarray}
\nonumber g_{ee}^{\ton\ttw}(\br_1,\br_2)&=&
        g_{ee,dir}^{\ton\ttw}(\br_1,\br_2)+
        2\delta_{\ton\ttw}g_{ee,exc}^{\ton\ttw}(\br_1,\br_2)+ 
        2\delta_{\ton\ttw}g_{ee,excj}^{\ton\ttw}(\br_1,\br_2)\\ 
        &=&N_{ee}^{\ton\ttw}(\br_1,\br_2)+X^{\ton\ttw}_{ee}(\br_1,\br_2)\,\,,\\
\nonumber g_{ee,dir}^{\ton\ttw}(\br_1,\br_2)&=&g_{dd}^{\ton\ttw}(\br_1,\br_2)
\Big[N_{ee}^{\ton\ttw}(\br_1,\br_2)+E_{ee,dir}^{\ton\ttw}(\br_1,\br_2)\\
\nonumber &+&\Big(N_{ed}^{\ton\ttw}(\br_1,\br_2)+E_{ed}^{\ton\ttw}
(\br_1,\br_2)\Big) \\  & &
\Big(N_{de}^{\ton\ttw}(\br_1,\br_2)+E_{de}^{\ton\ttw}(\br_1,\br_2)\Big)\Big]
\,\,,\\
\nonumber 
g_{ee,exc}^{\ton\ttw}(\br_1,\br_2)&=&g_{dd}^{\ton\ttw}(\br_1,\br_2) \Big[
           E_{ee,exc}^{\ton\ttw}(\br_1,\br_2) \\
\nonumber 
&-&\Big(N_{cc}^{\ton}(\br_1,\br_2)
           +E_{cc}^{\ton}(\br_1,\br_2)-\rho_0^{\ton}(\br_1,\br_2)\Big) \\
             &&\Big(N_{cc}^{\ttw}(\br_1,\br_2)+
           E_{cc}^{\ttw}(\br_1,\br_2)-\rho_0^{\ttw}(\br_1,\br_2)\Big) \Big]\,\,,\\
\nonumber
          g_{ee,excj}^{\ton\ttw}(\br_1,\br_2)&=&g_{dd}^{\ton\ttw}(\br_1,\br_2) \Big[
          E_{ee,excj}^{\ton\ttw}(\br_1,\br_2) \\
\nonumber 
&-& \Big(N_{ccj}^{\ton}(\br_1,\br_2)+
           E_{ccj}^{\ton}(\br_1,\br_2)-\rho_{0j}^{\ton}(\br_1,\br_2)\Big)\\
           &&\Big(N_{ccj}^{\ttw}(\br_1,\br_2)+
           E_{ccj}^{\ttw}(\br_1,\br_2)-\rho_{0j}^{\ttw}(\br_1,\br_2)\Big) \Big]\,\,,
\end{eqnarray}
In the above expression the subindex $dir$ denotes the case when the
two external particles are linked to different statistical loops, and
$exc(j)$ when they are connected to the same loop. In this last case,
a spin-isospin exchange operator is present, whose trace in the
isospin space is $2\delta_{\ton\ttw}$.  In the $exc(j)$ case, we
further distiguish the parallel spin case, $exc$, from the
antiparallel one, $excj$.

Finally, in the construction of the exchange parts, we have to
introduce the contribution of diagrams with open statistical loops.
These open loops appear in both the calculation of the energy and of
the OBDM.  We have to distinguish, also in this situation, the case
where the external particles have parallel or antiparallel spins. In
either case, the exchange loops may be combined with the two kinds of
dynamical correlations, therefore we define:
\begin{eqnarray}
\nonumber 
g_{cc(j)}^{\ton}(\br_1,\br_2)&=&g_{dd}^{\ton\ton}(\br_1,\br_2) \\
& &  \Big[N_{cc(j)}^{\ton}(\br_1,\br_2)+E_{cc(j)}^{\ton}(\br_1,\br_2)
          -\rho_{0(j)}^{\ton}(\br_1,\br_2)\Big] \,\,,\\
\nonumber
g_{w_cc(j)}^{\ton}(\br_1,\br_2)&=&g_{wd}^{\ton\ton}(\br_1,\br_2) \\ &&
          \Big[N_{w_cc(j)}^{\ton}(\br_1,\br_2)+E_{w_cc(j)}^{\ton}(\br_1,\br_2)
          -\rho_{0(j)}^{\ton}(\br_1,\br_2)\Big] \,\,,\\
\nonumber
g_{w_cw_c(j)}^{\ton}(\br_1,\br_2)&=&g_{ww}^{\ton\ton}(\br_1,\br_2) \\ &&
\Big[N_{w_cw_c(j)}^{\ton}(\br_1,\br_2)+E_{w_cw_c(j)}^{\ton}(\br_1,\br_2)
          -\rho_{0(j)}^{\ton}(\br_1,\br_2)\Big] ,
\end{eqnarray}
where with $(j)$ we have indicated the possible presence of the label
$j$. For all these six partial distribution functions we can write:
\begin{equation}
g_{mn(j)}^{\ton}(\br_1,\br_2)=N_{mn(j)}^{\ton}(\br_1,\br_2)+
X_{mn(j)}^{\ton}(\br_1,\br_2)
-\rho_{0(j)}^{\ton}(\br_1,\br_2)\,\,.
\end{equation}

The RFHNC expressions of all the nodal diagrams which do not have 
open statistical loops, can be written in a compact form as:
\begin{eqnarray}
\nonumber 
N_{mn}^{\ton\ttw}(\br_1,\br_2)& = &\sum_{\tth=p,n}\sum_{m'n'} \\ 
& &  \Big(X_{mm'}^{\ton\tth}(\br_1,\br_3)V_{m'n'}^{\tth}(\br_3) \Big|
N_{n'n}^{\tth\ttw}(\br_3,\br_2)+X_{n'n}^{\tth\ttw}(\br_3,\br_2)\Big) \,\,,
\end{eqnarray}
where $m,n=d,w,e$. In the previous equation, due to the diagrammatic
rules, the sums are limited to the values $(m'n')=dd,de,ed$.
Furthermore we have defined:
\begin{equation}
V_{mn}^{\tth}(i)= \left\{
\begin{array}{ccc} C^{\tth}(i) & {\rm for}& (mn)=dd  \\
 C_{d}^{\tth}(i) & & {\rm otherwise}  \end{array} \right. \,\,.
\label{eq:asoc-vert}
\end{equation}

The expressions of the nodal diagrams with open statistical loops in
the external particles, such as the diagrams of Fig.
\ref{fig:fin-ncc}, follow the classification presented in Eq.
(\ref{eq:fin-ncc}):
\begin{eqnarray}
\nonumber
N_{mn}^{(x)\ton}(\br_1,\br_2)&=&
  \Big(X_{mc}^{\ton}(\br_1,\br_3)C_{d}^{\ton}(\br_3)\Big| 
g_{cn}^{\ton}(\br_3,\br_2)\Big)\\
&-&\Big(X_{mcj}^{\ton}(\br_1,\br_3)C_{d}^{\ton}(\br_3)\Big| 
g_{cnj}^{\ton}(\br_3,\br_2)\Big)\,\,,\\
\nonumber
N_{mn}^{(\rho)\ton}(\br_1,\br_2)&=&
  -\Big(\rho_0^{\ton}(\br_1,\br_3)C_{d}^{\ton}(\br_3)\Big|
   N_{cn}^{(x)\ton}(\br_3,\br_2)+X_{cn}^{\ton}(\br_3,\br_2)\Big)\\
   \nonumber &-&\Big(\rho_0^{\ton}(\br_1,\br_3)(C_{d}^{\ton}(\br_3)-1)\Big|
   N_{cn}^{(\rho)\ton}(\br_3,\br_2)-\rho_0^{\ton}(\br_3,\br_2)\Big)\\
   \nonumber &+&\Big(\rho_{0j}^{\ton}(\br_1,\br_3)C_{d}^{\ton}(\br_3)\Big|
   N_{cnj}^{(x)\ton}(\br_3,\br_2)+X_{cnj}^{\ton}(\br_3,\br_2)\Big)\\
   &+&\Big(\rho_{0j}^{\ton}(\br_1,\br_3)(C_{d}^{\ton}(\br_3)-1)\Big|
   N_{cnj}^{(\rho)\ton}(\br_3,\br_2)-\rho_{0j}^{\ton}(\br_3,\br_2)\Big) ,
\label{eq:asoc-nrt1}
\\
\nonumber N_{mnj}^{(x)\ton}(\br_1,\br_2)&=&
   \Big(X_{mcj}^{\ton}(\br_1,\br_3)C_{d}^{\ton}(\br_3)\Big| 
g_{cn}^{\ton}(\br_3,\br_2)\Big)\\
&+&\Big(X_{mc}^{\ton}(\br_1,\br_3)C_{d}^{\ton}(\br_3)\Big| 
g_{cnj}^{\ton}(\br_3,\br_2)\Big)\,\,,\\
\nonumber N_{mnj}^{(\rho)\ton}(\br_1,\br_2)&=&
    -\Big(\rho_{0j}^{\ton}(\br_1,\br_3)C_{d}^{\ton}(\br_3)\Big|
     N_{cn}^{(x)\ton}(\br_3,\br_2)+X_{cn}^{\ton}(\br_3,\br_2)\Big)\\
\nonumber &-&\Big(\rho_{0j}^{\ton}(\br_1,\br_3)(C_{d}^{\ton}(\br_3)-1)\Big|
     N_{cn}^{(\rho)\ton}(\br_3,\br_2)-\rho_0^{\ton}(\br_3,\br_2)\Big)\\
\nonumber &-&\Big(\rho_0^{\ton}(\br_1,\br_3)C_{d}^{\ton}(\br_3)\Big|
     N_{cnj}^{(x)\ton}(\br_3,\br_2)+X_{cnj}^{\ton}(\br_3,\br_2)\Big)\\
&-&\Big(\rho_0^{\ton}(\br_1,\br_3)(C_{d}^{\ton}(\br_3)-1)\Big|
     N_{cnj}^{(\rho)\ton}(\br_3,\br_2)-\rho_{0j}^{\ton}(\br_3,\br_2)\Big) ,
\label{eq:asoc-nrt2}
\end{eqnarray}
where $m,n=c,w_c$. In the above equation, 
we have used the definitions:
\[ 
N_{mn(j)}^{(x) \ton}(\br_1,\br_2)=N_{mn(j)}^{xx \ton}(\br_1,\br_2)+
N_{mn(j)}^{x\rho \ton}(\br_1,\br_2)
\,\,,
\]
and 
\[
N_{mn(j)}^{(\rho) \ton}(\br_1,\br_2)=N_{mn(j)}^{\rho x \ton}(\br_1,\br_2)+
N_{mn(j)}^{\rho \rho \ton}(\br_1,\br_2)
\,\,. 
\]
It is worth to point out that Eq. (\ref{eq:asoc-nrt1}) indicates 
that $N_{mn(j)}^{(\rho) \ton}(\br_1,\br_2)$ does not depend on $m$.

The results of Sect. \ref{sec:fin-vertex} indicate that the vertex
corrections are given by:
\begin{eqnarray}
C_{m}^{\ton}(\br_1)&=&\exp[U_{m}^{\ton}(\br_1)]
\label{eq:asoc-vxe}
\,\,,\\
C^{\ton}(\br_1)&=&
C_{d}^{\ton}(\br_1)[U_{e}^{\ton}(\br_1)+\rho_0^{\ton}(\br_1)]
=\rho(\br_1)
\label{eq:asoc-vxd}
\,\,,
\end{eqnarray}
with $m=d,w$. In order to simplify the writing of the RFHNC expressions 
for $U_{d,w,e}$, we have defined the quantity:
\beq
S_{mn}^{\ton\ttw}(\br_1,\br_2) \equiv  \frac{1}{2}N_{mn}^{\ton\ttw}(\br_1,\br_2)+
E_{mn}^{\ton\ttw}(\br_1,\br_2) \,\,.
\label{eq:asoc-s}
\eeq
and we obtain:
\begin{eqnarray}
\nonumber U_{m}^{\ton}(\br_1)&=&
    \int d{\bf r}_{2}\sum_{\ttw=p,n} 
    \Big\{C^{\ttw}(\br_2)\Big[X_{md}^{\ton\ttw}(\br_1,\br_2)-E_{md}^{\ton\ttw}(\br_1,\br_2)
\\ \nonumber & & -
    S_{md}^{\ton\ttw}(\br_1,\br_2)(g_{md}^{\ton\ttw}(\br_1,\br_2)-1)\Big]\\
\nonumber &&+C_{d}^{\ttw}(\br_2)\Big[X_{me}^{\ton\ttw}(\br_1,\br_2)-
    E_{me}^{\ton\ttw}(\br_1,\br_2) \\  \nonumber & &
-S_{me}^{\ton\ttw}(\br_1,\br_2)
    (g_{md}^{\ton\ttw}(\br_1,\br_2)-1)\\
    &&-S_{md}^{\ton\ttw}(\br_1,\br_2)g_{me}^{\ton\ttw}(\br_1,\br_2)\Big]\Big\}
    +E_{m}^{\ton}(\br_1) 
\label{eq:asoc-ne}
\,\,,\\
\nonumber U_{e}^{\ton}(\br_1)&=& 
    \int d{\bf r}_{2}\sum_{\ttw=p,n}\Big\{C^{\ttw}(\br_2)
    \Big[X_{ed}^{\ton\ttw}(\br_1,\br_2)-E_{ed}^{\ton\ttw}(\br_1,\br_2)\\
\nonumber && -S_{ed}^{\ton\ttw}(\br_1,\br_2)
    (g_{dd}^{\ton\ttw}(\br_1,\br_2)-1)-S_{dd}^{\ton\ttw}(\br_1,\br_2)
     g^{\ton\ttw}_{ed}(\br_1,\br_2)\Big]\\
\nonumber &&+C_{d}^{\ttw}(\br_2)\Big[X_{ee}^{\ton\ttw}(\br_1,\br_2)
-E_{ee}^{\ton\ttw}(\br_1,\br_2) \\
\nonumber &&-S_{ee}^{\ton\ttw}(\br_1,\br_2)
    (g_{dd}^{\ton\ttw}(\br_1,\br_2)-1)-S_{ed}^{\ton\ttw}(\br_1,\br_2)g_{de}^{\ton\ttw}(\br_1,\br_2)
    \\ \nonumber &&-
    S_{de}^{\ton\ttw}(\br_1,\br_2)g_{ed}^{\ton\ttw}(\br_1,\br_2)
-S_{dd}^{\ton\ttw}(\br_1,\br_2)g_{ee}^{\ton\ttw}(\br_1,\br_2)\Big]\\
\nonumber &&+2\delta_{\ton\ttw}\Big[C_{d}^{\ttw}(\br_2)
    [S_{cc}^{\ton}(\br_1,\br_2)g_{cc}^{\ttw}(\br_1,\br_2)+S_{ccj}^{\ton}(\br_1,\br_2)
    g_{ccj}^{\ttw}(\br_1,\br_2)]\\
\nonumber && -\rho_0^{\ton}(\br_1,\br_2)
    (N_{cc}^{(\rho)\ttw}(\br_1,\br_2)-\rho_0^{\ttw}(\br_1,\br_2))
\\ & & -\rho_{0j}^{\ton}(\br_1,\br_2)(
    N_{ccj}^{(\rho)\ttw}(\br_1,\br_2)-\rho_{0j}^{\ttw}(\br_1,\br_2))\Big]\Big\}
    +E_{e}^{\ton}(\br_1) \,\,,
\label{eq:asoc-nend}
\end{eqnarray}
with $m=w,d$. 
The expressions (\ref{eq:asoc-gdd}-\ref{eq:asoc-nend}) form the close
set of RFHNC equations, valid for the scalar part of the correlation.
In terms of these quantities and defining other useful ones, we can
express the TBDF and the OBDM as:
\begin{eqnarray}
\rho_2^{1,\ton \ttw}(\br_1,\br_2) & = & 
\rho_{2,dir}^{\ton \ttw}(\br_1,\br_2) + 2 \delta_{\ton \ttw}
\Big(
\rho_{2,exc}^{\ton \ttw}(\br_1,\br_2)+ 
\rho_{2,excj}^{\ton \ttw}(\br_1,\br_2) \Big) \\
\nonumber
\rho_{2,dir}^{\ton \ttw} (\br_1,\br_2) & = &  C^{\ton}(\br_1)
\bigl(C^{\ttw}(\br_2) g_{dd}^{\ton \ttw}(\br_1,\br_2)+
 C_{d}^{\ttw}(\br_2) g_{de}^{\ton \ttw}(\br_1,\br_2)\bigr)\\
 &+&C_{d}^{\ton}(\br_1)
\bigl(C^{\ttw}(\br_2) g_{ed}^{\ton \ttw}(\br_1,\br_2) + 
C_{d}^{\ttw}(\br_2) g_{ee,dir}^{\ton \ttw}(\br_1,\br_2) \bigr)
\label{eq:aene-dens1}
\,\,,
\\
\rho_{2,exc}^{\ton \ttw} (\br_1,\br_2) & = &
 C_{d}^{\ton} (\br_1)C_{d}^{\ttw} (\br_2)
g_{ee,exc}^{\ton \ttw}(\br_1,\br_2)
\,\,,
\\
\rho_{2,excj}^{\ton \ttw} (\br_1,\br_2) & = &
 C_{d}^{\ton} (\br_1)C_{d}^{\ttw} (\br_2)
g_{ee,excj}^{\ton \ttw}(\br_1,\br_2)
\,\,,
\label{eq:aene-dens} \\
\rho^{\ton} (\br_1,\br_2) & = &-2
 C_{w}^{\ton} (\br_1)C_{w}^{\ton} (\br_2)
g_{w_cw_c}^{\ton \ton}(\br_1,\br_2)
\end{eqnarray}

We discuss now the case of state-dependent correlations, in the SOC
approximation.  We have mentioned in Sect. \ref{sec:fin-stdc}, the
need of separating the spin and isospin dependence of the TBDF in order
to describe nuclei not saturated in isospin.  In these systems, the
contribution of the linear isospin operators is not zero, therefore
we distiguish the proton and neutron dependence of the various
RFHNC/SOC terms.  This affects the chain equations in the calculation
of the nodal $N_{xy,p}^{\ton\ttw}(\br_1,\br_2)$ functions.  In order
to generate these chains, we have to consider the following folding
products: those of a function $X_{xy,p}(\br_1,\br_3)$ with another
function $X_{xy,q}(\br_3,\br_2)$, those of $X_{xy,p}(\br_1,\br_3)$
with $N_{xy,q}(\br_1,\br_3)$, and those of $X_{xy,p}(\br_1,\br_3)$,
or $N_{xy,q}(\br_1,\br_3)$, with $\rho_0(\br_3,\br_2)$. These
combinations are present also in isospin saturated systems.  In
addition, we should also consider that the action of a single isospin
operator on a single external point of the nodal function, produces a
non zero contribution.

We give in the following the expressions of the vertex corrections of
the nodal terms in SOC approximation, and also those of the various
terms of the TBDF. In these expressions we use the index $k$ which
can assume the values 1,2 and 3 and the index $l$ which can be 0 or 1.
The $p$, $q$ and $r$ indexes indicate the operator channels and can
assume values from 1 up to 6.

The expressions of the nodal functions without open statistical loops
can be written as:
\begin{eqnarray}
\nonumber
N_{mn,2k_{1}-1}^{\ton\ttw}(\br_1,\br_2)&=&
     \sum_{k_{2}k_{3}=1}^{3}\sum_{\tth=p,n}
     \Big[N_{mn,2k_{1}-1,2k_{2}-1,2k_{3}-1}^{\ton\ttw\tth}(\br_1,\br_2)\\
\nonumber &+& (2\delta_{\ton\tth}-1)
          N_{mn,2k_{1}-1,2k_{2},2k_{3}-1}^{\ton\ttw\tth}(\br_1,\br_2)\\
          &+& (2\delta_{\ttw\tth}-1) 
          N_{mn,2k_{1}-1,2k_{2}-1,2k_{3}}^{\ton\ttw\tth}(\br_1,\br_2)\Big]\,\,,
\\
          N_{mn,2k_{1}}^{\ton\ttw}(\br_1,\br_2)&=&\sum_{k_{2},k_{3}=1}^{3}
          \sum_{\tth=p,n}N_{mn,2k_{1},2k_{2},2k_{3}}^{\ton\ttw\tth}(\br_1,\br_2) \,\,,
\label{eq:asoc-soc1}
\end{eqnarray}
where we have defined:
\begin{eqnarray}
\hspace*{-5mm}
\nonumber
N_{mn,pqr}^{\ton\ttw\tth}(\br_1,\br_2)&=&\Big(X_{md,q}^{\ton\tth}(\br_1,\br_3)
\zeta_{132}^{k_{2}k_{3}k_{1}}C_{qr}^{\tth}(\br_3)\Big|
    X_{dn,r}^{\tth\ttw}(\br_3,\br_2)+N_{dn,r}^{\tth\ttw}(\br_3,\br_2)\Big)\\
\nonumber&+&\Big(X_{me,q}^{\ton\tth}(\br_1,\br_3)
    \zeta_{132}^{k_{2}k_{3}k_{1}}C_{d,qr}^{\tth}(\br_3)\Big|
    X_{dn,r}^{\tth\ttw}(\br_3,\br_2)+N_{dn,r}^{\tth\ttw}(\br_3,\br_2)\Big)\\
\nonumber 
    &+& \Big(X_{md,q}^{\ton\tth}(\br_1,\br_3)
    \zeta_{132}^{k_{2}k_{3}k_{1}}C_{d,qr}^{\tth}(\br_3)\Big|
    X_{en,r}^{\tth\ttw}(\br_3,\br_2)+N_{en,r}^{\tth\ttw}(\br_3,\br_2)\Big) .
\\
\end{eqnarray}
In the above equations we have defined
$p=2k_{1}-1+l_{1}$, $q=2k_{2}-1+l_{2}$, and 
$r=2k_{3}-1+l_{3}$. The sub-indexes $m$ and $n$ indicate the type of
link with the two external points, specifically $m,n=d,w,e$.

Since in the calculations of $ww$ diagrams, we must include the
isospin trace, for this case, we substitute Eq.(\ref{eq:asoc-soc1})
with:
\begin{equation}
N_{ww,2k_{1}}^{\ton\ton}(\br_1,\br_2)=\sum_{k_{2},k_{3}=1}^{3}
\sum_{\tth=p,n}\chi_2^{\ton \tth}
N_{ww,2k_{1},2k_{2},2k_{3}}^{\ton\ton\tth}(\br_1,\br_2) \,\,.
\end{equation}

The expressions of the nodal diagrams with open statistical loops
are:
\begin{eqnarray}
N_{mn,p}^{\ton}(\br_1,\br_2)&=&N_{mn,p}^{(x)\ton}(\br_1,\br_2)+
N_{mn,p}^{(\rho)\ton}(\br_1,\br_2)\hspace{0.1in} p=1,\ldots,6 
\label{eq:asoc-nmne}
\,\,,\\
N_{mnj,p}^{\ton}(\br_1,\br_2)&=&
N_{mnj,p}^{(x)\ton}(\br_1,\br_2)+N_{mnj,p}^{(\rho)\ton}(\br_1,\br_2) 
\hspace{0.1in} p=1,2
\label{eq:asoc-nmnej}
\,\,,
\end{eqnarray}
with $m,n=c,w_c$. We have used a symbology analogous to that of
Eqs. (\ref{eq:asoc-nrt1} - \ref{eq:asoc-nrt2}), to indicate the parallel
spin case, Eq. (\ref{eq:asoc-nmne}), and the antiparallel one, Eq.
(\ref{eq:asoc-nmnej}). In this last case, we have considered only the
contribution of the first two channels of the interaction, $p=1,2$, in
order to simplify our calculations.  This is a good approximation
since the relevance of the antiparallel loops is small. The explicit
expressions of the nodal functions present in the Eqs.
(\ref{eq:asoc-nmne}) and (\ref{eq:asoc-nmnej}) are:
\begin{eqnarray}
\nonumber
N_{mn(j),2k_{1}-1}^{(y)\ton}(\br_1,\br_2)&=&
 \sum_{k_{2},k_{3}=1}^{3}\sum_{\tth=p,n}
 \Big[N_{mn(j),2k_{1}-1,2k_{2}-1,2k_{3}-1}^{(y)\ton\tth}(\br_1,\br_2)\\
\nonumber &&+(2\delta_{\ton\tth}-1)
  \Big(N_{mn(j),2k_{1}-1,2k_{2},2k_{3}-1}^{(y)\ton\tth}(\br_1,\br_2)+\\
  &&N_{mn(j),2k_{1}-1,2k_{2}-1,2k_{3}}^{(y)\ton\tth}(\br_1,\br_2)\Big)\Big] \,\,,
\label{eq:asoc-ncc1}
\\
N_{mn(j),2k_{1}}^{(y)\ton}(\br_1,\br_2)&=&
 \sum_{k_{2},k_{3}=1}^{3}\sum_{\tth=p,n}
  N_{mn(j),2k_{1},2k_{2},2k_{3}}^{(y)\ton\tth}(\br_1,\br_2) \,\,,
\label{eq:asoc-ncc2}
\end{eqnarray}
where $y=x,\rho$. The other terms are:
\begin{eqnarray}
\nonumber
N_{mn,pqr}^{(x)\ton\tth}(\br_1,\br_2)&=&
  \Big(X_{mc,q}^{\ton}(\br_1,\br_3)\zeta_{132}^{k_{2}k_{3}k_{1}}
   C_{d,qr}^{\tth}(\br_3)\frac{\Delta^{k_{3}}}{2}\Big|
   g_{cn}^{\tth}(\br_3,\br_2)\Big)\\ \nonumber 
   &+&(1-\delta_{r,1})\Big(X_{mc}^{\ton}(\br_1,\br_3)\frac{\Delta^{k_{2}}}{2}
   \zeta^{k_{2}k_{3}k_{1}}_{132}C_{d,qr}^{\tth}(\br_3)\Big|
   g_{cn,r}^{\tth}(\br_3,\br_2)\Big) \\ \nonumber 
   &-& \delta_{k_1 1} \delta_{k_2 1} \delta_{k_3 1} \Big[
 \Big(X_{mcj,q}^{\ton}(\br_1,\br_3)C_{d,qr}^{\tth}(\br_3)\frac{1}{2}\Big|
     g_{cnj}^{\tth}(\br_3,\br_2)\Big)\\
     &+&(1-\delta_{r,1})\Big(X_{mcj}^{\ton}(\br_1,\br_3)C_{d,qr}^{\tth}(\br_3)\frac{1}{2}
\Big| g_{cnj,r}^{\tth}(\br_3,\br_2)\Big) \Big]\,\,,\\
\nonumber
N_{mn,pqr}^{(\rho)\ton\tth}(\br_1,\br_2)&=&
   -\Big(\rho_{0}^{\ton}(\br_1,\br_3)\frac{\Delta^{k_{2}}}{2}
   \zeta^{k_{2}k_{3}k_{1}}_{132}C_{d,qr}^{\tth}(\br_3)\Big|
    X_{cn,r}^{\tth}(\br_3,\br_2)\Big)\\
\nonumber 
   &-&\Big(\rho_{0}^{\ton}(\br_1,\br_3)\frac{\Delta^{k_{2}}}{2}
   \zeta^{k_{2}k_{3}k_{1}}_{132}C_{d,qr}^{\tth}(\br_3)\Big|
    N_{cn,r}^{(x)\tth}(\br_3,\br_2)\Big)\\
    \nonumber 
    &-&\Big(\rho_0^{\ton}(\br_1,\br_3)\frac{\Delta^{k_{2}}}{2}
    \zeta^{k_{2}k_{3}k_{1}}_{132}(C_{d,qr}^{\tth}(\br_3)-1)
    \Big|N_{cn,r}^{(\rho)\tth}(\br_3,\br_2)\Big)\\ \nonumber
    &+&\delta_{r,1}\Big(\rho_0^{\ton}(\br_1,\br_3)\frac{\Delta^{k_{2}}}{2}
    \zeta^{k_{2}k_{3}k_{1}}_{132}(C_{d,qr}^{\tth}(\br_3)-1)
    \Big|\rho_0^{\tth}(\br_3,\br_2)\Big) \\
\nonumber &+&\frac{\delta_{k_1 1} \delta_{k_2 1} \delta_{k_3 1}} 2 \Big[
     \Big(\rho_{0j}^{\ton}(\br_1,\br_3)C_{d,qr}^{\tth}(\br_3)\Big|
     X_{cnj,r}^{\tth}(\br_3,\br_2)\Big)\\ \nonumber 
     &+&\Big(\rho_{0j}^{\ton}(\br_1,\br_3)C_{d,qr}^{\tth}(\br_3)\Big|
     N_{cnj,r}^{\tth}(\br_3,\br_2)\Big)\\ \nonumber 
     &+&
\Big(\rho_{0j}^{\ton}(\br_1,\br_3)(C_{d,qr}^{\tth}(\br_3)-1)\Big|
N_{cnj,r}^{\tth}(\br_3,\br_2) \Big) \\
     &-&\delta_{r,1}
\Big(\rho_{0j}^{\ton}(\br_1,\br_3)(C_{d,qr}^{\tth}(\br_3)-1)\Big|
\rho_{0j}^{\tth}(\br_3,\br_2)\Big)\Big] ,\\
\nonumber N_{mnj,pst}^{(x)\ton\tth}(\br_1,\br_2)&=&
    \Big(X_{mc,s}^{\ton}(\br_1,\br_3)C_{d,st}^{\tth}(\br_3)\frac{1}{2}\Big|
     g_{cnj}^{\tth}(\br_3,\br_2)\Big)\\
\nonumber &+&(1-\delta_{t,1})\Big(X_{mc}^{\ton}(\br_1,\br_3)
     C_{d,st}^{\tth}(\br_3)\frac{1}{2}\Big|
     g_{cnj,t}^{\tth}(\br_3,\br_2)\Big)\\
\nonumber &+&\Big(X_{mcj,s}^{\ton}(\br_1,\br_3)
           C_{d,st}^{\tth}(\br_3)\frac{1}{2}\Big|
           g_{cn}^{\tth}(\br_3,\br_2)\Big)\\
           &+&(1-\delta_{t,1})\Big(X_{mcj}^{\ton}(\br_1,\br_3)C_{d,st}^{\tth}(\br_3)
           \frac{1}{2}\Big|
           g_{cn,t}^{\tth}(\br_3,\br_2)\Big)\,\,,\\
\nonumber N_{mnj,pst}^{(\rho)\ton\tth}(\br_1,\br_2)&=&
-\Big(\rho_0^{\ton}(\br_1,\br_3)C_{d,st}^{\tth}(\br_3)\frac{1}{2}\Big|
X_{cnj,t}^{\tth}(\br_3,\br_2)+N_{cnj,t}^{(x)\tth}(\br_3,\br_2)\Big)\\
\nonumber &-&\Big(\rho_0^{\ton}(\br_1,\br_3)(C_{d,st}^{\tth}(\br_3)-1)
\frac{1}{2}\Big| N_{cnj,t}^{(\rho)\tth}(\br_3,\br_2)\Big)\\ 
\nonumber &+&\delta_{t,1}
\Big(\rho_0^{\ton}(\br_1,\br_3)(C_{d,st}^{\tth}(\br_3)-1)
\frac{1}{2}\Big|\rho_{0j}^{\tth}(\br_3,\br_2)\Big)\\ 
\nonumber 
&-&\Big(\rho_{0j}^{\ton}(\br_1,\br_3)C_{d,st}^{\tth}(\br_3)\frac{1}{2}\Big|
          X_{cn,t}^{\tth}(\br_3,\br_2)+N_{cn,t}^{(x)\tth}(\br_3,\br_2)\Big)\\
\nonumber 
          &-& \Big(\rho_{0j}^{\ton}(\br_1,\br_3)(C_{d,st}^{\tth}(\br_3)-1)
\frac{1}{2}\Big|
          N_{cn,t}^{(\rho)\tth}(\br_3,\br_2)\Big) \\
          &+& \delta_{t,1}
\Big(\rho_{0j}^{\ton}(\br_1,\br_3)(C_{d,st}^{\tth}(\br_3)-1)
\frac{1}{2}\Big|\rho_0^{\tth}(\br_3,\br_2)\Big) \,.
\label{eq:asoc-nccend}
\end{eqnarray}
In the above equations the symbols $s$ and $t$ can assume the values 1
and 2.

In the calculations of $w_cw_c$ diagrams, we must include the
isospin trace, therefore we use:
\begin{equation}
N_{w_cw_c,2k_{1}}^{\ton}(\br_1,\br_2)=\sum_{k_{2},k_{3}=1}^{3}
\sum_{\tth=p,n}\chi_2^{\ton \tth}
N_{w_cw_c,2k_{1},2k_{2},2k_{3}}^{\ton\tth}(\br_1,\br_2) \,\,.
\end{equation}

The expressions of the TBDFs are rather similar to those of the
symmetric nuclear matter case. We define the quantities:
\begin{eqnarray}
h_{dd,p}^{\ton\ttw}(\br_1,\br_2) & = &  \frac{2f_{p}(r_{12})}
    {f_{1}(r_{12})}+N_{dd,p}^{\ton\ttw}(\br_1,\br_2) \,\,, \\
h_{wd,p}^{\ton\ttw}(\br_1,\br_2) & = &  \frac{f_{p}(r_{12})}
    {f_{1}(r_{12})}+N_{wd,p}^{\ton\ttw}(\br_1,\br_2) \,\,, \\
h_{ww,p}^{\ton\ttw}(\br_1,\br_2) & = &  N_{ww,p}^{\ton\ttw}(\br_1,\br_2) \,\,, 
\end{eqnarray}
with $p=2k-1+l>1$. We obtain:
\begin{eqnarray}
\nonumber g_{mn,p}^{\ton\ttw}(\br_1,\br_2)&=&
    g_{mn}^{\ton\ttw}(\br_1,\br_2)h_{mn,p}^{\ton\ttw}(\br_1,\br_2) \\
\nonumber
    &=&X_{mn,p}^{\ton\ttw}(\br_1,\br_2)+N_{mn,p}^{\ton\ttw}(\br_1,\br_2)\,\,,\\
\nonumber
g_{me,p}^{\ton\ttw}(\br_1,\br_2)&=&
    g_{me}^{\ton\ttw}(\br_1,\br_2)h_{md,p}^{\ton\ttw}(\br_1,\br_2)+
    g_{md}^{\ton\ttw}(\br_1,\br_2)N_{me,p}^{\ton\ttw}(\br_1,\br_2)\\
    &=& X_{me,p}^{\ton\ttw}(\br_1,\br_2)+N_{me,p}^{\ton\ttw}(\br_1,\br_2)\,\,,\\
\nonumber g_{ee,p}^{\ton\ttw}(\br_1,\br_2)&=&g_{ee,dir,p}^{\ton\ttw}(\br_1,\br_2)+
            g_{ee,exc,p}^{\ton\ttw}(\br_1,\br_2)+g_{ee,excj,p}^{\ton\ttw}(\br_1,\br_2)\\
            &=&X_{ee,p}^{\ton\ttw}(\br_1,\br_2)+N_{ee,p}^{\ton\ttw}(\br_1,\br_2)\,\,,\\
\nonumber g_{ee,dir,p}^{\ton\ttw}(\br_1,\br_2)&=&
          g_{ee,dir}^{\ton\ttw}(\br_1,\br_2)h_{dd,p}^{\ton\ttw}(\br_1,\br_2) \\
\nonumber
&+& g_{dd}^{\ton\ttw}(\br_1,\br_2)N_{ee,p}^{\ton\ttw}(\br_1,\br_2)\\
\nonumber 
&+& g_{de}^{\ton\ttw}(\br_1,\br_2)N_{ed,p}^{\ton\ttw}(\br_1,\br_2) \\
&+& g_{ed}^{\ton\ttw}(\br_1,\br_2)N_{de,p}^{\ton\ttw}(\br_1,\br_2)\,\,,\\
 g_{ee,exc,p}^{\ton\ttw}(\br_1,\br_2)&=&\Delta^{k}
             g_{ee,exc}^{\ton\ttw}(\br_1,\br_2)\,\,,\\
g_{ee,excj,p}^{\ton\ttw}(\br_1,\br_2)&=&\Delta^{k}
              g_{ee,excj}^{\ton\ttw}(\br_1,\br_2)\,\,,
\end{eqnarray}
with $m,n=d,w$.
 
For the other diagrams we can write:
\begin{eqnarray}
\nonumber 
g_{cc(j),p}^{\ton}(\br_1,\br_2)&=&
g_{cc(j)}^{\ton}(\br_1,\br_2)h_{dd,p}^{\ton\ton}(\br_1,\br_2)
\\ &+& 
g_{dd}^{\ton\ton}(\br_1,\br_2)N_{cc(j),p}^{\ton}(\br_1,\br_2) \,\,,\\ 
\nonumber 
g_{w_cc(j),p}^{\ton}(\br_1,\br_2)&=&
g_{w_cc(j)}^{\ton}(\br_1,\br_2)h_{wd,p}^{\ton\ton}(\br_1,\br_2)
\\ &+& 
g_{wd}^{\ton\ton}(\br_1,\br_2)N_{w_cc(j),p}^{\ton}(\br_1,\br_2) \,\,,\\ 
\nonumber 
g_{w_cw_c(j),p}^{\ton}(\br_1,\br_2)&=&
g_{w_cw_c(j)}^{\ton}(\br_1,\br_2)h_{ww,p}^{\ton\ton}(\br_1,\br_2)
\\ &+& 
g_{ww}^{\ton\ton}(\br_1,\br_2)N_{w_cw_c(j),p}^{\ton}(\br_1,\br_2) \,\,.
\label{eq:asoc-wcwc}
\end{eqnarray}

All the TBDFs can be expressed in terms of nodal and non nodal
diagrams, and for all of them, we can write:
\begin{equation}
 g_{mn(j),p}^{\ton}(\br_1,\br_2)= 
X_{mn(j),p}^{\ton}(\br_1,\br_2)+N_{mn(j),p}^{\ton}(\br_1,\br_2) \,\,.  
\end{equation}

The expressions of the vertex corrections for the operator dependent
part of the correlations are:
\begin{eqnarray}
C^{\ton}_{m,pq}(\br_1)&=&C_{m}^{\ton}(\br_1)\Big[1+
\delta_{pq,11}U_{m,SOC}^{\ton}(\br_1)\Big]
\label{eq:asoc-vxte}
\,\,, \\
\nonumber 
C^{\ton}_{pq}(\br_1) &=& 
C_{d,pq}^{\ton}(\br_1)\Big[\rho_0^{\ton}(\br_1)+
U_{e}^{\ton}(\br_1)\Big] \\
&+&  \delta_{pq,11}C_{d}^{\ton}(\br_1)\Big[U^{\ton}_{e,SOC}(\br_1)+
              U^{\ton}_{ej,SOC}(\br_1)\Big]
\label{eq:asoc-cd} 
\,\,,
\end{eqnarray}
with $m=w,d$ and where we have used:
\begin{eqnarray}
\nonumber
U_{m,SOC}^{\ton}(\br_1)&=&
\sum_{k_{1}=1}^{3}A^{k}\sum_{\ttw=p,n}
\Bigg\{
             (1-\delta_{k_{1}1})U_{m,2k_{1}-1,2k_{1}-1}^{\ton\ttw}(\br_1)\\
\nonumber  &+&\chi_{1}^{\ton\ttw}\Big[U_{m,2k_{1}-1,2k_{1}}^{\ton\ttw}(\br_1)+
              U_{m,2k_{1},2k_{1}-1}^{\ton\ttw}(\br_1)\Big]\\
           &+&\chi_{2}^{\ton\ttw}U_{m,2k_{1},2k_{1}}^{\ton\ttw}(\br_1)
\Bigg\}
\label{eq:asoc-u1} 
\,\,,
\\
\nonumber
U_{ej,SOC}^{\ton}(\br_1)&=& \sum_{k_{1}k_{2}=1}^{3}I^{k_{1}k_{2}2}
\sum_{\ttw=p,n}\Bigg\{
              U_{ej,2k_{1}-1,2k_{2}-1}^{\ton\ttw}(\br_1)\\
\nonumber  &+&\chi_{1}^{\ton\ttw}\Big[U_{ej,2k_{1}-1,2k_{2}}^{\ton\ttw}(\br_1)+
              U_{ej,2k_{1},2k_{2}-1}^{\ton\ttw}(\br_1)\Big]\\
           &+&\chi_{2}^{\ton\ttw}U_{ej,2k_{1},2k_{2}}^{\ton\ttw}(\br_1)
\Bigg\}
\,\,,
\end{eqnarray}
and we have considered $m=d,w,e$, the relation  
$p=2k_{1}-1+l_{1}$ and $q=2k_{2}-1+l_{2}$.
The expressions of the $U$ coefficients are:
\begin{eqnarray}\nonumber U_{d,pq}^{\ton\ttw}(\br_1)&=&\int d{\bf r}_{2}
          h_p^{\ton \ttw} (\br_1,\br_2) 
          \Bigg\{\Big[g_{dd}^{\ton\ttw}(\br_1,\br_2)C_{pq}^{\ttw}(\br_2)
\\ \nonumber &+&
         g_{de}^{\ton\ttw}(\br_1,\br_2)C_{d,pq}^{\ttw}(\br_2)\Big]
         h_{q}^{\ton\ttw}(\br_1,\br_2)\\
         &+& g_{dd}^{\ton\ttw}(\br_1,\br_2)C_{d,pq}^{\ttw}(\br_2)
         (1-\delta_{q,1})N_{de,q}^{\ton\ttw}(\br_1,\br_2)\Bigg\}
\,\,, \\
\nonumber U_{w,pq}^{\ton\ttw}(\br_1)&=&\int d{\bf r}_{2}
          h_{w,p}^{\ton \ttw} (\br_1,\br_2)(1-\delta_{q,1})            
          \Bigg\{\Big[g_{wd}^{\ton\ttw}(\br_1,\br_2)C_{pq}^{\ttw}(\br_2)
\\ \nonumber &+&
         g_{we}^{\ton\ttw}(\br_1,\br_2)C_{d,pq}^{\ttw}(\br_2)\Big]
         N_{wd,q}^{\ton\ttw}(\br_1,\br_2)\\
         &+& g_{wd}^{\ton\ttw}(\br_1,\br_2)C_{d,pq}^{\ttw}(\br_2)
         N_{we,q}^{\ton\ttw}(\br_1,\br_2)\Bigg\} 
\,\,, \\
\nonumber U_{e,pq}^{\ton\ttw}(\br_1)&=&\int d{\bf r}_{2}
          h_p^{\ton \ttw} (\br_1,\br_2) 
\Bigg\{ \Big[g_{ed}^{\ton\ttw}(\br_1,\br_2)C_{pq}^{\ttw}(\br_2)
\\ \nonumber &+&
              g_{ee,dir}^{\ton\ttw}(\br_1,\br_2)C_{d,pq}^{\ttw}(\br_2)\Big]
               h_{q}^{\ton \ttw} (\br_1,\br_2)\\
\nonumber  &+&
g_{dd}^{\ton\ttw}(\br_1,\br_2) C_{pq}^{\ttw}(\br_2) (1-\delta_{q,1}) 
N_{ed,q}^{\ton\ttw}(\br_1,\br_2)+ g_{dd}^{\ton\ttw}(\br_1,\br_2) 
C_{d,pq}^{\ttw} (\br_2)
 \\ 
\nonumber
&\times& (1-\delta_{q,1}) 
\Big[ N_{ee,q}^{\ton\ttw} (\br_1,\br_2)
+N_{de,q}^{\ton\ttw}(\br_1,\br_2) N_{ed}^{\ton\ttw}(\br_1,\br_2)
\\ \nonumber &&
               +N_{ed,q}^{\ton\ttw}(\br_1,\br_2) N_{de}^{\ton\ttw}(\br_1,\br_2)
\Big]\Bigg\} \\
\nonumber
              &+&  (1-\delta_{p,1}) 
\Delta^{k_2}\int d{\bf r}_{2}
\Bigg\{ h_p^{\ton \ttw}(\br_1,\br_2) 
             2 g_{ee,exc}^{\ton\ttw}(\br_1,\br_2) C_{d,pq}^{\ttw}(\br_2)
\\
\nonumber
&-& N^{\ton}_{cc,p}(\br_1,\br_2) g_{cc}^{\ttw}(\br_1,\br_2) 
C_{d,pq}^{\ttw}(\br_2) \\
&+& N^{(\rho) \ton}_{cc,p}(\br_2,\br_1)  
\Big[ N^{(\rho) \ttw}_{cc}(\br_2,\br_1)  
        - \rho^{\ttw}_{0}(\br_1,\br_2) \Big] \Bigg\}
\,\,,
\\
\nonumber
U_{ej,pq}^{\ton\ttw}&=&
(1-\delta_{p,1}) 
\Delta^{k_2}\int d{\bf r}_{2}
\Bigg\{
h_{p}^{\ton \ttw} (\br_1,\br_2) \,2\, 
g_{ee,excj}^{\ton\ttw}(\br_1,\br_2) C_{d,pq}^{\ttw}(\br_2) \\
\nonumber
&-& N^{\ton}_{cc,p}(\br_1,\br_2) g_{ccj}^{\ttw}(\br_1,\br_2) 
C_{d,pq}^{\ttw}(\br_2) \\
&+& N^{(\rho) \ton}_{cc,p}(\br_2,\br_1)  
\Big[ N^{(\rho) \ttw}_{ccj}(\br_2,\br_1) 
 - \rho^{\ttw}_{0j}(\br_1,\br_2) \Big] \Bigg\}
\label{eq:asoc-socend}
\end{eqnarray}
In the above equations we have used the functions 
\begin{eqnarray}
\nonumber 
h_{q}^{\ton\ttw}(\br_1,\br_2) & = &\frac{f_{q}(r_{12})} {f_{1}(r_{12})} +
(1- \delta_{q,1})N_{dd,q}^{\ton\ttw}(\br_1,\br_2)
\,\,, \\
\nonumber 
h_{w,q}^{\ton\ttw}(\br_1,\br_2) & = &\frac{f_{q}(r_{12})} {f_{1}(r_{12})} +
(1- \delta_{q,1})N_{wd,q}^{\ton\ttw}(\br_1,\br_2)
\,\,. 
\end{eqnarray}
In addition we remark that, since there is no contribution in the case 
$p=q=1$, we have 
\[
C^{\ttw}_{m,pq}(\br_2) = C^{\ttw}_{m,22}(\br_2)
\,\,.
\]

\section{The isospin matrix elements}
\label{sec:app-iso}
In Sect.  \ref{sec:fin-tiso} we have shown how to calculate the
expectation value of the diagram A of Fig. \ref{fig:fin-tiso} by using
the properties of the Pauli matrices.  We found a recursive relation
which allowed us to express a set of $n$ pairs of isospin operators as
a scalar term plus a single isospin operator pair, Eq.
(\ref{eq:fin-isorec1}). This recursive relation was
used to evaluate the expectation value of a set of $n$ isospin
operator pairs between two external points, Eq.  (\ref{eq:fin-chi}).

In this Appendix we calculate the isospin expectation value for the
situations represented by the B and C diagrams of Fig.
\ref{fig:fin-tiso}.  We call these diagrams vertex correction and
nodal diagram, respectively.

The starting point is a general expression for the expectation
value of a product of isospin operators between the external points 1
and 2, and a generic internal point, we label it as 3. Since this
expression is symmetric in the external points,
we write it in a general manner and we understand that the $i$ and $j$ 
points can be either 1 or 2. We define the expectation value:
\begin{eqnarray}
\nonumber
 {\cal T}_{l_1 l_2 l_3 l_4 l_5}^{\ton \ttw \tth}(i,j) & \equiv &  
\chi^+_{\ton} (1) \chi^+_{\ttw} (2) \chi^+_{\tth} (3)
(\btau_1 \cdot \btau_2)^{l_1}
(\btau_i \cdot \btau_3)^{l_2}
\\
\nonumber
&&
(\btau_1 \cdot \btau_2)^{l_3}
(\btau_j \cdot \btau_3)^{l_4}
(\btau_1 \cdot \btau_2)^{l_5} 
\chi_{\ton} (1 ) \chi_{\ttw} (2) \chi_{\tth} (3) \\
&=& b_1^\prime + b_2^\prime \delta_{\ton \ttw}+ 
b_3^\prime \delta_{\ton \tth} + b_4^\prime \delta_{\ttw \tth}
+ b_5^\prime \delta_{\ton \ttw}\delta_{\ton \tth}\delta_{\ttw
\tth} \,\,,
\label{eq:aiso-calt}
\end{eqnarray}
where the $b'_{1,5}$ coefficients are real numbers depending on
$l_{1,...,5}$ which indicate the number of isospin pairs appearing in
the expression.  For the properties of the Kronecker's $\delta$ symbol
we have that:
\beq
\delta_{\ton \ttw}\delta_{\ton \tth}\delta_{\ttw \tth} =
\frac 1 2 \left(\delta_{\ton \ttw}+ 
\delta_{\ton \tth} + \delta_{\ttw \tth} -1 \right) \,\,,
\eeq
therefore we obtain:
\beq
{\cal T}_{l_1 l_2 l_3 l_4 l_5}^{\ton \ttw \tth}(i,j)=
b_1 + b_2 \delta_{\ton \ttw}+ 
b_3 \delta_{\ton \tth} + b_4 \delta_{\ttw \tth} \,\,,
\label{eq:aiso-calt1}
\eeq
where we have defined $b_1 = b_1^\prime -b_5^\prime /2$ and $b_k =
b_k^\prime +b_5^\prime /2$ for $k=2,3,4$.  By exchanging the
coordinates 1 and 2 in Eq. (\ref{eq:aiso-calt}) we find that:
\beq
 {\cal T}_{l_1 l_2 l_3 l_4 l_5}^{\ton \ttw \tth}(3-i,3-j) = 
 {\cal T}_{l_1 l_2 l_3 l_4 l_5}^{\ttw \ton \tth}(i,j) \,\,.
\label{eq:aiso-sime}
\eeq

Eq. (\ref{eq:aiso-calt}) represents the most general expression for
all the cases we want to treat. All the possible combinations of
isospin operator pairs of our calculations, can be reconducted to this
expression, by appropriately redefining the values of the $l$ powers,
and those of the $i$ and $j$ coordinates.  The structure of the
sequence of operator pairs can be interpreted as follows. The isospin
pairs with exponent $l_1$ and $l_5$ represent the operators coming
from the dynamical correlations, and, eventually those coming from a
statistical correlation line applied to the points 1 and 2.  The pair
with $l_3$ represents the operator of the interaction. The isospin
pairs with power $l_2$ and $l_4$ are associated to the operators of
the vertex correction in the case $j=i$, and to the nodal diagram in
the case $j=3-i$. Since we work in SOC approximation, there is only a
single pair of isospin operators acting on these particles, therefore
$l_2,l_4=0,1$. On the other hand, there are no limitations on the
values that the other $l$ indexes can assume. We discuss below the
three possible cases, compatible with the SOC approximation.

When $l_2=l_4=0$ we have a structure analogous to that treated in 
Sect. \ref{sec:fin-tiso} and therefore:
\beq
{\cal T}_{l_1 0 l_3 0 l_5}^{\ton \ttw \tth}(i,j) =
 \chi_{l_1+l_3+l_5}^{\ton \ttw} \,\,,
\eeq
with $\chi_{l_1+l_3+l_5}^{\ton \ttw}$ given by Eq. (\ref{eq:fin-chi}). 

For the case $l_2 \ne l_4$, obviously one of the $l$ indexes is $1$
and the other one is $0$.  The expression of Eq. (\ref{eq:aiso-calt})
can be written, with an appropriated redefinition of the power
indexes, to an expression of the kind:
\begin{eqnarray}
\nonumber
T_{l_1 l_2}^{\ton \ttw \tth}(i) & = &
\chi_{\ton}^+ (1) \chi_{\ttw}^+ (2) \chi_{\tth}^+ (3) \\
&~&
(\btau_1 \cdot \btau_2)^{l_1}
(\btau_i \cdot \btau_3)
(\btau_1 \cdot \btau_2)^{l_2}
\chi_{\ton} (1 ) \chi_{\ttw} (2) \chi_{\tth} (3) \,\,.
\label{eq:aiso-capt}
\end{eqnarray}
These coefficients are related to the ${\cal T}$ coefficients by the 
relations:
\begin{eqnarray}
{\cal T}_{l_1 0 l_3 1 l_5}^{\ton \ttw \tth}(i,j) & = &  
T_{l_1+l_3 l_5}^{\ton \ttw \tth}(j)  \,\,,
\label{app:aiso-tt1}
\\
{\cal T}_{l_1 1 l_3 0 l_5}^{\ton \ttw \tth}(i,j) & =  & 
T_{l_1 l_3+l_5}^{\ton \ttw \tth}(i)  \,\,.
\label{app:aiso-tt2}
\end{eqnarray}

By using the recursive relation (\ref{eq:fin-isorec2}), we find for 
$T_{l_1 l_2}^{\ton \ttw \tth}(i)$ the following expression:
\begin{eqnarray}
 \nonumber  
T_{l_1 l_2}^{\ton \ttw \tth}(i) 
& = & a_{l_1} a_{l_2} T_{0 0}^{\ton \ttw \tth}(i)+ 
\left[ a_{l_1}(1- a_{l_2}) + (1-a_{l_1}) a_{l_2} \right]
T_{1 0}^{\ton \ttw \tth}(i)+ \\ 
& & (1-a_{l_1})(1- a_{l_2}) T_{1 1}^{\ton \ttw \tth}(i) \,\,.
\label{eq:aiso-chi1}
\end{eqnarray}

%%%%%%%%%%%%%%%%%%%%%%%%%%%%%%%
% Table isospin traces 1
%%%%%%%%%%%%%%%%%%%%%%%%%%%%%%%
\begin{table}[ht]
\begin{center}
\begin{tabular}{|cc|rrrr|}
\hline
 & & \multicolumn{4}{c|}{$T_{l_1 l_2}^{\ton \ttw \tth}(1)$} \\
$l_1$ & $l_2$ &
$b_1$ & $b_2$ & $b_3$ & $b_4$  \\
\hline
0 & 0 & --1 & 0 & 2 & 0   \\
1 & 0 & --1 & 0 & 0 & 2   \\
1 & 1 & --1 & 0 & --2 & 4   \\
\hline
\end{tabular}
\caption{\small The values of the $b$ coefficients of
 Eq. (\ref{eq:aiso-calt1}) used to calculate the basic  
 $T_{l_1l_2}^{\ton\ttw\tth}(1)$ terms of Eq. (\ref{eq:aiso-chi1}). 
}
\label{tab:aiso-chi1} 
\end{center}
\end{table}
%%%%%%%%%%%%%%%%%%%%%%%%%%%%%%%%%%%%

The values of the $b$ coefficients for each term of Eq.
(\ref{eq:aiso-chi1}) are given in Tab. \ref{tab:aiso-chi1} and have 
been calculated by using the expression:
\beq 
\btau_i \cdot \btau_j \chi_{t_i}(i) \chi_{t_j}(j) =
2\chi_{t_j}(i) \chi_{t_i}(j)-\chi_{t_i}(i) \chi_{t_j}(j)
\,\,.
\eeq
The values presented in Tab. \ref{tab:aiso-chi1} refer to the $i=1$
case.  By using Eq. (\ref{eq:aiso-sime}) we have:
\[
T_{l_1 l_2}^{\ton \ttw \tth}(2) 
=T_{l_1 l_2}^{\ttw \ton \tth}(1)
\,\,.
\]

The last case we have to analyze is: $l_2=l_4=1$.  In this case, we can
write:
\begin{eqnarray}
{\cal T}_{l_1 1 l_3 1 l_5}^{\ton \ttw \tth}(i,j) & = & 
a_{l_1} a_{l_3} 
\left[ a_{l_5} {\cal T}_{0 1 0 1 0}^{\ton \ttw \tth}(i,j) + 
(1 -a_{l_5}) {\cal T}_{0 1 0 1 1}^{\ton \ttw \tth}(i,j) \right]+ 
\label{eq:aiso-etar}
\\ & & 
(1-a_{l_1}) a_{l_3} 
\left[ a_{l_5} {\cal T}_{1 1 0 1 0}^{\ton \ttw \tth}(i,j)+
(1 -a_{l_5}) {\cal T}_{1 1 0 1 1}^{\ton \ttw \tth}(i,j) \right]+ 
\nonumber \\ & & 
a_{l_1}(1- a_{l_3}) 
\left[ a_{l_5} {\cal T}_{0 1 1 1 0}^{\ton \ttw \tth}(i,j)+
(1 -a_{l_5}) {\cal T}_{0 1 1 1 1}^{\ton \ttw \tth}(i,j) \right] + 
\nonumber \\ & & \nonumber
(1-a_{l_1})(1- a_{l_3}) 
\left[ a_{l_5} {\cal T}_{1 1 1 1 0}^{\ton \ttw \tth}(i,j) + 
(1 -a_{l_5}) {\cal T}_{1 1 1 1 1}^{\ton \ttw \tth}(i,j) \right] \,\,. 
\end{eqnarray}
The values of the matrix elements ${\cal T}$ of the previous equations
are given in Tab. \ref{tab:aiso-chi2} for $i=1$ and $j=1,2$.  The
other cases are calculated using Eq. (\ref{eq:aiso-sime}).

%%%%%%%%%%%%%%%%%%%%%%%%%%
% Table isospin traces 2
%%%%%%%%%%%%%%%%%%%%%%%%%%
\begin{table}[htb]
\begin{center}
\begin{tabular}{|ccc|rrrr|rrrr|}
\hline
 & & & \multicolumn{4}{c|}{${\cal T}_{l_1 1 l_3 1 l_5}^{\ton \ttw \tth}(1,1)$}
& \multicolumn{4}{c|}{${\cal T}_{l_1 1 l_3 1 l_5}^{\ton \ttw \tth}(1,2)$} \\
$l_1$ & $l_3$ & $l_5$ &
$b_1$ & $b_2$ & $b_3$ & $b_4$ & 
$b_1$ & $b_2$ & $b_3$ & $b_4$  \\
\hline
0 & 0 & 0 & 5 & 0 & --4 & 0 & --1 & 2 & 0 & 0   \\
1 & 0 & 0 & --1 & 6 & 0 & --4 & 5 & --4 & 4 & --4  \\
0 & 1 & 0 &  --1 & --2 & 0 & 4 & 5 & 4 & --4 & --4  \\
0 & 0 & 1 & --1 & 6 & 0 & --4 & 5 & --4 & --4 & 4  \\
1 & 1 & 0 &  --7 & 4 & 4 & 0 & 11 & --2 & 0 & --8  \\ 
1 & 0 & 1 & 17 & --12 & 4 & --8 & --13 & 14 & 0 & 0  \\
0 & 1 & 1 &  --7 & 4 & 4 & 0 & 11 & --2 & --8 & 0  \\
1 & 1 & 1 & 11 & --14 & 8 & --4 & --7 & 16 & --4 & --4 \\
\hline
\end{tabular}
\caption{\small 
 The values of the $b$ coefficients of
 Eq. (\ref{eq:aiso-calt1}) used to calculate the basic  
 ${\cal T}_{l_1 l_2 l_3 l_4 l_5}^{\ton\ttw\tth}(1,j)$ 
 terms of Eq. (\ref{eq:aiso-etar}) with $j=1,2$.
 }
\label{tab:aiso-chi2}
\end{center}
\end{table}
%%%%%%%%%%%%%%%%%%%%%%%%%%

In order to simplify the notation in the calculation of the energy, we
rename the matrix elements in the isospin space:
\begin{eqnarray} 
\chi_{l_1 l_2 l_3 l_4 l_5}^{\ton \ttw \tth}(i) & = & 
{\cal T}_{l_1 l_2 l_3 l_4 l_5}^{\ton \ttw \tth}(i,i) 
\label{eq:aiso-chi} \\
\eta_{l_1 l_2 l_3 l_4 l_5}^{\ton \ttw \tth}(i) & = & 
{\cal T}_{l_1 L_2 l_3 L_4 l_5}^{\ton \ttw \tth}(i,3-i) 
\label{eq:aiso-eta}
\end{eqnarray}
In the above definitions, the limitations related to the SOC
approximation in the possibilities of linking isospin exchange
operators, have been considered by defining $L_2 =
\delta_{i1}l_2+\delta_{i2}l_4$ and $L_4 =
\delta_{i1}l_4+\delta_{i2}l_2$. This implies that the isospin term of
Eq. (\ref{eq:aiso-calt}) with exponent $l_2$ acts on the pair 13, and
the term with exponent $l_4$ on the pair 23.

\section{The uncorrelated one-body densities}
\label{sec:app-obdt}
In this appendix we present the general expressions of the
uncorrelated one-body densities involved in the calculation of the
kinetic energy for systems not saturated in isospin.  We use a set of
single particle wave functions expressed as indicated by Eq.
(\ref{eq:fin-spwf}).

The one-body density $\rho_{T1}^{\ton}$ is given by:
\begin{eqnarray}
\nonumber
\rho^{\ton}_{T1}(\br_1)&=&
\sum_{nljm}\Big( \phi^{\ton *}_{nljm}(\br_1)\nabla_{1}^{2}
\phi^{\ton}_{nljm}(\br_1)
-\nabla_{1}\phi^{\ton *}_{nljm}(\br_1)\cdot\nabla_{1}
\phi^{\ton}_{nljm}(\br_1) \Big)\\
\nonumber
&=&\frac{1}{4\pi}\sum_{nlj}(2j+1)\Bigg[R^{\ton}_{nlj}(r_1)\Big(
D^{\ton}_{nlj}(r_1)-\frac{l(l+1)}{r_1^2}R^{\ton}_{nlj}(r_1)
\Big)\\
&&-[R^{\ton \prime}_{nlj}(r_1)]^2\Bigg] \,\,,
\end{eqnarray}
where we have carried out the trace in the spin space and we have defined:
\begin{equation}
D^{\ton}_{nlj}(r_1)=
R^{\ton \prime \prime}_{nlj}(r_1)+\frac{2}{r_1}R^{\ton \prime}_{nlj}(r_1)-
\frac{l(l+1)}{r_1^2}R^{\ton}_{nlj}(r_1).
\label{eq:app-dnlj}
\end{equation}

For the one-body density matrices $\rho_{T2}$ and $\rho_{T2,j}$ we find:
\begin{eqnarray}
\nonumber
\rho_{T2}^{\ton\ttw}(\br_1,\br_2)&=&
\rho_0^{\ton}(\br_1,\br_2)\nabla_{1}^{2}
\rho_0^{\ttw}(\br_1,\br_2)-\nabla_{1}
\rho_0^{\ton}(\br_1,\br_2)\cdot
\nabla_{1}\rho_0^{\ttw}(\br_1,\br_2)\\
\nonumber
&=&\frac{1}{4(4\pi)^2}\sum_{nljn'l'j'}(2j+1)(2j'+1)R^{\ton}_{nlj}
(r_2)R^{\ttw}_{n'l'j'}(r_2) \Bigg\{\\
\nonumber
&&\Big[R^{\ton}_{nlj}(r_1)D^{\ttw}_{n'l'j'}(r_1)
-R^{\ton \prime}_{nlj}(r_1)R^{\ttw \prime}_{n'l'j'}(r_1)\Big]
P_{l}(\cos{\theta})P_{l'}(\cos{\theta})\\
&&-\frac{\sin^2{\theta}}{r_1^2}R^{\ton}_{nlj}(r_1)
R^{\ttw}_{n'l'j'}(r_1)P^\prime_{l}(\cos{\theta})P^\prime_{l'}(\cos{\theta})
\Bigg\}\,\,,\\
\nonumber
\rho_{T2,j}^{\ton\ttw}(\br_1,\br_2)&=&
\frac{1}{(4\pi)^2}\sum_{nljn'l'j'}(-1)^{j+j'-l-l'-1}
R^{\ton}_{nlj}(r_2)R^{\ttw}_{n'l'j'}(r_2) \Bigg\{\\
\nonumber
&&\Big[R^{\ton}_{nlj}(r_1)D^{\ttw}_{n'l'j'}(r_1)
-R^{\ton \prime}_{nlj}(r_1)R^{\ttw \prime}_{n'l'j'}(r_1)\Big]
Q_{l}(\cos{\theta})Q_{l'}(\cos{\theta})\\
&&-\frac{\sin^2(\theta)}{r_1^2}R^{\ton}_{nlj}(r_1)R^{\ttw}_{n'l'j'}(r_1)
Q'_{l}(\cos{\theta})Q'_{l'}(\cos{\theta})\Bigg\} \,\,,
\end{eqnarray}
where $P_{l}$ are Legendre polynomials, $\theta$ is the angle between
the vectors $\br_1$ and $\br_2$ and we have defined:
\begin{eqnarray*}
Q_{l}(\cos{\theta})&=&\sin{\theta}P'_{l}(\cos{\theta}) \,\,,\\
Q'_{l}(\cos{\theta})&=&\frac{1}{\sin{\theta}}\Big(\cos{\theta}
P'_{l}(\cos{\theta})-l(l+1)P_{l}(\cos{\theta})\Big).
\end{eqnarray*}

For the $\rho_{T3}$ densities we have:
\begin{eqnarray}
\nonumber
\rho_{T3}^{\ton}(\br_1,\br_2)&=&2\nabla_1^2\rho^{\ton}_{0}(\br_1,\br_2) \\
&=&\frac{1}{4\pi}\sum_{nlj}(2j+1)R^{\ton}_{nlj}(r_2)D^{\ton}_{nlj}(r_1)
P_{l}(\cos{\theta}) \,\,,\\ 
\nonumber
\rho_{T3,j}^{\ton}(\br_1,\br_2)&=&
2\nabla_1^2\rho_{0j}^{\ton}(\br_1,\br_2) \\
&=&\frac{1}{2\pi}\sum_{nlj}(-1)^{j-l-\frac{1}{2}}R^{\ton}_{nlj}(r_2)
D^{\ton}_{nlj}(r_1)Q_{l}(\cos{\theta})\,\,.
\end{eqnarray}

The last density that appears in the calculation of the
center of mass energy can be written as:
\begin{eqnarray}
\nonumber
\rho_{T4}^{\ton}(\br_1,\br_2) &=&
\rho_{T6}^{\ton}(\br_1,\br_2) -
\rho_{0}^{\ton}(\br_1,\br_2) 
\rho_{T5}^{\ton}(\br_1,\br_2)\\
&-& \rho_{0j}^{\ton}(\br_1,\br_2)
\rho_{T5,j}^{\ton}(\br_1,\br_2) \,\,,
\end{eqnarray}
where we have defined:
\begin{eqnarray}
\nonumber
\rho_{T6}^{\ton}(\br_1,\br_2) & = & 2 \Big(
\nabla_1 \rho_{0}^{\ton}(\br_1,\br_2) \cdot
\nabla_2 \rho_{0}^{\ton}(\br_1,\br_2) + \\  & & 
\nabla_1 \rho_{0j}^{\ton}(\br_1,\br_2) \cdot
\nabla_2 \rho_{0j}^{\ton}(\br_1,\br_2) \Big) 
\label{eq:app-rhot6}
\,\,, \\
\rho_{T5,(j)}^{\ton}(\br_1,\br_2) & = & 2 
\nabla_1 \cdot \nabla_2 \rho_{0(j)}^{\ton}(\br_1,\br_2)\,\,.
\label{eq:app-rhot5}
\end{eqnarray}

The explicit expressions of the above defined quantities are:
\begin{eqnarray}
\rho_{T5}^{\ton}(\br_1,\br_2)& = & \frac 1 {4\pi}
\sum_{nlj} (2j+1) \Biggl[ R_{nlj}^{\ton\prime}(r_1)
R_{nlj}^{\ton\prime} (r_2)
\cos \theta P_{l}(\cos \theta)+ \\
& & \left( 
R_{nlj}^{\ton \prime} (r_2) \frac{R_{nlj}^{\ton} (r_1)}{r_1}+
R_{nlj}^{\ton \prime} (r_1) \frac{R_{nlj}^{\ton} (r_2)}{r_2} \right)
\sin^2 \theta P_l^\prime(\cos \theta)+
 \nonumber \\
& &  \frac{R_{nlj}^{\ton} (r_1)R_{nlj}^{\ton} (r_2)}{r_1r_2}
\Big(\sin^2 \theta P_l^\prime(\cos \theta)+ 
l(l+1) \cos \theta P_{l}(\cos \theta)\Big) \Biggr] \nonumber ,\\
\rho_{T5,j}^{\ton}(\br_1,\br_2)& = & \frac 1 {2\pi}
\sum_{nlj} (-1)^{j-l-1/2} \Biggl[ R_{nlj}^{\ton\prime}(r_1)
R_{nlj}^{\ton\prime} (r_2) \cos \theta Q_{l}(\cos \theta)+  \\
& & \left( 
R_{nlj}^{\ton \prime} (r_2) \frac{R_{nlj}^{\ton} (r_1)}{r_1}+
R_{nlj}^{\ton \prime} (r_1) \frac{R_{nlj}^{\ton} (r_2)}{r_2} \right)
\sin^2 \theta Q_l^\prime(\cos \theta) +\nonumber \\
& & \frac{R_{nlj}^{\ton} (r_1)R_{nlj}^{\ton} (r_2)}{r_1r_2}
\Bigg\{ \sin^2 \theta Q_{l}^\prime(\cos \theta)+\nonumber \\ & & 
\Big( l(l+1)-
\frac 1 {\sin^2 \theta} \Big) \cos \theta Q_l(\cos \theta) \Bigg\}
 \Biggr] \nonumber \,\,,\\
\rho_{T6}^{\ton}(\br_1,\br_2)& = & \frac 1 {2(4\pi)^2}
\sum_{nlj \atop n'l'j'} (2j+1)(2j'+1)R_{nlj}^{\ton} (r_2)
R_{n'l'j'}^{\ton} (r_1) \\
& & \Biggl\{ \cos \theta \Biggl[ R_{nlj}^{\ton\prime} (r_2)
R_{n'l'j'}^{\ton\prime} (r_1)
P_l(\cos \theta)P_{l'}(\cos \theta)-  \nonumber \\
& & \sin^2 \theta \frac{R_{nlj}^{\ton} (r_2)R_{n'l'j'}^{\ton} (r_1)}
{r_1 r_2} P_l^\prime(\cos \theta)P_{l'}^\prime(\cos \theta)
\Biggr]+ \nonumber \\
& &  \sin^2 \theta \Biggl[
\frac{R_{nlj}^{\ton} (r_2)}{r_2}R_{n'l'j'}^{\ton\prime} (r_1)
 P_l^\prime(\cos \theta)P_{l'}(\cos \theta)+ \nonumber \\
 & & R_{nlj}^{\ton\prime} (r_2)\frac{R_{n'l'j'}^{\ton} (r_1)}
{r_1} P_l(\cos \theta)P_{l'}^\prime(\cos \theta) \Biggr] \Biggr\}+
 \nonumber \\
& & \frac 1 {2(2\pi)^2}
\sum_{nlj \atop n'l'j'} (-1)^{j+j'-l-l'-1}R_{nlj}^{\ton} (r_2)
R_{n'l'j'}^{\ton} (r_1) \nonumber \\
& & \Biggl\{ \cos \theta \Biggl[ R_{nlj}^{\ton\prime} (r_2)
R_{n'l'j'}^{\ton\prime} (r_1)
Q_l(\cos \theta)Q_{l'}(\cos \theta)-  \nonumber \\
& & \sin^2 \theta \frac{R_{nlj}^{\ton} (r_2)R_{n'l'j'}^{\ton} (r_1)}
{r_1 r_2} Q_l^\prime(\cos \theta)Q_{l'}^\prime(\cos \theta)
\Biggr]+ \nonumber \\
& &  \sin^2 \theta \Biggl[
\frac{R_{nlj}^{\ton} (r_2)}{r_2}R_{n'l'j'}^{\ton\prime} (r_1)
 Q_l^\prime(\cos \theta)Q_{l'}(\cos \theta)+ \nonumber \\
 & & R_{nlj}^{\ton\prime} (r_2)\frac{R_{n'l'j'}^{\ton} (r_1)}
{r_1} Q_l(\cos \theta)Q_{l'}^\prime(\cos \theta) \Biggr] \Biggr\}
 \nonumber \,\,. 
\end{eqnarray}

\section{The expressions of the energy expectation value}
\label{sec:app-ene}
The {\em interaction energy} defined in Sect. \ref{sec:ene} as
$T_F+V_2 \equiv W$, can be expressed in terms of:
\begin{eqnarray}
\nonumber
H_{JF}^{p,q,r}(r_{12}) & = &
\frac{1}{f_{1}^2(r_{12})}
\biggl\{-{{\hbar^2}\over{2m}}\delta_{q,1}
\Bigl[ f_p(r_{12})\nabla^2 f_r(r_{12})-
 \nabla f_p(r_{12})\cdot\nabla f_r(r_{12})
\\ \nonumber & & 
-\frac 6 {r_{12}^2}f_p(r_{12})f_r(r_{12})
(\delta_{r,5}+\delta_{r,6})(1+\delta_{p,5}+\delta_{p,6})
\Bigr]\\
& & + f_p(r_{12})v^q(r_{12})f_r(r_{12})\biggr\}
\,\,.
\label{eq:aene-hjf}
\end{eqnarray}

In the following, we shall use the separation of the spin and isospin
operators as was done in Eqs. (\ref{eq:inf-separ}) and
(\ref{eq:fin-stcorr}).  We shall identify the various isospin
independent operators by using the $k$ and $u$ indexes, which can
assume the values 1,2, and 3. The $l$ indexes can assume the values 0
and 1.

As in Sect. \ref{sec:ene}, the expectation
value of the interaction energy is calculated in four parts, see Eq.
(\ref{eq:ene-W}) and Fig. \ref{fig:ene-W}. The first contribution,
called $W_0$, is given by:
\begin{eqnarray}
\hspace*{-4mm}
W_0 & = & \frac 1 2 \int d{\bf r}_1 d{\bf r}_2
H_{JF}^{2k_1-1+l_1,2k_2-1+l_2,2k_3-1+l_3}(r_{12}) \biggl[ 
\\ \nonumber & & 
I^{k_1 k_2 k_3} A^{k_3} 
\rho_{2,dir}^{\ton \ttw} (\br_1,\br_2) 
\chi_{l_1+l_2+l_3}^{\ton \ttw} \nonumber \\
\nonumber &+& 
I^{k_4 k_1 k_5} I^{k_2 k_3 k_5} A^{k_5} \Delta^{k_4} 
\rho_{2,exc}^{\ton \ttw} (\br_1,\br_2) 
\chi_{l_1+l_2+l_3+l_4}^{\ton \ttw} \\
&+& 
I^{k_4 k_1 k_5} I^{k_2 k_3 k_6}I^{k_5 k_6 2} \Delta^{k_4} 
\rho_{2,excj}^{\ton \ttw} (\br_1,\br_2) 
\chi_{l_1+l_2+l_3+l_4}^{\ton \ttw}
\biggr]
\label{eq:aene-w0} 
\,\,,
\end{eqnarray}
and corresponds to the case when $p>1$ operators act between the
external particles only.  These operators are associated to $f_p$,
$v^q$ and $f_r$ functions in Eq.  (\ref{eq:aene-hjf}) for the direct
terms. In addition, we should recall that the exchange terms have
additional spin and isospin dependent operators because of the
presence of the exchange operator $\Pi^{\sigma\tau} (1,2)$, see Eq.
(\ref{eq:inf-siexch}).  By definition, this operator may act on the bra or 
on the ket.  In our conventions we make it act on the ket so it is always
to the left of the rest of the operators. The sequence of the three
operators we have mentioned above, plus the exchange operators, are 
present in all the terms we are going to analyze, therefore we
shall always use the same set of indexes. Furthermore, we should point
out that in Eq.  (\ref{eq:aene-w0}) a sum on all the $k,l$ and $t$
indexes is understood.  This convention will be used in all the
equations of this appendix.

The expressions of the densities used in Eq. (\ref{eq:aene-w0}) are
given in Eqs.(\ref{eq:aene-dens1}-\ref{eq:aene-dens}) defined in
Appendix \ref{sec:app-fhncsoc}, and include only state--independent
vertex corrections. In this way, the $\rho$ functions consider all the
direct and exchange central dressing linked to the external points.
As discussed in Sect. \ref{sec:finite}, $A^{k=1,2,3}=1,3,6$ and
$\Delta^k=1-\delta_{k,3}$. The values of $I^{k_1k_2k_3}$ and
$J^{k_1k_2k_3}$ are given by Eqs.  (\ref{eq:fin-iijk}) and
(\ref{eq:fin-jijk}).  The $\chi^{\ton\ttw}_{l}$ functions give the
isospin traces and their values are given by Eq. (\ref{eq:fin-chi}).

%%%%%%%%%%%%%%%%%%%%%%%%%%%%%%%%%%%%%%
% table spin trace 0
%%%%%%%%%%%%%%%%%%%%%%%%%%%%%%%%%%%%%%
\vskip 0.5 cm 
\begin{table}[htb]
\begin{center}
\begin{tabular}{|l|l|}
\hline
$u_1 k_1 k_2 k_3 u_2$ & 
$I^{k_1k_2k_3} A^{k_3} A^{u_1} \delta_{u_1 u_2}$ \\
$u_1 k_1 k_2 u_2 k_3$ & 
$I^{k_1k_2k_3} A^{k_3}  (1+ D_{k_3 u_1}) A^{u_1} \delta_{u_1 u_2}$ \\
$k_1 u_1 k_2 k_3 u_2$ & 
$I^{k_1k_2k_3} A^{k_3}  (1+ D_{k_1 u_1}) A^{u_1} \delta_{u_1 u_2}$ \\
$k_1 u_1 k_2  u_2 k_3$  & 
$I^{k_1k_2k_3} A^{k_3}  (1+ D_{k_2 u_1}) A^{u_1} \delta_{u_1 u_2}$ \\
\hline
\end{tabular}
\caption{\small Spin traces of 
the operators in Eq. (\ref{eq:aene-wstr}).}
\label{tab:aene-wst0}
\end{center}
\end{table}
%%%%%%%%%%%%%%%%%%%%%%%%%%%%%%%%%%%%%%%%%%%%%%%%%%%%%

We now discuss the effects produced by one SOR linked to one of the
interacting particles. This is the $W_s$ diagram of Fig.
\ref{fig:ene-W}.  The operator structure that we must analyze, for the
direct case, is:
\begin{eqnarray}
\nonumber
&~&\frac 1 2 \{k_1 , u_1 \} k_2 \frac 1 2 \{k_3 , u_2 \} = \\
&~&
\frac 1 4 \left( u_1  k_1 k_2 k_3 u_2 + u_1 k_1 k_2 u_2 k_3 + 
k_1 u_1 k_2 k_3 u_2 +
 k_1 u_1 k_2 u_2k_3 \right)
\,\,,
\label{eq:aene-wstr}
\end{eqnarray}
In the above equation, the $k$ and $u$ indicate the operators $P^k$
and $P^u$, Eq. (\ref{eq:inf-separ}).  The $k_1$, $k_2$ and $k_3$
operators act on the pair of particles 1 and 2. The $u_1$ and $u_2$
operators act, instead, on the particles pair 1 and 3, or 2 and 3. The
symbol $\{,\}$ indicates the anticommutator. The values of the traces
of the various terms of Eq. (\ref{eq:aene-wstr}), are given in Tab.
\ref{tab:aene-wst0}.

%
%%%%%%%%%%%%%%%%%%%%%%%%%%%%%%%%%%%%%%
% table spin trace 1
%%%%%%%%%%%%%%%%%%%%%%%%%%%%%%%%%%%%%%
\vskip 0.5 cm 
\begin{table}[htb]
\begin{center}
\begin{tabular}{|l|l|}
\hline
$u_1 k_1 k_2 k_3 u_2$ & 
$I^{k_4k_1k_5} I^{k_2k_3k_5} A^{k_5}  (1+ D_{k_4 u_1}) 
A^{u_1} \delta_{u_1 u_2}$ \\
$u_1 k_1 k_2 u_2 k_3$ & 
$I^{k_4k_1k_5} I^{k_2k_3k_5} A^{k_5}  (1+ D_{k_5 u_1}) 
A^{u_1} \delta_{u_1 u_2}$ \\
$k_1 u_1 k_2 k_3 u_2$ & 
$I^{k_4k_1k_5} I^{k_2k_3k_5} A^{k_5}  (1+ D_{k_5 u_1}) 
A^{u_1} \delta_{u_1 u_2}$ \\
$k_1 u_1 k_2  u_2 k_3$  & 
$I^{k_4k_1k_5} I^{k_2k_3k_5} A^{k_5}  (1+ D_{k_2 u_1}) 
A^{u_1} \delta_{u_1 u_2}$ \\
\hline
\end{tabular}
\caption{\small Spin traces for the  parallel spins case
 of the operators of Eq. (\ref{eq:aene-wstr}).}
\label{tab:aene-wst}
\end{center}
\end{table}
%
%%%%%%%%%%%%%%%%%%%%%%%%%%%%%%%%%%%%%%%%%%%%%%%%%%%%
% table spin trace 2 
%%%%%%%%%%%%%%%%%%%%%%%%%%%%%%%%%%%%%%%%%%%%%%%%%%%%
\begin{table}[htb]
\begin{center}
\begin{tabular}{|l|l|}
\hline
$k_4u_1 k_1 k_2 k_3 u_2$ & 
$I^{k_4k_1k_5} I^{k_2k_3k_6}I^{k_5k_62} (1+ D_{k_4 u_1}) A^{u_1} 
\delta_{u_1 u_2}$ \\
$k_4u_1 k_1 k_2 u_2 k_3$ & 
$I^{k_4k_1k_5} I^{k_2k_3k_6}I^{k_5k_62}  (1+ D_{k_5 u_1}) A^{u_1} 
\delta_{u_1 u_2}$ \\
$k_4k_1 u_1 k_2 k_3 u_2$ & 
$I^{k_4k_1k_5} I^{k_2k_3k_6}I^{k_5k_62}  (1+ D_{k_5 u_1}) A^{u_1} 
\delta_{u_1 u_2}$ \\
$k_4k_1 u_1 k_2  u_2 k_3$  & 
$I^{k_4k_1k_5} I^{k_2k_3k_6}I^{k_5k_62}  (1+ D_{k_2 u_1}) A^{u_1} 
\delta_{u_1 u_2}$ \\
\hline
\end{tabular}
\caption{\small Spin traces for the case 
of antiparallel spins between particles 1 and 2
in Eq. (\ref{eq:aene-wstr}).} 
\label{tab:aene-wst2}
\end{center}
\end{table}
%%%%%%%%%%%%%%%%%%%%%%%%%%%%%%%%%%%%%%%%%%%%%%%%%%%%
%%%%%%%%%%%%%%%%%%%%%%%%%%%%%%%%%%%%%%%%%%%%%%%%%%%%
% table spin trace 3
%%%%%%%%%%%%%%%%%%%%%%%%%%%%%%%%%%%%%%%%%%%%%%%%%%%%
\begin{table}[htb]
\begin{center}
\begin{tabular}{|l|l|}
\hline
$u_1 k_1 k_2 k_3 u_2$ & 
$I^{k_1k_2k_3} A^{k_3} I^{u_1u_22}$ \\
$u_1 k_1 k_2 u_2 k_3$ & 
$I^{k_1k_2k_3} A^{k_3} I^{u_1u_22}(1+ D_{k_3 u_1})$ \\
$u_1k_1 u_2 k_2 k_3$ & 
$I^{k_1k_2k_3} A^{k_3} I^{u_1u_22}(1+ D_{k_1 u_1})$ \\
$u_1u_2 k_1 k_2  k_3$  & 
$I^{k_1k_2k_3} A^{k_3} I^{u_1u_22}$ \\
\hline
\end{tabular}
\caption{Spin traces for the case of 
antiparallel spins between the  
particles in 1 and 3 or 2 and 3 in Eq. (\ref{eq:aene-wstr}).}
\label{tab:aene-wst3}
\end{center}
\end{table}
%%%%%%%%%%%%%%%%%%%%%%%%%%%%%%%%%%%%%%%%%%%%%%%%%%%%

In the exchange case, an additional $k_4$ operator must be included on
the left hand side, following our conventions.  Since we work with
single particle wave functions expressed in a $jj$ coupling scheme,
antiparallel spin terms are usually contributing. For this reason in
the exchange case, we have to distinguish the parallel and
antiparallel spin situations. For the case of parallel spins between
the interacting points, by following Ref. \cite{pan79}, we obtain
results given in Tab. \ref{tab:aene-wst}.

For the antiparallel spin term, we have to distinguish the case when
the antiparallel spins are those of interacting points, whose results
are given in Tab. \ref{tab:aene-wst2}, from the case when they are
those of the 1 and 3 particles, or those of the 2 and 3 particles. In
this last case $u_1$ must be an exchange operator acting on the left
hand side of the operator product. This produces the results given in
Tab. \ref{tab:aene-wst3}.

In the Tabs. \ref{tab:aene-wst}, \ref{tab:aene-wst2},
\ref{tab:aene-wst3}, we have used the following values of the $D$
terms:
\begin{equation}
D_{k_1,k_2} = \left( \begin{array}{rrr}
 0 & 0 & 0 \\ 0 & -4/3 & -4/3 \\0 & -4/3 & -4/3 \\
\end{array} \right)
\,\,,
\label{eq:aene-dk1k2}
\end{equation}
which corresponds to the odd values of $E_{pq}$ defined in Tab.
\ref{tab:inf-dpq} of Appendix~\ref{sec:app-matrix}.
 
All the isospin parts of the above operators can be written by using
the $\chi_{l_1 l_2 l_3 l_4 l_5}^{\ton \ttw \tth}(i)$, function defined
in Eq. (\ref{eq:aiso-chi}).  The contribution of the SOR to the
interaction energy is:
\begin{eqnarray}
\nonumber 
\hspace*{-3mm}
W_s & = & \frac 1 8 \int d{\bf r}_1 d{\bf r}_2
H_{JF}^{2k_1-1+l_1,2k_2-1+l_2,2k_3-1+l_3}(r_{12}) \Biggl( 
I^{k_1 k_2 k_3} A^{k_3} \\ & \biggr\{ & 
\rho_{2,dir}^{\ton \ttw} (\br_1,\br_2)  
\left[
M_{d,l_1, l_2, l_3}^{\ton \ttw k_1 k_2 k_3} (\br_1)+
M_{d,l_1, l_2, l_3}^{\ton \ttw k_1 k_2 k_3} (\br_2) \right] 
\nonumber \\
&+& \left[ g_{dd}^{\ton \ttw} (\br_1,\br_2) C_{22}^{\ttw}(\br_2)+
g_{de}^{\ton \ttw} (\br_1,\br_2) C_{d,22}^{\ttw}(\br_2) \right] 
C_{d,22}^{\ton} (\br_1)  M_{e,l_1, l_2, l_3}^{\ton \ttw k_1 k_2 k_3} (\br_1)
\nonumber \\ 
&+& \left[ g_{dd}^{\ton \ttw} (\br_1,\br_2) C_{22}^{\ton}(\br_1)+
g_{ed}^{\ton \ttw} (\br_1,\br_2) C_{d,22}^{\ton}(\br_1) \right]
C_{d,22}^{\ttw} (\br_2)  
M_{e,l_1, l_2, l_3}^{\ton \ttw k_1 k_2 k_3,} (\br_2) \nonumber \\
&+& \left[ g_{dd}^{\ton \ttw} (\br_1,\br_2) C_{22}^{\ttw}(\br_2)+
g_{de}^{\ton \ttw} (\br_1,\br_2) C_{d,22}^{\ttw}(\br_2) \right] 
C_{d,22}^{\ton} (\br_1)  M_{ej,l_1, l_2, l_3}^{\ton \ttw k_1 k_2 k_3} (\br_1)
\nonumber \\ 
&+& \left[ g_{dd}^{\ton \ttw} (\br_1,\br_2) C_{22}^{\ton}(\br_1)+
g_{ed}^{\ton \ttw} (\br_1,\br_2) C_{d,22}^{\ton}(\br_1) \right]
C_{d,22}^{\ttw} (\br_2)  
M_{ej,l_1, l_2, l_3}^{\ton \ttw k_1 k_2 k_3} (\br_2)
\biggr\} \nonumber \\
&+& 
I^{k_4 k_1 k_5} I^{k_2 k_3 k_5} A^{k_5} \Delta^{k_4}  
\rho_{2,exc}^{\ton \ttw} (\br_1,\br_2) 
\left[
M_{d,l_1, l_2, l_3,l_4}^{\ton \ttw k_2 k_4 k_5} (\br_1)+
M_{d,l_1, l_2, l_3,l_4}^{\ton \ttw k_2 k_4 k_5} (\br_2) 
\right]
\nonumber \\
\nonumber  &+& 
I^{k_4 k_1 k_5} I^{k_2 k_3 k_6}I^{k_5k_6 2} \Delta^{k_4}  
\rho_{2,excj}^{\ton \ttw} (\br_1,\br_2) \\
& & \left[
M_{d,l_1, l_2, l_3,l_4}^{\ton \ttw k_2 k_4 k_5} (\br_1)+
M_{d,l_1, l_2, l_3,l_4}^{\ton \ttw k_2 k_4 k_5} (\br_2)
\right]
\Biggr) \,\,, 
\end{eqnarray}
where we have defined 
\begin{eqnarray}
\nonumber
M_{m,l_1, l_2, l_3}^{\ton \ttw k_1 k_2 k_3} (\br_i) & = & 
M_{m,0, l_1+l_2+l_3, 0}^{\ton \ttw u_1} (\br_i) +
(1+D_{k_3 u_1}) M_{m,0, l_1+l_2, l_3}^{\ton \ttw u_1} (\br_i)  
 \\ \nonumber 
&+&  (1+D_{k_1 u_1}) M_{m,l_1, l_2+l_3, 0}^{\ton \ttw u_1} (\br_i) 
\\
&+&
(1+D_{k_2 u_1}) M_{m,l_1, l_2, l_3}^{\ton \ttw u_1} (\br_i) 
\label{eq:aene-m1}
\,\,,
\\
\nonumber 
M_{ej,l_1, l_2, l_3}^{\ton \ttw k_1 k_2 k_3} (\br_i) & = & 
M_{ej,0, 0,l_1+l_2+l_3}^{\ton \ttw u_1u_2} (\br_i) +
(1+D_{k_3 u_1}) M_{ej,0, l_1+l_2, l_3}^{\ton \ttw u_1u_2} (\br_i) 
\\ 
&+&  (1+D_{k_1 u_1}) M_{ej,0, l_1, l_2+l_3}^{\ton \ttw u_1u_2} (\br_i) +
M_{ej,0,l_1+l_2+l_3,0}^{\ton \ttw u_1u_2} (\br_i) 
\,\,,
\\
\nonumber  
M_{d,l_1, l_2, l_3,l_4 }^{\ton \ttw k_2 k_4 k_5} (\br_i) & = & 
(1+D_{k_5 u_1}) \bigl(M_{d,l_4+l_1, l_2+l_3, 0}^{\ton \ttw u_1} (\br_i) +
 M_{d,l_4, l_1+l_2, l_3}^{\ton \ttw u_1} (\br_i) \bigr)
\\ \nonumber  &+& 
(1+D_{k_4 u_1}) M_{d,l_4, l_1+l_2+l_3, 0}^{\ton \ttw u_1} (\br_i)
\\ 
&+& 
(1+D_{k_2 u_1}) M_{d,l_4+l_1, l_2, l_3}^{\ton \ttw u_1} (\br_i)
\,,
\end{eqnarray}

\begin{eqnarray}
M_{m,l_1, l_2, l_3}^{\ton \ttw u_1} (\br_i) & = & A^{u_1} 
 \biggl[
(1 - \delta_{u_1,1})
\chi_{l_1 0 l_2 0 l_3}^{\ton \ttw \tth}(i)
U_{m,2u_1-1,2u_1-1}^{\mu \tth} (\br_i) \nonumber \\ 
&+& \chi_{l_1 1 l_2 0 l_3}^{\ton \ttw \tth}(i)
U_{m,2u_1,2u_1-1}^{\mu \tth} (\br_i) + 
\chi_{l_1 0 l_2 1 l_3}^{\ton \ttw \tth}(i)
U_{m,2u_1-1,2u_1}^{\mu \tth} (\br_i)  \nonumber \\ 
&+& \chi_{l_1 1 l_2 1l_3}^{\ton \ttw \tth}(i) 
U_{m,2u_1,2u_1}^{\mu \tth} (\br_i)  \biggr]
\,\,,
\end{eqnarray}
\begin{eqnarray}
\hspace*{-8mm}
M_{ej,l_1, l_2, l_3}^{\ton \ttw u_1u_2} (\br_i) & = &  
{\displaystyle I^{u_1u_22} \biggl[ } 
\chi_{l_1 0 l_2 0 l_3}^{\ton \ttw \tth}(i)
U_{ej,2u_1-1,2u_2-1}^{\mu \tth} (\br_i)+ 
 \chi_{l_1 1 l_2 0 l_3}^{\ton \ttw \tth}(i)
U_{ej,2u_1,2u_2-1}^{\mu \tth} (\br_i)  \nonumber \\
&+& \chi_{l_1 0 l_2 1 l_3}^{\ton \ttw \tth}(i)
U_{ej,2u_1-1,2u_2}^{\mu \tth} (\br_i) + 
\chi_{l_1 1 l_2 1l_3}^{\ton \ttw \tth}(i) 
U_{ej,2u_1,2u_2}^{\mu \tth} (\br_i)  \biggr]
\label{eq:aene-m2}
\,\,,
\end{eqnarray}
In the above equations we have considered that $i=1,2$, $m=d,e$ , and
$\mu = \ton$ for $i=1$ and $\mu = \ttw$ for $i=2$. The expressions of
the $U_{m,pq}^{\ton \ttw}(\br_i)$ and $U_{ej,pq}^{\ton \ttw}(\br_i)$ functions
are given by Eqs. (\ref{eq:asoc-socend}).

The structure of the $W_c$ term is more involved.  We calculate
separately the various terms according to the direct or exchange
nature of the correlations reaching the external points.
\begin{equation}
W_c = W_c (dd) + W_c (ed) + W_c (de) + W_c (ee) + W_c (cc)
\,\,.
\label{eq:aene-wc}
\end{equation}
The operator structure that we must analyze in the $dd$ direct case
is:
\begin{eqnarray}
\nonumber
& & \frac 1 4 
\biggl[ \frac 1 4 \{k_1 , u_1 \} k_2 \{k_3 , u_2 \}  +
\frac 1 4 \{k_1 , u_2 \} k_2 \{k_3 , u_1 \}+\\
\nonumber &&
\frac 1 6 \Big(  \{ \{ u_1, u_2 \}, k_1 \} + u_1 k_1 u_2 + u_2 k_1 u_1 
\Big) k_2k_3 +
\\ & & 
\frac 1 6 k_1k_2 
\Big( \{ \{ u_1, u_2 \}, k_3 \} + u_1 k_3 u_2 + u_2 k_3 u_1 
\Big)
\biggr]
\,\,,
\label{eq:aene-wctr}
\end{eqnarray}
where the various symbols have the same meaning as in Eq. 
(\ref{eq:aene-wstr}).

%%%%%%%%%%%%%%%%%%%%
% Table
%%%%%%%%%%%%%%%%%%%
\begin{table}[t]
\begin{center}
\begin{tabular}{|l|l|l|}
\hline
 & \multicolumn{1}{c|}{Direct} & \multicolumn{1}{c|}{Exchange} \\
\hline
$u_1 k_1 k_2 k_3 u_2$ & 
$I^{k_1k_2k_6} I^{k_3 k_5 k_6} A^{k_6}$ & $I^{k_1k_2k_6} I^{k_6k_3k_7} J^{k_4 k_5k_7}$ \\
$u_1 k_1 k_2 u_2 k_3$ & 
$I^{k_1k_2k_6} J^{k_3 k_5 k_6}$ & $I^{k_1k_2k_6} I^{k_3k_4k_7} J^{k_7 k_5 k_6}$ \\
$k_1 u_1 k_2 k_3 u_2$ & 
$I^{k_2k_3k_6} J^{k_1 k_5 k_6}$ & $I^{k_2k_3k_6} I^{k_4k_1k_7} J^{k_7 k_5 k_6}$ \\
$k_1 u_1 k_2  u_2 k_3$ & 
$I^{k_3k_1k_6} J^{k_2 k_5 k_6}$ & $I^{k_3k_4k_6} I^{k_6k_1k_7} J^{k_2 k_5 k_7}$ \\
$u_1 u_2 k_1 k_2 k_3$ & 
$I^{k_1k_2k_6} I^{k_3 k_5 k_6} A^{k_6}$ & $I^{k_4k_5k_6} I^{k_1k_2k_7} I^{k_6 k_7 k_3} A^{k_3} $ \\
$k_1 u_1 u_2  k_2 k_3$ & 
$I^{k_1k_2k_6} I^{k_3 k_5 k_6} A^{k_6}$ & $I^{k_4k_5k_6} I^{k_1k_2k_7} I^{k_6 k_7 k_3} A^{k_3} $ \\
$u_1 k_1 u_2 k_2 k_3$ & 
$I^{k_2k_3k_6} J^{k_1 k_5 k_6}$ & $I^{k_2k_3k_6} I^{k_6k_4k_7} J^{k_1 k_5 k_7} $ \\
$k_1 k_2 k_3 u_1 u_2$ & 
$I^{k_1k_2k_6} I^{k_3 k_5 k_6} A^{k_6}$ & $I^{k_4k_5k_6} I^{k_1k_2k_7} I^{k_6 k_7 k_3} A^{k_3} $ \\
$k_1 k_2 u_1 u_2 k_3$ & 
$I^{k_1k_2k_6} I^{k_3 k_5 k_6} A^{k_6}$ & $I^{k_4k_5k_6} I^{k_1k_2k_7} I^{k_6 k_7 k_3} A^{k_3} $ \\
$k_1 k_2 u_1  k_3 u_2 $ & 
$I^{k_1k_2k_6} J^{k_3 k_5 k_6}$ & $I^{k_4k_1k_6} I^{k_6k_2k_7} J^{k_3 k_5 k_7} $ \\
\hline
\end{tabular}
\end{center}
\caption{\small The traces obtained for the $W_{c}(dd)$ term.}
\label{tab:aene-ddtraces}
\end{table}
%%%%%%%%%%%%%%%%%%%%%%%%%%%%%%%%
%

We calculate the value of this term by using the result of Eq.
(\ref{eq:inf-lpqrdef1}), and the results are given in Tab.
\ref{tab:aene-ddtraces}.  In this table, the factors $\zeta_{132}^{u_1
  u_2 k_5}$ are not present, since they will be included in the
expressions of the nodal diagrams terms $N$ which we shall define in
analogy to Eq. (\ref{eq:inf-nodalop}).  The value of the spin traces
are the same if we exchange $u_1$ and $u_2$.  For this reason, we have
only shown the results when $u_1$ is on the left hand side of $u_2$.
The isospin traces do not have this property and their values are
given by the coefficients $\eta_{l_1 l_2 l_3 l_4 l_5}^{\ton \ttw
  \tth}(i)$ defined in Eq. (\ref{eq:aiso-eta}).  Since we have
observed that the exchange contribution from antiparallel spins is
much smaller than that from the parallel ones, we have neglected the
contribution of the antiparallel spins in the SOC terms.

By using the above definitions we can write:
\begin{eqnarray}
\nonumber \hspace*{-5mm}
W_c (dd) & = & \frac{1}{24} \int d{\bf r}_1 d{\bf r}_2
H_{JF}^{2k_1-1+l_1,2k_2-1+l_2,2k_3-1+l_3}(r_{12}) 
\Biggl( 
\rho_{2,dir}^{\ton \ttw} (\br_1,\br_2) \biggl[ \\
 \nonumber & & 
I^{k_1 k_2 k_6} I^{k_3 k_5 k_6} A^{k_6} 
M_{1,dd,l_1, l_2, l_3}^{\ton \ttw k_5} (\br_1,\br_2) 
\\ \nonumber &+&
I^{k_1 k_2 k_6} J^{k_3 k_5 k_6} 
M_{2,dd,l_1, l_2, l_3}^{\ton \ttw k_5} (\br_1,\br_2) \\
\nonumber
&+& 
I^{k_2 k_3 k_6} J^{k_1 k_5 k_6} 
M_{3,dd,l_1, l_2, l_3}^{\ton \ttw k_5} (\br_1,\br_2) +
\frac 3 2 I^{k_3 k_1 k_6} J^{k_2 k_5 k_6} 
M_{dd,l_1, l_2, l_3}^{\ton \ttw k_5} (\br_1,\br_2) \biggr] \\
\nonumber
&+& \Delta^{k_4}  
\rho_{2,exc}^{\ton \ttw} (\br_1,\br_2)  \biggl\{ 
\frac 3 2 \biggl[ I^{k_1 k_2 k_6} I^{k_6 k_3 k_7} J^{k_4 k_5 k_7}
M_{dd,l_4, l_1+l_2+l_3, 0}^{\ton \ttw k_5} (\br_1,\br_2) \\
\nonumber  &+& 
I^{k_1 k_2 k_6} I^{k_3 k_4 k_7} J^{k_7 k_5 k_6}
M_{dd,l_4, l_1+l_2, l_3}^{\ton \ttw k_5} (\br_1,\br_2) \\
\nonumber &+& 
I^{k_2 k_3 k_6} I^{k_4 k_1 k_7} J^{k_7 k_5 k_6}
M_{dd,l_4+l_1, l_2+l_3, 0}^{\ton \ttw k_5} (\br_1,\br_2) \\
\nonumber &+& 
I^{k_3 k_4 k_6} I^{k_6 k_1 k_7} J^{k_2 k_5 k_7}
M_{dd,l_4+l_1, l_2, l_3}^{\ton \ttw k_5} (\br_1,\br_2) \biggr] \\
\nonumber &+& 
I^{k_4 k_5 k_6} I^{k_1 k_2 k_7} I^{k_6 k_7 k_3} A^{k_3}
M_{4,dd,l_1, l_2, l_3, l_4}^{\ton \ttw k_5} (\br_1,\br_2) \\
\nonumber &+& 
I^{k_2 k_3 k_6} I^{k_6 k_4 k_7} J^{k_1 k_5 k_7}
M_{dd,l_4, l_1, l_2+l_3}^{\ton \ttw k_5} (\br_1,\br_2)\\
 &+& 
I^{k_4 k_1 k_6} I^{k_6 k_2 k_7} J^{k_3 k_5 k_7}
M_{dd,l_4+l_1+l_2, l_3, 0}^{\ton \ttw k_5} (\br_1,\br_2) 
\biggr\} \Biggr)
\,\,,
\end{eqnarray}
where we have defined
\begin{eqnarray}
M_{dd,l_1, l_2, l_3}^{\ton \ttw k_5} (\br_1,\br_2) & = &
\frac 1 2 \left[ M_{dd,l_1, l_2, l_3}^{\ton \ttw k_5} (\br_1,\br_2;1) +
M_{dd,l_1, l_2, l_3}^{\ton \ttw k_5} (\br_1,\br_2;2) \right]
,
\\
\nonumber
M_{1,dd,l_1, l_2, l_3}^{\ton \ttw k_5} (\br_1,\br_2) 
& = & \frac 3 2 M_{dd,0, l_1+l_2+l_3, 0}^{\ton \ttw k_5} (\br_1,\br_2) +
M_{dd,0, 0, l_1+l_2+l_3}^{\ton \ttw k_5} (\br_1,\br_2) + 
\\ \nonumber & & 
M_{dd,l_1, 0, l_2+l_3}^{\ton \ttw k_5} (\br_1,\br_2) +  
M_{dd,l_1+l_2+l_3, 0, 0}^{\ton \ttw k_5} (\br_1,\br_2) +
\\ & & 
M_{dd,l_1+l_2, 0, l_3}^{\ton \ttw k_5} (\br_1,\br_2)
\,\,,
\\
M_{2,dd,l_1, l_2, l_3}^{\ton \ttw k_5} (\br_1,\br_2) 
& = & \frac 3 2 M_{dd,0, l_1+l_2, l_3}^{\ton \ttw k_5} (\br_1,\br_2) +
M_{dd,l_1+l_2, l_3, 0}^{\ton \ttw k_5} (\br_1,\br_2) 
\,\,,
\\ 
M_{3,dd,l_1, l_2, l_3}^{\ton \ttw k_5} (\br_1,\br_2) 
& = & \frac 3 2 M_{dd,l_1, l_2+l_3, 0}^{\ton \ttw k_5} (\br_1,\br_2) +
M_{dd,0, l_1, l_2+l_3}^{\ton \ttw k_5} (\br_1,\br_2) 
\,\,,
\\ 
\nonumber
\hspace*{-5mm}
M_{4,dd,l_1, l_2, l_3, l_4}^{\ton \ttw k_5} (\br_1,\br_2) 
& = & M_{dd,l_4, 0, l_1+l_2+l_3}^{\ton \ttw k_5} (\br_1,\br_2) +
M_{dd,l_4+l_1, 0, l_2+l_3}^{\ton \ttw k_5} (\br_1,\br_2)+ \\ 
& & M_{dd,l_4+l_1+l_2+l_3, 0, 0}^{\ton \ttw k_5} (\br_1,\br_2) +
M_{dd,l_4+l_1+l_2, 0, l_3}^{\ton \ttw k_5} (\br_1,\br_2),
\end{eqnarray}
and
\begin{eqnarray}
\nonumber
\hspace*{-1cm}
M_{mn,l_1, l_2, l_3}^{\ton \ttw k_5} (\br_1,\br_2;i)
& =& \sum_{u_1,u_2=1}^3 \sum_{\tth=p,n} \biggl[
\eta_{l_1, 1, l_2, 1, l_3}^{\ton \ttw \tth}(i)
N_{mn,2k_5,2u_1,2u_2}^{\ton \ttw \tth}  (\br_1,\br_2)  \\
\nonumber &+& 
\eta_{l_1, 1, l_2, 0, l_3}^{\ton \ttw \tth}(i)
N_{mn,2k_5-1,2u_1,2u_2-1}^{\ton \ttw \tth}  (\br_1,\br_2) \\
\nonumber &+& 
\eta_{l_1, 0, l_2, 1, l_3}^{\ton \ttw \tth}(i)
N_{mn,2k_5-1,2u_1-1,2u_2}^{\ton \ttw \tth}  (\br_1,\br_2)\\
&+& 
(1 - \delta_{k_5 1})
\eta_{l_1, 0, l_2, 0, l_3}^{\ton \ttw \tth}(i)
N_{mn,2k_5-1,2u_1-1,2u_2-1}^{\ton \ttw \tth}  (\br_1,\br_2)\biggr]  
\,\,,
\label{eq:aene-mmn}
\end{eqnarray}
The $mn$ labels indicate that this last equation is valid not only for
the $dd$ case but also for the $ed$, $de$ and $ee$ ones.

So far, we have calculated diagrams where only dynamical correlations
reach the external points 1 and 2. The statistical correlations,
labeled with $e$, are treated by using the spin-isospin exchange
operator. In the $ed$ part, the operator structure is:
\begin{equation}
\frac 1 2  u_1 \biggl[ \frac 1 2 k_1  k_2 \{k_3 , u_2 \}  +
\frac 1 2 \{k_1 , u_2 \} k_2  k_3
\biggr] 
\,\,,
\end{equation}
since $u_1$ is the spin-isospin exchange operator acting always on the
left hand side.  The various terms have already been calculated in the
evaluation of the $dd$ part (see Tab. \ref{tab:aene-ddtraces}). We
obtain:
\begin{eqnarray}
\hspace*{-7mm}
W_c (ed) & = &\frac{1}{8} \int d{\bf r}_1 d{\bf r}_2
H_{JF}^{2k_1-1+l_1,2k_2-1+l_2,2k_3-1+l_3}(r_{12})\cdot \\
\nonumber
&&\bigl( g_{dd}^{\ton \ttw} (\br_1,\br_2)  C_{22}^{\ttw} (\br_2) +
g_{de}^{\ton \ttw} (\br_1,\br_2)  C_{d,22}^{\ttw} (\br_2) \bigr) 
C_{d,22}^{\ton} (\br_1)  
\\ & \biggr\{ & 
I^{k_1 k_2 k_6} I^{k_3 k_5 k_6} A^{k_6} \Bigl[
M_{ed,0, l_1+l_2+l_3, 0}^{\ton \ttw k_5} (\br_1,\br_2;1) +
M_{ed,0, 0, l_1+l_2+l_3}^{\ton \ttw k_5} (\br_1,\br_2;1) \Bigr]
\nonumber  \\ &+& 
I^{k_1 k_2 k_6} J^{k_3 k_5 k_6} 
M_{ed,0, l_1+l_2, l_3}^{\ton \ttw k_5} (\br_1,\br_2;1) 
\nonumber  \\ &+& 
I^{k_2 k_3 k_6} J^{k_1 k_5 k_6} 
M_{ed,0, l_1, l_2+l_3}^{\ton \ttw k_5} (\br_1,\br_2;1) \biggr\} 
\,\,.
\nonumber 
\end{eqnarray}

The $de$ term of Eq. (\ref{eq:aene-wc}) has the same operator
structure as the $ed$ term when $u_1$ and $u_2$ are interchanged. This
does not change the spin traces, therefore the result is:
\begin{eqnarray}
\hspace*{-7mm}
W_c (de) & = & \frac{1}{8} \int d{\bf r}_1 d{\bf r}_2
H_{JF}^{2k_1-1+l_1,2k_2-1+l_2,2k_3-1+l_3}(r_{12})\\ 
&&\Bigl[ 
g_{dd}^{\ton \ttw} (\br_1,\br_2)  C_{22}^{\ton} (\br_1) +
g_{ed}^{\ton \ttw} (\br_1,\br_2)  C_{d,22}^{\ton} (\br_1) \Bigr] 
C_{d,22}^{\ttw} (\br_2)   
\nonumber  \\ & \biggr\{ & 
I^{k_1 k_2 k_6} I^{k_3 k_5 k_6} A^{k_6} \Bigl[
M_{de,0, l_1+l_2+l_3, 0}^{\ton \ttw k_5} (\br_1,\br_2;2) +
M_{de,0, 0, l_1+l_2+l_3}^{\ton \ttw k_5} (\br_1,\br_2;2) \Bigr]
\nonumber  \\ &+& 
I^{k_1 k_2 k_6} J^{k_3 k_5 k_6} 
M_{de,0, l_1+l_2, l_3}^{\ton \ttw k_5} (\br_1,\br_2;2)
\nonumber  \\ &+& 
I^{k_2 k_3 k_6} J^{k_1 k_5 k_6} 
M_{de,0, l_1, l_2+l_3}^{\ton \ttw k_5} (\br_1,\br_2;2) \biggr\} 
\nonumber
\,\,.
\end{eqnarray}

Since $u_1$ and $u_2$ are spin-isospin exchange operators, in the $ee$
term, the only possible ordering of operators is $1/2 \{u_1 , u_2 \}
k_1 k_2 k_3$.  We obtain:
\begin{eqnarray}
\hspace*{-3mm}
W_c (ee) & = & \frac{1}{2} \int d{\bf r}_1 d{\bf r}_2
H_{JF}^{2k_1-1+l_1,2k_2-1+l_2,2k_3-1+l_3}(r_{12}) 
\nonumber  \\ 
& & 
C_{d,22}^{\ton} (\br_1)  
g_{dd}^{\ton \ttw} (\br_1,\br_2)  C_{d,22}^{\ttw} (\br_2) 
I^{k_1 k_2 k_6} I^{k_3 k_5 k_6} A^{k_6}  \nonumber \\ & & 
\frac 1 2 \left[ M_{ee,l_1 l_2 l_3}^{\ton \ttw k_5} (\br_1,\br_2;1)
+ M_{ee,0,0,l_1+l_2+l_3}^{\ton \ttw k_5} (\br_1,\br_2;2) \right]
\,\,.
\end{eqnarray}

To calculate the $W_{c}(cc)$ term of the eq. (\ref{eq:aene-wc}) we
found it useful to consider separately the situations where the $p>1$
operators appear on the left, or on the right hand 
sides of the folding
integrals. Specifically we define non-nodal diagrams as:
\begin{eqnarray}
\nonumber
X_{cc,p}^{Z \ton} (\br_1,\br_2) &=& 
\left[ 2 f_1 (r_{12}) f_p (r_{12})
+N_{dd,p}^{\ton \ton} (\br_1,\br_2) \right]  g_{cc}^{\ton} (\br_1,\br_2)+ \\
&~&
\left[ g_{dd}^{\ton \ton} (\br_1,\br_2) - 1 \right]
N_{cc,p}^{Z \ton} (\br_1,\br_2)
\,\,,
\label{eq:aene-xleftright}
\end{eqnarray}
where the label $Z$ can be $L$ (for left) and $R$ (for right).
By using Eqs. (\ref{eq:asoc-ncc1}) and (\ref{eq:asoc-ncc2})
we define the left and right nodal diagrams as:
\begin{eqnarray}
\nonumber
&~&
N_{cc,pqr}^{L \ton \tth} (\br_1,\br_2) = \\
&~&
\biggl(X_{cc,q}^{L \ton} (1,3) \zeta_{132}^{k_2 k_3 k_1} 
C_{d,qr}^{\tth} (3) \frac{\Delta^{k_3}} 2  \Bigl| X_{cc}^{\tth} (3,2) +
N_{cc}^{\tth} (3,2) - \rho_0^{\tth} (3,2) \biggl) 
\,\,,
\label{eq:aene-nodleft} 
\\
\nonumber
&~&
N_{cc,pqr}^{R \ton \tth} (\br_1,\br_2) = \\
&~&
\biggl(X_{cc}^{\ton} (1,3)  \frac{\Delta^{k_2}} 2 \zeta_{132}^{k_2 k_3 k_1} 
C_{d,qr}^{\tth} (3)  \Bigl| X_{cc,r}^{R \tth} (3,2) +
N_{cc,r}^{R \tth} (3,2) \biggl) 
\,\,.
\label{eq:aene-nodright}
\end{eqnarray}

The above equations
(\ref{eq:aene-xleftright}),(\ref{eq:aene-nodleft}),
(\ref{eq:aene-nodright}) form a set of hypernetted equations which can
be solved iteratively.  For example, one may start by setting the
nodal diagrams equal to zero in Eq. (\ref{eq:aene-xleftright}).  The
$(cc)$ nodal diagrams to be used in the evaluation of $W_{c}(cc)$ are
those where the left and right nodal diagrams are subtracted:
\begin{equation}
N_{cc,p}^{int, \ton} (\br_1,\br_2) = 
N_{cc,p}^{\ton} (\br_1,\br_2) - N_{cc,p}^{R \ton} (\br_1,\br_2)
- N_{cc,p}^{L \ton} (\br_1,\br_2)
\,\,.
\end{equation}

The operator structure of the spin-tensor terms of the $R$ diagrams is:
\begin{equation}
\frac 1 4 \{ k_4,u_1 \} \biggl[ \frac 1 2 \{k_1 , u_2 \} k_2 k_3  +
\frac 1 2 k_1 k_2  \{k_3 , u_2 \}
\biggr]
\,\,. 
\label{eq:aene-wcct}
\end{equation}
The terms related to the diagrams $L$ are obtained by exchanging $u_1$
and $u_2$.
The various possibilities are given in Tab. \ref{tab:aene-wccc}.
%%%%%%%%%%%%%%%%%%%%%%%%%%%
% Table
%%%%%%%%%%%%%%%%%%%%%%%%%%%
\begin{table}[htb]
\begin{center}
\begin{tabular}{|l|l|}
\hline
  & \multicolumn{1}{c|}{Traces} \\
\hline
$k_4 u_1 u_2 k_1 k_2 k_3$ & $I^{k_2k_3k_6} I^{k_6k_4k_7} I^{k_1 k_7 k_5} 
A^{k_5}$ \\
$k_4 u_1 k_1 u_2 k_2 k_3$ & $I^{k_2k_3k_6} I^{k_6k_4k_7} J^{k_1 k_7 k_5}$ \\
$k_4 u_1 k_1 k_2 u_2 k_3$ & $I^{k_1k_2k_6} I^{k_3k_4k_7} J^{k_7 k_6 k_5}$ \\
$k_4 u_1 k_1 k_2 k_3 u_2$ & $I^{k_2k_3k_6} I^{k_1k_6k_7} J^{k_4 k_7 k_5}$ \\
$u_1 k_4 u_2 k_1 k_2 k_3$ &  $I^{k_2k_3k_6} I^{k_1k_6k_7} J^{k_4 k_7 k_5}$ \\
$u_1 k_4 k_1 u_2 k_2 k_3$ & $I^{k_2k_3k_6} I^{k_4k_1k_7} J^{k_6 k_7 k_5}$ \\
$u_1 k_4 k_1 k_2 u_2 k_3$ & $I^{k_1k_2k_6} I^{k_4k_6k_7} J^{k_7 k_3 k_5}$ \\
$u_1 k_4 k_1 k_2 k_3 u_2$ & $I^{k_2k_3k_6} I^{k_4k_1k_7} I^{k_6 k_7 k_5} 
A^{k_5}$ \\
\hline
\end{tabular}
\caption{\small Tensor-spin traces for the Eq. (\ref{eq:aene-wcct}).} 
\label{tab:aene-wccc}
\end{center}
\end{table}
%%%%%%%%%%%%%%%%%%%%%%%%%%%%%%5

By putting together the various terms we obtain 
\begin{eqnarray}
\hspace*{-4mm}
W_c (cc) & = & - \frac{1}{8} \int d{\bf r}_1 d{\bf r}_2
H_{JF}^{2k_1-1+l_1,2k_2-1+l_2,2k_3-1+l_3}(r_{12}) 
\nonumber \\ & & 
C_{d,22}^{\ton} (\br_1)  
g_{cc}^{\ton} (\br_1,\br_2)  C_{d,22}^{\ttw} (\br_2) \Delta^{k_4}
\nonumber  \\ 
&  \biggl[ & 
8 I^{k_2 k_3 k_6} I^{k_1 k_6 k_7} J^{k_7 k_4 k_5} 
\chi_{l_1+l_2+l_3+l_4} N_{cc,2k_5-1+l_5}^{int, \ttw} (\br_1,\br_2)  
\nonumber  \\ &+& 
I^{k_2 k_3 k_6} I^{k_6 k_4 k_7} I^{k_1 k_7 k_5} A^{k_5} 
M_{cc,l_4, 0, l_1+l_2+l_3}^{ \ttw k_5} (\br_1,\br_2)
\nonumber  \\ &+& 
I^{k_2 k_3 k_6} I^{k_6 k_4 k_7} J^{k_1 k_7 k_5} 
M_{cc,l_4, l_1, l_2+l_3}^{\ttw k_5} (\br_1,\br_2)
\nonumber  \\ &+& 
I^{k_1 k_2 k_6} I^{k_3 k_4 k_7} J^{k_7 k_6 k_5} 
M_{cc,l_4, l_1+l_2, l_3}^{ \ttw k_5} (\br_1,\br_2)
\nonumber  \\ 
&+& 
I^{k_2 k_3 k_6} I^{k_1 k_6 k_7} J^{k_4 k_7 k_5} 
M_{cc,0, l_4, l_1+l_2+l_3}^{ \ttw k_5} (\br_1,\br_2) 
\nonumber  \\ &+& 
I^{k_2 k_3 k_6} I^{k_1 k_6 k_7} J^{k_4 k_7 k_5} 
M_{cc,l_4, l_1+l_2+l_3, 0}^{ \ttw k_5} (\br_1,\br_2) 
\nonumber  \\ &+& 
I^{k_2 k_3 k_6} I^{k_4 k_1 k_7} J^{k_6 k_7 k_5} 
M_{cc,0, l_4+l_1, l_2+l_3}^{ \ttw k_5} (\br_1,\br_2)
\nonumber  \\ &+& 
I^{k_1 k_2 k_6} I^{k_4 k_6 k_7} J^{k_7 k_3 k_5} 
M_{cc,0, l_4+l_1+l_2, l_3}^{ \ttw k_5} (\br_1,\br_2)
\nonumber  \\ &+& 
I^{k_2 k_3 k_6} I^{k_4 k_1 k_7} I^{k_6 k_7 k_5} A^{k_5} 
M_{cc,0, l_4+l_1+l_2+l_3, 0}^{ \ttw k_5} (\br_1,\br_2)
\biggr]
\,\,,
\end{eqnarray}
with
\begin{equation}
M_{cc,l_1, l_2, l_3}^{\ton k_5} (\br_1,\br_2) = 
\frac 1 2 \left[ M_{cc,l_1, l_2, l_3}^{R \ton k_5} (\br_1,\br_2;1)
+ M_{cc,l_1, l_2, l_3}^{L \ton k_5} (\br_1,\br_2;2) \right]
\,\,.
\end{equation}
In the above equation the $R$ and $L$ functions are defined as in Eq.
(\ref{eq:aene-mmn}) by substituting the nodal diagrams $N$ with the
left and right nodal diagrams of Eqs. (\ref{eq:aene-nodleft}) and
(\ref{eq:aene-nodright}).

We give now the expression of the kinetic energy terms.
For the $T_\phi^{(1)}$ term we obtain:
\begin{equation}
T_\phi^{(1)} = - {{\hbar^2}\over{4m}} 
\int d{\bf r}_1 \rho_{T1}^{\ton} ({\bf r}_1) C_{d,11}^{\ton} ({\bf r}_1)
\,\,,
\end{equation}
where $\rho_{T1}^{\ton} ({\bf r}_1)$ 
has been defined in Eq. (\ref{eq:ene-rhot1}) and 
a sum on $\ton$ is understood.

As indicated in Sect. \ref{sec:ene-kinv6}, we separate the remaining
terms in three parts:
\begin{equation}
T_{\phi}^{(n)} = T_{\phi,0}^{(n)} + T_{\phi,s}^{(n)}
+ T_{\phi,c}^{(n)} \ \ \ \ n=2,3
\,\,.
\label{eq:aene-enecin}
\end{equation}
In order to express the above quantity in a closed form,
we define the function:
\begin{equation}
\hspace*{-8mm}
h_{p,r}^{\ton \ttw} (\br_1,\br_2) = 
\left[ \frac{f_p (r_{12}) f_r (r_{12})}
{f_1^2(r_{12})} g_{dd}^{\ton \ttw} (\br_1,\br_2) - \delta_{p,1} \delta_{r,1}
\right] C_{d,22}^{\ton} (\br_1) C_{d,22}^{\ttw} (\br_2)
\,\,,
\end{equation}
where we have included the scalar dressing of the correlation operator
acting on the external particles. We can see an analogy with the
kinetic energy part in Eq. (\ref{eq:aene-hjf}).  In addition, we have
to consider a spin-isospin exchange operator acting on the external
particles. By associating the indexes $k_1$ and $l_1$ to $p$, $k_3$
and $l_3$ to $r$ and $k_4$ and $l_4$ to the exchange operator, we can
use the results obtained for the traces of the interaction energy in
the case $q=1$, corresponding to $k_2=1$ and $l_2=0$.

Consequently, for $T_{\phi,0}^{(n)}$ we get:
\begin{eqnarray}
T_{\phi,0}^{(2)} & = & 
\frac{\hbar^2}{4m} \int d{\bf r}_1 d{\bf r}_2
\biggl\{ h_{2k_1-1+l_1,2k_3-1+l_3}^{\ton \ttw} (\br_1,\br_2) \Delta^{k_4}
\chi_{l_1+l_3+l_4}^{\ton \ttw} \\
\nonumber & & 
\Bigl[I^{k_4 k_1 k_3} A^{k_3}\rho_{T2}^{\ton \ttw} (\br_1,\br_2) +
I^{k_4 k_1 k_5} I^{k_5 k_3 2}\rho_{T2,j}^{\ton \ttw} (\br_1,\br_2) \Bigr]+ \\
\nonumber & &
2\delta_{\ton \ttw} 
\Bigl[\rho_{T2}^{\ton \ttw} (\br_1,\br_2) +
\rho_{T2,j}^{\ton \ttw} (\br_1,\br_2) \Bigr]
C_{d,22}^{\ton} (\br_1) \Bigl[ C_{d,22}^{\ttw} (\br_2) -1 \Bigr] \biggr\}
\,\,, \\
T_{\phi,0}^{(3)} & = & -\frac{\hbar^2}{2m} \int d{\bf r}_1 d{\bf r}_2
\biggl( h_{2k_1-1+l_1,2k_3-1+l_3}^{\ton \ttw} (\br_1,\br_2) 
\Delta^{k_4} \chi_{l_1+l_3+l_4}^{\ton \ttw} \\
\nonumber & & 
\Bigl[I^{k_4 k_1 k_3} A^{k_3}\rho_{T3}^{\ton \ttw} (\br_1,\br_2) 
N_{cc}^{\ttw}(\br_1,\br_2) + \\ \nonumber & & 
I^{k_4 k_1 k_5} I^{k_5 k_3 2}\rho_{T3,j}^{\ton \ttw} (\br_1,\br_2) 
N_{ccj}^{\ttw}(\br_1,\br_2) \Bigr]+ 
2\delta_{\ton \ttw} C_{d,22}^{\ton} (\br_1) \Bigl\{ \\
\nonumber & & 
\Bigl[\rho_{T3}^{\ton \ttw} (\br_1,\br_2)N_{cc}^{(x) \ttw}(\br_2,\br_1) +
\rho_{T3,j}^{\ton \ttw} (\br_1,\br_2) N_{ccj}^{(x) \ttw}(\br_2,\br_1)  \Bigr]
C_{d,22}^{\ttw} (\br_2)+ \\ & &  \nonumber
\Bigl[\rho_{T3}^{\ton \ttw} (\br_1,\br_2)N_{cc}^{(\rho) \ttw}(\br_2,\br_1) +
\rho_{T3,j}^{\ton \ttw} (\br_1,\br_2) N_{ccj}^{(\rho) \ttw}(\br_2,\br_1) \Bigr]
\\ \nonumber & & 
\Bigl[ C_{d,22}^{\ttw} (\br_2) -1 \Bigr]
\Bigr\} \biggr)\,\,, 
\end{eqnarray}
where the one-body densities $\rho_{T2}$, $\rho_{T2,j}$, $\rho_{T3}$
and $\rho_{T3,j}$ have been defined in Appendix \ref{sec:app-obdt}.  A
comparison with Eq. (\ref{eq:aene-w0}) shows that we have only the
exchange terms.  We have substituted $\rho_{2,exc(j)}$ with
$\rho_{Tn,(j)}$ and we have used $I^{1 k_3 k_5}=\delta_{k_3 k_5}$. The
term without operators that appears in the above equations is caused by
the different vertex correction that must be used when no dynamical
correlations reach the particle $2$.

Using the property mentioned above for the $I$ matrices, we get the
following expressions for the SOR contributions:
\begin{eqnarray}
T_{\phi,s}^{(2)} & = & \frac{\hbar^2}{4m} \int d{\bf r}_1 d{\bf r}_2
\biggl\{ \frac 1 4 h_{2k_1-1+l_1,2k_3-1+l_3}^{\ton \ttw} (\br_1,\br_2) 
\Delta^{k_4} \\
\nonumber & & 
\Bigl[I^{k_4 k_1 k_3} A^{k_3}\rho_{T2}^{\ton \ttw} (\br_1,\br_2) +
I^{k_4 k_1 k_5} I^{k_5 k_3 2}\rho_{T2,j}^{\ton \ttw} (\br_1,\br_2) \Bigr] \\
\nonumber & & 
\Bigl[ M_{d,l_1, 0,l_3, l_4}^{\ton \ttw 1 k_4 k_3}(\br_1)
+  M_{d,l_1, 0,l_3, l_4}^{\ton \ttw 1 k_4 k_3}(\br_2) \Bigr]+ \\
\nonumber & &
 2 \delta_{\ton \ttw} \Bigl[\rho_{T2}^{\ton \ttw} (\br_1,\br_2) +
\rho_{T2,j}^{\ton \ttw} (\br_1,\br_2) \Bigr] C_{d,22}^{\ton} (\br_1)
\\  \nonumber & & \biggl[
U_{d,SOC}^{\ton} (\br_1) \Bigl[ C_{d,22}^{\ttw} (\br_2) -1 \Bigr]+ 
C_{d,22}^{\ttw} (\br_2) U_{d,SOC}^{\ttw} (\br_2) \biggr]  
\biggr\} \,\,, \\
T_{\phi,s}^{(3)} & = & -\frac{\hbar^2}{2m} \int d{\bf r}_1 d{\bf r}_2
\Biggl[ \frac 1 4 h_{2k_1-1+l_1,2k_3-1+l_3}^{\ton \ttw} (\br_1,\br_2) 
\Delta^{k_4} \\
\nonumber & &   
\Bigl[I^{k_4 k_1 k_3} A^{k_3}\rho_{T3}^{\ton \ttw} (\br_1,\br_2) 
N_{cc}^{\ttw}(\br_1,\br_2) + \\ \nonumber & & 
I^{k_4 k_1 k_5} I^{k_5 k_3 2}\rho_{T3,j}^{\ton \ttw} (\br_1,\br_2) 
N_{ccj}^{\ttw}(\br_1,\br_2) \Bigr] \\
\nonumber & & 
\Bigl[ M_{d,l_1, 0,l_3, l_4}^{\ton \ttw 1 k_4 k_3}(\br_1)
+  M_{d,l_1, 0,l_3, l_4}^{\ton \ttw 1 k_4 k_3}(\br_2) \Bigr]+
2\delta_{\ton \ttw} C_{d,22}^{\ton} (\br_1) \Biggl( \\
\nonumber & & \Bigl[\rho_{T3}^{\ton \ttw} (\br_1,\br_2) 
N_{cc}^{\ttw} (\br_1,\br_2) +
\rho_{T3,j}^{\ton \ttw} (\br_1,\br_2)  N_{ccj}^{\ttw} (\br_1,\br_2)\Bigr]
\\ \nonumber
& &  C_{d,22}^{\ttw} (\br_2) U_{d,SOC}^{\ttw} (\br_2) + 
U_{d,SOC}^{\ton} (\br_1) \biggl\{ 
\\ \nonumber & & 
\Bigl[\rho_{T3}^{\ton \ttw} (\br_1,\br_2) 
N_{cc}^{(x)\ttw} (\br_2,\br_1) +
\rho_{T3,j}^{\ton \ttw} (\br_1,\br_2)  N_{ccj}^{(x)\ttw} (\br_2,\br_1)\Bigr]
C_{d,22}^{\ttw} (\br_2) + \\ \nonumber & & 
\Bigl[\rho_{T3}^{\ton \ttw} (\br_1,\br_2) 
N_{cc}^{(\rho)\ttw} (\br_2,\br_1) +
\rho_{T3,j}^{\ton \ttw} (\br_1,\br_2)  N_{ccj}^{(\rho)\ttw} (\br_2,\br_1)\Bigr]
\\ \nonumber & & \Bigl[ C_{d,22}^{\ttw} (\br_2) -1 \Bigr] 
\biggr\} \Biggr) \Biggr]\,. 
\end{eqnarray}

In the evaluation of the $T_{\phi,c}$ terms we neglect 
$T_{\phi,c}^{(3)}$. This approximation is justified by the fact that
the contribution of the $T_{\phi,0,s}^{(3)}$ terms are
much smaller than those of the $T_{\phi,0,s}^{(2)}$ terms.
The structure of the $T_{\phi,c}^{(2)}$ term is analogous to that of
the exchange case of $W_c(dd)$ when $k_2=1$ and $l_2=0$. In this case,
we use $J^{1k_5k_7}=A^{k_5} \delta_{k_5 k_7}$, and obtain the expression:
\begin{eqnarray}
\nonumber
\hspace*{-7mm}
T_{\phi,c}^{(2)} & = & - \frac{\hbar^2}{48 m} \int d{\bf r}_1 d{\bf r}_2
\rho_{T2}^{\ton \ttw}(\br_1,\br_2) f_{2k_1-1+l_1}(r_{12}) 
 f_{2k_3-1+l_3}(r_{12})\cdot 
\\ \nonumber & &
\Delta^{k_4}  C_{d,22}^{\ton}(\br_1) C_{d,22}^{\ttw}(\br_2)
g_{dd}^{\ton \ttw} (\br_1,\br_2) \Biggl\{ 
\frac 3 2 \biggl[ I^{k_1 k_3 k_6} J^{k_4 k_5 k_6}
M_{dd,l_4, l_1+l_3, 0}^{\ton \ttw k_5} (\br_1,\br_2)\\ 
\nonumber &+&
I^{k_3 k_4 k_6} J^{k_6 k_5 k_1}
M_{dd,l_4, l_1, l_3}^{\ton \ttw k_5} (\br_1,\br_2) + 
I^{k_4 k_1 k_6} J^{k_6 k_5 k_3}
M_{dd,l_4+l_1, l_3, 0 }^{\ton \ttw k_5} (\br_1,\br_2)\\
\nonumber &+& I^{k_4 k_3 k_6} I^{k_6 k_1 k_5} A^{k_5}
M_{dd,l_4+l_1, 0, l_3}^{\ton \ttw k_5} (\br_1,\br_2) \biggr] +
I^{k_3 k_4 k_6} J^{k_1 k_5 k_6}
M_{dd,l_4, l_1, l_3}^{\ton \ttw k_5}(\br_1,\br_2) \\
&+&  I^{k_4 k_5 k_6} I^{k_6 k_1 k_3} A^{k_3}
M_{4,dd,l_1, 0, l_3, l_4}^{\ton \ttw k_5}(\br_1,\br_2) \\
&+& I^{k_4 k_1 k_6} J^{k_3 k_5 k_6}
M_{dd,l_4+l_1, l_3, 0}^{\ton \ttw k_5} (\br_1,\br_2) 
\Biggr\}
\,\,.
\nonumber
\end{eqnarray}

We present now the contribution of the spin-orbit terms.  In our
calculations we use correlations up to the tensor channels, therefore
the presence of spin-orbit operators is only due to the interaction.
We consider only the case of spin-orbit operators acting on the
external particles. In this case we write:
\begin{eqnarray}
\nonumber
<v^{7+l_2}>_{0}&=&-9\int d\br_1d\br_2 \frac{f_{5+l_1}(r_{12})
v^{7+l_2}(r_{12})f_{5+l_3}(r_{12})}
{f_{1}^2(r_{12})}\\
&&\biggl\{\rho_{2,dir}^{\ton\ttw} (\br_1,\br_2)\chi^{\ton\ttw}_{l_1+l_2+l_3}+
\label{eq:aene-ls} \\
\nonumber
&&2\Bigl[\rho_{2,exc}^{\ton\ttw} (\br_1,\br_2)
+ \frac 1 3 \rho_{2,excj}^{\ton\ttw} (\br_1,\br_2) \Bigr]
\Bigl[\chi^{\ton\ttw}_{l_1+l_2+l_3}+
\chi^{\ton\ttw}_{l_1+l_2+l_3+1}\Bigr]\biggr\}
\,\,,
\end{eqnarray}
where we have used the relation:
\begin{equation}
C \Big[
P_{12}^{k_1} ({\bf L} \cdot {\bf S})_{12}P_{12}^{k_3} \Big]
=-18 \delta_{k_1 3}
\delta_{k_3 3} \,\,.
\end{equation}

Finally, we analyze the expressions used to calculate the contribution to 
the energy of the three-body potential, Eq.(\ref{eq:ene-v3}). 
The diagrams we consider in this calculation have been presented in 
Fig. \ref{fig:ene-3b}. We start by considering the $v^R_{123}$ term
of the three-body force,  Eq. (\ref{eq:ene-v3vr}), which is a 
scalar function. The diagram (3.1) of Fig. \ref{fig:ene-3b}, is the
leading term, and its contribution can be expressed as:
\begin{equation}
 <v^{R}_{123}>_{3.1}=\frac{1}{6}
  \int d{\bf r}_{1}d{\bf r}_{2}d{\bf r}_{3}v^{R}_{123}
   \rho_{3}^{\ton\ttw\tth}(\br_1,\br_2,\br_3) \,\,,
\end{equation}
where a sum on the $t$ indexes is understood, $\rho_3$ is the
three-body density that can be written as:
\begin{eqnarray}
\nonumber & &
\rho_{3}^{\ton\ttw\tth}(\br_1,\br_2,\br_3) = 
\rho_{3,dir}^{\ton\ttw\tth}(\br_1,\br_2,\br_3)
+\delta_{\ton \ttw}\delta_{\ton \tth}
\rho_{3,exc}^{\ton\ttw\tth}(\br_1,\br_2,\br_3) \\
\nonumber & & \hspace*{-5mm}
 =\sum_{mm^\prime,nn^\prime,ll^\prime=dd,de,ed} \hspace*{-5mm}
  g_{ml^\prime}^{\ton\tth}(\br_1,\br_3)V_{l l^\prime}^{\tth}(\br_3)
  g_{ln^\prime }^{\tth\ttw}(\br_3,\br_2)
  V_{nn^\prime}^{\ttw}(\br_2) g_{lm^\prime}^{\ttw\ton}(\br_2,\br_1)
  V_{mm^\prime }^{\ton}(\br_1)\\
  &&+2\delta_{\ton \ttw}\delta_{\ton \tth}
g_{cc}^{\tth}(\br_1,\br_3)V_{cc}^{\tth}(\br_3)g_{cc}^{\ttw}(\br_3,\br_2)
  V_{cc}^{\ttw}(\br_2)g_{cc}^{\ton}(\br_2,\br_1)V_{cc}^{\ton}(\br_1) \,,
\label{eq:aene-vr31}
\end{eqnarray}
where $V_{mn}^{t_i}(\br_i)$ has been defined in Eq.
(\ref{eq:asoc-vert}). We have separated the direct and the
exchange parts of the three-body density, and we have neglected the 
contribution of the Abe diagrams \cite{abe59}. These Abe diagrams
are simple non-nodal ones with three external points that play the same role
as elementary diagrams in the TBDF.

The expressions of the diagram (3.2) is:
\begin{eqnarray}
\nonumber 
   <v^{R}_{123}>_{3.2}&=&
    \frac{1}{2}\int d{\bf r}_{1}d{\bf r}_{2}d{\bf r}_{3}
     \frac{f_{2k_{1}-1+l_{1}}(r_{12})
     f_{2k_{2}-1+l_{2}}(r_{12})}{[f_{1}(r_{12})]^{2}}\\
\label{eq:aene-vr32}
     & &\Bigl[A^{k_{1}}\delta_{k_1,k_2}\chi^{\ton\ttw}_{l_{1}+l_{2}}
     \rho_{3,dir}^{\ton\ttw\tth}(\br_1,\br_2,\br_3)+ \\
     \nonumber 
     & &\Delta^{k_3}I^{k_{3}k_{1}k_{2}}A^{k_{2}}
     \chi^{\ton\ttw}_{l_{1}+l_{2}+l_{3}}
     \rho_{3,exc}^{\ton\ttw\tth}(\br_1,\br_2,\br_3)\Bigr] \,\,.
\end{eqnarray}
where, as in all the previous equations, a sum on the operator
channels, here indicated by the indexes $k$ and $l$, is understood.
As indicated in diagram (3.2) of Fig. \ref{fig:ene-3b}, the various
operators act between the 1 and 2 particles. In this respect, we
should point out that the scalar operator term, i. e. the case
$k_1=k_2=1$ and $l_1=l_2=0$, is not considered in Eq.
(\ref{eq:aene-vr32}), since this would produce a double counting with
the contribution of the (3.1) diagram, Eq. (\ref{eq:aene-vr31}).

We consider now the contribution of the $v^{2\pi}_{123}$ term of the
three-body force, Eq. (\ref{eq:ene-v3v2pi}). We define first two
effective potentials:
\begin{eqnarray}
\nonumber v_{eff,mn}^{k_3,\ton\ttw}(\br_1,\br_2)&=&
\sum_{k_1,k_2=2}^3 \sum_{\tth}4A_{2\pi}\int d{\bf r}_{3}
\Big[g_{md}^{\ton\tth}(\br_1,\br_3)C^{\tth}_{22}(\br_3)
g_{dn}^{\tth\ttw}(\br_3,\br_2)\\
\nonumber 
&+& g_{md}^{\ton\tth}(\br_1,\br_3)C^{\tth}_{d,22}(\br_3)
g_{en}^{\tth\ttw}(\br_3,\br_2)\\ \nonumber 
&+&g_{me}^{\ton\tth}(\br_1,\br_3)C_{d,22}^{\tth}(\br_3)
g_{dn}^{\tth\ttw}(\br_3,\br_2)\Big] \\
& & \zeta^{k_{1}k_{2}k_{3}}_{132}X^{k_{1}}(r_{13})X^{k_{2}}(r_{32})\\
\nonumber 
   v_{eff,cc}^{k_3,\ton}(\br_1,\br_2)&=& \sum_{k_1,k_2=2}^3
   \sum_{\tth}4A_{2\pi}\int d{\bf r}_{3}
\nonumber          
g_{cc}^{\ton}(\br_1,\br_3)C^{\tth}_{d,22}(\br_3)
g_{cc}^{\tth}(\br_3,\br_2)\\
& &\zeta^{k_{1}k_{2}k_{3}}_{132}X^{k_{1}}(r_{13})X^{k_{2}}(r_{32})
\end{eqnarray}
with $m,n=d,e$ and $X^k(r)$ defined in Eq. (\ref{eq:ene-x}). 
With the help of the above defined quantities we
express the contribution of $v^{2\pi}_{123}$, related to the (2.1)
diagram of Fig. \ref{fig:ene-3b}, as:
\begin{eqnarray}
\nonumber \hspace*{-4mm} <v^{2\pi}_{123}>_{2.1}&=&
 \frac{1}{2}\sum_{\ton\ttw}\int d{\bf r}_{1}
 \int d{\bf r}_{2}\frac{f_{2k_{1}-1+l_{1}}(r_{12})
  f_{2k_{2}-1+l_{2}}(r_{12})}{f_{1}^2(r_{12})}\\
\nonumber
  & \cdot&I^{k_{1}k_{3}k_{2}}A^{k_{2}}
  \chi^{\ton\ttw}_{l_{1}+l_{2}+1}\bigg\{ \\
\nonumber 
 & & v_{eff,dd}^{k_{3},\ton\ttw}(\br_1,\br_2)
\Big[g^{\ton\ttw}_{dd}(\br_1,\br_2)C_{22}^{\ton}(\br_1)C_{22}^{\ttw}(\br_2)\\
\nonumber  &+&
    g^{\ton\ttw}_{de}(\br_1,\br_2)C_{22}^{\ton}(\br_1)C_{d,22}^{\ttw}(\br_2)+
g^{\ton\ttw}_{ed}(\br_1,\br_2)
            C_{d,22}^{\ton}(\br_1)C_{22}^{\ttw}(\br_2)\\
\nonumber  &+&    g^{\ton\ttw}_{ee}(\br_1,\br_2) 
              C_{d,22}^{\ton}(\br_1)C_{d,22}^{\ttw}(\br_2)\Big]\\
\nonumber   
   &+&v_{eff,de}^{k_{3},\ton\ttw}\Big(\br_1,\br_2)C_{d,22}^{\ttw}(\br_2)
    \Big[g^{\ton\ttw}_{dd}(\br_1,\br_2)
     C_{22}^{\ton}(\br_1)\\
\nonumber  &+&
    g^{\ton\ttw}_{ed}(\br_1,\br_2)
     C_{d,22}^{\ton}(\br_1)\Big]\\
\nonumber  
    &+&v_{eff,ed}^{k_{3},\ton\ttw}(\br_1,\br_2)C_{d,22}^{\ton}(\br_1)
 \Big[g^{\ton\ttw}_{dd}(\br_1,\br_2)C_{22}^{\ttw}(\br_2)\\
\nonumber  &+&
       g^{\ton\ttw}_{de}(\br_1,\br_2)C_{d,22}^{\ttw}(\br_2)\Big]\\
\nonumber  
    &+&v_{eff,ee}^{k_{3},\ton\ttw}(\br_1,\br_2)
     g^{\ton\ttw}_{dd}(\br_1,\br_2)
      C_{d,22}^{\ton}(\br_1)C_{d,22}^{\ttw}(\br_2)\bigg\}\\
\nonumber  
    &-&\frac{1}{2}\int d{\bf r}_{1}
    \int d{\bf r}_{2}f_{2k_{1}-1+l_{1}}(r_{12})
    f_{2k_{2}-1+l_{2}}(r_{12})
    g_{dd}^{\ton\ttw}(\br_1,\br_2)\\
\nonumber  
     &\cdot&      C_{d,22}^{\ton}(\br_1)C^{\ttw}_{d,22}(\br_2)
     I^{k_{1}k_{3}k_{4}}I^{k_{4}k_{2}k_{5}}A^{k_{5}}
     (\chi^{\ton\ttw}_{l_{1}+l_{2}+1}+
     \chi^{\ton\ttw}_{l_{1}+l_{2}+2})\Delta^{k_{5}}\\
\nonumber  
      & & \bigg\{v_{eff,dd}^{k_3,\ton\ttw}(\br_1,\br_2)
     \Big[N_{cc}^{\ton}(\br_1,\br_2)-\rho^{\ton}_{0}(\br_1,\br_2)\Big] \\
\nonumber & & 
     [N_{cc}^{\ttw}(\br_1,\br_2)-\rho^{\ttw}_{0}(\br_1,\br_2)]\\
     &+&  2v_{eff,cc}^{k_3,\ton}(\br_1,\br_2)
    \Big [N_{cc}^{\ttw}(\br_1,\br_2)-\rho^{\ttw}_{0}(\br_1,\br_2)\Big]
\bigg\} \,\,,
\end{eqnarray}
In the above expression, we have considered only the 
contribution of the anticommutator term.

In the (2.2) diagram of Fig. \ref{fig:ene-3b}, the operators act on the
13 and 23 particle pairs. In this case, the operator structure is:
\begin{eqnarray}
\nonumber
&\frac{1}{4}\Bigl[\frac{1}{2}\{O_{13}^{p},O_{23}^{s_{1}}\}
            (O_{13}^{q}O_{23}^{s_{2}})_{\pm}+
  \frac{1}{2}(O_{13}^{q},O_{23}^{s_{2}})_{\pm}
  \{O_{13}^{p}O_{23}^{s_{1}}\}&\\
 & +O_{13}^{p}(O_{13}^{q},O_{23}^{s_{2}})_{\pm}O_{23}^{s_{1}}+
  O_{23}^{s_{1}}(O_{13}^{q},O_{23}^{s_{2}})_{\pm}O_{13}^{p}
\Bigr]
&
\label{eq:aene-v3tr}
\end{eqnarray}
where we have defined:
\begin{eqnarray*}
& (O_{13}^{q},O_{23}^{s_{2}})_{\pm}=O_{13}^{q}O_{23}^{s_{2}}\pm
  O_{23}^{s_{2}} O_{13}^{q}&\\
\end{eqnarray*}

The contributions of Eq. (\ref{eq:aene-v3tr}), with $p=2k_3 -1+l_1$,
$q=2k_1 $, $s_1=2k_4 -1+l_2$ and $s_2=2k_2$, are given in Tab.
\ref{tab:aene-v3t}, where we used the matrix $R^{k_1k_2k_3k_4}$
 defined as \cite{Wir80}:
\begin{eqnarray}
R^{k_1k_2k_3k_4}& = & C(P^{k_1}_{13}P^{k_2}_{23}
P^{k_3}_{13}P^{k_4}_{23}) \\
\nonumber & = & A^{k_1} \delta_{k_1 k_3}A^{k_2} \delta_{k_2 k_4}
\Big( 1 + D_{k_2 k_3} \Big) 
+12 P_2(\hat{r}_{13}\cdot \hat{r}_{23}) \times \\
\nonumber & & \Big[ 1 - (1-\delta_{k_1 3})(1-\delta_{k_3 3}) \Big]
\Big[1 -(1-\delta_{k_2 3})(1-\delta_{k_4 3}) \Big] 
 \,\,.
\label{eq:aene-rks}
\end{eqnarray} 
The plus signs in the table correspond to the anticommutator terms
and the minus signs to the commutator ones.  

%%%%%%%%%%%%%%%%%%%%%%%%%%%%%%%%%%%%%%%%%%%5
% Table
%%%%%%%%%%%%%%%%%%%%%%%%%%%%%%%%%%%%%%%%%%%5
\begin{table}[tb]
\begin{center}
\begin{tabular}{|l|l|}
\hline
 & \multicolumn{1}{c|}{Traces} \\
\hline
$k_3 k_4 (k_1 , k_2 )_{\pm} $ & $R^{k_3 k_4 k_1 k_2}\pm$ 
                                $ A^{k_3}\delta_{k_3 k_1}$
                                $ A^{k_4}\delta_{k_2 k_4}$ \\
$k_4 k_3 (k_1 , k_2 )_{\pm}  $ & $A^{k_3}\delta_{k_3 k_1}$
                                $ A^{k_4}\delta_{k_2 k_4}\pm$
                                $ R^{k_4 k_3 k_2 k_1} $ \\
$(k_1 , k_2 )_{\pm} k_3 k_4 $ & $R^{k_1 k_2 k_3 k_4}\pm$
                                $ A^{k_3}\delta_{k_3 k_1}$
                                $ A^{k_4}\delta_{k_2 k_4}$ \\
$(k_1 , k_2 )_{\pm} k_4 k_3 $ & $A^{k_3}\delta_{k_3 k_1}$
                                $ A^{k_4}\delta_{k_2 k_4}\pm$
                                $ R^{k_1 k_2 k_3 k_4} $ \\
$k_3 (k_1 , k_2 )_{\pm} k_4  $ & $A^{k_3}\delta_{k_3 k_1}$
                                $ A^{k_4}\delta_{k_2 k_4}\pm$
                                $ R^{k_3 k_2 k_1 k_4}$ \\
$k_4 (k_1 , k_2 )_{\pm} k_3 $ & $R^{k_4 k_1 k_2 k_3}\pm$
                                $ A^{k_3}\delta_{k_3 k_1}$
                                $ A^{k_4}\delta_{k_2 k_4}$ \\
\hline
\end{tabular}
\caption{\small Tensor-spin traces of the operator of
Eq. (\ref{eq:aene-v3tr}).}
\label{tab:aene-v3t}
\end{center} 
\end{table}
%%%%%%%%%%%%%%%%%%%%%%%%%%%%%%%%%%%%%%%%%%%%%%%%%%%%%%%%%%%%%%%%%%%%%%%%

When the isospin traces are included we obtain:
\begin{eqnarray}
\nonumber
\Omega_{k_1 k_2 k_3 k_4, l_1 l_2}^{\ton\ttw\tth \pm}
&=&
\frac{1}{8} \Big(R^{k_3 k_4 k_1 k_2}
\pm A^{k_3}\delta_{k_3 k_1}
 A^{k_4}\delta_{k_4 k_2} \Big)
\\ \nonumber &~&
 \Bigl[\chi^{\ton\tth\ttw}_{l_1 l_2 1 1 0}(2)\pm
                         \chi^{\ton\tth\ttw}_{l_1 l_2 0 1 1}(2)\pm
                         \chi^{\ton\tth\ttw}_{0 l_2 l_1 +1 1 0}(2)\\
\nonumber
  &~&+\chi^{\ton\tth\ttw}_{0 l_2 l_1 1 1}(2)+
                         \chi^{\ton\tth\ttw}_{1 1 l_1 l_2 0}(2)\pm
                         \chi^{\ton\tth\ttw}_{0 1 l_1 +1 l_2 0}(2)\pm
                         \chi^{\ton\tth\ttw}_{1 1 0 l_2 l_1}(2)\\
\nonumber
 &~&+\chi^{\ton\tth\ttw}_{0 1 1 l_2 l_1}(2)\pm 2
                         \chi^{\ton\tth\ttw}_{l_1 +1  1 0 l_2 0}(2)+
                         2\chi^{\ton\tth\ttw}_{l_1 1 1 l_2 0}(2)+2
                         \chi^{\ton\tth\ttw}_{0 l_2 1 1 l_1}(2)\\
 &~& \pm 2 \chi^{\ton\tth\ttw}_{0 l_2 0 1 l_1 +1}(2) \Bigr]
\label{eq:aene-omega}
\end{eqnarray}
where we have defined the 
$\Omega_{k_1 k_2 k_3 k_4, l_1 l_2}^{\ton\ttw\tth \pm}$ 
symbols which allows us to express 
the contribution of diagram (2.2) of Fig. \ref{fig:ene-3b} as:
\begin{eqnarray}
\nonumber \hspace*{-3mm} <v^{2\pi}_{123}>_{2.2}&=&
   A_{2\pi}\int d{\bf r}_{1}d{\bf r}_{2}d{\bf r}_{3}
   \frac{f_{2k_{3}-1+l_1}(r_{13})}{f_{1}(r_{13})}
   X^{k_{1}}(r_{13})\frac{f_{2k_{4}-1+l_2}(r_{23})}
   {f_{1}(r_{23})}X^{k_{2}}(r_{32})\\
   &&
    \rho_{3}^{\ton\ttw\tth}(\br_1,\br_2,\br_3) \Bigr(
\Omega_{k_1 k_2 k_3 k_4,l_1 l_2}^{\ton\ttw\tth+}+
\frac 1 4 \Omega_{k_1 k_2 k_3 k_4,l_1 l_2}^{\ton\ttw\tth-} \Bigl)\,,
\end{eqnarray} 
where the sum over all the $k$, $l$ and $t$ indexes is understood. 

In the diagram (2.3) of Fig. \ref{fig:ene-3b}, the operator dependent
correlations act on the 1,2 and 1,3 particle pairs. This implies that
the only contribution comes from the commutator term of Eq.
(\ref{eq:ene-v3v2pi}). In this case, the operator structure is:
\begin{eqnarray}
\nonumber
&\frac{1}{4}\Biggl(\frac{1}{2}\{O_{13}^{p},O_{12}^{s_{1}}\}
            [O_{13}^{q},O_{23}^{s_{2}}]+
  \frac{1}{2}[O_{13}^{q},O_{23}^{s_{2}}]
  \{O_{13}^{p},O_{12}^{s_{1}}\}&\\
 & +O_{13}^{p}[O_{13}^{q},O_{23}^{s_{2}}]O_{12}^{s_{1}}+
  O_{12}^{s_{1}}[O_{13}^{q},O_{23}^{s_{2}}]O_{13}^{p}\Biggr)&
\label{eq:ene-v3ttr}
\end{eqnarray}

The spin and  tensor traces of this set of operators are 
given in Tab. \ref{tab:aene-v3tt}, where we have not written 
a function $\zeta_{123}^{k_4 k_2 k_5}$ that is present in all the
traces.
%%%%%%%%%%%%%%%%%%%%%%%%%%%%%%%%%%%%%%%%%%%%%%%%%%%%%%%%
% Table
%%%%%%%%%%%%%%%%%%%%%%%%%%%%%%%%%%%%%%%%%%%%%%%%%%%%%%%%
\begin{table}[t]
\begin{center}
\begin{tabular}{|l|l|}
\hline
 & \multicolumn{1}{c|}{Traces} \\
\hline
$k_3 k_4 [k_1 , k_2 ] $ & $J^{k_3 k_1 k_5}-$ 
                                $ I^{k_3 k_1 k_5}A^{k_5}$\\
$k_4 k_3 [k_1 , k_2 ]  $ & $-J^{k_3 k_1 k_5}+$ 
                                $ I^{k_3 k_1 k_5}A^{k_5}$\\
$[k_1 , k_2 ] k_3 k_4 $ & $J^{k_3 k_1 k_5}-$ 
                                $ I^{k_3 k_1 k_5}A^{k_5}$\\
$[k_1 , k_2 ] k_4 k_3 $ & $-J^{k_3 k_1 k_5}+$ 
                                $ I^{k_3 k_1 k_5}A^{k_5}$\\
$k_3 [k_1 , k_2 ] k_4  $ & $-J^{k_3 k_1 k_5}+$ 
                                $ I^{k_3 k_1 k_5}A^{k_5}$\\
$k_4 [k_1 , k_2 ] k_3 $ & $J^{k_3 k_1 k_5}-$ 
                                $ I^{k_3 k_1 k_5}A^{k_5}$\\
\hline
\end{tabular}
\caption{\small Tensor-spin traces of the operators of
Eq. (\ref{eq:ene-v3ttr}).}
\label{tab:aene-v3tt}
\end{center} 
\end{table}
%%%%%%%%%%%%%%%%%%%%%%%%%%%%%%%%%%%%%%%%%%%%%%%%%%%%%%%%%

After considering the isospin traces we obtain the expression:
\begin{eqnarray}
\nonumber
&\frac{1}{8}(J^{k_3 k_1 k_5 }- I^{k_3 k_1 k_5}A^{k_5})
                        \Big[\eta^{\ton\tth\ttw}_{l_1 l_2 1 1 0}(1)-     
                    \eta^{\ton\tth\ttw}_{l_1 l_2 0 1 1}(1)-
                         \eta^{\ton\tth\ttw}_{0 l_2 l_1 +1 1 0}(1)&\\
\nonumber
                         &+\eta^{\ton\tth\ttw}_{0 l_2 l_1 1 1}(1)+
                         \eta^{\ton\tth\ttw}_{1 l_2 l_1 1 0}(2)-
                         \eta^{\ton\tth\ttw}_{1 l_2 0 1 l_1}(2)-
                         \eta^{\ton\tth\ttw}_{0 l_2 l_1+1 1 0}(2)&\\
\nonumber
                         &+\eta^{\ton\tth\ttw}_{0 l_2 1 1 l_1}(2)-2
                         \eta^{\ton\tth\ttw}_{l_1 +1  l_2 0 1 0}(2)+
                         2\eta^{\ton\tth\ttw}_{l_1 l_2 1 1 0}(2)+2
                         \eta^{\ton\tth\ttw}_{0 l_2 1 1 l_1}(1)&\\
                         &- 2 \eta^{\ton\tth\ttw}_{0 l_2 0 1 l_1 +1}(1)\Big]&
\label{eq:aene-xi}
\end{eqnarray}
Calling $\Xi_{k_1 k_3 k_5,l_1 l_2}^{\ton\ttw\tth}$ 
the above trace, we can write: 
\begin{eqnarray}
\nonumber <v^{2\pi}_{123}>_{2.3}&=&
   \frac{A_{2\pi}} 4\int d{\bf r}_{1}d{\bf r}_{2}d{\bf r}_{3}
   \frac{f_{2k_{3}-1+l_1}(r_{13})}{f_{1}(r_{13})}
   X^{k_{1}}(r_{13})
   \frac{f_{2k_{4}-1+l_2}(r_{12})}{f_{1}(r_{12})} \\ && X^{k_{2}}(r_{32}) 
   \zeta_{123}^{k_4 k_2 k_5} \rho_{3}^{\ton\ttw\tth}(\br_1,\br_2,\br_3) \, 
    \Xi_{k_1 k_3 k_5, l_1 l_2}^{\ton\ttw\tth}
\end{eqnarray}

\section{The Euler procedure}
\label{sec:app-eul}
In this Appendix we present the method used to fix the correlation
function. We have named this method {\sl Euler procedure}.  The
starting point is the calculation of the hamiltonian mean value,
considering only cluster terms up to the second order.  In this case,
the contribution to the $W$ terms of Eq. (\ref{eq:ene-W}) is given
only by the $W_{0}$ term, since the other terms are produced by
clusters of higher order. In analogy, only the terms
$T^{(1,2)}_{\Phi}$ contribute to the $T_{\Phi}$ term of Eq.
(\ref{eq:ene-Tphi}). Therefore, the total energy in the two-body
cluster approximation is given by:
\begin{equation}
E_{2}=W_{2}+T^{(1)}_{\Phi,2}+T^{(2)}_{\Phi,2} \,\,,
\label{eq:aeul-e2}
\end{equation}
where the expression of the various terms are:
\begin{eqnarray}
\nonumber
W_{2}&=&\frac{1}{2}\int d\br_1d\br_2 
H_{JF}^{2k_1-1+l_1,2k_2-1+l_2,2k_3-1+l_3}(r_{12})\\
\nonumber
&&\Bigg\{\rho_0^{s_1s_2 \ton}(\br_1,\br_1)
\rho_0^{s_3s_4 \ttw}(\br_2,\br_2)
\chi^{+}_{s_1}(1)\chi^{+}_{s_3}(2)P_{12}^{k_1}P_{12}^{k_2}P_{12}^{k_3}
\chi_{s_2}(1)\chi_{s_4}(2)\\
\nonumber
&&\chi^{+}_{\ton}(1)\chi^{+}_{\ttw}(2)
(\btau_1\cdot\btau_2)^{l_1+l_2+l_3}
\chi_{\ton}(1)\chi_{\ttw}(2)\\
\nonumber
&&-\rho_0^{s_1s_2 \ton}(\br_1,\br_2)
\rho_0^{s_3s_4 \ttw}(\br_2,\br_1)\chi^{+}_{s_1}(1)\chi^{+}_{s_3}(2)
P_{12}^{k_1}P_{12}^{k_2}P_{12}^{k_3}\chi_{s_4}(1)\chi_{s_2}(2)\\
&&\chi^{+}_{\ton}(1)
\chi_{\ttw}^{+}(2)(\btau_1\cdot\btau_2)^{l_1+l_2+l_3}
\chi_{\ttw}(1)\chi_{\ton}(2)\Bigg\}
\,\,,
\\
\nonumber
T^{(1)}_{\Phi,2}&=&-\frac{\hbar^2}{4m}\int d\br_1
\rho^{s_1s_2 \ton}_{T1}(\br_1,\br_1)\int d\br_2
\rho_0^{s_3s_4 \ttw}(\br_2,\br_2)
\\ \nonumber
&&\Big(f_{2k_1-1+l_1}(r_{12})f_{2k_2-1+l_2}(r_{12})
-\delta_{2k_1-1+l_1,1}\delta_{2k_2-1+l_2,1}\Big)\\
\nonumber
&&\chi^{+}_{s_1}(1)\chi^{+}_{s_3}(2)P_{12}^{k_1}P_{12}^{k_2}
\chi_{s_2}(1)\chi_{s_4}(2)\\
&&\chi^{+}_{\ton}(1)\chi^{+}_{\ttw}(2)(\btau_1\cdot\btau_2)^{l_1+l_2}
\chi_{\ton}(1)\chi_{\ttw}(2)
\,\,,
\\
\nonumber
T_{\Phi,2}^{(2)}&=&\frac{\hbar^2}{4m}\int
d\br_1d\br_2\rho^{s_1s_2s_3s_4 \ton \ttw}_{T2}(\br_1,\br_2)
\\ \nonumber 
&&\Big(f_{2k_1-1+l_1}(r_{12})f_{2k_2-1+l_2}(r_{12})
-\delta_{2k_1-1+l_1,1}\delta_{2k_2-1+l_2,1}\Big)\\
\nonumber
&&\chi_{s_1}^{+}(1)\chi_{s_3}^{+}(2)
P_{12}^{k_1}P_{12}^{k_2}\chi_{s_4}(1)\chi_{s_2}(2)\\
&&\chi_{\ton}^{+}(1)\chi_{\ttw}^{+}(2)
(\btau_1\cdot\btau_2)^{l_1+l_2}
\chi_{\ttw}(1)\chi_{\ton}(2)
\,\,.
\end{eqnarray}
As in the calculation of the energy expectation values, the $k$
indexes can assume the values 1, 2 and 3, and the $l$ indexes the
values 0 and 1. The $s = \pm 1/2$ indexes indicate the spin third
component and the $t = \pm 1/2$ the isospin third component. A sum on
all the third components of spin and isospin and on the repeated
indexes is understood in the above equations, and in the following
ones.

The other quantities of the above equations are the two-body
correlation functions $f_p$, $H_{JF}^{pqr}$
defined in Eq. (\ref{eq:aene-hjf}) and the isospin independent $P_{12}^k$
operators, see Eq. (\ref{eq:inf-operators}). We use the expressions of the
one-body densities (\ref{eq:ene-rhot1}) and (\ref{eq:ene-rhot2}) where
the spin dependence is explicit:
\begin{eqnarray}
\rho_{T1}^{\ton}(\br_1)&=&
\sum_{s_1,s_2} \rho_{T1}^{s_1s_2 \ton}(\br_1,\br_1) 
\chi_{s_1}^*(1)\chi_{s_2}(1)
\,\,,
\\
\nonumber
\rho_{T2}^{s_1s_2s_3s_4\ton \ttw}(\br_1,\br_2)&=&
\rho_0^{s_1s_2 \ton}(\br_1,\br_2)\nabla_{1}^{2}
\rho_0^{s_3s_4 \ttw}(\br_2,\br_1)\\
&&-\nabla_{1}\rho_0^{s_1s_2 \ton}(\br_1,\br_2)\cdot
\nabla_{1}\rho_0^{s_3s_4 \ttw}(\br_2,\br_1)
\,\,.
\end{eqnarray}

We obtain the optimal correlation functions $f_p$ by solving the 
Euler-Lagrange equation:
\begin{equation}
\frac{\delta \left(E_{2}-C_{2}\right)}{\delta f_p}=0
\,\,,
\label{eq:aeul-var2}
\end{equation}
where we have indicated with $C_2$ the contributions of the
constraints.  The expression of $C_2$ is analogous to that of $W_2$
after substituting $H_{JF}^{pqr}$ with $f_p \lambda_q f_r$,
$\lambda_q$ being the Lagrange multipliers.  The values of these
multipliers, are fixed by imposing the conditions that the various
terms of the two-body correlation function assume their asymptotic
values after a certain internucleon distance $d_p$, called healing
distance:
\begin{eqnarray}
f_1(r \ge d_1) & = & 1  \,\,,
\\
f_{p>1}(r \ge d_p) & = & 0
\,\,,
\end{eqnarray}
In addition we impose the condition:
\beq
\left. \frac{\partial f_p}{\partial r}\right|_{r=d_p}=0
\,\,.
\eeq

Since we use correlation functions composed by six operator channels,
the minimization procedure (\ref{eq:aeul-var2}) should be applied to
fix the values of six healing distances $d_p$. This produces a system
of six interconnected differential equations. It is possible to
separate these equations by using a representation of the Euler
equations in terms of the total spin and isospin $S$ and $T$ of the
correlated nucleon pair. We use the projection operators:
\begin{eqnarray}
\Pi^{S}_{12}&=&\frac{1}{4}
\bigl[2S+1+(-1)^{S+1}\bsigma_1\cdot\bsigma_2 \bigr]
\,\,,
\\
\Pi^{T}_{12}&=&\frac{1}{4}
\bigl[ 2T+1+(-1)^{T+1}\btau_1\cdot\btau_2  \bigr]
\,\,,
\end{eqnarray}
with $S,T=0,1$. In the ($T$,$S$) representation we can write:
\begin{equation}
\sum_{p=1}^6x_{p}O^{p}_{12}=
\sum_{S,T=0,1}\left( x_{TS}+ \delta_{S,1} x_{Tt} S_{12}
\right)\Pi_{12}^{T} \Pi_{12}^{S}
\,\,,
\label{eq:aeul-sdcorr}
\end{equation}
where $x_p$ can be the scalar part of the correlation, $f_p$, that of
the interaction $v^p$, or the  Langrange multiplier $\lambda_p$.
The relation between the expressions of these quantities in the
two representations is:
\begin{eqnarray}
x_{ST}& =&  x_1 + (4T-3)x_2+(4S-3)x_3+(4T-3)(4S-3)x_4 \,\,,\\
x_{tT}& =&  x_5 + (4T-3)x_6 \,\,.
\end{eqnarray}

As discussed in Sect. \ref{sec:res-corr}, in our calculations we have
used only two healing distances, one for the four central
channels, $d_{p=1,2,3,4}=d_c$, and the other one for the two tensor
channels, $d_{p=5,6}=d_t$. In terms of these quantities, we can
rewrite the boundary conditions as:
\begin{eqnarray}
f_{ST}(r \ge d_c) = 1   & \hspace*{2cm} & 
\left. \frac{\partial f_{ST}}{\partial r}\right|_{r=d_c} = 0
\,\,,
\\
f_{Tt}(r \ge d_t) =  0  & \hspace*{2cm} & 
\left. \frac{\partial f_{Tt}}{\partial r}\right|_{r=d_t}  =  0
\,\,.
\end{eqnarray}

After some algebra, and by using the properties of the Pauli matrices,
we obtain for $E_2$ the following expression:
\begin{eqnarray}
\nonumber
E_{2}&=&\sum_{s_1s_2s_3s_4\ton\ttw}
\Bigg\{\int d\br_1d\br_2\Bigg[ \frac{1}{2} v_{TS}(r_{12})
f_{TS}^2(r_{12})\\
\nonumber
&&-\frac{\hbar^2}{4m} 
\Big(f_{TS}(r_{12})\nabla^2f_{TS}(r_{12})
-\left(\nabla f_{TS}(r_{12})\right)^2\Big)\\
\nonumber && + 8 \delta_{S,1}\Bigg( \left(\frac{1}{2} v_{T1}(r_{12})
-v_{Tt}(r_{12})\right)f_{Tt}^2(r_{12})
+v_{Tt}(r_{12})f_{T1}(r_{12})f_{Tt}(r_{12}) \\
\nonumber
&&-\frac{\hbar^2}{4m} 
\Big(f_{Tt}(r_{12})\left(\nabla^2f_{Tt}(r_{12})
-\frac{12}{r_{12}^2}f_{Tt}(r_{12})\right)
-\left(\nabla f_{Tt}(r_{12})\right)^2\Big)\Bigg)\Bigg]\\
\nonumber
&&\Big[\rho_0^{s_1s_2 \ton}(\br_1,\br_1)
\rho_0^{s_3s_4 \ttw}(\br_2,\br_2)
-(-1)^{T+S}\rho_0^{s_1s_2 \ton}(\br_1,\br_2)
\rho_0^{s_3s_4 \ttw}(\br_1,\br_2)\Big]\\
\nonumber
&&-\frac{\hbar^2}{4m}\int d\br_1\int d\br_2
\Big(f_{TS}^2(r_{12})+8\delta_{S,1}f_{Tt}^2(r_{12}) -1\Big)\\
\nonumber
&&
\Big[\rho^{s_1s_2 \ton}_{T1}(\br_1,\br_1)
\rho_0^{s_3s_4 \ttw}(\br_2,\br_2)
-(-1)^{T+S} \rho^{s_1s_2s_3s_4 \ton \ttw}_{T2}(\br_1,\br_2) \Big]\Bigg\}\\
&&\frac{1}{2}\Big(\delta_{s_1s_2}\delta_{s_3s_4}+(-1)^{S+1}
\delta_{s_1s_4}\delta_{s_2s_3}\Big)\frac{1}{2}
\Big(1+(-1)^{T+1}\delta_{\ton\ttw}\Big)
\,\,.
\label{eq:aeul-ene22}
\end{eqnarray}

In order to make the variation on $E_2$, we found convenient to
rewrite the above expression as a function of two new quantities
$P_{TS}(\br_1,\br_2)$ and $Q_{TS}(\br_1,\br_2)$, defined as:
\begin{eqnarray}
\nonumber
P_{TS}(\br_1,\br_2)&=&\frac{4}{2T+1}
\Bigg[\delta_{T1}\Big[
\rho_0^{p}(\br_1)\rho_0^{p}(\br_2)+
\rho_0^{n}(\br_1)\rho_0^{n}(\br_2)\\
\nonumber
&&+(-1)^{S}4\Big(\rho_0^{p}(\br_1,\br_2)\rho_0^{p}(\br_1,\br_2)+
\rho_0^{n}(\br_1,\br_2)\rho_0^{n}(\br_1,\br_2)\Big)\Big]\\
\nonumber
&&+\frac{1}{2}\Big[\rho_0^{p}(\br_1)\rho_0^{n}(\br_2)+
\rho_0^{n}(\br_1)\rho_0^{p}(\br_2)\\
\nonumber
&&-(-1)^{T+S}4\Big(\rho_0^{p}(\br_1,\br_2)\rho_0^{n}(\br_1,\br_2)+
\rho_0^{n}(\br_1,\br_2)\rho_0^{p}(\br_1,\br_2)\Big)\Big]
\Bigg]\\
\nonumber
&&+\frac{16(-1)^T}{(2S+1)(2T+1)}
\Big[ \\ \nonumber & & 
\delta_{T1}\Big(\rho_{0j}^{p}(\br_1,\br_2)\rho_{0j}^{n}(\br_2,\br_1)+
\rho_{0j}^{n}(\br_1,\br_2)\rho_{0j}^{p}(\br_2,\br_1)\Big)\\
&&+\frac{1}{2}\Big(\rho_{0j}^{p}(\br_1,\br_2)\rho_{0j}^{n}(\br_2,\br_1)
+\rho_{0j}^{n}(\br_1,\br_2)\rho_{0j}^{p}(\br_2,\br_1)\Big)\Big]
\,\,,
\\
\nonumber
Q_{TS}(\br_1,\br_2)&=&-\frac{\hbar^2}{4m}\sum_{\ton\ttw=p,n}
\frac 1 2 \Big(1 + (-1)^{T+1}\delta_{t_1,t_2} \Big) \\ 
\nonumber &&
\Bigg[\frac{4}{2T+1} \Big( \rho_{T1}^{t_1}(\br_1) \rho_0^{t_2}(\br_2)  
 -4 (-1)^{T+S}
\Big(\rho_0^{\ton}(\br_1,\br_2)
\nabla^2_{1}\rho_0^{\ttw}(\br_2,\br_1) \\
\nonumber &&
-\nabla_{1}\rho_0^{\ton}(\br_1,\br_2)\cdot \nabla_{1}
\rho_0^{\ttw}(\br_2,\br_1)
\Big)\Big)\Big]\\
&&+\frac{16(-1)^T}{(2S+1)(2T+1)}
\Big(\rho_{0j}^{\ton}(\br_1,\br_2)\nabla_{1}^2
\rho_{0j}^{\ttw}(\br_2,\br_1)
\\  \nonumber &&
-\nabla_{1}\rho_{0j}^{\ton}(\br_1,\br_2) \cdot
\nabla_{1}\rho_{0j}^{\ttw}(\br_2,\br_1)\Big)\Bigg]
+\frac{1}{2}P_{TS}(\br_1,\br_2)v_{TS}(r_{12})
\,,
\end{eqnarray}
where we have explicitly written the sum over the spin and the
isospin. In order to obtain a quantity depending only on the relative
distance $r_{12}$ between the particles $1$ and $2$ we integrate Eq.
(\ref{eq:aeul-ene22}) over $r_1$ and $r_2$ by keeping fixed the value
of $r_{12}$.  We define the quantity:
\begin{equation}
\tilde{P}_{TS}=\int d\br_2P_{TS}(\br_1,\br_2)=
\frac{2\pi}{r_{12}}\int_{0}^{\infty} r_2dr_2\int_{|r_2-r_{12}|}
^{|r_2+r_{12}|}r_1dr_1P_{TS}(\br_1,\br_2)
\end{equation}
and an analogous expression for $\tilde{Q}_{TS}$, therefore we write
$E_2$  as:
\begin{eqnarray}
E_{2} & = &\int
d\br_{12}\Bigg[
\tilde{Q}_{TS}f_{TS}^2-\frac{\hbar^2}{4m}
\tilde{P}_{TS}\Big(f_{TS}
\nabla^2f_{TS}-\left(\nabla f_{TS}\right)^2 \Big)
\\ \nonumber & &
+ 8 \delta_{S,1} \Bigg(\left(\tilde{Q}_{T1}-\tilde{P}_{T1}v_{Tt}
+\frac{3\hbar^2}{mr_{12}^2}\tilde{P}_{T1}\right)f_{Tt}^2+
\tilde{P}_{T1}v_{T}f_{T1}f_{Tt}
\\ & & \nonumber 
-\frac{\hbar^2}{4m}
\tilde{P}_{T1}\Big(f_{Tt}
\nabla^2f_{Tt}-\left(\nabla f_{Tt}\right)^2 \Big)\Bigg)
+\tilde{Q}_{TS}+\frac{1}{2}\tilde{P}_{TS}v_{TS}\Bigg]
\,\,.
\label{eq:aeul-eps2}
\end{eqnarray}
where we must understand that all the functions and the operators act
on the $\br_{12}$ coordinate.  The expression of the variation of
$E_2$ with respect to $f_{TS}$ and $f_{Tt}$ is:
\begin{eqnarray}
\nonumber 
\delta \left(E_{2}-C_{2} \right) &=&\int d\br_{12}
\bigg[\delta f_{TS}P_{TS}^{1/2}\Big\{
-\frac{\hbar^2}{m}\nabla^2F_{TS}+\left(\tilde{V}_{TS}-\lambda_{TS}\right)F_{TS}
\\ \nonumber &&
+\delta_{S,1}\left(v_{Tt}-\lambda_{Tt}\right)F_{Tt} \Big\}
+8 \delta_{S,1}\delta f_{Tt}P_{T1}^{1/2}\Big\{
-\frac{\hbar^2}{m}\nabla^2F_{Tt}+
\\ && \nonumber
\left(\tilde{V}_{T1}-\lambda_{T1}-2v_{Tt} +2\lambda_{Tt}+ 
\frac{6\hbar^2}{mr_{12}^2} \right)F_{Tt}+
\\ && 
\left(v_{T1}-\lambda_{T1}\right)F_{T1} \Big\} \bigg]=0
\,\,,
\label{eq:aeul-vareps2}
\end{eqnarray}
where we have defined 
\begin{eqnarray}
F_{TS}& =& f_{TS}\tilde{P}_{TS}^{1/2} \,\,, \\ 
F_{Tt}& =& f_{Tt}\tilde{P}_{T1}^{1/2} \,\,, \\
\tilde{V}_{TS} & = & \frac{1}{\tilde{P}^{TS}}
\left[2\tilde{Q}_{TS}
+\frac{\hbar^2}{4m}\left(\nabla^2\tilde{P}_{TS}
-\frac{(\nabla\tilde{P}_{ST})^2}{\tilde{P}_{TS}}
\right)\right] \,\,.
\end{eqnarray}

The fact that Eq. (\ref{eq:aeul-vareps2}) has to be valid for every
variation of $f_{TS}$ or $f_{Tt}$, implies that both expressions
included in the braces has to be zero.  By imposing this condition we
obtain the Euler-Lagrange equations:
\begin{eqnarray}
\nonumber 
-\frac{\hbar^2}{m}\nabla^2F_{TS}&+&\left(\tilde{V}_{TS}-\lambda_{TS}\right)
F_{TS}
+\delta_{S,1}\left(v_{Tt}-\lambda_{Tt}\right)F_{Tt} =  0
\,\,,\\ \nonumber 
\hspace*{-1cm}
-\frac{\hbar^2}{m}\nabla^2F_{Tt}&+&
\left(\tilde{V}_{T1}-\lambda_{T1}-2v_{Tt} +2\lambda_{Tt}+ 
\frac{6\hbar^2}{mr_{12}^2} \right)F_{Tt}
\\ &+&
\left(v_{T1}-\lambda_{T1}\right)F_{T1} = 0
\,\,.
\label{eq:aeul-euler}
\end{eqnarray} 
The expressions (\ref{eq:aeul-euler}) represent a system of
differential equations, the two equations corresponding to $S=0$ are
not coupled, while the other equations are coupled. The solution of
the above equations gives the optimal value for $f_{TS}$ and $f_{Tt}$
and by using Eq. (\ref{eq:aeul-sdcorr}) we obtain the correlation
functions $f_{p}$, in the representation used in the FHNC
calculations.

In the case of infinite systems, nuclear and neutron matter, these
equations have been generalized to include the spin--orbit channels in
the correlation with $v_8$ \cite{pan79} and $v_{14}$ \cite{lag81b}
potentials.

\section{Acronyms}
\label{sec:acronyms}
\begin{tabular}{ll}
AFDMC      & Auxiliary Field Diffusion Monte Carlo \\
AV8'       & Argonne $v'_{8}$ nucleon-nucleon potential \\
AV18       & Argonne $v_{18}$ nucleon-nucleon potential \\
CBF        & Correlated Basis Function\\
FHNC       & Fermi HyperNetted Chain\\
FHNC/SOC   & Fermi HyperNetted Chain/Single Operator Chain\\
% FP:       & Friedman-Pandharipande interaction\\
GFMC       & Green's Function Monte Carlo \\
HF         & Hartree-Fock\\
HNC        & HyperNetted Chain\\
% IP         & Interacting Particles \\
IPM        & Independent Particle Model\\
MBCF       & Many-Body Correlation Function\\
MF         & Mean Field \\ 
NN         & Nucleon-Nucleon \\
OBDF       & One-Body Distribution Function \\
OBDM       & One-Body Density Matrix\\
RFHNC      & Renormalized Fermi HyperNetted Chain\\
QCD        & Quantum ChromoDynamics \\
SOC        & Single Operator Chain\\
SOR        & Single Operator Ring\\
SRC        & Short Range Correlations\\
TBCF       & Two-Body Correlation Function\\
TBDF       & Two-Body Distribution Function\\
TBDM       & Two-Body Density Matrix\\
U14        & Urbana $v_{14}$ nucleon-nucleon potential \\
UVII       & Urbana VII three-nucleon interaction \\
UIX        & Urbana IX  three-nucleon interaction \\
\end{tabular}

\section{Symbols}
\label{sec:symbols}

\begin{tabular}{lll}
Symbol     & Meaning & Note \\
           &         &       \\
$\Big(\Big|\Big)$ & folding product & Eq. (\ref{eq:inf-folprod}) \\
$< |>$ & Clebsch-Gordan coefficient &  \\
$A$        & number of particles & \\
$A^k$      & $B^{2k-1}$ & \\
$a^\alpha_{0}$ , 
$a^\alpha_{ls}$ & parameters of the WS potential &  Eq. (\ref{eq:fin-wspot}) \\
$B^p$      & $C$-trace of two operators & Tab. \ref{tab:inf-ap}\\
$C(\rot)$ & sum of the composite diagrams & Eq. (\ref{eq:inf-composite})\\
$C(O_{12}^p)$ & $C$-trace & Sect. \ref{sec:inf-traces} \\
$C^t_m(1)$   & vertex corrections & 
Eq. (\ref{eq:asoc-vxe}-\ref{eq:asoc-vxd}) \\
$C^t_{m,pq}(1)$  & vertex corrections & 
Eq. (\ref{eq:asoc-vxte}-\ref{eq:asoc-cd}) \\
$cc$       & sub-index for cyclic-cyclic & Sect. \ref{sec:inf-fermion} \\
$D_{k_1 k_2}$   & $E_{2k_1-1 2k_2-1}$ & \\ 
$D^t_{nlj}(r)$   & & Eq. (\ref{eq:app-dnlj}) \\ 
$dd$       & sub-index for dynamical-dynamical &
           Sect. \ref{sec:inf-fermion} \\
$de$       & sub-index for dynamical-exchange & Sect. \ref{sec:inf-fermion} \\
$E(\rot)$  & contribution of elementary diagrams & \\
$E_{pq}$   & matrix used in SOC calculations &
           Tab. \ref{tab:inf-dpq}\\ 
$ee$       & sub-index for exchange-exchage &
           Sect. \ref{sec:inf-fermion} \\ 
$F(x_1,....,x_A)$ & many-body correlation function & Eq. (\ref{eq:in-trial}) \\
${\cal F}(1...A)$ & operator dependent correlation function &
 Eq. (\ref{eq:inf-correl}) \\
$f(r)$            & two-body correlation function  &
           Eq. (\ref{eq:in-jastrow}) \\
$f_p(r)$      & state dependent correlation function  &
           Eq. (\ref{eq:inf-stcorr}) \\
$g(x_1,x_2)$  & two-body distribution function &
           Eq. (\ref{eq:inf-tbdf})\\
$g^{\ton \ttw}_{p}(1,2)$ & State dependent TBDF & 
Eq. (\ref{eq:inf-sdtbdf}) \\
$H_{ij}$   & state dependent part of the TBCF  &
           Eq. (\ref{eq:inf-defhij}) \\
$H^{pqr}_{JF}(\rot)$ & & Eq. (\ref{eq:aene-hjf}) \\ 
\end{tabular}

\newpage

\begin{tabular}{lll}
$h(r)$     & two-body h-function  & Eq. (\ref{eq:inf-hdef}) \\
$h_c(r)$   & used in SOC Eqs.     & Eq. (\ref{eq:inf-hc}) \\
$h_p(r)$   & used in SOC Eqs.     & Eq. (\ref{eq:inf-hp}) \\
$I^{ijk}$ & spin part of $K^{pqr}$ & 
Eq. (\ref{eq:fin-iijk}) \\
$J^{ijk}$ & spin part of $L^{pqr}$  & 
Eq. (\ref{eq:fin-jijk}) \\
$K^{pqr}$  & used in the product of two operators &
           Tab. \ref{tab:inf-kpqr}\\ 
$L(r_{12})$ & used in SOC Eqs.     & Eq. (\ref{eq:inf-bigl}) \\
$L^{pqr}$ & used in the product of four operators & 
Tab. \ref{tab:inf-lpqr} \\
$\ell(x)$  & Slater function & Eq. (\ref{eq:inf-slatfun}) \\
$M^{\ton \ttw  k_{1} k_{2} k_{3} }_{x, l_{1},l_{2},l_{3}}(i)$
& &  Eq. (\ref{eq:aene-m1}-\ref{eq:aene-m2})\\
$m_\pi$ & pion mass & Eq. (\ref{eq:fin-wspot}) \\
$N$        & number of neutrons & \\
$N(\rot)$  & contribution of nodal diagrams & \\
$N^{x x}_{cc}$ , $N^{x \rho}_{cc}$ , $N^{\rho \rho}_{cc}$  & & 
Eq. (\ref{eq:asoc-ncc1}-\ref{eq:asoc-nccend})\\
${\cal N}^t_{nljm}(r)$ & used in the quasi-hole wave function & 
Eq.(\ref{eq:res-caln})\\
$n^t(k)$ & momentum distribution & Eq.(\ref{eq:res-md})\\
$O^p_{ij}$  & interaction and correlation operators &
Eq. (\ref{eq:inf-operators}) \\
$P_l(cos \theta)$ & Legendre polynomials & \\
$P^t_1$ & proton / neutron projector operator & Eq. (\ref{eq:fin-obd}) \\
$P^k_{ij}$  & $O^{2k-1}_{ij}$ &
Eq. (\ref{eq:inf-separ}) \\
$R^t_{0}$, $R^t_{ls}$ , $R_C$ 
& parameters of the WS potential &  
Eqs. (\ref{eq:fin-wspot}, \ref{eq:fin-coulpot}) \\
$R^t_{nlj}(r)$ & radial part of $\phi(x)$ & 
Eq. (\ref{eq:fin-spwf}) \\
$R^{ k_{1} k_{2} k_{3} k_{4} }$ & spin $C$-traces & Eq. (\ref{eq:aene-rks}) \\
$\br_i$    & spatial coordinate of the particle $i$ & \\
$S(r_{12})$ &  &  Eq. (\ref{eq:asoc-s}) \\
$S_{ij}$ & tensor operator & Eq. (\ref{eq:inf-tensor}) \\
$S_1^{p,n}$ & one-body sum rule &  Eq. (\ref{eq:res-s1})\\ 
$S_2^{pp,pn,nn}$ & two-body sum rule &  
Eqs. (\ref{eq:res-s2}) \\ 
$S_{2,\sigma}$ & spin sum rule & Eq. (\ref{eq:res-s2sigma}) \\
\end{tabular}

\newpage
 
\begin{tabular}{lll}
$S_{nlj}^t$ & spectroscopic factor & Eq. (\ref{eq:res-sf}) \\
${\cal S}$ & symmetrizer operator & \\
$T^{(i)}_\phi$, $T_F$, $T_{c.m.}$ &
kinetic energy terms & 
Eqs. (\ref{eq:ene-Tphi}-\ref{eq:ene-cm})\\
$T_{l_1 l_2}^{\ton \ttw \tth}(i)$ &  
used in the isospin traces &
Eq. (\ref{eq:aiso-capt}) \\
${\cal T}_{l_1 l_2 l_3 l_4 l_5}^{\ton \ttw \tth}(i)$  
& used in the isospin traces &
Eq. (\ref{eq:aiso-calt}) \\
$U^t(r)$ & mean-field potential & 
        Eqs. (\ref{eq:fin-obham},\ref{eq:fin-wspot}) \\
$U^t_x(1)$ &  & 
Eqs. (\ref{eq:asoc-ne}-\ref{eq:asoc-nend}) \\
$U^{\ton \ttw}_{m,pq}(1)$ &  & 
Eqs. (\ref{eq:asoc-u1}-\ref{eq:asoc-socend}) \\
$V$ & volume of the system \\
$V^t_{0}$ ,$V^t_{ls}$ & parameter of the WS potential &  
Eq. (\ref{eq:fin-wspot}) \\
$v^p(\rot)$ & scalar parts of the interaction & \\
$v_{123}^{2\pi}$ & part of the three-body force & Eq. (\ref{eq:ene-v3v2pi}) \\
$v_{123}^R$ & part of the three-body force & Eq. (\ref{eq:ene-v3vr}) \\
$X(\rot)$  & contribution of non-nodal diagrams &   
Eq. (\ref{eq:inf-hnc2})\\
$X_{ij}$ & & Eq. (\ref{eq:ene-x}) \\
${\cal X}^t_{nljm}$ & used in the quasi-hole states  & \\
$x_i$      & generalized coordinate of  particle $i$ & \\
$ Y_{l\mu}(\Omega_i)$ & spherical harmonics & \\
${\bf Y}^m_{lj}$ & spin spherical harmonics & Eq. (\ref{eq:fin-spwf}) \\
$W_0$ $W_s$ $W_c$  $W_{cs}$ & interaction energy terms & 
Eq. (\ref{eq:ene-W}) \\ 
$Z$        & number of protons  &  \\

$\Delta^k$ & $\Gamma^{2k-1}$ &  \\
$\Delta_p(1,...,p)$ & sub-determinant & Eq. (\ref{eq:inf-subdet}) \\
$\eta_{l_1 l_2 l_3 l_4 l_5}^{\ton \ttw \tth}(i)$ &
used in the isospin traces
& Eq. (\ref{eq:aiso-eta}) \\
$\Gamma^p$ & matrix used in SOC Eqs. & Tab. \ref{tab:inf-delta} \\
$\nu$           & spin-isospin degeneracy &
                  Eq. (\ref{eq:inf-spwave}) \\ 
$\xi^{pqr}_{123}$ & function used to calculate SOR &
           Eq. (\ref{eq:inf-xiprop}) \\
$\Xi^{\ton \ttw \tth}_{k_1 k_3 k_5, l_1 l_2}$ & & 
Eq. (\ref{eq:aene-xi})\\
$\zeta^{k_1k_2k_3}_{123}$ & spin  part of $\xi^{pqr}_{123}$  &
           Eq. (\ref{eq:inf-xiprop}) \\
\end{tabular}

\newpage 

\begin{tabular}{lll}
$\Pi^{\sigma,\tau}$ & spin-isospin exchange operator &  
Eq. (\ref{eq:inf-siexch}) \\
$\rho$          & particle density & \\
$\rho_{T1,..,T4}$ & kinetic energy densities & 
Eqs. (\ref{eq:ene-rhot1}-\ref{eq:ene-rhot4})\\
$\rho_{T5,T6}$ & kinetic energy densities & 
Eqs. (\ref{eq:app-rhot6}, \ref{eq:app-rhot5})\\
$\rho^t(\br_1)$      & OBDF & Eq. (\ref{eq:fin-obd})\\
$\rho^t_0$      & parallel spin OBDM & 
                 Eqs. (\ref{eq:fin-ropar},\ref{eq:fin-rhoz})\\
$\rho^t_{0j}$   & antiparallel spin OBDM & 
                 Eqs. (\ref{eq:fin-roantipar},\ref{eq:fin-rhoz})\\
$\rho^{q, \ton \ttw}_2(1,2)$  & operator dependent TBDF &
Eq. (\ref{eq:fin-tbdm}) \\
$\rho^{q, \ton \ttw}_{2,dir,exc,exj}(1,2)$ & & 
Eq. (\ref{eq:aene-dens1} - \ref{eq:aene-dens})\\
$\rho^{s,s';t}(\br_1,\br_2)$ & OBDM & Eq. (\ref{eq:res-obdm})\\
$\Phi(1,...,A)$ & independent particle wave function & \\
$\phi(x)$ & single particle wave function & \\
$\phi^{t,NO}_{nlj}(x)$ & natural orbit & Eq. (\ref{eq:res-no})\\
$\chi_{s,t}$ & Pauli spinors & Eq. (\ref{eq:inf-spwave}) \\
$\chi_n^{\ton \ttw}$ & used in the isospin traces
& Eq. (\ref{eq:fin-chi}) \\
$\Psi(1,...,A)$ & correlated many-body wave function & \\
$\psi^{t}_{nljm}(x)$ & quasi-hole wave function & Eq. (\ref{eq:res-qhfun})\\
$\Omega_{i}$ & the polar angles $\theta_i$ and $\phi_i$ & \\
$\Omega_{k_1 k_2 k_3 k_4, l_1 l_2}^{\ton\ttw\tth \pm}$ &
& Eq. (\ref{eq:aene-omega})  \\
\end{tabular}
\newpage
\end{document}